\documentclass[prd,twocolumn,10pt,nofootinbib,superscriptaddress,
floatfix,aps
]{revtex4-1}

\pdfoutput=1
\usepackage{latexsym}
\usepackage{mathtools}
\usepackage{amsmath}
\usepackage{amssymb}
\usepackage{cancel}
\usepackage{epsfig,amsmath,graphics}
\usepackage{epstopdf}
\usepackage{verbatim}
\usepackage{hyperref}
\usepackage{graphicx}
\usepackage{subfigure}
\usepackage{color}
\usepackage{enumerate}
\usepackage{multirow}
\usepackage{booktabs}
\usepackage{dcolumn}
\newcolumntype{d}[1]{D{.}{.}{#1}}

\newcommand{\invcm}{\text{cm}^{-1}}
\newcommand{\eV}{\text{eV}}

\newcommand{\GeV}{\text{GeV}}

\newcommand{\angstrom}{\text{\AA}}
\newcommand{\s}{\text{s}}

\newcommand{\rad}{\text{rad}}

\definecolor{deepblue}{rgb}{0.2,0.2,0.8}

\definecolor{deepred}{rgb}{0.8,0.2,0.2}

\usepackage{bm}
\newcommand{\vect}[1]{\boldsymbol{\mathbf{#1}}}

\makeatletter
\newcommand{\pushright}[1]{\ifmeasuring@#1\else\omit\hfill$\displaystyle#1$\fi\ignorespaces}
\newcommand{\pushleft}[1]{\ifmeasuring@#1\else\omit$\displaystyle#1$\hfill\fi\ignorespaces}
\makeatother

\begin{document}

\title{Resonant absorption of bosonic dark matter in molecules}

\author{Asimina Arvanitaki}
\email{aarvanitaki@perimeterinstitute.ca}
\affiliation{Perimeter Institute for Theoretical Physics, Waterloo, Ontario N2L 2Y5, Canada}
\author{Savas Dimopoulos} 
\email{savas@stanford.edu}
\affiliation{Stanford Institute for Theoretical Physics, Stanford University, Stanford, California 94305, USA}
\author{Ken Van Tilburg} 
\email{kenvt@ias.edu}
\email{kenvt@nyu.edu}
\affiliation{School of Natural Sciences, Institute for Advanced Study, Princeton, NJ 08540, USA}
\affiliation{Center for Cosmology and Particle Physics, Department of Physics, New York University, New York, NY 10003}
\date{\today}

\begin{abstract}
We propose a new class of bosonic dark matter (DM) detectors based on resonant absorption onto a gas of small polyatomic molecules. Bosonic DM acts on the molecules as a narrow-band perturbation, like an intense but weakly coupled laser.
The excited molecules emit the absorbed energy into fluorescence photons that are picked up by sensitive photodetectors with low dark count rates. This setup is sensitive to any DM candidate that couples to electrons, photons, and nuclei, and may improve on current searches by several orders of magnitude in coupling for DM masses between 0.2~eV and 20~eV. This type of detector has excellent intrinsic energy resolution, along with several control variables---pressure, temperature, external electromagnetic fields, molecular species/isotopes---that allow for powerful background rejection methods as well as precision studies of a potential DM signal. 
The proposed experiment does not require usage of novel exotic materials or futuristic technologies, relying instead on the well-established field of molecular spectroscopy, and on recent advances in single-photon detection.
Cooperative radiation effects, which arise due to the large spatial coherence of the nonrelativistic DM field in certain detector geometries, can tightly focus the DM-induced fluorescence photons in a direction that depends on the DM's velocity, possibly permitting a detailed reconstruction of the full 3D velocity distribution in our Galactic neighborhood, as well as further background rejection.

\end{abstract}

\maketitle

\tableofcontents

\section{Introduction}\label{sec:introduction}

Dark matter (DM), a form of nonrelativistic matter that amounts to 25\% of the energy budget of the universe but does not appear to emit light, is by now the conservative option to explain a wealth of astrophysical and cosmological data that can otherwise not be accommodated for with the known interactions and particles in the Standard Model (SM). The motion of stars in galaxies, the velocity dispersion of galaxies in clusters, gravitational lensing by galaxy clusters, temperature anisotropies in the cosmic microwave background, baryon acoustic oscillation measurements, and early-universe structure formation; all point to a new form of matter that is largely inert save for its gravitational interactions.
Many questions remain unanswered: What are the properties---mass, spin, parity---of the dark matter particle(s)?  What are its nongravitational interactions, if any? How is it produced?

While there are many possible answers to the first two questions, the number of dark matter candidates dwindles once you focus on the ones with a realistic production mechanism. One such great DM candidate is the so-called WIMP, a type of particle which may be produced with the correct relic abundance in the early Universe through the thermal freeze-out mechanism, provided it has a mass and interaction strength close to the electroweak scale. Searches for WIMPs are still in full swing, but previous iterations of both direct and indirect detection experiments have come up empty, ruling out the simplest implementations of this paradigm.

Ultralight, weakly interacting bosons constitute another large category of DM candidates with a natural production mechanism. Light spin-0 particles generically appear as relics of inflation through field misalignment, while massive spin-1 particles acquire an abundance set by quantum fluctuations during the last inflationary e-fold, among other possible production channels. As long as they are sufficiently weakly coupled such that they never reach thermal equilibrium throughout the cosmological evolution, bosons as light as $10^{-21}$~eV can be cold DM, i.e.~behave as an inert, pressureless, nonrelativistic fluid.

These bosons can arise in many theories beyond the Standard Model. The most famous example is the QCD axion, a light remnant in a class of theories that can explain the smallness of the neutron's electric dipole moment. Topological complexity in string compactifications naturally gives rise to a plenitude of bosonic states, such as axions and spin-1 fields, also sometimes known as ``dark'' or ``hidden'' photons. These states may easily have extremely weak couplings to the SM, as well as very small masses. Scalar fields associated with the shape and size of extra dimensions, as well as those that determine fundamental constants in our vacuum, often called moduli, can also couple very weakly and be extremely light. All of these states are associated with the same ingredients that give rise to the string landscape. 

When bosons lighter than 15~eV make up a significant fraction of the local DM energy density, their number density is so large that there are many of them per de Broglie wavelength volume. When that happens, their superposition can be described as a classical field oscillating at a frequency set by the mass, and a coherence time determined by the inverse energy spread, roughly $10^6$ periods of oscillation. This field also exhibits macroscopic spatial coherence on a length scale of order its deBroglie wavelength, $10^3$ times larger than its Compton wavelength. The amplitude of the field oscillation is proportional to $\sqrt{\rho_\text{DM}}$, where $\rho_\text{DM}$ is the local DM density.

In this work, we take advantage of this behavior to propose a novel class of DM detectors. We describe how DM can act as a laser that resonantly excites transitions in molecules when its mass closely matches the transition energy, thus utilizing the DM's temporal coherence. Resonant absorption of DM can excite molecules to a higher-energy state that is otherwise not thermally occupied. This excited internal molecular state decays via emission of a photon, which eventually impinges onto a sensitive photodetector. Our techniques are applicable to DM masses between $0.2~\eV$ and $20~\eV$, and can probe a variety of DM candidates, including axions, dark photons, and moduli.
Two experimental configurations are shown in fig.~\ref{fig:setup}. Molecular gas is placed in a container capable of supporting moderately high pressures. In the ``bulk'' configuration, a fraction of the container walls are  instrumented with large-area photodetectors, and the rest of the container boundary is lined with an optically reflective coating to retain the isotropic fluorescence.  The second, ``stack'' container exploits the spatial coherence of DM to focus the fluorescence onto a much smaller photodetector.

As we will show, the proposed setups have great intrinsic energy resolution and other advantages which allow for efficient background rejection and signal discrimination. In sec.~\ref{sec:theoryoverview}, we review the dynamics of a two-level system under influence of a nonrelativistic wave, and the types of molecular states and transitions that can be excited by bosonic DM. Section~\ref{sec:expsetup} contains a more detailed description of our experimental setup and strategy, as well as a discussion of backgrounds and signal discrimination techniques. We provide estimates for the sensitivity of our setup to scalar, pseudoscalar, and vector DM candidates in sec.~\ref{sec:sensitivity}. Finally, we compare our methods to other proposals from the literature in sec.~\ref{sec:discussion}. Appendix~\ref{sec:fullquantum} shows that the calculations performed in the semi-classical approximation throughout this work give---on average---the correct results.


\section{Theoretical overview}\label{sec:theoryoverview}

In this section, we give an overview of the relevant resonant absorption theory (sec.~\ref{sec:resonance}), cooperative radiation effects (sec.~\ref{sec:cooperation}), the energy eigenstates of diatomic molecules (sec.~\ref{sec:states}), and the types of the transitions between them given certain operator structures of the dark-matter interactions (sec.~\ref{sec:transitions}). Some of this material is not new; we provide it merely to set up notation and give a self-contained review. We use natural units with $\hbar = c = k_B = 1$ throughout.


\subsection{Resonant excitation of a two-level system}\label{sec:resonance}
Consider first a single molecule with two internal energy eigenstates $|0\rangle$ and $|1\rangle$ of an unperturbed Hamiltonian $H_0$ under which they have a relative energy splitting $\omega_0$ with $|0\rangle$ the lower-energy state. We may parametrize the most general state of this system as
\begin{align}
|\Psi\rangle = \sin\left(\frac{\theta}{2}\right) |0\rangle + \cos\left(\frac{\theta}{2}\right) e^{-i\varphi}|1\rangle,
\end{align}
a two-parameter space in a superposition angle $\theta$ and a relative phase $\varphi$ which jointly define the surface of the unit (Bloch) sphere. The unperturbed time-evolution equations of this system are simply $\dot{\theta} = 0$ and $\dot{\varphi} = \omega_0$, determining the interaction-picture states $|0'\rangle = |0\rangle$ and $|1'\rangle = e^{-i\omega_0 t} |1\rangle$. We furthermore assume the excited state has a radiative decay rate of $\gamma_0$ due to spontaneous emission of photons, which will drive $\theta$ to $\pi$.

We are interested in the dynamics of this system influenced by a weakly perturbing DM wave, in particular resonant absorption, which may be effected by a harmonic interaction Hamiltonian $\delta H(t)$ with a small but nonzero matrix element
\begin{align}
\langle  1 | \delta H(t) | 0 \rangle = \Omega e^{-i\alpha_1} \cos(\omega t + \alpha_2). \label{eq:rabi2}
\end{align}
with $\alpha_1$ an extracted phase such that the Rabi frequency $\Omega$ is real, and $\alpha_2$ an arbitrary phase. In sec.~\ref{sec:transitions}, we develop the necessary tools to calculate the dark matter's Rabi frequency $\Omega$ given a nonrelativistic interaction Hamiltonian with nucleons, electrons, and photons, for the molecular states classified in sec.~\ref{sec:states}. Here, we first derive the absorption rate given a certain $\Omega$, allowing an estimate of the minimum detectable $\delta \Omega$ later in sec.~\ref{sec:expsetup}.

Our treatment is valid for any two molecular levels, but the reader may keep in mind the analogous systems in NMR or ESR, of a spin-$1/2$ particle with gyromagnetic ratio $\gamma$ in a static magnetic field $B_0 \hat{\vect{z}}$, as well as an oscillating one $\vect{\delta B} \cos(\omega t + \alpha_2)$. There exists an exact mapping of the spin projection states $|0\rangle \cong |\downarrow\rangle$ and $|1\rangle \cong |\uparrow\rangle$ (with the interaction picture being the rotating frame), the Larmor precession frequency $\omega_0 = \gamma B_0$ and the Rabi frequency $\Omega = \gamma |\vect{\delta B} \times \hat{\vect{z}}|$, and the position on the Bloch sphere denoting the direction of the spin expectation value. Often the classical interpretation of precessing spins gives a useful intuition about the dynamics of the system.

Suppose the system starts in the ground state $|0\rangle$ at $t = 0$, at which point the DM wave is turned on. In the interaction picture, where $\delta H(t)$ contains a term proportional to $|1\rangle \langle 0| = e^{i\omega_0 t} |1'\rangle\langle 0'|$, the state vector $|\psi(t)\rangle$ evolves as (for short times $t>0$):
\begin{align}
|\Psi(t)'\rangle &= e^{-i \int_0^t \delta H'(t') dt'} |0'\rangle \label{eq:rabi1}\\
& \simeq |0'\rangle -i e^{-i (\alpha_1 + \alpha_2)} \frac{\Omega}{2} \frac{e^{i(\omega_0 - \omega)t}-1}{i(\omega_0 - \omega)} |1'\rangle. \nonumber
\end{align}
In the second line, we used perturbation theory first-order in $\delta H(t)$, and discarded rapidly oscillating terms of frequency $\omega_0 + \omega$, which give small corrections upon integration. For times $t \ll 1/\max\lbrace|\omega_0-\omega|, \gamma_{0}\rbrace$, notice that the system rotates into a partly excited state with $\theta(t) = \pi -\Omega t$, and that the phase of the DM wave is imprinted onto the molecule as $\varphi(t) = \pi/2 + \alpha_1 + \alpha_2+ \omega t$. 

Heuristically, the state vector keeps precessing into a more excited state at angular velocity $\dot{\theta} = -\Omega$ until $t \sim 1/|\omega_0 - \omega|$ when the DM wave and the molecule dephase relative to each other, or until $t \sim \gamma_0^{-1}$, the $1/e$-lifetime of the excited state, whichever is shorter, so that the maximum excitation probability is $|\langle 1|\Psi'(t)\rangle|^2 \sim \Omega^2 /\max\lbrace|\omega_0-\omega|, \gamma_{0}\rbrace$. At late times, we therefore expect an equilibrium to be reached between DM absorption and photon emission, each at a rate of $\gamma_0 |\langle 1|\Psi'(t)\rangle|^2$. 
This intuitive result is also borne out by the fully-quantized treatment in ref.~\cite{km1976}, where the absorption rate $\Gamma^{(1)}_\text{abs}(\omega)$ for a single two-level system at late times $t \gg \gamma_0^{-1}$ was found to asymptote to:
\begin{align}
\Gamma^{(1)}_\text{abs}(\omega) = \gamma_0 \frac{\Omega^2/\gamma_0^2}{1 + 4 (\omega_0 - \omega)^2/\gamma_0^2} + \mathcal{O}\left( \Omega^4,e^{-\gamma_0 t/2} \right),\label{eq:Gamma1}
\end{align}
for an excitation field in a coherent state. In App.~\ref{sec:fullquantum}, we show that over integration times of interest, eq.~\ref{eq:Gamma1} gives the correct result even when the field mode is not in a coherent state, as well as when the average occupation number in the mode becomes so low such that the semiclassical approxation breaks down.

One would naively think that extending this result to $N = n p_0 V$ molecules, of number density $n$ in a volume $V$, and probability $p_0$ of occupying the state $|0\rangle$, would be trivial, with a total absorption rate of $N \Gamma_\text{abs}^{(1)}(\omega)$. However, this is \emph{not} correct, because it ignores \emph{cooperative effects} due to the fact that the molecules interact with common excitation and radiation fields. Firstly, the nonrelativistic DM wave is expected to be phase coherent over lengths of order $1/m v_0$ with $v_0 \sim 10^{-3}$ a measure of the DM's local velocity dispersion, so that roughly the same phase $\varphi$ is imprinted on the molecules within a sphere of this radius. Secondly, the radiation from nearby phase-matched molecules generally interferes, constructively so for separations close to an integer number of photon wavelengths $2\pi/\omega$, and destructively so for separations equal to a half-integer number of wavelengths.

In the next subsection, sec.~\ref{sec:cooperation}, we explore these cooperative effects in more detail. In general, we find that the total radiative width of $N$ molecules is not simply $N \gamma_0$, but $N \bar{r} \gamma_0$, with $\bar{r}$ an ``average cooperation number'' that can easily be much larger than unity.  
This leads to the surprising result that the radiative width \emph{per molecule} is
\begin{align}
\gamma_\text{rad} = \bar{r} \gamma_0, \label{eq:superrad}
\end{align} 
and depends on the molecular density, on the DM's phase coherence structure, and on the container size and geometry. For example, for a rectangular container of thickness $R_z > 1/m$ and where the other two dimensions are very large compared to the DM's de Broglie wavelength, we derive in sec.~\ref{sec:cooperation} that, for noninteracting molecules,
\begin{align}
\bar{r} \simeq 1 + \frac{8 \pi n p_0}{m^4 R_z} \approx 1 + \frac{5.1 \times 10^6}{m R_z} \left(\frac{1~\text{eV}}{m}\right)^3 \left(\frac{n p_0}{n_0}\right),\label{eq:barrslab}
\end{align}
with $n_0$ the number density at standard atmospheric conditions with a gas pressure $P =  1~\text{bar}$ and temperature $T = 273~\text{K}$.  We defined $p_0$ as the thermal occupation probability of the state $|0\rangle$. Indeed, we find $\bar{r} \gg 1$ for all but the largest and most dilute containers (before taking into account decoherence). For a vertical stack of slabs, the average cooperation number can be even larger by an amount $\bar{S}$, which in an optimal arrangement is equal to the number of slabs in the stack within a coherence length.

An electric dipole transition has a spontaneous emission rate $\gamma_0 = |\mu_{1,0}|^2 \omega_0^3 / 3\pi$, where $\mu_{1,0}$ is the electric dipole matrix element between $|0\rangle$ and $|1\rangle$. In a stack of thin slabs, this then implies a radiative width at large $\bar{r}$ of:
\begin{align}
\gamma_\text{rad} &\simeq \frac{8 n p_0  \bar{S} |\mu_{1,0}|^2}{3 m R_z} \label{eq:gammarad} \\
&\approx 5.5 \times 10^{10}~\text{Hz} \frac{p_0 \bar{S}}{m R_z}\left(\frac{n}{n_0}\right) \left(\frac{ |\mu_{1,0}|}{0.1 e a_0} \right)^2. \nonumber
\end{align}
For a single slab, $\bar{S} = 1$ by definition; for a large stack of slabs, it can be as large as $\bar{S} \sim 1/v_0$ (see sec.~\ref{sec:cooperation} for details).

Finally, we are led to a total steady-state absorption/emission rate of:
\begin{align}
\Gamma^{(N)}_\text{abs}(\omega,\omega_0) \simeq N \gamma_\text{rad} \frac{\Omega^2/\gamma_\text{rad}^2}{1 + 4 (\omega_0 - \omega)^2/\gamma_\text{rad}^2} . \label{eq:Gammaabsrad}
\end{align}
We note that on-resonance, when $|\omega-\omega_0| \ll \gamma_\text{rad}$, the absorption rate is density independent, while the width of the resonance scales linearly with number density, because $\gamma_\text{rad}\propto \bar{r} \propto n$. Off-resonance, when $|\omega-\omega_0| \gg \gamma_\text{rad}$, the rate of DM absorption and subsequent photon emission scales as $\propto n^2$, where it can be regarded as a coherent conversion of DM into photons. 

Besides radiative broadening, there are several other sources of resonance broadening, all of which we have ignored up until now. In an initial DM search---rather than a verification procedure of a signal---the only other broadening mechanism of importance is collisional broadening, as it is the only other one enhanced at high number densities. We will treat both the DM signal's frequency width and Doppler broadening in sec.~\ref{sec:discrimination}, and nonradiative quenching in sec.~\ref{sec:photodetection}. Inhomogenous broadening is always negligible. The effects of molecular collisions can be complex, but are often well-modeled by the ``impact approximation'', whereby molecules are unperturbed between collisions, but any state gets dephased to a new random phase between $0$ and $2\pi$ when it collides elastically with other molecules~\cite{landau1980statistical}. (Inelastic collisions, where internal energy is exchanged between the molecules, typically occur less frequently.) Under these assumptions, effects from collisions lead to a Lorentzian lineshape $g_0^\text{col}(\omega_0',\omega_0) = \frac{2}{\pi \gamma_\text{col}} \frac{1}{1+{(\omega_0' - \omega_0)^2}/{\gamma_\text{col}^2}}$, with the mean collision rate per molecule 
\begin{align}
\gamma_\text{col} = n \sigma_\text{col} \sqrt{\frac{2T}{M_\text{mol}}} \approx 8.9 \times 10^{9}~\text{Hz} \label{eq:gammacol}
\end{align}
in terms of an average collision cross-section $\sigma_\text{col}$, the temperature $T$, and the total molecular mass $M_\text{mol}$~\cite{corney1978atomic,loudon2000quantum}. The size of the elastic collision  cross-section as defined above is typically somewhat larger than the geometric size of the molecule. In the numerical estimate of eq.~\ref{eq:gammacol}, we used $\sigma_\text{col} \approx 10^2~\angstrom^2$ as a typical benchmark value, and also assumed $M_\text{mol} = 40 m_p$ and a number density $n \equiv P/T = n_0$ at standard atmospheric conditions, with a pressure of $P=1~\text{bar}$ and temperature $T = 273~\text{K}$. In the NMR/ESR analogy, $\gamma_\text{col}^{-1}$ is the equivalent of the transverse spin relaxation time $T_2$, the timescale over which a pair of spins dephase due to magnetic dipole interactions.

We can fold in the effects of collisions into the total absorption rate formula, via the convolution $\Gamma_\text{abs}(\omega,\omega_0) = \int_0^\infty {\rm d}\omega_0' g_0^\text{col}(\omega_0',\omega_0) \Gamma_\text{abs}^{(N)}(\omega,\omega_0')$, which yields
\begin{align}
\Gamma_\text{abs}(\omega,\omega_0) \simeq N \gamma \frac{\Omega^2/\gamma^2}{1 + 4 (\omega_0 - \omega)^2/\gamma^2}, \label{eq:Gammaabscol}
\end{align}
with $\gamma \equiv \gamma_\text{rad} + 2 \gamma_\text{col}$. We should also note that the dephasing caused by the collisions also modifies the radiative width to $\gamma_\text{rad} \simeq \gamma_0 + \eta_\text{coh} (\bar{r}-1)\gamma_0$ with $\eta_\text{coh}$ given by eq.~\ref{eq:etacoh1}. Even so, it is still roughly true that $\gamma \propto n$, so that the absorption rate is density-independent on resonance, and scales as $n^2$ off-resonance.

A molecular sample in the gas phase will be at finite temperature, so typically a number of states will have appreciable occupation probability $p_0$. The initial thermal state usually has nonnegligible support over several rotational states (sometimes also excited vibrational states) of the molecule, each occupied with the Boltzmann probability
\begin{align}
p_{0,j} = \frac{e^{-E_j/T}}{\sum_k e^{-E_k/T}},
\end{align}
with $j,k$ labeling all possible molecular states, each with energy $E_j$. One can view the total molecular population as several independent subpopulations $j$ with total numbers $N_{j} = p_{0,j} nV$. 
Furthermore, each subpopulation will be sensitive to dark-matter excitations at a \emph{set of transition energies} $\lbrace \omega_{0,j} \rbrace$ (not just one $\omega_0$ as assumed above) far above the temperature $T$. This set of transition energies will be different for each subpopulation. The dense discretuum of states makes gas-phase molecules an effective multimode resonant system.



\subsection{Cooperative radiation}\label{sec:cooperation}

The increased per-molecule radiative width of eq.~\ref{eq:superrad} is a phenomenon known as cooperative radiation, closely related to the phenomenon of superradiance first described at length in ref.~\cite{d1954}.
Cooperative effects are particularly dramatic for the emission following absorption of bosonic dark matter rather than photons, due to the DM's nonrelativistic nature. 

This type of cooperative emission can be understood in classical wave mechanics, as it is simply due to the principle of superposition. Going back to the analogy between the two-level quantum molecular system and a precessing spin in a magnetic field, if the spins in an NMR/ESR sample are precessing in phase, they produce an effective magnetic dipole moment $N$ times larger than that of a single spin.  The radiated power then scales as $N^2$, leading to a radiation reaction force per spin that is  linearly proportional to $N$, in direct analogy to the Abraham-Lorentz force on an oscillating charge. This damping force leads to a characteristic damping rate---and thus emission rate---proportional to the density of spins, just as in eqs.~\ref{eq:superrad}~and~\ref{eq:barrslab}.

In the rest of this section, we perform a semiclassical computation of the cooperation number $\bar{r}$, a dimensionless number which parametrizes to what extent the radiation of different molecules interferes. For independent emitters, $\bar{r}$ is defined to be unity. However, constructive (or destructive) interference  means that $\bar{r} = 1 + \mathcal{O}(n)$ in general. We will find below that $\bar{r}$ depends on the phase coherence structure of the excitation source, as well as on the container geometry. In addition, for certain container geometries, the coherent emission can be highly directional.

Suppose we have a container of volume $V$ filled with $N$ molecules at discrete positions $\vect{x}' \in V$ at large and uniform density $n = N/V$. The molecules are coherently excited by dark matter, which causes each of them to radiate a field at a spacetime point $(t,\vect{x})$ of the form:
\begin{align}
A(t,\vect{x},\vect{x}') = \frac{q}{4\pi |\vect{x}-\vect{x}'|} \cos\left[\omega \left(t - |\vect{x}-\vect{x}'|\right) + \alpha(\vect{v},\vect{x}')\right]. \label{eq:radfield}
\end{align}
In reality, the molecules will of course emit photons; as a proxy, we use massless scalar radiation to illustrate the essential physics. Also for simplicity, we will ignore the kinetic-energy contributions to the frequency of the wave, and thus take $\omega = m$.  In eq.~\ref{eq:radfield}, $q$ is an effective ``charge'' set by the coupling of the molecule to DM as well as to the radiation field. The phase $\alpha(\vect{v},\vect{x}')$ of the DM wave is random: for a single velocity component $\vect{v}$ of the whole DM ensemble it can be written as
\begin{align}
\alpha(\vect{v},\vect{x}') = \alpha_{\vect{v}} - m \vect{v} \cdot \vect{x}'.
\end{align}
The DM wave is, however, an incoherent superposition of many waves of different $\vect{v}$ and randomly distributed initial phase $\alpha_{\vect{v}}$. We expect the DM to be approximately virialized in our Galaxy, with the kinetic energy having a Maxwell-Boltzmann probability distribution, which translates into a velocity probability density
\begin{align}
f(\vect{v}) = \left( \frac{1}{\pi v_0^2}\right)^{3/2} \exp\left[-\frac{(\vect{v} - \vect{v}_{\text{lab}})^2}{v_0^2}\right],\label{eq:fv}
\end{align}
where $v_0 \approx 235~\text{km/s}$ is a measure of the DM velocity dispersion~\cite{mcmillan2010uncertainty,kerr1986review,reid2009trigonometric}. We have boosted to the laboratory frame which moves through the Galactic rest frame at velocity $\vect{v}_\text{lab}$. This relative velocity $\vect{v}_\text{lab} = \vect{v}_{\bigodot} + \vect{v}_{\bigoplus}$ is a vector sum of the Sun's velocity $v_{\bigodot} \approx 220~\text{km/s}$ in the DM's rest frame, plus the Earth's orbital velocity $v_{\bigoplus} \approx 30~\text{km/s}$ around the Sun~\cite{schonrich2010local,mignard2000local}. More accurately, this distribution should have a velocity cutoff at the Galactic escape velocity $v_\text{esc} \approx 550~\text{km/s}$~\cite{smith2007rave}, but we shall ignore this complication, as it will not significantly affect the results below.

The expected field at $(t,\vect{x})$ from a single molecule at $\vect{x}'$ is zero trivially
\begin{align}
&\langle A(t,\vect{x},\vect{x}') \rangle_{\vect{v},\alpha} \equiv  \\ 
& \int \text{d}^3\vect{v} f(\vect{v})\int_0^{2\pi}\frac{\text{d}\alpha_{\vect{v}}}{2\pi}  \cos\left[m \left(t - |\vect{x}-\vect{x}'|\right) + \alpha(\vect{v},\vect{x}')\right] = 0 \nonumber
\end{align}
because the phase average of $\cos[\alpha + \dots]$ returns zero. Note that averaging over the initial DM phases $\alpha_{\vect{v}}$ automatically takes care of time averaging. The same must thus also be true for the total field from all molecules in the volume $V$:
\begin{align}
\langle A_\text{tot}(t,\vect{x}) \rangle_{\vect{v},\alpha} \equiv \sum_{\vect{x}' \in V} \langle A(t,\vect{x},\vect{x}') \rangle_{\vect{v},\alpha} = 0. 
\end{align}

The total emitted energy density $m^2 \langle A_\text{tot}^2 \rangle$ does not vanish, as the square of the total field has the  expectation value:
\begin{widetext}
\begin{align}
\langle A_\text{tot}(t,\vect{x})^2 \rangle_{\vect{v},\alpha} &=  \Big\langle \Big(\sum_{\vect{x}'} A(t,\vect{x},\vect{x}') \Big) \Big( \sum_{\vect{y}'} A(t,\vect{x},\vect{y}') \Big) \Big\rangle_{\vect{v},\alpha} \nonumber \\
&= \sum_{\vect{x}'} \Big\langle A(t,\vect{x},\vect{x}')^2 \Big\rangle_{\vect{v},\alpha} + \sum_{\vect{x}'}\sum_{\vect{y}' \neq \vect{x}'} \Big\langle A(t,\vect{x},\vect{x}')A(t,\vect{x},\vect{y}') \Big\rangle_{\vect{v},\alpha} \nonumber \\
&\simeq \int_V \text{d}^3 \vect{x}' n(\vect{x}') \frac{q^2}{2(4\pi)^2 L^2}  + \iint_V \text{d}^3\vect{x}' \text{d}^3\vect{y}' n(\vect{x}') n(\vect{y}') \frac{q^2}{2(4\pi)^2 L^2} \int \text{d}^3\vect{v} f(\vect{v}) \cos\left[m (\hat{\vect{x}} - \vect{v}) \cdot (\vect{x}' - \vect{y}') \right] \nonumber \\
&\equiv \frac{q^2}{2(4\pi)^2 L^2} n V r(\hat{\vect{x}}). \label{eq:intensity}
\end{align}
\end{widetext}
In the second line, we have separated independent radiation terms and ``cooperative'' radiation terms, the latter of which capture interference effects. To get to the third line, we replaced sums with integrals as $\sum_{\vect{x}'} \to \int_V \text{d}^3\vect{x}' n(\vect{x}')$, used eq.~\ref{eq:radfield}, and averaged over time/phase. We also took the far-field limit, with $|\vect{x}-\vect{x}'| \simeq L - \hat{\vect{x}} \cdot \vect{x}'$ (which gets simplified further to just $L$ in the denominator factors) for all $\vect{x}' \in V$, and where $\hat{\vect{x}}$ is a unit vector along the line-of-sight direction. In the last line, we wrote the answer as the square of the total field from a single molecule $\frac{q^2}{2(4\pi)^2 L^2}$, times the number of molecules $n V$, times a ``directional cooperation number'':
\begin{align}
r(\hat{\vect{x}}) = 1 + \frac{n}{V} \iint_V \text{d}^3\vect{x}' \text{d}^3\vect{y}' g(\hat{\vect{x}}, \vect{x}'-\vect{y}'). \label{eq:rdir}
\end{align}
The correlation function $ g(\hat{\vect{x}}, \vect{x}'-\vect{y}')$ itself is defined as
\begin{align}
g(\hat{\vect{x}}, \vect{x}' - \vect{y}') &\equiv \int \text{d}^3\vect{v} f(\vect{v}) \cos\left[m (\hat{\vect{x}} - \vect{v} )\cdot (\vect{x}' - \vect{y}')  \right]. \label{eq:gdef}
\end{align}
For the Boltzmann distribution of eq.~\ref{eq:fv}, we can evaluate this exactly:
\begin{align}
g_B(&\hat{\vect{x}}, \vect{x}' - \vect{y}') = \label{eq:gBoltz}\\
 &\exp\left[-\frac{m^2 v_0^2}{4} (\vect{x}'-\vect{y}')^2\right] \cos\left[m (\hat{\vect{x}}-\vect{v}_{\text{lab}}) \cdot (\vect{x}' - \vect{y}')\right].  \nonumber
\end{align}
One can attribute the rapidly oscillating cosine factor mostly to the photon path length difference between $|\vect{x}-\vect{x}'|$ and $|\vect{x}-\vect{y}'|$, with a longer-range modulation proportional to $v_\text{lab}$ from the coherent phase imprinted on the molecules by DM.
The much more smoothly varying gaussian exponential factor quantifies the phase correlation between the states of two molecules with a separation of $|\vect{x}'-\vect{y}'|$ and a coherence length of $2/mv_0$, equal to 0.5~mm for $m = 1~\eV$. A similar correlation function was found to describe signals of bosonic dark matter in pairs of detectors located at different points in spacetime~\cite{derevianko2016detecting}.

A directional cooperation number $r(\hat{\vect{x}})$ greater than unity means that the radiation constructively interferes in the direction $\hat{\vect{x}}$. Take for example the simplest and also most spectacular case, in which all $N$ molecules are located within a DM's Compton wavelength of each other, $|\vect{x}'-\vect{y}'| \ll 1/m, \forall \vect{x}',\vect{y}' \in V$. In that case, the correlation function $g_B(\hat{\vect{x}}, \vect{x}' - \vect{y}')$ is unity for all directions and all separations, so $r(\hat{\vect{x}}) = N$, with perfect constructive interference.


Another instructive example is that of a rectangular prism with sides of lengths $R_x$, $R_y$, and $R_z$ aligned with the $x$, $y$, and $z$ axes, respectively. For $f(\vect{v})$ as in eq.~\ref{eq:fv}, we can calculate $r(\theta,\varphi)$, the cooperation number for radiation at an angle $\theta$ relative to the $z$-axis, and at an angle $\varphi$ in the $xy$-plane, as:
\begin{align}
r(\theta,\varphi) = 1 + \frac{n}{m^3}\frac{I[\tilde{R}_x,\beta_x,v_0]  I[\tilde{R}_y,\beta_y,v_0]I[\tilde{R}_z,\beta_z,v_0]}{\tilde{R}_x \tilde{R}_y \tilde{R}_z}. 
\end{align}
Above, we have defined the dimensionless lengths $\tilde{R}_i \equiv m R_i$, and the quantities $\beta_x \equiv \sin \theta \cos \varphi - v_{\text{lab},x}$, $\beta_y \equiv \sin \theta \sin \varphi - v_{\text{lab},y}$, and $\beta_z \equiv \cos \theta - v_{\text{lab},z}$. We also defined the integral function
\begin{align}
I[\tilde{R},\beta,v_0] &\equiv \iint_0^{\tilde{R}} \text{d}x' \text{d}y' e^{-\frac{v_0^2}{4} (x'-y')^2 + i \beta (x'-y')} \nonumber\\
&\simeq \begin{cases}
\tilde{R}^2, \hfill |\beta| \ll v_0~\&~\tilde{R} \ll v_0^{-1}\\
\frac{2\sqrt{\pi}\tilde{R}}{v_0}, \hfill |\beta| \ll v_0~\&~\tilde{R} \gg v_0^{-1} \\
 2-2 e^{-\frac{v_0^2}{4}\tilde{R}^2} \cos \tilde{R}. \qquad |\beta-1| \ll 1   \end{cases}
\end{align}
The limiting cases quoted in the second line are correct up to $\mathcal{O}(v_0)$ fractional error or better. 
Around $|\beta| = 0$ and for large $\tilde{R}$, the integral function falls of very steeply as $I \propto \exp(-\beta^2/v_0^2)$.

\begin{figure}
\includegraphics[width=0.4\textwidth, trim = 0 3.6cm 0 3.6cm, clip]{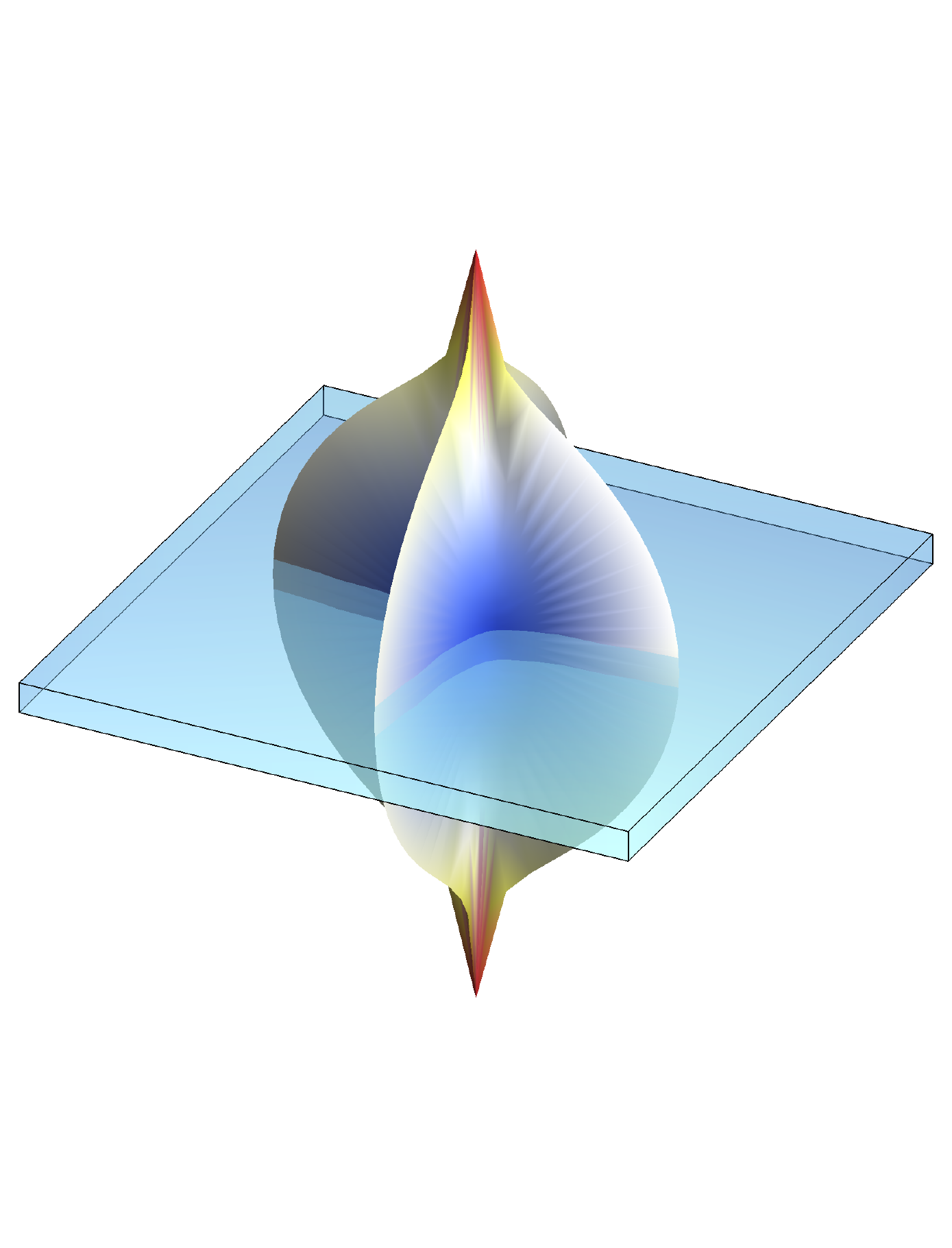}
\caption{Angular dependence of the logarithmic intensity emitted by a molecular gas at large cooperation number $\bar{r} \gg 1$ in a slab-shaped container with dimensions $\tilde{R}_x = \tilde{R}_y = 10^3$ and $\tilde{R}_z = 1$, whose orientation is depicted by the blue prism. The intensity is exponentially peaked in the directions close to the normal vector of each of the faces, and is largest by far along the $z$-direction, perpendicular to the two largest faces of the container. For illustrative purposes, we picked $v_0 = 10^{-2}$ and $v_\text{lab} = 0$, while we also exaggerated the thickness of the slab.}\label{fig:intensity-slab}
\end{figure}

For a prism smaller than a $1/m$ on all sides, we recover the previous result of direction-independent coherent emission with $r(\theta,\varphi) = 1 + n R_x R_y R_z$, up to $\mathcal{O}(1/N)$ fractional corrections. More surprising effects occur when some of the dimensions become large. For a ``slab" with  e.g.~$R_x,R_y \gg 1/mv_0$, we find to a good approximation:
\begin{align}
r(\theta,\varphi) = 1 + \frac{n}{m^3} \frac{8 \pi}{v_0^2} \frac{1- e^{-\frac{v_0^2}{4}\tilde{R}_z^2} \cos \tilde{R}_z}{\tilde{R}_z}  \exp\left[ - \frac{\beta_x^2 + \beta_y^2}{v_0^2} \right]
\end{align}
for $\theta \ll 1$. The full angular dependence of the radiation from a single slab is depicted in fig.~\ref{fig:intensity-slab}. We find that the coherent piece of the emission is \emph{highly focused}, with 84\% of the coherent radiation contained in a cone of angular radius $v_0 \approx 0.78 \times 10^{-4}~\text{rad}$, and 99.5\% within twice that opening angle. For random angles $\theta,\varphi \sim 1$ and $R_z \gtrsim 1/m$, one would typically find a much smaller cooperation number $r(\theta,\varphi) \sim 1+  n/(m^6 R_x R_y R_z)$. The coherent radiation cone is very nearly perpendicular to the slab, with the center of the cone $(\theta_c, \varphi_c)$ slightly offset from $\theta = 0$ by an amount determined by the lab velocity through the DM halo:
\begin{align}
\theta_c &= \text{arcsin} \sqrt{v_{\text{lab},x}^2 + v_{\text{lab}_y}^2}, \label{eq:center1}\\
\varphi_c &= \text{arccot} \frac{v_{\text{lab},x}}{v_{\text{lab},y}}. \label{eq:center2}
\end{align}
The dependence of the emission cone's size and center on the DM's velocity dispersion $v_0$ and average velocity $\vect{v}_\text{lab}$ is illustrated in fig.~\ref{fig:intensity-slab}.
In this way, a measurement of the angular distribution of the coherent radiation constitutes a measurement of the DM velocity distribution, including precise directional information!

\begin{figure}
\includegraphics[width=0.3\textwidth]{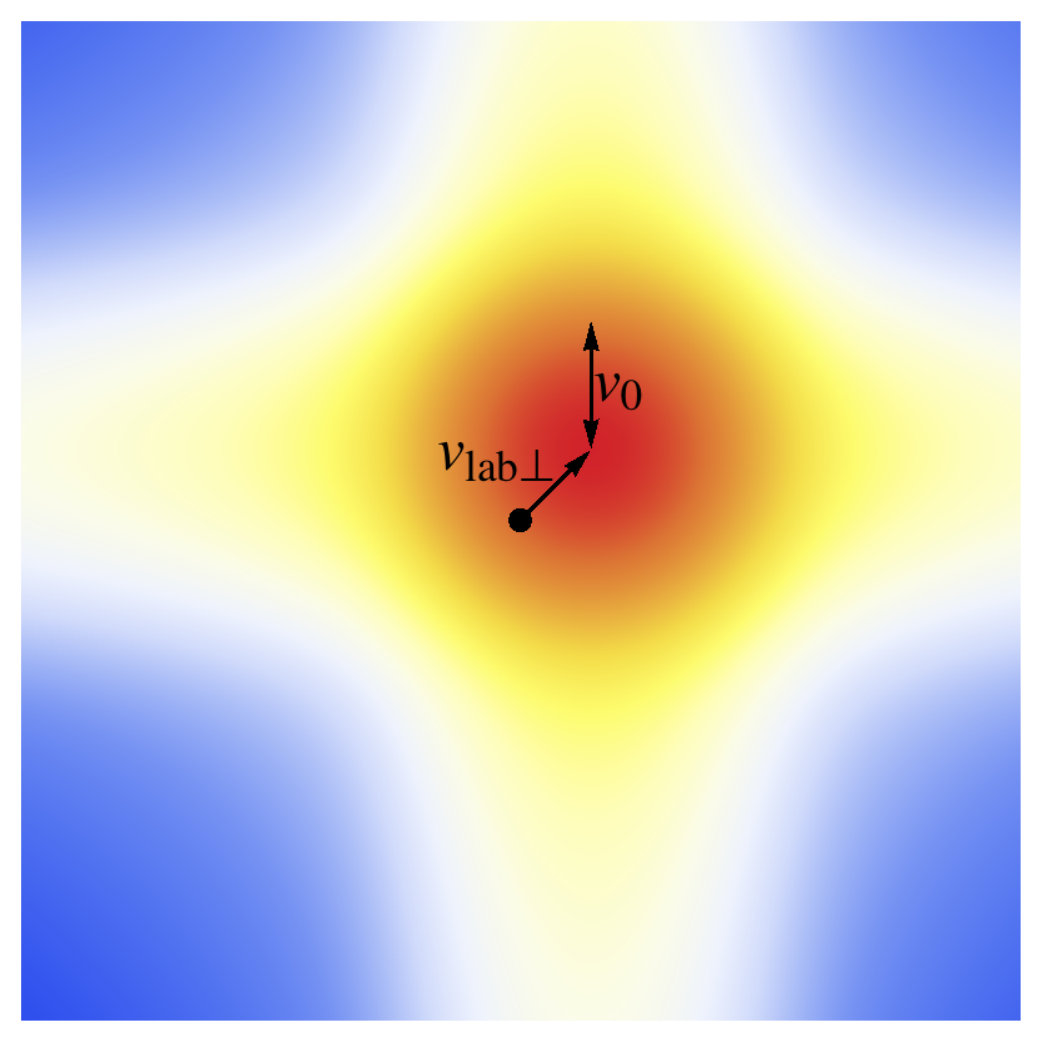}
\caption{Density plot of the logarithmic intensity as measured by a horizontal screen placed far above a thin slab in the $xy$-plane, zoomed in on the directions close to the normal vector defined by $\theta =0$, which pierces the screen at the location of the dot shown. The angular radius of the main emission region is the expected DM velocity dispersion measure $v_0 \approx 0.78 \times 10^{-4}~\text{rad}$. The vector shows the angular offset of the cone's center to the normal; its magnitude and direction are given by the projection $\vect{v}_{\text{lab}\perp}$ of the average DM velocity $\vect{v}_\text{lab}$ onto the $xy$-plane in the lab frame. }\label{fig:intensity-cube-dir}
\end{figure}

For a thin slab, the coherent radiation can dominate the isotropic, incoherent fluorescence. The appropriate measure is the average cooperation number
\begin{align}
\bar{r} = \frac{1}{4\pi} \int_0^\pi \text{d}\theta \int_0^{2\pi}\text{d}\varphi \, r(\theta,\varphi).
\end{align}
When $\bar{r} \gg 1$, most of the outgoing radiation will be coherent. More precisely, the fraction of the radiation that is coherent is $1-\bar{r}^{-1}$. For the thin slab, we find:
\begin{align}
\bar{r} =  1 + \frac{4 \pi n}{m^3}  \frac{1- e^{-\frac{v_0^2}{4}\tilde{R}_z^2} \cos \tilde{R}_z}{\tilde{R}_z}. \label{eq:barr1}
\end{align}
For the small $v_0$ under consideration, the second fraction has a maximum value of $0.72$ at $\tilde{R}_z \approx 2.3$, with subleading local maxima of $\lbrace 0.21, 0.13,0.091,\dots,\sim 2/\tilde{R}_z\rbrace$ at $\tilde{R}_z \approx \lbrace 9.2,15.6,21.9,\dots, \rbrace$. 
Numerically, we then find that the coherent emission can easily dominate for a gas (in absence of decoherence due to e.g.~collisions) in a slab-like container at standard atmospheric conditions and assuming (locally) optimal thickness $R_z$, as we have previously shown in eq.~\ref{eq:barrslab}.

The fact that the cooperation number in eq.~\ref{eq:barr1} is inversely proportional to the thickness at large $R_z$ indicates that cooperative radiation is a surface effect rather than a bulk effect. We expect the scaling of $\bar{r}$ inversely proportional to the (smallest) linear size $R$ of the container to hold for any simply-connected, convex container volume for $R \gg 1/mv_0$. We furthermore expect this scaling and the overall value of $\bar{r}$ to be quite insensitive to the particular shape of the velocity distribution $f(\vect{v})$.

For disjoint container volumes, it is clear that constructive interference of the radiation from any pair of molecules in separate containers can be ignored if the containers are separated by more than a coherence length $1/mv_0$. When the containers are closer together, it is more appropriate to replace the sums $\sum_{\vect{x}'\in V}$ with integrals $\int_V \text{d}^3\vect{x}' n(\vect{x}')$ and a piecewise uniform density $n(\vect{x}')$, such that
\begin{align}
r(\hat{\vect{x}}) = 1 + \frac{\iint_V \text{d}^3\vect{x}'\text{d}^3\vect{y}' n(\vect{x}')n(\vect{y}') g(\hat{\vect{x}},\vect{x}'-\vect{y}')}{\int_V \text{d}^3\vect{x}' n(\vect{x}')}. \label{eq:rdisjoint}
\end{align}
Disjoint container volumes may thus be arranged such that they interfere with each other in a controlled way.
An important example is that of a stack of $N_\text{s}$ identical thin slabs (from the previous example) placed at a regular intervals along the $z$-axis, their centers a distance $L$ apart. When the spatial periodicity along the $z$-axis is an integer number of wavelengths $L = k 2 \pi /m, \, k = 1,2,\dots$, then the radiation from each slab constructively interferes with that of another, at least in the $z$-direction and for slabs within a coherence length. The full calculation of the directional cooperation number for such an arrangement is quite tedious, but can be done numerically with eq.~\ref{eq:rdisjoint}. Roughly, the outcome is that the cooperation number of the stack is enhanced relative to $r(\theta,\varphi)$ for the single slab calculated in eq.~\ref{eq:barr1} by an amount $S(\theta,\varphi)$:
\begin{align}
\frac{r_{\text{stack}}(\theta,\varphi)}{r(\theta,\varphi)} \equiv S(\theta,\varphi) \sim \begin{cases} N_\text{s},  & N_\text{S} L \ll 1/m v_0 \\
1/kv_0. & N_\text{S} L \gg 1/mv_0 \end{cases}
\end{align}
At best, the average cooperation number can be enhanced by about a factor of $\bar{S} \sim 1/v_0$ for a judiciously chosen slab separation $L = 2\pi/m$.

The above analysis has so far ignored decoherence, as it has implicitly assumed that each identical molecular system has retained its internal-state phase imprinted on it by the DM wave. Elastic collisions between the molecules will scramble the molecular phases, destroying any constructive interference in the radiation and driving $r \to 1$ if it occurs at too large a rate. We estimate the coherent part of the radiative rate to be suppressed by the coherence efficiency
\begin{align}
\eta_\text{coh} \simeq \frac{(\bar{r} - 1) \gamma_0}{\bar{r} \gamma_0 + 2\gamma_\text{col}}, \label{eq:etacoh1}
\end{align}
in the presence of collisional broadening of width $\gamma_{\text{col}}$ as in eq.~\ref{eq:gammacol}. This efficiency factor quantifies the fraction of the radiation emitted coherently (i.e.~by the second term in eq.~\ref{eq:intensity}) with angular dependence proportional to $r(\hat{\vect{x}})-1$. The rest will be emitted isotropically. To focus the majority of the total radiative emission, one needs $\eta_\text{coh} >1/2$ or $(\bar{r}-1)\gamma_0 > \gamma_0 + 2\gamma_\text{col}$.

\subsection{Energy eigenstates of diatomic molecules}\label{sec:states}

In this section, we provide a short but self-contained review of the types of molecular states, focusing in particular on the simplest system---the diatomic molecule---for pedagogical reasons. The classification below is a prerequisite to illustrate the different types of transitions that can be induced by dark matter; readers familiar with this material can skip ahead to sec.~\ref{sec:transitions}. Our abridged treatment below is almost entirely based on ref.~\cite{struve1989fundamentals}.

The full Hamiltonian of a neutral, diatomic molecule in absence of external fields is:
\begin{align}
H_0 = & \sum_{N=1}^2 \frac{-\vect{\nabla}_{N}^2}{2 M_{N}} + \sum_{n=1}^{Z_1 + Z_2} \frac{-\vect{\nabla}_{e_n}^2}{2 m_e} + \frac{\alpha Z_1 Z_2}{|\vect{R}_1 - \vect{R_2}|}  \label{eq:H0}\\
 & + \sum_{n<m}^{Z_1 + Z_2} \frac{\alpha}{|\vect{r}_{e_n}-\vect{r}_{e_m}|} - \sum_{n=1}^{Z_1+Z_2} \sum_{N=1}^2 \frac{\alpha Z_N}{|\vect{r}_{e_n}-\vect{R}_N|} . \nonumber
\end{align}
Here $M_1$, $M_2$ are the two nuclear masses and the electron mass, $\alpha$ the fine structure constant, $Z_1$ and $Z_2$ the nuclear charges, $-i\vect{\nabla}_N$ ($\vect{R}_N$) and $-i\vect{\nabla}_{e_n}$ ($\vect{r}_{e_n}$) the momentum (position) operators for the $N$th nucleus and $n$th electron. We neglect spin-orbit coupling and relativistic corrections for the rest of this discussion; the results in this subsection and sec.~\ref{sec:transitions} are only valid insofar as $Z_N \ll 1/\alpha$.

Due to the large mass splitting of the nuclear masses and the electron mass---$m_e/M_N$ ranges from $10^{-3}$ to $10^{-5}$ for light and heavy nuclei---the Hamiltonian $H_0$ is separable into ``fast'' electronic motion and ``slow'' nuclear motion (the Born-Oppenheimer approximation). Its internal energy (i.e.~discounting translational motion) eigenstates $|\Psi_k\rangle$ with energy $E_{k}$ can be written as
\begin{align}
H_0|\Psi_k\rangle &= E_{k} |\Psi_k\rangle \\
& \simeq \left[E^\text{el}_k + E^\text{vib}_k + E^\text{rot}_k \right] |\chi^\text{el}_k \rangle |\psi^{\text{vib}}_k \rangle |Y^{\text{rot}}_k\rangle \nonumber
\end{align}
with the wavefunction $| \Psi_k \rangle$  factorized into one for electronic motion $|\chi^\text{el}_k\rangle$, one for nuclear vibrational motion $|\psi^\text{vib}_k\rangle$, and one for nuclear rotational motion $|Y^\text{rot}_k\rangle$. 
The separability is manifested in the fact that transitions in the rotational and vibrational states leave the electronic state unaltered to a good approximation, and that the vibrational and rotational motion are largely factorized from each other. Different electronic states generally have different effective vibrational and rotational Hamiltonians, however.

Integrating out the electronic motion gives rise to a Hamiltonian of the form
\begin{align}
H^\text{cm,rot,vib} = \frac{-\vect{\nabla}^2_{\vect{R}_\text{cm}}}{2M_\text{mol}} + \frac{-\vect{\nabla}^2_{\vect{R}}}{2M}  + U(R),
\end{align}
where we wrote the nuclear kinetic energies in terms of the center-of-mass momentum operator $-i \vect{\nabla}_{\vect{R}_\text{cm}}$ conjugate to $\vect{R}_\text{cm} \equiv (M_1 \vect{R}_1 + M_2 \vect{R}_2)/(M_1 + M_2)$, and the relative momentum operator $-i \vect{\nabla}_{\vect{R}}$ conjugate to the internuclear separation vector $\vect{R} \equiv R_2 - R_1$. We also defined the total molecular mass $M_\text{mol}\equiv M_1 + M_2$ and the reduced nuclear mass $M \equiv M_1 M_2 / (M_1+M_2)$. The relative kinetic energy can be written out as:
\begin{align}
\frac{-\vect{\nabla}^2_{\vect{R}}}{2M} = \frac{-1}{2M}\left[\frac{1}{R^2} \partial_R \left( R^2 \partial_R \right)- \frac{\vect{J}^2}{R^2} \right],
\end{align}
where $\vect{J}^2$ is the molecule's angular momentum operator
\begin{align}
\vect{J}^2 = -\left[ \frac{1}{\sin \theta} \partial_\theta \left(\sin \theta \partial_\theta \right)+ \frac{1}{\sin^2\theta} \partial_\phi^2 \right].
\end{align}
Parametrically, electronic energy splittings are of order $\Delta E^\text{el} \sim \alpha^2 m_e$, and so is the binding energy $U(\infty) - U(R_e)$. The equilibrium radius $R_e$ is of order the Bohr radius $ a_0 = 1/(\alpha m_e)$. Vibrational energy splittings turn out to be of order $\Delta E^{\text{vib}} \sim \alpha^2 m_e (m_e/M)^{1/2}$, while rotational energy splittings are yet lower at order $\Delta E^\text{rot} \sim \alpha^2 m_e (m_e/M)$, as we will see below. This hierarchy of scales is important, as it endows the molecules with a large set of absorption lines that is finely spaced and uniformly distributed in terms of transition energy, as we will explore further in sec.~\ref{sec:transitions}.


\subsubsection{Rotational states}
In anticipation of the hierarchy of vibrational and rotational energies, we can take $R \simeq R_e$ constant and study just the rotational spectrum (``rigid-rotor'' approximation). At fixed $R$, the Hamiltonian simplifies to
\begin{align}
H^\text{rot} = B_e \vect{J}^2,
\end{align}
where the energy scale $B_e = 1/(2 M R_e^2)$ is half the inverse moment of inertia of the molecule. The energy eigenstates of this operator are just the spherical harmonics $|JM\rangle \equiv |Y_{JM}(\theta,\phi)\rangle$ with $J = 0,~1,~,2,~\dots$ and $M = -J,~-J+1,~\dots,~+J$ which have the $(2J+1)$-fold degenerate energies $E^\text{rot}_J = B_e J(J+1)$. [These results hold true if the electronic level has zero spin $S$ and zero orbital angular momentum $\Lambda$ projected along the $z$-axis (see sec.~\ref{sec:electronic} for definitions). For $\Lambda = 0$ but $S \neq 0$, each $J$ level gets split into (up to) $2 S + 1$ sublevels, ``Hund's case (b)''. For $\Lambda \neq 0$, ``Hund's case (a)'', the $J$ levels get split into two of opposite reflection symmetry $\sigma_v$, and $J$ has a lower bound of $|\Omega|\equiv |\Lambda + \Sigma|$, where $\Sigma$ is the component of the electron spin projected onto the molecular axis.]

For example, the ground state $J=0$ and the first excited state $J=1$ are split by $2B_e \simeq 1/(M R_e^2) \sim \alpha^2 m_e (m_e/M)$, after approximating $R_e \sim a_0 = 1/(\alpha m_e)$. The rotational constant $B_e$ typically varies by $\mathcal{O}(1)$ for different electronic states, as they stabilize the nuclei at different equilibrium separations. The variation of $B_e$ among different vibrational states is much weaker, and can be modeled by a correction factor $\alpha_e$:
\begin{align}
E^\text{rot}_{v,J} = B_e J(J+1) - \alpha_e \left(v+\frac{1}{2}\right) J(J+1),
\end{align}
with $v$ the vibrational quantum number (see below). The minus sign indicates that higher vibrational states typically have \emph{lower} expectation values $\langle 1/R^2 \rangle$. From naive dimensional analysis, $\alpha_e/B_e \sim (m_e/M)^{1/2}$. Other corrections to the rigid-rotor approximation include centrifugal distortion, usually denoted by another correction term of the form $D_e J^2(J+1)^2$ which is even smaller ($D_e/B_e \sim m_e/M$) and will be ignored hereafter.


\subsubsection{Vibrational states}
Allowing for a dynamical internuclear separation $R$, we can integrate in the vibrational spectrum. Taking the energy eigenstate wavefunctions to be $\langle R \theta \phi | \psi^\text{vib} Y^\text{rot}\rangle  = \tilde{\psi}_v(R) Y_{JM}(\theta,\phi)/R$, we find that the radial wavefunction $\tilde{\psi}_v(R)$ obeys the Schr\"odinger equation
\begin{align}
\left[-\frac{1}{2M} \frac{d^2}{dR^2} + U(R) + \frac{J(J+1)}{2M R^2} \right] \tilde{\psi}_v(R) = E^\text{vib}_v.
\end{align}
At leading order, we can ignore the $R$-dependence of the third term, the centrifugal potential, since it is typically a small correction to $U(R)$.  In that case, the vibrational energy eigenvalues and eigenfunctions become $J$-independent. For a bound diatomic, the potential $U(R)$ has a minimum at some $R = R_e$, near which it can be approximated by a harmonic oscillator potential:
\begin{align}
U(R) \simeq \frac{M \omega_e^2}{2} (R - R_e)^2,
\end{align}
which has eigenstates $|v\rangle$ with energies
\begin{align}
E^\text{vib}_v = \omega_e \left(v + \frac{1}{2}\right)
\end{align}
for $v = 0,~1,~2,~\dots$. By naive dimensional analysis (NDA), $U(R)$ must vary by an amount of $\mathcal{O}(\alpha^2 m_e)$ over distances of $\mathcal{O}(a_0)$, so the effective spring constant $k_e \equiv M \omega_e^2$ must be $\mathcal{O}(\alpha^4 m_e^3)$. It follows that vibrational energy splittings are $\omega_e \sim \alpha^2 m_e (m_e/M)^{1/2}$.

Anharmonicities in the vibrational potential can be modeled by the Morse potential:
\begin{align}
\frac{M \omega_e^2}{2} (R-R_e)^2 \rightarrow D_e\left[1-e^{-a(R-R_e)}\right]^2, \label{eq:Morsepotential}
\end{align}
which has closed-form eigenfunctions with the exact eigenvalues:
\begin{align}
E^\text{vib}_v = \omega_e \left(v+\frac{1}{2}\right) - \omega_e x_e\left(v+\frac{1}{2}\right)^2.
\end{align}
where $\omega_e^2 \equiv 2 a^2 D_e / M$ and $\omega_e x_e \equiv \omega_e^2 / 4 D_e$. In other words, one recovers a pure harmonic oscillator for dissociation energy $D_e \to \infty$ while keeping the spring constant fixed at $k_e = 2 a^2 D_e$.


\subsubsection{Electronic states}\label{sec:electronic}

The electronic structure and wavefunctions are more complicated than those of vibration-rotation modes. Nevertheless, the electronic levels can be classified according to their symmetry structure. 

Because of the mass hierarchy between the electron and the nuclei, we can take the internuclear separation $\vect{R} = R \hat{\vect{z}}$ as parametrically fixed to point along the $z$-axis. (Note that we are thus temporarily adopting molecule-fixed coordinates that are co-rotating with the molecule, rather than the space-fixed coordinates of the previous section.) One can find the electronic wavefunction and energy parametrically as a function of $R$, with the minimum of energy being $T_e$ at the equilibrium radius $R_e$ of the level under consideration. Knowing this parametric potential energy function $U(R)$ also allows one to compute the vibration-rotation quantities like e.g.~$\omega_e$ and $B_e$, which in general depend on the electronic level. In the Born-Oppenheimer approximation with factorizable motion of the electrons and the nuclei, we can consistently expand the energy eigenvalues in those of the electronic Hamiltonian with rovibrational fine structure for each electronic level (the total level labeled by $k$):
\begin{align}
E^\text{el}_k = ~& T_{e,k} + \omega_{e,k} \left(v_k+\frac{1}{2}\right) - \omega_{e,k} x_{e,k} (v_k+1/2)^2 \\
& + B_{e,k} J_k (J_k + 1) - \alpha_{e,k} \left(v_k+\frac{1}{2}\right) J_k(J_k+1). \nonumber
\end{align}
Conventionally, the $T_{e,k}$ are measured relative to the minimum of the lowest electronic state, which is taken to have $T_{e,k} = 0$.

A diatomic molecule exhibits cylindrical symmetry, so the projection of the electronic angular momentum $L_z$ around the molecular axis is a good quantum number when the spin-orbit coupling is small. Hence we can classify electronic energy eigenstates according to their (molecule-fixed) $\phi$ dependence:
\begin{align}
L_z |\chi^\text{el}\rangle = L_z e^{\pm i \Lambda \phi} |\chi^{\prime \text{el}}\rangle = \pm \Lambda e^{\pm i \Lambda \phi} |\chi^{\prime \text{el}}\rangle  = \pm \Lambda |\chi_\text{el}\rangle \nonumber
\end{align}
States with $\Lambda = 0, 1, 2, 3, \dots$ are conventionally denoted by the letters $\Sigma,\Pi,\Delta,\Phi,\dots$. Reflections around any plane through the molecular axis are also symmetries of the system, so eigenstates with $\Lambda = 0$ can be labeled by whether their wavefunctions are even $(+)$ or odd $(-)$ under such reflections, which are denoted by the operator $\sigma_v$. In absence of spin-orbit coupling, total spin $\vect{S}$ and spin projection $S_z$ onto the molecular axis are also good symmetries, with quantum numbers $S$ and $\Sigma$ that can be half-integers. Finally, homonuclear diatomics (those with identical atoms) give rise to potentials for the electrons that are symmetric about the molecule-fixed inversion $i$, which takes $(x,y,z)\to(-x,-y,-z)$. Hence the electronic states can be categorized into those that are even $(g)$ and odd $(u)$ under $i$. For example, a homonuclear diatomic state with $S = \Lambda = 0$ that is even under $\sigma_v$ and $i$ would be denoted by ${}^1 \Sigma_g^+$, whereas a heteronuclear diatomic state with spin $S$ and orbital angular momentum $\Lambda = 1$ would be denoted by ${}^{2S+1}\Pi$. Finally, we note that the electronic wavefunction must return to a sum of separated-atom wavefunctions in the limit $R \gg R_e$, so there exists a one-to-one correspondence between molecular orbitals and combinations of atomic orbitals. For the purposes of this discussion, we shall ignore complications due to spin-orbit coupling and nuclear spin symmetries; more details can be found in~\cite{struve1989fundamentals}.

\begin{figure}\begin{center}
\includegraphics[width=0.48\textwidth]{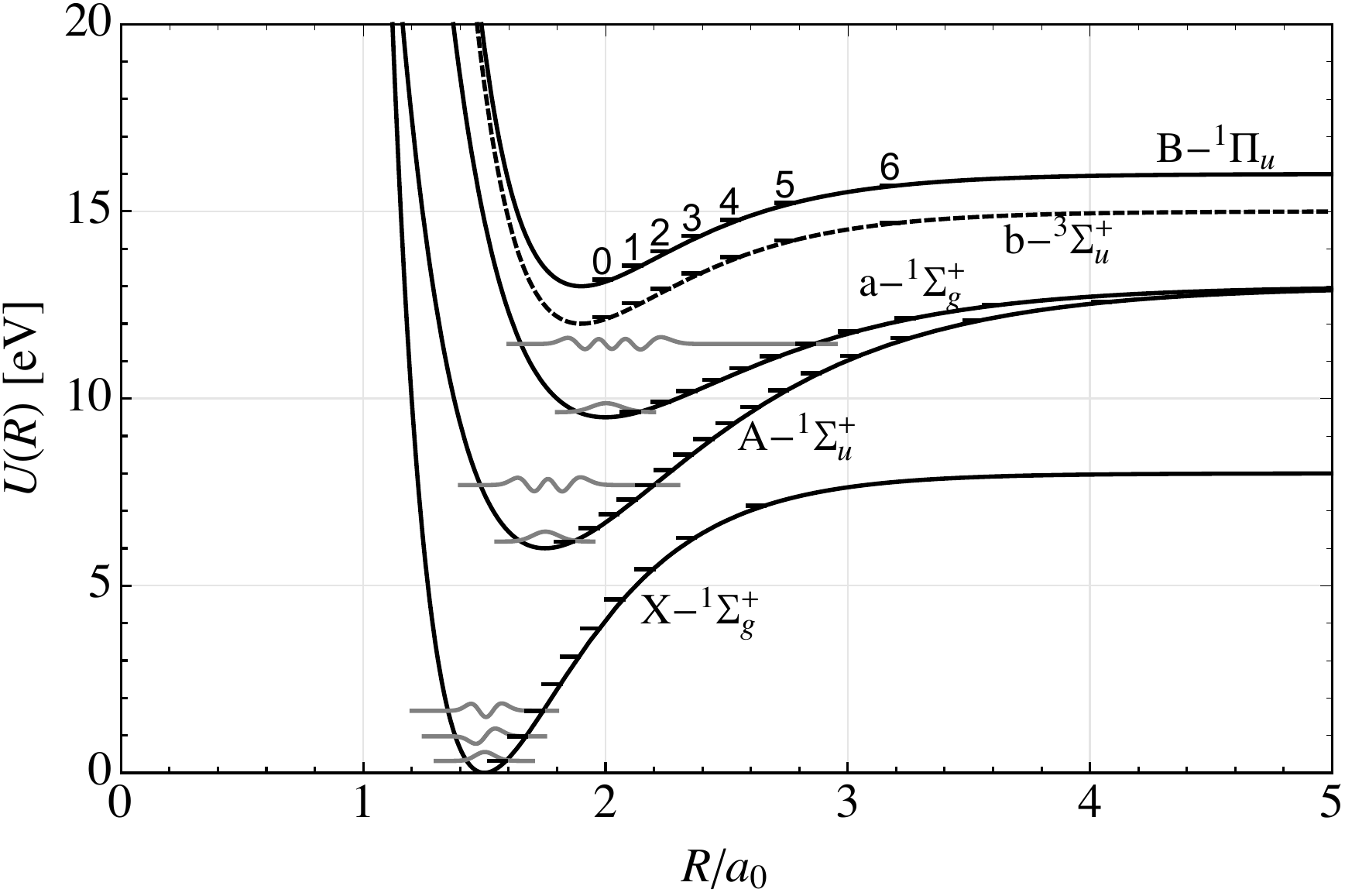}
\end{center}
\caption{Electronic potential energy $U(R)$ curves for a (hypothetical) homonuclear diatomic molecule as a function of internuclear separation $R$ for five different electronic states, labeled by their quantum numbers, $\text{X-}{}^1\Sigma_g^+$,  $\text{A-}{}^1\Sigma_u^+$,  $\text{a-}{}^1\Sigma_g^+$,  $\text{b-}{}^3\Sigma_u^+$,  and $\text{B-}{}^1\Pi_u$, respectively. Ticks in the potential well of each curve indicate vibrational energy levels, e.g.~labeled from $v=0,...,6$ in the fourth excited electronic state $\text{B-}{}^1\Pi_u$. Exemplary vibrational wavefunctions $\psi_v(R)$ are plotted in gray (with arbitrary vertical units) for a few vibrational states, including the lowest three of the ground electronic state. Rotational energy splittings are not shown. The ground state $\text{X-}{}^1\Sigma_g^+$ can be excited to the first excited state $\text{a-}{}^1\Sigma_g^+$ by a monopole operator, and to $\text{A-}{}^1\Sigma_u^+$ and $\text{B-}{}^1\Pi_u$ by a dipole operator. A spin-dipole operator can also cause transitions to the spin-triplet state $\text{b-}{}^3\Sigma_u^+$, in addition to those caused by a regular dipole operator.}\label{fig:UR}
\end{figure}

In figure~\ref{fig:UR}, we depict a typical set of potential energy curves $U(R)$ for five different electronic states in a hypothetical diatomic molecule. Insofar as the relative nuclear motion factorizes from the electronic motion, one can then compute the vibrational wavefunctions $\psi_v(R)$ in the potential well defined by $U(R)$ for each electronic state separately. In the Born-Oppenheimer approximation, vibrational states can be excited within one electronic level. However, transitions between different electronic levels are generally accompanied by a change in nuclear motion---and thus a change in vibrational quantum number. The probability that, in an electronic transition from $|\chi^\text{el}_i \rangle \to |\chi^\text{el}_f \rangle$, the vibrational state $|v'_i\rangle$ in the initial electronic level changes to the vibrational state $|v''_f\rangle$ of the final electronic level, is determined by the so-called Franck-Condon (FC) factor:
\begin{align}
 \left| \langle v''_f | v'_i \rangle \right|^2 = \left| \int_0^{\infty} dR \, \psi_{v''_f}^*(R) \psi_{v'_i}(R)\right|^2.\label{eq:FC}
\end{align}
This factor largely controls the relative absorption rates (and radiative emission probabilities) among the vibrational states. In the case of fig.~\ref{fig:UR} for example, if the electronic state is excited from $\chi^\text{el}_i = {}^1 \Sigma_g^+$ to the first-excited state $\chi^\text{el}_f = {}^1 \Sigma_u^+$, the vibrational quantum number $v'_i = 0$ is much more likely to change to $v''_f = 4$ than to $v''_f = 0$ because of the larger vibrational wavefunction overlap (which can be read off visually from the wavefunctions in fig.~\ref{fig:UR}). Tabulations of these FC factors have been computed and calibrated against measurement for a vast number of molecules; we will reference them where used in the text.
 In general, there are no selection rules controlling changes in vibrational quantum number in electronic molecular transitions, greatly enhancing the number of potential absorption lines, and thus accounting for the much richer spectroscopy of diatomic and polyatomic gases than that of monoatomics. This increased complexity does \emph{not} come at the cost of calculability nor spectral resolution for sufficiently small molecules. In table~\ref{tab:molecules}, we list the defining spectral characteristics of several of the electronic energy levels used throughout the text.
 
\begin{table*}[t]
\bgroup
\def\arraystretch{1.5}
\centering
  \begin{tabular}{l l | d{6.2} d{4.5} d{3.5} d{1.5} d{1.6} d{1.5} d{3.3}}
    molecule & $\chi^\text{el}$ & \multicolumn{1}{c}{$T_e$ [$\invcm$]} & \multicolumn{1}{c}{$\omega_{e}$ [$\invcm$]} & \multicolumn{1}{c}{$\omega_e x_e$ [$\invcm$]} & \multicolumn{1}{c}{$B_e$ [$\invcm$]} & \multicolumn{1}{c}{$\alpha_e$ [$\invcm$]} & \multicolumn{1}{c}{$R_e$ [\angstrom]} & \multicolumn{1}{c}{$T_b$ [K]} \\
    \hline{}
    \multirow{5}{*}{${}^{1}\text{H}{}^{1}\text{H}$} & c-${}^3 \Pi_u$ & 95938 & 2466.8 & 63.51 & 31.07 & 1.42 & 1.037 &  \\
						    & EF-${}^1 \Sigma_g^+$ & 100082.3 & 2588.9 & 130.5 & 32.68 & 1.818 & 1.011 & \\ 
    						    & C-${}^1 \Pi$ & 100089.8 & 2443.77 & 69.524 & 31.362 & 1.664 & 1.0327 &  \\
    						    & B-${}^1 \Sigma_u^+$ & 91700 & 1358.09 & 20.888 & 20.015 & 1.184 & 1.2928 & \\
						    & X-${}^1 \Sigma_g^+$ & 0 & 4401.21 & 121.33 & 60.853 & 3.062 & 0.74144 & 20.28 \\
    \hline{}
    \multirow{3}{*}{${}^{1}\text{H}{}^{2}\text{H}$} & C-${}^1 \Pi$ & 100092.9 & 2119.6 & 53.31 & 23.522 & 1.096 & 1.0329 &  \\
    						    & B-${}^1 \Sigma_u^+$ & 91698.3 & 1177.16 & 15.59 & 15.071 & 0.820 & 1.2904 & \\
						    & X-${}^1 \Sigma_g^+$ & 0 & 3813.1 & 91.65 & 45.655 & 1.986 & 0.74142 &  \\
    \hline{}
    ${}^{2}\text{H}{}^{2}\text{H}$ & X-${}^1 \Sigma_g^+$ & 0 & 3115.50 & 61.82 & 30.443 & 1.0786 & 0.74152 &  \\
    \hline
    ${}^{16}\text{O}{}^{16}\text{O}$ & X-${}^3 \Sigma_g^-$ & 0 &  1580.19 & 11.98 & 1.44563 & 0.0159 & 1.20752 & 90.19 \\
    \hline
    \multirow{4}{*}{${}^{12}\text{C}^{16}\text{O}$} & A-${}^1 \Pi$ & 65075.7 & 1518.2 & 19.40 & 1.6115 & 0.0232 & 1.2353 &  \\
    						    & a'-${}^3 \Sigma^+$ & 55825.4 & 1228.60 & 10.468 & 1.3446 & 0.0189 & 1.3523 &  \\
    						    & a-${}^3 \Pi$ & 48686.70 & 1743.4 & 14.3 & 1.69124 & 0.01904 & 1.20574 & \\
						    & X-${}^1 \Sigma^+$ & 0 & 2169.81358 & 13.28831 & 1.93128 & 0.0175 & 1.12832 & 81.65\\
    \hline
    ${}^{12}\text{C}^{18}\text{O}$ & X-${}^1 \Sigma^+$ & 0 & 2117.5 & 12.66 & 1.839 & 0.0163 & 1.128 & \\
    \hline
    \multirow{2}{*}{${}^{14}\text{N}^{14}\text{N}$} 	& A-${}^3 \Sigma_u^+$& 50203.6 & 1460.64 & 13.87 & 1.4546 & 0.0180 &1.2866 \\
							& X-${}^1\Sigma_g^+$ & 0 & 2358.57 & 14.324 & 1.99824 & 0.017318 & 1.09768 & 77.355 \\
    \hline
    ${}^1\text{H}^{35}\text{Cl}$ & X-${}^1 \Sigma^+$ & 0 & 2990.946 & 51.8 & 10.59341 & 0.30718 & 1.27455 & 188.10 \\
    \hline
    \multirow{2}{*}{${}^{127}\text{I}{}^{127}\text{I}$} & B-${}^3 \Pi_{0^+u}$ & 15769.01 & 125.69 & 0.764 & 0.02903 & 0.000158 & 3.024 & \\
					  		& X-${}^1 \Sigma_g^+$ & 0 & 214.50 & 0.614 & 0.03737 & 0.000113 & 2.666 & 457.4\\
  \end{tabular}
  \caption{Spectroscopic properties of select diatomic molecules in the ground electronic state $\chi^\text{el} = \text{X}$, as well as in a few exemplary excited electronic states $\chi^\text{el} = A,a,B,b,\dots$ taken from Ref.~\cite{huber1979g}. Energetic quantities such as the electronic excitation energy $T_e$, vibrational energy splitting $\omega_e$ and anharmonic correction $\omega_e x_e$, rotational constant $B_e$, and vibration-rotation correction constant $\alpha_e$ are expressed in the conventional inverse photon-equivalent-wavelength units of $\invcm$; conversion to units of energy is achieved via the substitution $1~\invcm \leftrightarrow 1.23967 \times 10^{-4}~\eV$. We also quote the equilibrium radius $R_e$ and boiling point temperature (for ground states) at standard atmospheric pressure $P = 1~\text{bar}$.  }\label{tab:molecules}
  \egroup
\end{table*}


\subsection{Types of dark-matter induced transitions}\label{sec:transitions}

In this section, we classify the types of transitions according to the operator structure of $\delta H$, and derive how they act on nuclear and electronic wavefunctions. For each type, we derive the corresponding selection rules for rotational, vibrational, and electronic transitions, and give estimates of the expected transition matrix elements in small diatomic molecules.


\subsubsection{Monopole transitions}
The simplest operator structure occurs when the perturbing Hamiltonian $\delta H$ is spherically symmetric in all respects, in the sense that it acts trivially on the angular-momentum eigenstates of the entire molecule, the electronic configuration, and all spin degrees of freedom. In that case, $\delta H$ must be diagonal in these angular eigenstate bases, so any such ``monopole'' transition obeys the selection rules:
\begin{align}
\Delta J = \Delta M = \Delta \Lambda = \Delta S = \Delta \Sigma = \Delta \Omega = 0. \label{eq:SelMon1}
\end{align}
We expect these rules to be obeyed insofar as spin-orbit coupling can be neglected, except for the conservation of total angular momentum ($\Delta J = 0$), which is exact. 

It follows immediately that pure rotational transitions cannot be induced by a monopole operator. Vibrational transitions can occur, for example by the operators
\begin{align}
\delta H^{0}_\text{I} \propto R-R_e, \quad \delta H^{0}_\text{II} \propto (R-R_e)^2, \quad \delta H^{0}_\text{III} \propto \vect{\nabla}_{\vect{R}}^2. \label{eq:HMon1}
\end{align}
These act trivially on the angular wavefunction, but can excite an initial ground vibrational state $|v_i = 0\rangle$ to a first- or second-excited vibrational state $|v_f\rangle$ via the matrix elements:
\begin{align}
\langle v_f = 1 | (R-R_e) | v_i = 0 \rangle & = \frac{-i}{(2M \omega_e)^{1/2}},  \label{eq:Mat1} \\
\langle v_f = 2 | (R-R_e)^2 | v_i = 0 \rangle & = \frac{1}{M \omega_e}, \label{eq:Mat2}  \\
\langle v_f = 2 | \frac{d^2}{dR^2} | v_i = 0 \rangle & = M \omega_e. \label{eq:Mat3}
\end{align}
In the limit of a pure harmonic oscillator ($\omega_e x_e \to 0$), these are the only nonzero, off-diagonal matrix elements connecting $|v_i=0\rangle$ to excited levels. In the presence of anharmonicities ($\omega_e x_e \neq 0$), these selection rules are weakly broken, and e.g.~$(R-R_e)$ can also excite the ground vibrational state to $v_f \ge 2$, with the matrix elements:
\begin{align}
|\langle v_f = 2 | (R-R_e) | v_i = 0 \rangle| & \simeq \frac{\left(\frac{1}{8}\right)^{1/2} \left(\frac{\omega_e x_e}{\omega_e}\right)^{1/2}}{(2M \omega_e)^{1/2}} , \label{eq:Mat1b} \\
|\langle v_f = 3 | (R-R_e) | v_i = 0 \rangle| & \simeq \frac{\left(\frac{1}{24}\right)^{1/2} \left(\frac{\omega_e x_e}{\omega_e}\right)}{(2M \omega_e)^{1/2}} , \label{eq:Mat1c} \\
|\langle v_f = 4 | (R-R_e) | v_i = 0 \rangle| & \simeq \frac{\left(\frac{3}{128}\right)^{1/2} \left(\frac{\omega_e x_e}{\omega_e}\right)^{3/2}}{(2M \omega_e)^{1/2}}  ,\label{eq:Mat1d} 
\end{align}
up to $\mathcal{O}(\omega_e x_e /\omega_e)^{1/2}$-suppressed fractional corrections. In gross sensitivity estimates using the matrix elements of eqs.~\ref{eq:Mat1}--\ref{eq:Mat1d}, we will often use the parametric estimates:
\begin{align}
M \omega_e \sim \frac{\alpha^4 m_e^3}{\omega_e}, \qquad \frac{\omega_e}{\omega_e x_e} \sim \frac{\alpha^2 m_e}{4\omega_e}, \label{eq:paramest}
\end{align}
typically correct up to a factor of 2, and ``solving'' for $M$ and $\omega_e x_e$ in favor of the vibrational frequency $\omega_e$.

Monopole vibrational transitions are not accompanied by a change in molecular rotation, and thus do not come with much rotational substructure in the possible transition energies (no dependence on $B_e$):
\begin{align}
&|v_i=0, J_i, M_i \rangle \to |v_f, J_i, M_i\rangle:  \\
&\hspace{5em} \omega_0 = \omega_e v_f - \omega_e x_e (v_f^2 + v_f) - \alpha_e v_f J_i(J_i+1). \nonumber
\end{align}
The only dependence on the rotational quantum number enters via the vibration-rotation coupling $\alpha_e$, which in the Born-Oppenheimer approximation is suppressed by a factor of $\mathcal{O}(m_e/M)^{3/2}$ relative to the vibrational splitting $\omega_e$.

Monopole operators that can cause electronic transitions include
\begin{align}
\delta H^{0}_\text{IV} & \propto \sum_n \vect{\nabla}_{e,n}^2, \label{eq:HMon2} \\
\delta H^{0}_\text{V} & \propto -\sum_{n,N} Z_k \frac{1}{\left|\vect{r}_{e,n} - \vect{R}_{N}\right|} + \sum_{n,m} \frac{1}{\left|\vect{r}_{e,n} - \vect{r}_{e,m}\right|}. \nonumber 
\end{align}
Both the inversion $i$ and the reflection $\sigma_v$ operations commute with these operators, so electronic transitions follow the selection rules
\begin{align}
g \leftrightarrow g, \quad u & \leftrightarrow u, \quad g \nleftrightarrow u, \quad \text{(homonuclear)}\\
+ \leftrightarrow +, \quad - & \leftrightarrow -, \quad + \nleftrightarrow -, \quad \text{(for $\Sigma \leftrightarrow \Sigma$)}
\end{align}
in addition to the ones mentioned previously in eq.~\ref{eq:SelMon1}. Typical sizes for transition matrix elements in nonrelativistic molecules can be estimated via NDA as
\begin{align}
&\Big| \langle \chi^\text{el}_f | -\sum_{n,N} Z_N \frac{1}{\left|\vect{r}_{e,n} - \vect{R}_{N}\right|} + \sum_{n,m} \frac{1}{\left|\vect{r}_{e,n} - \vect{r}_{e,m}\right|} | \chi^\text{el}_i\rangle \Big| \nonumber\\
&\hspace{9em} = \frac{1}{2 \alpha m_e}   \Big|\langle \chi^\text{el}_f |\sum_n \vect{\nabla}_{e,m}^2| \chi^\text{el}_i\rangle\Big| \sim \frac{\omega_0}{\alpha},\label{eq:Hmonvir}
\end{align}
with $R_e$ the equilibrium radius of the initial electronic state $|\chi^\text{el}_i\rangle$, usually an $\mathcal{O}(1)$ number times the Bohr radius $a_0 = 1/\alpha m_e$. The off-diagonal matrix elements of $\delta H^{0}_\text{IV}$ and $\delta H^{0}_\text{V}$ can be related as in eq.~\ref{eq:Hmonvir} because they are both component terms of $H_0$ in eq.~\ref{eq:H0}, which itself acts diagonally on energy eigenstates by construction. Monopole transitions from the ground electronic state to an excited electronic state can occur at many different energies due to rovibrational substructure:
\begin{align}
\omega_0 = & ~T_{e,f} + \left[\omega_{e,f} \left(v_f+\frac{1}{2}\right) - \omega_{e,i} \left(v_i+\frac{1}{2}\right)\right] \nonumber\\
&- \left[\omega_{e,f} x_{e,f} \left(v_f+\frac{1}{2}\right)^2  - \omega_{e,i} x_{e,i} \left(v_i+\frac{1}{2}\right)^2\right]\nonumber\\
& + \left[B_{e,f}-B_{e,i}\right] J_i(J_i+1) \nonumber\\
&- \left[\alpha_{e,f} \left(v_f+\frac{1}{2}\right)-\alpha_{e,i} \left(v_i+\frac{1}{2}\right)\right] J_i(J_i+1). \nonumber
\end{align}
We can see that even though $\Delta J = 0$, there is a significant amount of rotational substructure because the rotational constants $B_e$ of two different electronic states generally differ by an $\mathcal{O}(1)$ fractional amount. As explained around eq.~\ref{eq:FC}, there are no selection rules associated with changes in vibrational quantum number for electronic transitions, introducing a large amount of vibrational substructure, which also holds true for electronic transitions induced by other types of operators, to which we turn next.


\subsubsection{Dipole transitions} \label{sec:dipoletransitions}
Transitions caused by operators of the form
\begin{align}
\delta H^{1}_\text{I} \propto \hat{\vect{k}} \cdot \vect{R}, \qquad \delta H^{1}_\text{II} \propto \hat{\vect{k}} \cdot \sum_j \vect{r}_{e,j} \label{eq:HPosDip}
\end{align}
are perhaps the most familiar, since they include the ordinary electric dipole transitions from photon absorption. Here, the unit vector $\hat{\vect{k}}$ denotes a unit vector in a space-fixed (as opposed to molecule-fixed) direction, like the direction of a vector field or the DM velocity, that acts trivially on the wavefunction. The effective direction $\hat{\vect{k}}$ will in general vary in time and space, but only over scales of order the coherence time and length of the DM field, both much larger than the relevant temporal and spatial scales of the molecule for the energies under consideration.

The operator proportional to $\vect{R} = R \hat{\vect{R}}$ can induce pure rotational transitions through diagonal action of $\langle R \rangle = R_e$ on the vibrational state but nontrivial action of $\hat{\vect{R}}$ on the rotational state, such that  $\langle J_f M_f | \hat{\vect{k}} \cdot \hat{\vect{R}}|J_i M_i \rangle \neq 0$. 
These well-known matrix elements have selection rules $\Delta J = \pm 1$ as well as $\Delta M = \pm 1$ for $\hat{\vect{k}}\cdot \hat{\vect{z}}=0$ and $\Delta M = 0$ for $\hat{\vect{k}} = \hat{\vect{z}}$. The rotational transitions which \emph{increase} the internal energy are those with $\Delta J = +1$, and have transition energy:
\begin{align}
|J_i, M_i \rangle \to |J_i+ 1, M_f\rangle: \quad \omega_0 = 2 \left(B_e - \frac{1}{2}\alpha_e\right) (J_i+1). \label{eq:omega1R}
\end{align}
These pure rotational transitions are too low in energy for the experimental setup under consideration in this work, because $B_e$ (for any molecule) is lower than the thermal energy $T$ even at liquid nitrogen temperatures. However, they form an important part of the substructure for vibrational transitions.

The operator $\vect{R} = R \hat{\vect{R}}$ can also act nontrivially on the vibrational state, e.g.~$\langle v_f, J_f, M_f |\hat{\vect{k}} \cdot  \vect{R} | v_i , J_i, M_i \rangle = \langle v_f | (R-R_e) | v_i \rangle \langle J_f  M_f | \hat{\vect{k}} \cdot \hat{\vect{R}} | J_i  M_i \rangle$. The transition amplitudes are products of the vibrational matrix element in eq.~\ref{eq:Mat1} and the pure rotational matrix elements. Over long time scales, the unit vector $\hat{\vect{k}}$ will change direction, so we directionally average the rotational matrix elements according to:
\begin{align}
\left| \langle J_f | \hat{\vect{k}}\cdot \hat{\vect{R}}| J_i \rangle \right|^2_\text{avg} \equiv   \sum_{M_f,M_i}\sum_{j=x,y,z}  \frac{\left| \langle J_f M_f| \hat{R}_j | J_i M_i \rangle \right|^2}{3(2J_i + 1)} 
, \label{eq:dipmatrot}
\end{align}
with $\hat{R}_x$, $\hat{R}_y$, $\hat{R}_z$ unit vectors in the $x$, $y$, $z$ directions, respectively. 
These average square matrix elements can be computed to be $(J_i+1)/[3(2J_i+1)]$ for $\Delta J = +1$ transitions, and $J_i/[3(2J_i+1)]$ for $\Delta J = -1$ transitions; at large $J_i$, they tend to $1/6$.
Vibrational transitions induced by $H^{1}_\text{I}$ thus obey the selection rules $\Delta v = \pm 1$, $\Delta J = \pm 1$, and $\Delta M = 0, \pm 1$. Changes in the $M$ quantum number are from the components of $\hat{\vect{R}}$ in the $x$ and $y$ directions; the $z$-component of $\hat{\vect{R}}$ leaves $M$ unchanged. The allowed vibrational transitions and energies from the ground vibrational state are:
\begin{widetext}
\begin{align}
&| v_i = 0, J_i, M_i \rangle \to 
\begin{cases} 
|v_f = 1, J_f = J_i + 1, M_f \rangle: & \qquad \omega_0 = \omega_e - 2 \omega_e x_e + 2 \left(B_e - \frac{1}{2}\alpha_e\right) (J_i+1),  \\
|v_f = 1, J_f = J_i - 1, M_f \rangle: & \qquad \omega_0 = \omega_e - 2 \omega_e x_e - 2 \left(B_e - \frac{1}{2}\alpha_e\right) J_i. 
\label{eq:transPosDip}
\end{cases}
\end{align}
\end{widetext}
Higher-$\Delta v$ are only weakly allowed (cfr.~eqs.~\ref{eq:Mat1b}--\ref{eq:Mat1d}) and occur at energies that are roughly integer multiples of the first energy gap. To leading order in the Born-Oppenheimer approximation, the operator $H^{1}_\text{I}$ does not excite electronic transitions.

The operator $\delta H^{1}_\text{II} \propto \sum_n \vect{r}_{e_n}$ primarily causes transitions between electronic states. The components of $\vect{r}_{e_n}$ that point along the molecular axis are unaffected by $L_z$ rotations, are even under $\sigma_v$ reflections, and odd under the inversion $i$. Components of $\vect{r}_{e_n}$ transverse to the molecular axis transform as $\Lambda = 1$ states under $L_z$ rotations, and are odd under $\sigma_v$ reflections and the inversion $i$. We thus find the selection rules:
\begin{align}
& \Delta \Lambda = 0,\pm 1 \label{eq:sele1R1}\\
& + \leftrightarrow +, \quad - \leftrightarrow -, \quad - \nleftrightarrow + \quad \text{(for $\Sigma \leftrightarrow \Sigma$)}\label{eq:sele1R2}\\
& g \leftrightarrow u, \quad g \nleftrightarrow g, \quad u \nleftrightarrow u, \quad \text{(homonuclear)}\label{eq:sele1R3} \\
& \Delta S = \Delta \Sigma = 0. \label{eq:sele1R4}
\end{align}
The last line follows from the trivial action on the spin coordinates, so both the spin multiplicity $S$ and the projection of the spin onto the molecular axis $\Sigma$ are unchanged in absence of spin-orbit coupling. For transitions that are allowed by these selection rules, we can estimate the matrix elements with NDA to be parametrically of order:
\begin{align}
\Big|\langle \chi^\text{el}_f |\sum_n \vect{r}_{e_n}|  \chi^\text{el}_i \rangle\Big| \sim R_e. \label{eq:dipmatvir}
\end{align}
Alternatively, for $\delta H^{1}_\text{II}$ the above matrix elements can also be measured via absorption or emission intensities in electric dipole transitions.

As for any electronic transition, there are generally no selection rules for changes in vibrational quantum number. The total rotational quantum number $J$ has the selection rule $\Delta J = \pm 1$ for $\Delta \Lambda = 0$ transitions, whereas both $\Delta J = \pm 1$ and $\Delta J = 0$ are allowed for $\Delta \Lambda = \pm 1$ transitions (for most initial $J_i$). We refer the reader to \cite{struve1989fundamentals} for more details about rotational fine structure in electronic transitions. The bottom line is that the extra rotational fine structure makes the discretuum of transition energies for electronic dipole transitions even more rich as that of electronic monopole transitions, with lines occurring at energies
\begin{align}
\omega_0 = & ~T_{e,f} + \left[\omega_{e,f} \left(v_f+\frac{1}{2}\right) - \omega_{e,i} \left(v_i+\frac{1}{2}\right)\right] \nonumber\\
&- \left[\omega_{e,f} x_{e,f} \left(v_f+\frac{1}{2}\right)^2  - \omega_{e,i} x_{e,i} \left(v_i+\frac{1}{2}\right)^2\right]\nonumber\\
& + \left[B_{e,f}  - \alpha_{e,f} \left(v_f+\frac{1}{2}\right) \right] J_f(J_f+1) \nonumber\\
&- \left[B_{e,i} - \alpha_{e,i} \left(v_i+\frac{1}{2}\right)\right] J_i(J_i+1),
\end{align}
for transitions from the ground electronic state to one particular excited electronic state with excitation energy $T_{e,f}$.


\subsubsection{Spin-dipole transitions}

Dark matter may also interact via less-familiar operators which exhibit a simultaneous coupling to both the spin and the momentum of nucleons or electrons: 
\begin{align}
\delta H^{1\text{S}}_\text{I} \propto \vect{\sigma}_N \cdot \vect{\nabla}_{\vect{R}}, \qquad 
\delta H^{1\text{S}}_\text{II} \propto \sum_n \vect{\sigma}_{e,n} \cdot \vect{\nabla}_{e,n}.
\end{align}
Despite the wholly different operator structure and besides their action onto the spin coordinates, they are in many ways similar to the dipole operators of eq.~\ref{eq:HPosDip}, with analogous selection rules on the orbital wavefunction in absence of spin-orbit coupling. For simplicity of discussion, we will take $\vect{\sigma}_N$ to be the nuclear spin operator on only \emph{one} of the nuclear spins (e.g.~if the other one were spinless). For two spins, one would need to consider a larger spin Hilbert space, and use an interaction Hamiltonian of the form $\alpha_1 \vect{\sigma}_{1}\cdot \vect{\nabla}_1 + \alpha_2 \vect{\sigma}_{2} \cdot \vect{\nabla}_2 =  (-\alpha_1 \vect{\sigma}_{1} + \alpha_2 \vect{\sigma}_2)\cdot \vect{\nabla}_{\vect{R}} + \mathcal{O}(\vect{\nabla}_{\vect{R}_\text{cm}})$ with coefficients $\alpha_1, \alpha_2$.

The first operator can cause pure rotational transitions, as well as vibrational transitions with rotational substructure. In spherical coordinates, we have 
\begin{align}
\vect{\nabla}_{\vect{R}} = \hat{\vect{R}} \partial_R + \hat{\vect{\theta}} \frac{1}{R} \partial_\theta + \hat{\vect{\phi}} \frac{1}{R \sin \theta} \partial_\phi.
\end{align}
Let us now see how this operator acts on a given rovibrational eigenstate. We have:
\begin{align}
\langle R \theta \phi | \vect{\nabla}_{\vect{R}} | \psi^\text{vib} Y^\text{rot} \rangle = ~&\langle \theta \phi|\hat{\vect{R}}|J M \rangle \langle R | \left(\frac{d}{dR}-\frac{1}{R}\right) | v \rangle \nonumber \\
&+ \langle \theta \phi|\hat{\vect{\theta}}\partial_\theta |J M \rangle \langle R | \frac{1}{R} | v \rangle \nonumber\\
&+ \langle \theta \phi|\hat{\vect{\phi}}\frac{1}{\sin \theta}\partial_\phi|J M \rangle \langle R | \frac{1}{R} | v \rangle.\label{eq:psigmaaction}
\end{align}
where $\langle R | \psi^\text{vib}_v \rangle = \tilde{\psi}_v(R)/R$ with $\tilde{\psi}_v(R)$ QHO eigenstate wavefunctions, and $\langle \theta \phi | Y^\text{rot}_{JM} \rangle = Y_{JM}(\theta,\phi)$ are the spherical harmonics.

When the radial part of $\vect{\sigma}_N \cdot \vect{\nabla}_{\vect{R}}$ acts diagonally on the vibrational state, we have that $\langle v | 1/R | v \rangle = 1/R_e$ and $\langle v | d/dR | v \rangle =0$. In this case, the angular and nuclear-spin action can still cause pure rotational transitions along with nuclear spin transitions:
\begin{widetext}
\begin{alignat}{3}
\langle \Sigma_{N_f} J_f M_f | \vect{\sigma}_N \cdot \vect{\nabla}_{\vect{R}} | \Sigma_{N_i} J_i M_i \rangle = &+  \frac{1}{R_e} \langle \Sigma_{N_f} | \sigma_{N,x} | \Sigma_{N_i} \rangle && \langle J_f M_f | (-\sin \theta \cos \phi + \cos \theta \cos \phi \partial_\theta - \frac{\sin\phi}{\sin\theta} \partial_\phi)| J_i M_i \rangle \nonumber \\
&+ \frac{1}{R_e}  \langle \Sigma_{N_f} | \sigma_{N,y} | \Sigma_{N_i} \rangle && \langle J_f M_f | (-\sin \theta \sin \phi+\cos \theta \sin \phi \partial_\theta + \frac{\cos\phi}{\sin\theta} \partial_\phi)| J_i M_i \rangle \nonumber \\
&+ \frac{1}{R_e}  \langle \Sigma_{N_f} | \sigma_{N,z} | \Sigma_{N_i} \rangle && \langle J_f M_f |(\cos \theta -\sin \theta \partial_\theta)| J_i M_i \rangle.
\end{alignat}
\end{widetext}
Above we have assumed that the nucleus has a total spin $S_N$ equal to an integer or half-integer, and of which the component along the $z$-axis has the possible values $\Sigma_N = - S_N, -S_N + 1, \dots, + S_N$. The selection rules for these transitions are:
\begin{align}
\Delta J = \pm 1, \quad 
\begin{cases}
\Delta M = \pm 1 & ~ \& ~~~ \Delta \Sigma_{N} = \mp 1, \\
\Delta M =  0 & ~ \& ~~~ \Delta \Sigma_{N} =  0.\label{eq:selNProt}
\end{cases}
\end{align}
We distinguish between the spin-flip (from $\sigma_{N,x}$ and $\sigma_{N,y}$) and spin-preserving (from $\sigma_{N,z}$) transitions, as the former can receive linear Zeeman energy shifts in an external magnetic field. They otherwise have the same rotational energy splittings and selection rules as in eq.~\ref{eq:omega1R}. In a thermal state, the nuclear spins are in an unpolarized mixed state. The transition rate from an energy level with $J = J_i$ to one with $J = J_f$ is thus proportional to $|\langle J_f | \hat{ \vect{\sigma}} \cdot \vect{\nabla}_{\vect{R}} | J_i \rangle|^2_\text{avg}$, which can be found by averaging the square of the matrix element $\langle J_f M_f |\hat{ \vect{\sigma}} \cdot \vect{\nabla}_{\vect{R}} | J_i M_i \rangle$ over the spin directions $\hat{ \vect{\sigma}}$, averaging over $M_i$ and summing over the possible values of $M_f$. 
One finds that $|\langle J_f | \hat{ \vect{\sigma}} \cdot \vect{\nabla}_{\vect{R}} | J_i \rangle|^2_\text{avg}$ is equal to $(J_i-1)^2(J_i+1)/[3(2J_i-1)]$ for $\Delta J = +1$ transitions, and $J_i(J_i+2)^2/[3(2J_i-1)]$ for $\Delta J = -1$. They tend to $J_i^2/6$ at large $J_i$.

The operator $\vect{\nabla}_{\vect{R}}$ can also induce vibrational transitions through its off-diagonal action. As can be seen from eq.~\ref{eq:psigmaaction}, it acts on vibrational states $|v\rangle$ through the operator terms $d/dR$ and $1/R$. The former has a matrix element connecting the ground state to the first-excited state:
\begin{align}
\langle v_f = + 1 | \frac{d}{dR} | v_i = 0 \rangle = -i \left( \frac{M\omega_e}{2}\right)^{1/2}, \label{eq:Mat4}
\end{align}
all other off-diagonal matrix elements from the ground state being suppressed by the anharmonicity of the vibrational potential. 
The second operator term, $1/R$, has smaller vibrational matrix elements, as can be seen from expanding
\begin{align}
\frac{1}{R} = \frac{1}{R_e} + \frac{R-R_e}{R_e^2} + \dots
\end{align}
and using eq.~\ref{eq:Mat1} to find that the dominant $1/R$ matrix element is smaller than that of eq.~\ref{eq:Mat4} by a factor of $1/M\omega_e R_e^2 \sim \mathcal{O}(m_e/M)^{1/2}$. Therefore, the dominant vibrational transitions obey the same $\Delta v = \pm 1$ selection rules as in eq.~\ref{eq:transPosDip}. The accompanying rotational matrix elements are the same as well, cfr.~eq.~\ref{eq:dipmatrot}, as long as one remembers that changes in the orbital angular momentum projected onto any particular axis are correlated with simultaneous changes in the nuclear spin projected onto the same axis as in eq.~\ref{eq:selNProt}.

The operator $\vect{\sigma}_e \cdot \vect{\nabla}_e$ mostly excites electronic transitions, similar to $\delta H_\text{II}^{1}$ of the previous section. The operator $\vect{\nabla}_e$ has the same transformation properties under $L_z$, $i$, and $\sigma_v$ as $\vect{r}_e$, so $\delta H_\text{II}^{1\text{S}}$ obeys the same selection rules as $\delta H_\text{II}^{1}$ on changes in the orbital part of the electronic motion, which are already listed in eqs.~\ref{eq:sele1R1}--\ref{eq:sele1R3}. However, the nontrivial spin structure of $\delta H_\text{II}^{1\text{S}}$ means that it can induce both spin-preserving ($\Delta S = \Delta \Sigma = 0$) transitions---the only ones $\delta H_\text{II}^{1}$ can excite---as well as spin-flip transitions ($\Delta \Sigma = \pm 1$ and $\Delta S = 0, \pm 1$):
\begin{align}
&\Delta \Lambda = 0~\&~\Delta \Sigma = \Delta S = 0, \label{eq:sele1P1A} \\
&\Delta \Lambda = \pm 1 ~\&~ \Delta \Sigma = \mp 1,  \label{eq:sele1P1B} \\
& + \leftrightarrow +, \quad - \leftrightarrow -, \quad - \nleftrightarrow +, \quad \text{(for $\Sigma \leftrightarrow \Sigma$)}\label{eq:sele1P2}\\
& g \leftrightarrow u, \quad g \nleftrightarrow g, \quad u \nleftrightarrow u. \quad \text{(homonuclear)}\label{eq:sele1P3}
\end{align}
Such spin-flip transitions are highly suppressed in small molecules for the usual dipole transitions, for which a ground state of e.g.~${}^1 \Sigma$ can normally only be excited to higher spin-singlet states of symmetry ${}^1 \Sigma$ and ${}^1 \Pi$, whereas these spin-dipole transitions can excite the same ground state to the spin singlets ${}^1 \Sigma$ or the spin triplets ${}^3 \Pi$. For transitions that respect the selection rules of eqs.~\ref{eq:sele1P1A}--\ref{eq:sele1P3}, we expect electronic matrix elements of size:
\begin{align}
\Big|\langle \chi^\text{el}_f | \sum_n \frac{\vect{\nabla}_{e_n}}{m_e} | \chi^\text{el}_i\rangle\Big| = \omega_0 \Big|\langle  \chi^\text{el}_f | \sum_n \vect{r}_{e_n} | \chi^\text{el}_i \rangle\Big| \sim \omega_0 R_e, \label{eq:spindipmatel}
\end{align}
with $\omega_0 = |E_f - E_i|$ the transition energy.
The first equality follows by virtue of the identity $\vect{\nabla}_{e_n} = - m_e \left[ H_0, \vect{r}_{e_n} \right]$ for the nonrelativistic $H_0$ from eq.~\ref{eq:H0}. The matrix elements for the spin-dipole transitions can thus be inferred from those of the regular dipole transitions.


\section{Experimental setup}\label{sec:expsetup}

We describe the general detector requirements and molecular container configurations in sec.~\ref{sec:configuration}, along with a detailed discussion of signal detection in sec.~\ref{sec:photodetection}, background levels in sec.~\ref{sec:backgrounds}, and signal discrimination techniques in sec.~\ref{sec:discrimination}. 

\begin{table*}
\bgroup
\def\arraystretch{1.5}
\begin{tabular}{l | l | l}
 & \multicolumn{1}{c|}{Bulk} & \multicolumn{1}{|c}{Stack} \\
\hline
\multirow{6}{*}{Phase I} & $V = (0.3~\text{m})^3$, $P \sim 0.1~\text{bar}$, $T = 300~\text{K}$ & $A = \pi (0.3~\text{m})^2$, $D = 1~\text{mm}$, $P \sim 10~\text{bar}$, $T \sim 100~\text{K}$ \\
 & PMT, $\text{DCR} = 1~\text{Hz}$, $A_\text{det} = (0.3~\text{m})^2$, $\eta_\gamma = 0.3$ & MKID/TES, $\text{DCR} 
 \lesssim 10^{-5}~\text{Hz}$, $A_\text{det} = (0.3~\text{mm})^2$, $\eta_\gamma = 0.5$ \\
 & any electronic $\rightarrow$ intermediate & E1-allowed electronic \\
 &  - & E1-allowed vibrational \\
 & Stark/Zeeman tuning, $t_\text{shot} = 10^2~\text{s}$ & collisional broadening, $t_\text{shot} = 10^6~\text{s}$ \\
 & $\delta \Omega \approx 2.9 \times 10^{-7}~\text{rad s}^{-1}$ & $\delta \Omega \approx 9.4 \times 10^{-9}~\text{rad s}^{-1}$ \\
 \hline
\multirow{6}{*}{Phase II} & $V = (2~\text{m})^3$, $P \sim 0.1~\text{bar}$, $T \sim 100~\text{K}$ & $A = \pi (2~\text{m})^2$, $D = 100~\text{mm}$, $P \sim 10~\text{bar}$, $T \sim 100~\text{K}$\\
 & MKID, $\text{DCR} \lesssim 10^{-3}~\text{Hz}$, $A_\text{det} = (0.1~\text{m})^2$, $\eta_\gamma = 0.5$ & MKID/TES, $\text{DCR} \lesssim 10^{-7}~\text{Hz}$, $A_\text{det} = (2~\text{mm})^2$, $\eta_\gamma = 1$\\
 & any electronic $\rightarrow$ intermediate & E1-allowed electronic \\
 & any vibrational with optically thin fluorescence & E1-allowed vibrational\\
 & Stark/Zeeman tuning, $t_\text{shot} = 10^3~\text{s}$ & collisional broadening, $t_\text{shot} = 10^7~\text{s}$ \\
 & $\delta \Omega \approx 9.9 \times 10^{-10}~\text{rad s}^{-1}$ & $\delta \Omega \approx 1.8 \times 10^{-11}~\text{rad s}^{-1}$
\end{tabular}
\caption{Experimental configurations and their specifications for Phase~I prototypes, and ultimate Phase~II implementations. The six lines in each cell correspond to (1) thermodynamic variables, (2) photodetection parameters, (3) accessible electronic transition types, (4) accessible vibrational transition types, (5) frequency coverage strategy, and (6) Rabi frequency sensitivity at $\text{SNR} = 1$ over frequencies where BBR can be ignored, assuming collisional broadening dominates the absorption width. The frequency coverage strategy of the Bulk configuration can be used by the Stack setup, and vice versa, albeit typically at lower sensitivity in a DM search over a broad energy range. The four versions are abbreviated as BI, BII, SI, and SII.}\label{tab:configs}
\egroup
\end{table*}

\begin{figure}\begin{center}
\includegraphics[width=0.42\textwidth, , trim = 0cm 0.1cm 0 0.1cm , clip]{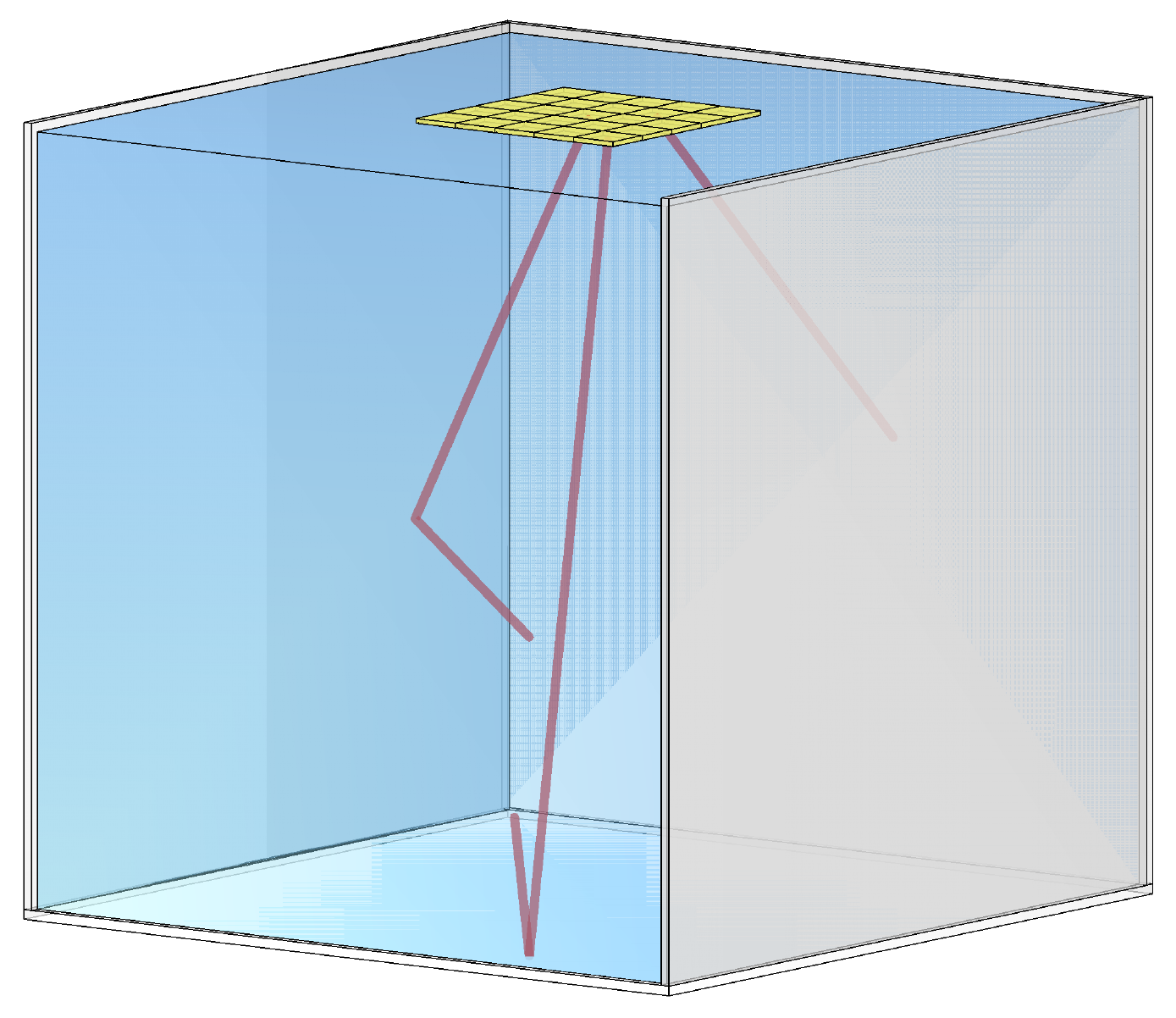}
\includegraphics[width=0.48\textwidth , trim = 0cm 0.7cm 0 2.2cm , clip]{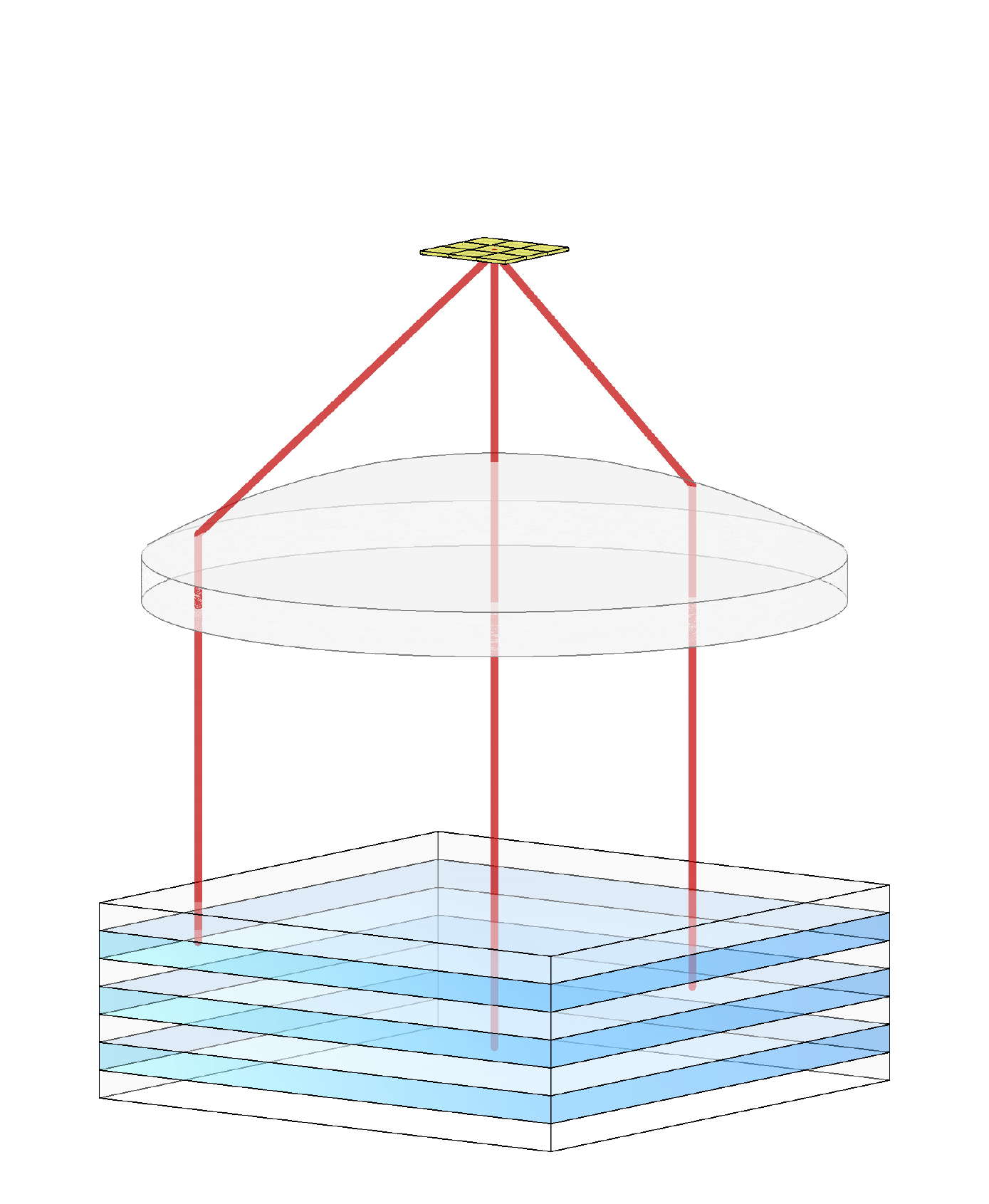}
\end{center}
\caption{Experimental setup: Bulk (top) and Stack (bottom) configurations. Molecular gas (depicted by light blue volumes) is pumped into containers capable of supporting pressures up to 10~bar. In the Bulk configuration, DM absorption events yield isotropic, single fluorescence photons, whose paths are indicated by thick red lines. Reflective coatings (shown as silver colored sheets) lining the container boundary retain the signal photons until they impinge onto a large-area photodetector, displayed as yellow tiles at the top. The Stack configuration features a pattern of alternating molecular density in the form of multiple slabs. This container geometry and the spatial coherence of DM can produce cooperatively emitted photons nearly perpendicular to the slabs, making it possible to focus them onto a tiny photodetector by a lens. The photon direction is sensitive to the DM velocity vector projected onto the plane defined by the slabs. For illustrative purposes, we do not show shielding, cooling, or electromagnetic field and pressure control systems in either setup. We also left out the reflective coatings on the front and top faces of the Bulk setup.}\label{fig:setup}
\end{figure}

\subsection{Configurations and search strategies} \label{sec:configuration}
We envision two configuration types to detect bosonic DM in the mass range between $0.2~\text{eV}$ and $20~\text{eV}$: one is a ``bulk'' detector volume, the other a layered set of slabs in a ``stack'' arrangement. We depict both configurations in fig.~\ref{fig:setup}, and summarize their specifications in table~\ref{tab:configs}, which will be explained below. We consider a prototype ``Phase~I'' and an optimistic, ultimate ``Phase~II'' of both the Bulk and Stack configuration.  We denote the four versions as BI, BII, SI, and SII. 

The molecules are kept in the gaseous phase so that they can be regarded as approximately independent subsystems with a discretuum of energy levels, each with a resonant response to near-monochromatic excitations; intermolecular interactions in the liquid and solid phases tend to produce a non-resonant continuum in all but a few special cases. Dark matter waves of the right frequency---i.e.~DM particles of the right mass---can excite molecules from their ground state(s) in thermal equilibrium to nonoccupied, higher-energy states, which in turn can fluoresce or cooperatively emit single photons. This radiation is to be read out by sensitive photodetectors, and serves as the signal in our setup. 

Two primary considerations are key in a DM detector design based on resonant absorption onto molecules in the gas phase: \emph{radiative efficiency} and \emph{frequency coverage}. We will first explain the relevant physics for both of these requirements, and then how the Bulk and Stack detector designs address each of them.

\begin{figure}
\includegraphics[width=0.48\textwidth]{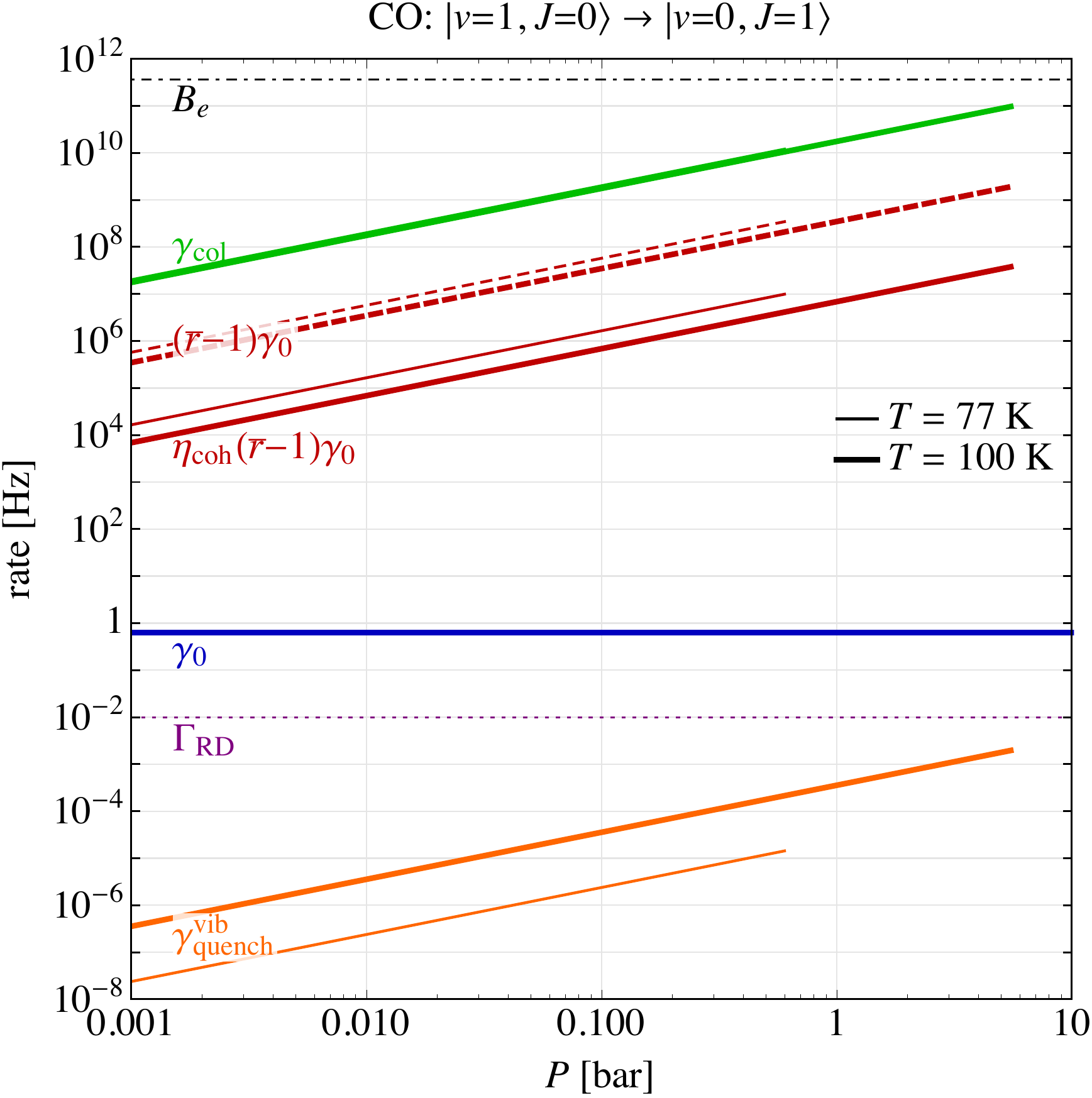}
\caption{Dynamic rates of the two-level subsystem $|v=0,J=1\rangle \leftrightarrow |v=1,J=0\rangle$ in carbon monoxide (CO), as a function of pressure $P$. Plotted are the collisional broadening rate $\gamma_\text{col}$ (green), the coherent emission rate in absence/presence of decoherence---$(\bar{r}-1)\gamma_0$ in dashed red / $\eta_\text{coh}(\bar{r}-1)\gamma_0$ in solid red---and the nonradiative quenching rate $\gamma_\text{quench}^\text{vib}$ (orange), for two temperatures $T = 100~\text{K}$ (thick) and $T = 77~\text{K}$ (thin). Also shown are the incoherent radiative rate $\gamma_0$ (blue), and a benchmark natural radioactive decay rate $\Gamma_\text{RD}$ (dashed purple) before an active veto. The collisional width $\gamma_\text{col}$ comes within a factor of 3 (30) from the rotational energy $B_e$ (dashed black) at the boiling point pressure of 5.5~bar (0.6~bar) at $T = 100~\text{K}$ ($T=77~\text{K}$). This shows that collisional broadening is an effective mechanism for frequency scanning. Notice that because the collisional rate dominates every other dynamical timescale, it reduces the coherent emission rate so that only 1 every $\mathcal{O}(1000)$ DM particles absorbed will produce a photon.}\label{fig:rate1}
\end{figure}

\textbf{Radiative efficiency---} Ideally, every DM absorption quantum leads to a detectable fluorescence photon, as opposed to heat or fluorescence photons that are difficult to detect (more on that in sec.~\ref{sec:photodetection}). The dominant channel for conversion of internal energy of small polyatomic molecules to heat is via two-body collisions wherein the excitation quanta in electronic or vibrational state are converted to rotational and/or translational kinetic energy, a process called radiative quenching. (Other nonradiative pathways through which the absorbed energy can dissipate include vibronic decays in large polyatomic molecules, and molecular dissociation in weakly bound ones.) By analogy to the collisional broadening rate in eq.~\ref{eq:gammacol}, we can thus parametrize the collisional quenching rate as:
\begin{align}
\gamma_\text{quench} = n \sigma_\text{quench} v_\text{mol} \label{eq:gammaquench}
\end{align}
with a quenching cross-section $\sigma_\text{quench}$ independent of density. For many excited electronic states, the inelastic, quenching cross-section is typically an order of magnitude smaller than the elastic cross-section $\sigma_\text{col}$~\cite{massey1949collisions}. Precise data on quenching cross-sections for excited electronic states is scarce, so we will take $\sigma_\text{quench}^\text{el} \sim 10~\angstrom^2$ as a benchmark value. Vibrational quanta are less easily quenched in diatomic molecules, an effect theoretically understood in the context of Landau-Teller theory~\cite{landauteller,massey1949collisions}, with a quenching cross-section that depends strongly on relative molecular velocities and thus temperature. Ref.~\cite{millikan1963systematics} found that a wide array of polyatomic systems obey the following empirical relation for the quenching rate $\gamma_\text{quench}^\text{vib}$ of vibrational quanta, to a precision of 50\%: 
\begin{align}
&\log_{10}\left( \frac{\gamma_\text{quench}^\text{vib}}{\text{Hz}} \frac{1~\text{bar}}{P}\right) \approx 8.00 \label{eq:gammaquenchvib} \\
&\hspace{3.6em} - 1.3 \times 10^2 \tilde{\mu}^{1/2}  \tilde{\omega}_e^{4/3} \left(\tilde{T}^{-1/3} - 1.5 \times 10^{-2} \tilde{\mu}^{1/4} \right), \nonumber
\end{align}
with the definitions of the dimensionless quantities $\tilde{\mu} \equiv \mu / m_p$, $\tilde{\omega}_e \equiv \omega_e / \text{eV}$, and $\tilde{T} = T/\text{K}$.
Above, $\mu$ is the reduced mass of the colliding pair of molecules, so $\mu = M_\text{mol}/2$ for a single molecular species of mass $M_\text{mol}$. At sufficiently low temperature $T$, the quenching rate becomes exponentially slow, as indicated by the orange line in fig.~\ref{fig:rate1}, with about a factor 10 decrease in the quenching rate of carbon monoxide by lowering the temperature from 100~K to 77~K. (The long relaxation time of the molecular sample means environmental background rejection of cosmic rays and radioactivity may become harder in some cases, as we discuss in sec.~\ref{sec:backgrounds}.)

We define the overall radiative efficiency $\eta_\text{rad}$ as the ratio of the fluorescence rate $\Gamma_\text{rad}$ to the DM absorption rate $\Gamma_\text{abs}$ from eq.~\ref{eq:Gammaabsrad}
\begin{align}
\Gamma_\text{rad} = \eta_\text{rad} \Gamma_\text{abs}.
\end{align}
In general, $\eta_\text{rad}$ is a complicated function of the DM energy $\omega$, as there are many possible target states that can be excited by dark matter, each having their own radiative properties. Given that $\Gamma_\text{abs}(\omega)$ is highly peaked for $\omega$ near any one out of a set of transition energies $\lbrace \omega_0 \rbrace$, we can take the overall radiative efficiency at $\omega$ to be that of the pair of states $|0\rangle$ and $|1\rangle$ with transition energy $\omega_0$ closest to $\omega$. The radiative efficiency around any such target state $|1\rangle$ can be written as:
\begin{align}
\eta_\text{rad} \simeq \frac{\gamma_0 + \sum_i \gamma_i}{\gamma_0 + \eta_\text{coh} (\bar{r}-1) \gamma_0 + \sum_i \gamma_i + \gamma_\text{quench}} + \eta_\text{coh}, \label{eq:etarad}
\end{align}
where $\gamma_0$ is the radiative width for the process $|1\rangle \to |0\rangle$ in vacuum, and $\gamma_i$ is the radiative width of $|1 \rangle \to |i\rangle$ with $|i\rangle \neq |0 \rangle$ any intermediate state with energy below that of $|1\rangle$. 

The fraction of absorption quanta that are coherently radiated is given by the second term in eq.~\ref{eq:etarad}, and is equal to:
\begin{align}
\eta_\text{coh} \simeq \frac{(\bar{r} - 1) \gamma_0}{(\bar{r} \gamma_0 + \sum_i \gamma_i + 2\gamma_\text{col})} \label{eq:etacoh2}
\end{align}
which is the generalization of eq.~\ref{eq:etacoh1} in the presence of other radiative channels. For large cooperation numbers $\bar{r}-1 \gg 1$ such that other radiative decays can be ignored, the condition for most of the absorbed to be coherently radiated is thus still $(\bar{r}-1)\gamma_0 \gg \gamma_\text{col}$.
Likewise, the fraction of fluorescence radiation to the intermediate state $|i\rangle$ is 
\begin{align}
f_i = \frac{\gamma_i}{\gamma_0 + \eta_\text{coh} (\bar{r}-1) \gamma_0 + \sum_i \gamma_i + \gamma_\text{quench}}.\label{eq:fi}
\end{align}
The sum over all of the intermediate-state branching ratios, $\sum_i f_i$, can become close to unity when the intermediate radiative decays dominate over quenching at low number density, and when the radiative decay rate $\gamma_0$ back to the initial state is small because of selection rules, low $\bar{r}$, and/or combinatoric factors.

\textbf{Frequency coverage---} Any one molecule can be regarded as a multimode resonator capable of absorbing DM particles with energies $\omega$ near any one of the set of transition energies $\lbrace \omega_0 \rbrace$. It is important that most of the ``frequency gaps'' among the adjacent $\omega_0$ can be efficiently covered, either via scanning the $\lbrace \omega_0 \rbrace$ by tuning some external variable like an external electromagnetic field, or by broadening each individual line by increasing the molecular number density.

\begin{figure}\begin{center}
\includegraphics[width=0.35\textwidth]{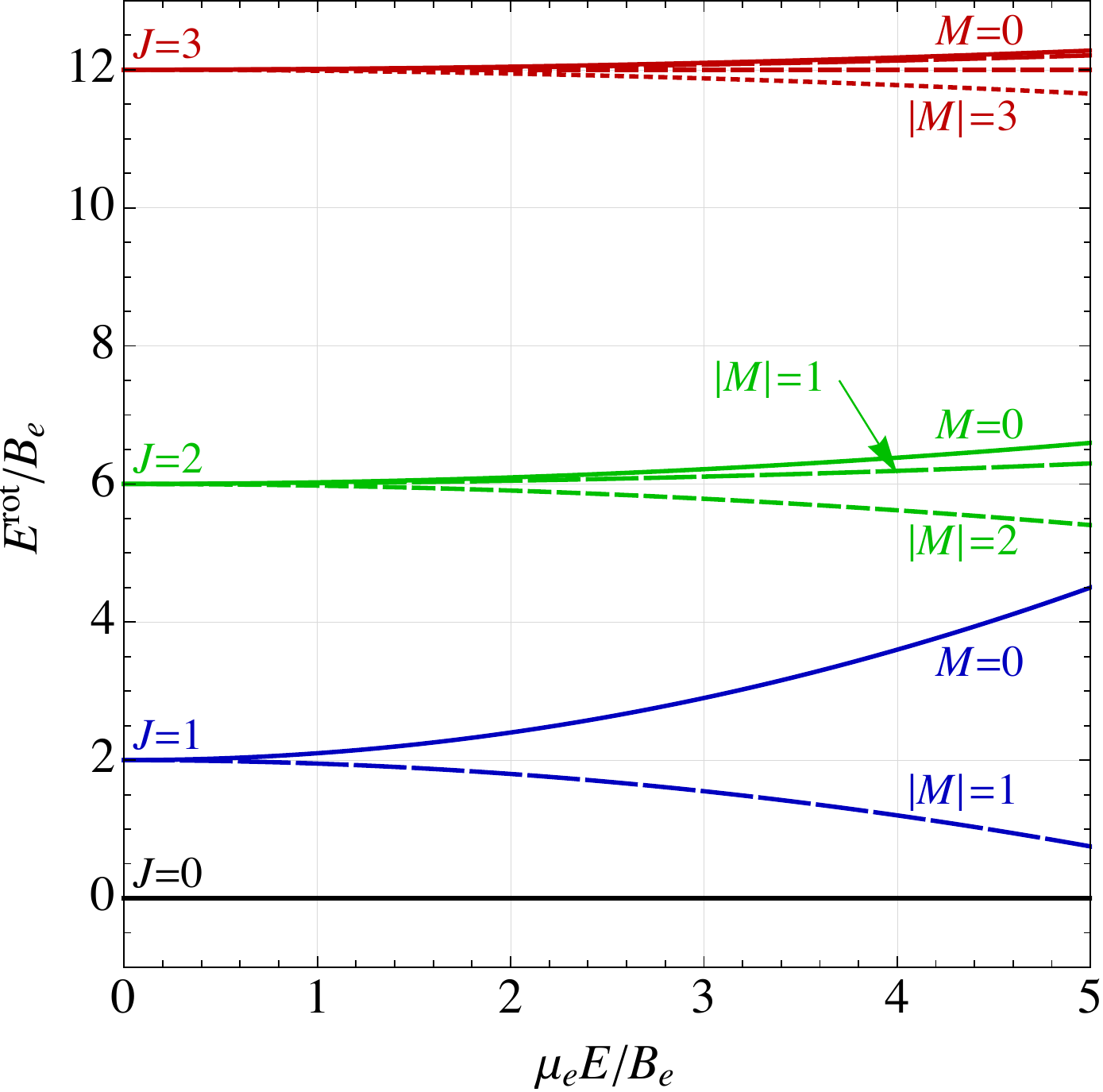}\\
\includegraphics[width=0.35\textwidth]{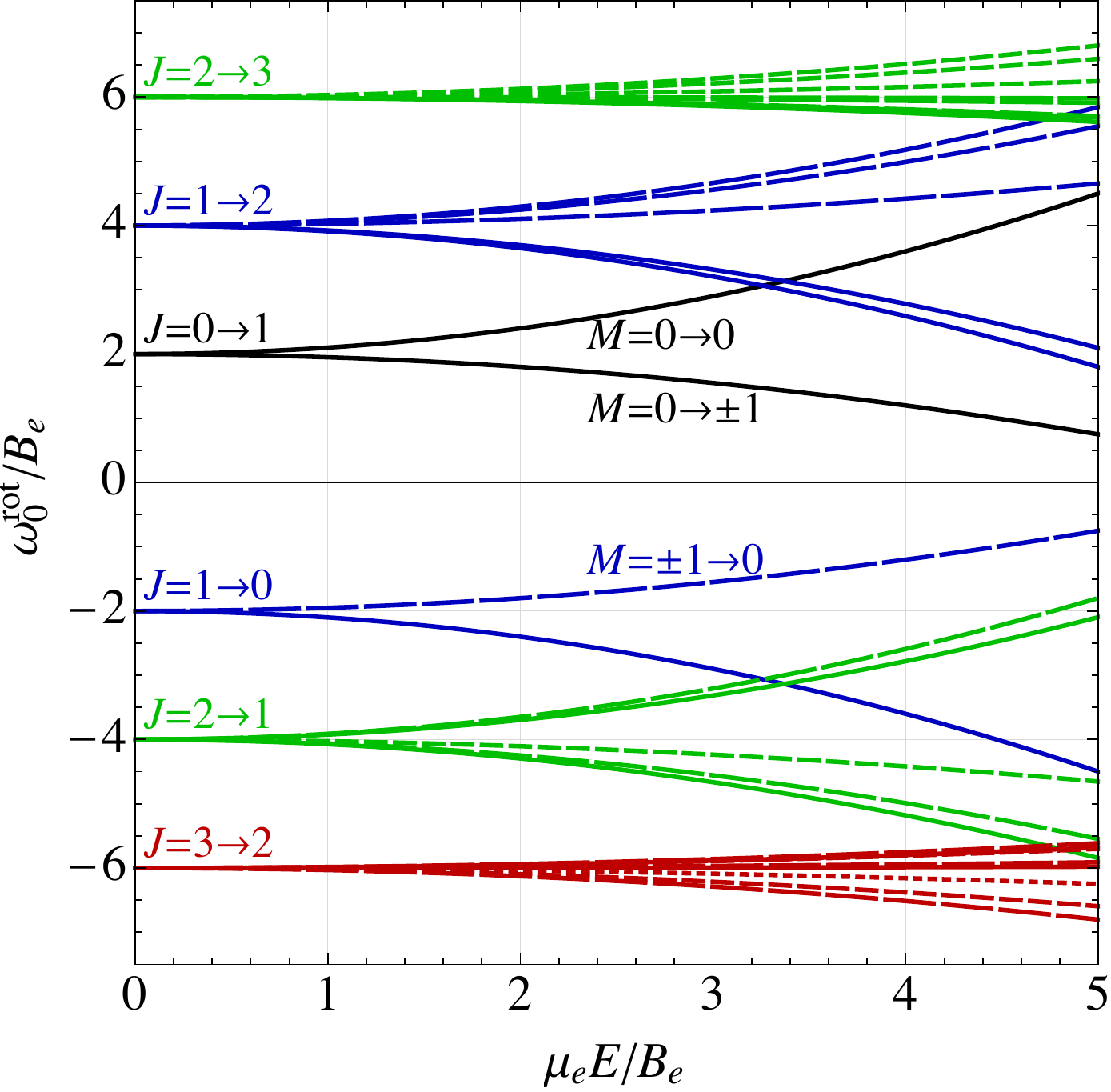}
\end{center}
\caption{Figures of Stark frequency scanning that show how electric fields of order $B_e/\mu_e$ can shift rotational transition energy splittings by an $\mathcal{O}(1)$ amount providing continuous coverage. Top panel: Rotational energy levels $E^\text{rot}$ with quantum numbers $(J,M)$ as a function of electric field $E = |\vect{E}|$ for a diatomic molecule with rotational energy constant $B_e$ and permanent electric dipole moment $\mu_e$. Bottom panel: Rotational energy splittings $\omega^\text{rot}_{0}$ for transitions with selection rules $\Delta J = \pm 1$ and $\Delta M = 0,\pm1$ (which includes E1-allowed transitions), as a function of electric field $E$. Color coding and dashing pattern indicate the initial state, and is consistent with the top panel. Only levels with $J \le 3$ are shown; Stark shifts become less dramatic with increasing $J$. }\label{fig:starktune}
\end{figure}

The splittings among the possible transition energies are typically of order the rotational energy constants of the ground vibrational state. A diatomic molecule has rotational energies $E_\text{rot} = B_e J(J+1),~J=0,1,2,\dots$ in its ground state, where the magnitude of the rotational constant $B_e$ is determined by the inverse moment of inertia, i.e.~the reduced mass $M$ of the diatomic times its mean square separation $\langle R^2 \rangle \simeq R_e^2$:
\begin{align}
B_e =  \frac{1}{2 M \langle R^2 \rangle} \approx 2.1 \times 10^{-4}~\text{eV} \left(\frac{10m_P}{M} \right)\left(\frac{\angstrom}{R_e} \right)^2. \label{eq:Benum}
\end{align}
For example, for dipole vibrational transitions, we expect a rotational fine structure with a splitting of $2B_e$, as derived in eq.~\ref{eq:transPosDip} and depicted in the lower panel of fig.~\ref{fig:starktune}. Most other types of transitions have similar fine structure splittings at the same order or lower. Polyatomic molecules composed out of more than two atoms typically have multiple rotational constants (one around each axis, unless it is a linear molecule) that are smaller, and thus exhibit an even richer fine structure and lower degeneracy. For example, $\text{SF}_6$ has rotational constants of $1.1 \times 10^{-5}~\text{eV}$ around all three axes. 

The splittings in rotational transition energies can be scanned by tuning the magnitude of an external electric field via the second-order Stark effect, if the molecule exhibits a permanent electric dipole moment $\vect{\mu}_e$ (in molecule-fixed coordinates, not laboratory-fixed coordinates of course). In an electric field $\vect{E}$, the energy eigenvalues $E^\text{rot}(J,M)$ of the combined rotational and Stark Hamiltonian are
\begin{align}
\frac{E^\text{rot}}{B_e} \simeq  J (J+1) + \frac{1}{2}\lambda^2 \frac{J(J+1) - 3 M^2}{J(J+1)(2J-1)(2J+3)},
\end{align}
valid up to $\mathcal{O}(\lambda/2)^4$ for $\lambda \equiv \mu_e |\vect{E}| / B_e$, a dependence plotted in the upper panel of fig.~\ref{fig:starktune}. Note that the $2J + 1$-fold degeneracy among levels with the same $J$ but different $M$ is now lifted into a two-fold degeneracy among levels with the same $(J,|M|)$. In the lower panel, we plot how the rotational transition energy splittings $\omega^\text{rot}_{0} = E^\text{rot}(J_f,M_f) - E^\text{rot}(J_i, M_i)$ depend on the ratio $\mu_e E / B_e$ for dipole vibrational transitions. An $\mathcal{O}(1)$ fraction of the rotational splittings can be ``scanned'' if the expansion coefficient 
\begin{align}
\frac{\mu_e E}{B_e} \approx 0.63 \left(\frac{\mu_e}{\text{D}} \right)\left(\frac{E}{3 \times 10^6~\text{V/m} }\right)\left(\frac{10^{-4}~\text{eV}}{B_e} \right)
\end{align}
becomes $\mathcal{O}(1)$. In the above numerical estimate, we took the electric field to be the breakdown electric field value (for large separations) of air at standard atmospheric conditions, and a dipole moment of $1~\text{D} \approx 0.39 e a_0$. 
Transitions involving an electronic spin flip can furthermore receive linear Zeeman corrections to their transition energy in an external magnetic field $B$, by an amount 
\begin{align}
\Delta \omega^\text{Zeeman}_{0} = g_e \mu_B B \approx 5.8 \times 10^{-4}~\text{eV} \left(\frac{g_e}{2} \right)\left(\frac{B}{5~\text{T}} \right) \label{eq:Zeeman}
\end{align}
with $g_{e}$ the effective g-factor and $\mu_{B}$ the Bohr magneton. For nuclear spin flips, the Zeeman shift is of less relevance due to the smallness of the nuclear gyromagnetic ratio. We thus conclude that it is possible to scan the gaps in the rotational fine structure of the absorption spectrum for molecules with small rotational splittings, which include heavy (large $M$) and weakly bound (large $\langle R^2 \rangle$) diatomics, or moderately large polyatomics (large moments of inertia, usually around several axes).

Another important way to achieve contiguous frequency coverage is to broaden all of the transition lines $\lbrace \omega_0 \rbrace$ at high number densities $n$. The full width in $\omega$ at half-maximum absorption rate in eq.~\ref{eq:Gammaabsrad} is
\begin{align}
\gamma = \gamma_\text{rad} + 2 \gamma_\text{col} \approx 1.1 \times 10^{-5}~\text{eV} \left( 1 + \frac{\gamma_\text{rad}}{2\gamma_\text{col}} \right)\left(\frac{n}{n_0}\right), \label{eq:gamma2}
\end{align}
where we took $T = 273~\text{K}$ and a molecular mass of $M_\text{mol} = 40m_P$ in the collisional width $2\gamma_\text{col}$ defined in eq.~\ref{eq:gammacol}. Collisional broadening alone can already bridge the gap between typical rotational splittings, cfr.~eq.~\ref{eq:Benum}, at moderate pressures of $10~\text{bar}$. The radiative width $\gamma_\text{rad} = \gamma_0 + \eta_\text{coh} (\bar{r}-1) \gamma_0 + \sum_i \gamma_i$ can increase the FWHM even more, especially if $\bar{r} \gtrsim n/m^3$ and $\gamma_0$ is an E1-allowed decay rate, cfr.~eq.~\ref{eq:stackineq}.

\textbf{Bulk configuration---} The Bulk detector configuration comprises of a large convex volume, $V= (0.3~\text{m})^3$ in Phase~I, $V= (2~\text{m})^3$ in Phase~II, filled with a single molecular species at any time.  A large-area photodetector array (PMT in Phase I, MKID array in Phase II) detects the signal photons. The rest of the container area is coated with a highly reflective layer, necessary to retain the fluorescence photons since they are emitted isotropically. Cooperative effects may be ignored for a bulk container, i.e.~$\eta_\text{coh}(\bar{r} -1) \ll 1$, as we showed in sec.~\ref{sec:cooperation}. Photons from the $|1\rangle \to |0\rangle$ fluorescence channel are at a danger of being reabsorbed elsewhere in the detector volume, and are thus more easily quenched after multiple re-absorptions and re-emissions (we discuss optical thickness issues in more detail in sec.~\ref{sec:photodetection}). Optical thickness is a problem whenever the matrix element of the $|0\rangle \leftrightarrow |1\rangle$ is E1-allowed and $\gamma_0$ is thus relatively large; on the other hand, if $|0\rangle \leftrightarrow |1\rangle$ is E1-forbidden, then $\gamma_0$ can usually only dominate over the quenching rate from eq.~\ref{eq:gammaquench} at very low number densities, potentially suppressing the absorption rate.

A Bulk detector configuration will therefore largely rely on radiative decays $|1\rangle \to |i\rangle$ where $|i\rangle$ is any intermediate state with negligible occupation probability in thermal equilibrium. These decays will produce fluorescence photons for which the medium is optically transparent. Hence, the detectable photon fluorescence rate $\Gamma_\text{rad}^\text{B}$ for a Bulk configuration is
\begin{align}
\Gamma_\text{rad}^\text{B}\left( \omega, \lbrace \omega_0 \rbrace \right) = \sum_i^\text{thin}f_i \Gamma_\text{abs} \left( \omega, \lbrace \omega_0 \rbrace \right),
\end{align}
where it is understood that $f_i$ from eq.~\ref{eq:fi} is to be evaluated for the target level $|1\rangle$ with $\omega_0$ closest to $\omega$, and that the sum is to be performed only for intermediate levels $|i\rangle$ that produce fluorescence photons for which the molecular medium is optically thin.

Frequency coverage for a DM search over a broad range of masses in a bulk detector can be achieved by performing several narrow searches, each with a different molecule. Covering gaps in the rotational fine structure of the absorption spectrum can be done in the two ways mentioned above: scanning at low density or broadening the lines at high number density. Within the scanning strategy, the optimal regime is to work at a low number density where $\gamma_\text{quench} \lesssim \sum^\text{thin}_i\gamma_i$, such that the fluorescence fraction is large, $\sum_i^\text{thin} f_i \approx 1$. The resulting narrow linewidth $\gamma$ means that many (on the order of $\gamma/B_e$) shots, each at a different set of $\lbrace \omega_0 \rbrace$, are needed for contiguous frequency coverage. Each shot can thus last a small fraction of the total integration time $t_\text{int}$ for the molecule: $t_\text{shot} \sim (\gamma/B_e) t_\text{int}$. At high number densities where $\gamma = 2\gamma_\text{col} \simeq B_e$, only one shot is needed, but at the cost of a lower fluorescence ratio $\sum_i^\text{thin} f_i \simeq \sum_i^\text{thin} \gamma_i / \gamma_\text{quench} \propto 1/n$. At finite background rate levels, the scanning method typically provides better optimum sensitivity. In the limit of zero background, they yield roughly equivalent sensitivity, though the broadening method has the advantage that it can be used for all transitions in all molecules (including those without permanent electromagnetic moments), and is experimentally more straightforward.

\textbf{Stack configuration---} To take full advantage of the cooperative radiation effects analyzed in sec.~\ref{sec:cooperation}, the detector volume should be composed out of a planar stack of slab-like containers made out of highly transparent material such as glass or silicon, as shown in the bottom diagram of fig.~\ref{fig:setup}. The signal photons are emitted nearly perpendicularly to the planes, and could thus be focused onto a much smaller photosensitive area by a lens or reflecting mirror (or an array thereof). Any deviation of the emission direction from the normal vector is of order the DM velocity divided by the speed of light, so a high-resolution photodetector such as an MKID could ``image'' the DM velocity distribution as in fig.~\ref{fig:intensity-cube-dir} as a function of time. 

Alternatively, an ``artificial stack'', without physical barriers between the slabs, may be effectively created by a standing electromagnetic wave pattern in a bulk container. At the antinodes of the standing waves, the ground state population could be depleted via resonant pumping to an intermediate level, while the molecules would be much less affected near the nodes of the standing waves. Even if no such suitable intermediate level is available, the standing wave pattern could create a spatially-dependent quadratic Stark shift, moving the molecules off-resonance at the antinodes and on-resonance at the nodes (or vice versa) at sufficiently small line widths. This artificial stack approach likely introduces additional complications; an evaluation of its feasibility is beyond the scope of this work.

The aim of the Stack design is to operate in a regime where the equality
\begin{align}
(\bar{r} - 1) \gamma_0 \sim \gamma_\text{col}, \label{eq:stackineq}
\end{align}
is roughly satisfied.
This accomplishes three goals simultaneously: radiation focusing, high radiative efficiency ($\eta_\text{coh} \approx 1$), and potential contiguous frequency coverage. The coherent, focused fluorescence rate is:
\begin{align}
\Gamma^\text{S}_\text{rad}(\omega,\lbrace \omega_0 \rbrace) = \eta_\text{coh} \Gamma_\text{abs}(\omega,\lbrace \omega_0 \rbrace). \label{eq:GammaSrad}
\end{align}
 The FWHM of the DM absorption lines will be as in eq.~\ref{eq:gamma2} with the radiative width dominating if eq.~\ref{eq:stackineq} is satisfied. Molecules with an E1-allowed transition with transition dipole moment $\vect{\mu}_{1,0} = \langle 1 | \vect{\mu}_e | 0 \rangle$ in a stacked-slab configuration have a cooperative radiative width as in eq.~\ref{eq:gammarad}, so we have that the ratio controlling the validity of eq.~\ref{eq:stackineq} is
\begin{align}
\frac{(\bar{r} - 1) \gamma_0}{\gamma_\text{col}} = \frac{8 |\vect{\mu}_{1,0}|^2 p_0 \bar{S}}{3 m R_z \sigma_\text{col} v_\text{mol}} \approx 6.3 \frac{p_0 \bar{S}}{mR_z},
\end{align}
for $\sigma_\text{col} = 10^2 ~\angstrom^2$, $T = 273~\text{K}$, and $M_\text{mol} = 40m_p$. 

The radiation focusing effect analyzed in sec.~\ref{sec:cooperation} has several important advantages. It allows for a smaller photosensitive area, permitting the use of highly sensitive, cryogenic photodetecors such as a TES or MKID without a prohibitively high cost. In addition, it isolates the signal from the mostly isotropic environmental backgrounds (see sec.~\ref{sec:backgrounds}). The direction of the coherent photons depends on the DM velocity and dispersion as shown in sec.~\ref{sec:cooperation}, yielding a spectacular intrinsic sensitivity to the direction of dark matter. Low-dark-count, 10-kilopixel MKIDs as large as $(10~\text{mm})^2$ have already been built, and would have an intrinsic DM velocity resolution of $\lesssim 10^{-5}$ (roughly given by the pixel size divided by the transverse size of slab) in the two dimensions parallel to the stack, i.e.~at the sub-percent level fractionally given the expected DM velocity at $10^{-3}$ the speed of light. Full 3D velocity information could be gleaned over long integration times because of Earth's rotation and orbit, and/or with two experimental setups.

\textbf{Sensitivity estimates---} We quantify the sensitivity of each detector version in terms of the smallest Rabi frequency $\delta \Omega$ it can detect at unity signal-to-noise ratio (SNR) after a shot time $t_\text{shot}$. A photodetector with an intrinsic dark count rate DCR within its bandwidth $\Delta \omega$ around an energy $\omega$ has a $\text{SNR} = 1$ sensitivity to detected photon counts at a rate of
\begin{align}
\delta \Gamma_\text{det}(\omega,\Delta \omega) \simeq \sqrt{\frac{\eta_\text{det} \Gamma_\text{bckg} + \text{DCR} + t_\text{shot}^{-1}}{t_\text{shot}}}, \label{eq:deltaGammadet}
\end{align}
where $\Gamma_\text{bckg}$ is the true photon background rate impinging on the photodetector with detection efficiency $\eta_\text{det}$, and in the same bandwidth $\Delta \omega$. If we define $\eta_\gamma$ as the probability that a radiated photon is detected by the photodetector, then the signal detection rate is
\begin{align}
\Gamma_\text{det} = \eta_\gamma \Gamma_\text{rad} = \eta_\gamma \eta_\text{rad} \Gamma_\text{abs}. \label{eq:Gammadet}
\end{align}
The minimum detectable Rabi frequency $\delta \Omega$ is the $\Omega$ for which $\Gamma_\text{det} = \delta \Gamma_\text{det}$. When the absorption rate at a given energy $\omega$ is dominated by a single resonance with transition energy $\omega_0$ closest to $\omega$, we can write:
\begin{align}
\delta \Omega \simeq \sqrt{\frac{\delta \Gamma_\text{det} \gamma [1 + 4(\omega-\omega_0)^2/\gamma^2]}{\eta_\gamma \eta_\text{rad} n V p_0}}. \label{eq:deltaRabi}
\end{align}
In table~\ref{tab:configs}, we have compiled the typical on-resonance sensitivity in terms of $\delta \Omega$ for $p_0 = 0.1$, $\gamma = 2 \gamma_\text{col}$, $\sigma_\text{col} = 10^2~\angstrom^2$, and $M_\text{mol}= 40m_P$ for each configuration and phase, assuming $\Gamma_\text{bckg} \ll \text{DCR} + t_\text{shot}^{-1} $. Comparison with the signal Rabi frequencies $\Omega$ in table~\ref{tab:transitions} allows one to estimate the signal-to-noise ratio for many DM candidates at a few benchmark frequencies. We shall see in sec.~\ref{sec:sensitivity} that the Phase~I prototypes already explore new parameter space of some DM candidates, and that Phase~II experiments will be capable of probing previously unexplored parameter space in \emph{all} of the DM models and couplings listed in table~\ref{tab:transitions}.

\begin{figure}
\includegraphics[width=0.48\textwidth]{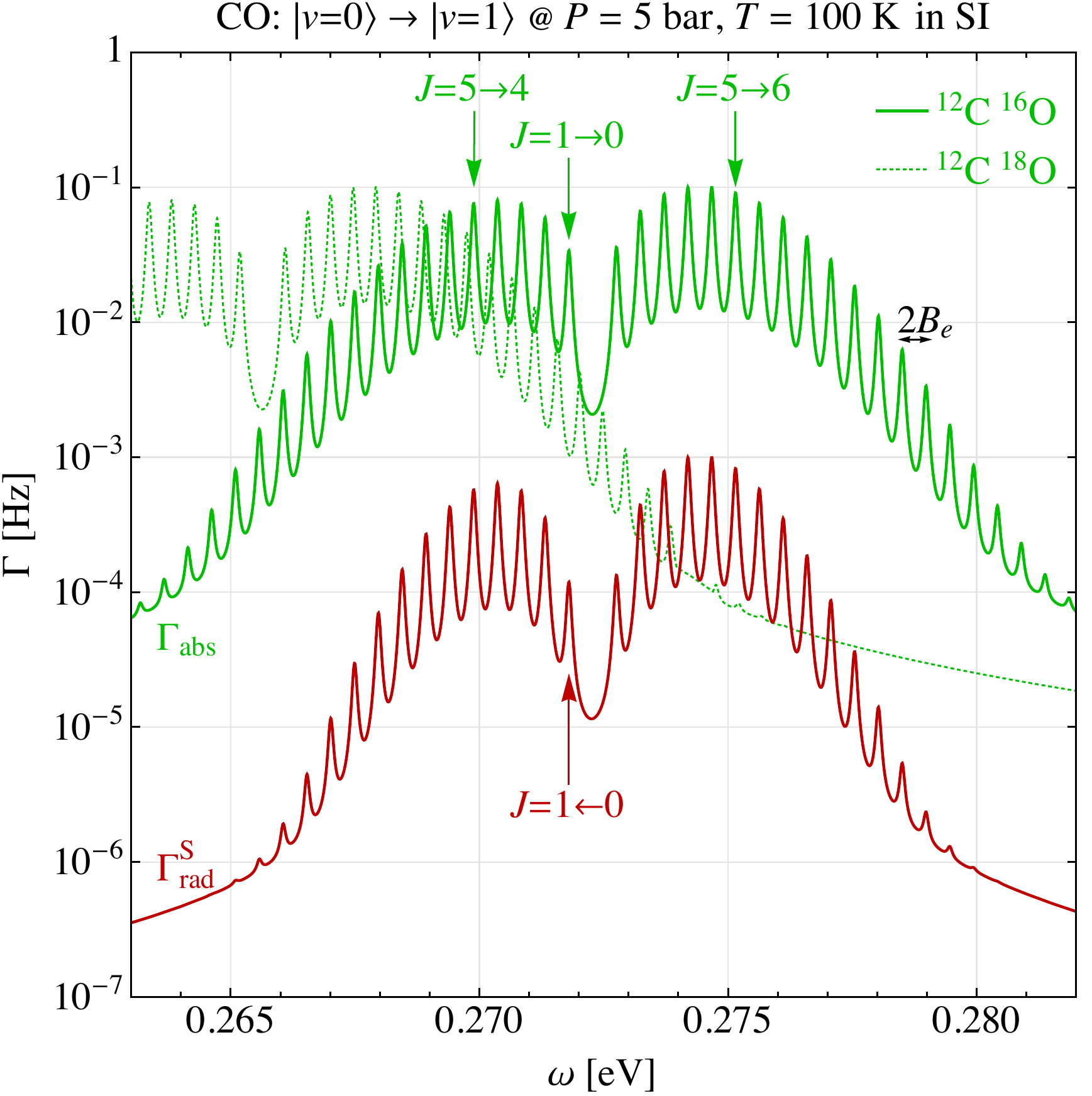}
\caption{
Rate of absorption $\Gamma_\text{abs}$ (solid green) and rate of focused, coherent fluorescence $\Gamma_\text{rad}^\text{S}$ (solid red) as a function of DM energy $\omega \simeq m$, in the Phase~I version of the Stack configuration, filled with carbon monoxide (the ${}^{12}\text{C}{}^{16}\text{O}$ isotopologue) at a pressure of 5~bar, temperature of 100~K, and ``stack enhancement parameter'' $\bar{S}/m R_z \sim 10$, cfr.~eq.~\ref{eq:gammarad}. The case study shown assumes a kinetically-mixed photon with $\epsilon=10^{-12}$, zoomed in on the energy range around CO's first vibrational absorption line $|v=0\rangle \to |v=1\rangle$ at $\omega_e = 0.27~\eV$, with rotational fine structure of splitting $2B_e$ clearly visible. This plot shows that the thermally occupied rotational levels of the ground state allows us to expand the frequency coverage of vibrational and electronic transitions.  In dashed green, we also show the absorption rate for ${}^{12}\text{C}{}^{18}\text{O}$, displaying the isotope shift on the rovibrational structure.
}\label{fig:rate2}
\end{figure}

In figures~\ref{fig:rate1}~and~\ref{fig:rate2}, we illustrate many of the considerations in this section so far with a simple case study, namely the first-excited vibrational level in carbon monoxide. Figure~\ref{fig:rate1} shows the inverse time scales of radiative and collisional dynamics as a function of pressure and temperature. We also show on the same axis the rotational energy constant $B_e$, and the rate $\Gamma_\text{RD}$ to which radioactive decays can be held in the large BII setup with low-contaminant materials (before any active veto). Figure~\ref{fig:rate2} shows in green the absorption rate $\Gamma_\text{abs}$ of a kinetically-mixed hidden photon DM particle with $\epsilon = 10^{-12}$ (see sec.~\ref{sec:vectors} for more details) for the ${}^{12}\text{C}{}^{16}\text{O}$ isotope in the SI setup. The resulting coherent radiation rate of eq.~\ref{eq:GammaSrad} is plotted in red, about 2--3 orders of magnitude below. This is because CO has a rather small transition electric dipole moment, and thus $\eta_\text{coh} \ll 1$ even with the assumed stack parameters of $\bar{S}/mR_z$. Molecules with stronger absorption strengths---such as e.g.~$\text{HCl}$, $\text{CO}_2$, and $\text{CH}_4$---can achieve $\eta_\text{coh}$ much closer to 1.

\subsection{Photodetection} \label{sec:photodetection}

In sec.~\ref{sec:resonance}, we calculated the absorption rate of the DM particles of energy $\omega$ given a Rabi frequency $\Omega$, or equivalently the rate at which molecular states with transition energy $\omega_0$ close to $\omega$ are excited. The signal detection consists of reading out the fluorescence photons spontaneously or coherently emitted from these excited levels, at a rate $\Gamma_\text{rad} = \eta_\text{rad}\Gamma_\text{abs}$. However, the photon detection rate is typically lower than this radiation rate $\Gamma_\text{rad}$, cfr.~\ref{eq:Gammadet}, by the overall detection efficiency factor
\begin{align}
\eta_\gamma = \eta_\text{refl}\eta_\text{trans}\eta_\text{thick}\eta_\text{det}. \label{eq:etas}
\end{align}
We describe our estimates for $\eta_\gamma$ below, and show that it can be $\mathcal{O}(1)$ in our setup.  We will finish this subsection with a brief summary of potential photodetectors and their specifications, including dark count rates.

Four main loss mechanisms are responsible for $\eta_\gamma \le 1$, each with their own efficiency factor $\eta_i$ as schematically indicated in eq.~\ref{eq:etas}.
They are due to, respectively, absorption onto the reflective walls of the container with probability $1-\eta_\text{refl}$, finite transmission probability $\eta_\text{trans}$ of glass elements in the optical path, re-absorption and subsequent fluorescence quenching (if the gas is optically thick) with probability $1-\eta_\text{thick}$, and the intrinsic quantum detection efficiency $\eta_\text{det}$ of the single-photon counter. In what follows, we discuss how each of the $\eta_i$ can be of order unity.

\textbf{Reflection efficiency---} Fluorescence photons may be absorbed by the (not perfectly) reflective walls of the gas container in the Bulk configuration. We envision that the container walls are coated with a material with high reflectance $\mathcal{R}$, which depends on the wavelength, polarization, and incidence angle of the incoming photon. Appropriately averaging over polarizations and incident angles for an effective wavelength-dependent reflectance $\bar{\mathcal{R}}(\omega)$, a photon can be reflected an expected $1/[1-\bar{\mathcal{R}}(\omega)]$ times before being absorbed onto the coating. The loss fraction of signal photons will be small $\eta_\text{refl} \approx 1$, as long as $A_\text{det}/{[V^{2/3} [1-\bar{\mathcal{R}}(\omega)]}\gtrsim 1$ for a detector volume $V$ with an aspect ratio near unity, and an area $A_\text{det}$ instrumented with photodetectors. For photon energies $\omega \lesssim 1.9~\eV$ or  wavelengths $\lambda_\gamma \gtrsim 650~\text{nm}$, silver has $1-\bar{\mathcal{R}}(\omega) \lesssim 10^{-2}$. For higher-energy light, aluminum is better with $1-\bar{\mathcal{R}}(\omega) \lesssim 10^{-1}$ for all $\lambda > 100~\text{nm}$. Dielectric coatings can achieve even much lower absorbance values $1-\bar{\mathcal{R}} \ll 10^{-2}$ in narrow energy ranges over all wavelengths of interest ($100~\text{nm}$--$10~\mu\text{m}$). High-reflectance coatings can thus allow for a small photodetector-instrumented area $A_\text{det}$, of linear size $1/\sqrt{1-\bar{\mathcal{R}}(\omega)}$ smaller than the bulk detector size, while keeping $\eta_\text{refl} \sim 1$. 

\textbf{Transmission efficiency---} A Stack detector configuration does not require reflective coatings, since the radiation is focused; in fact, it will typically need \emph{anti}-reflective (AR) coatings on the interfaces between the gas (with index of refraction close to 1) and the material separating the different slabs, such as glass or silicon, that will have a substantially higher index of refraction. Graded-index coatings can achieve $0.1\%$ reflection over a broad range of wavelengths, while even better performance may be expected over narrow wavelength ranges with thin-film interference coatings. Bulk absorption by the slab container material must also be considered. Synthetic quartz glasses can easily achieve absorption depths in excess of 1~\text{cm} in the wavelength range 180~nm--2.5~$\mu$m, while silicon exhibits this property for wavelengths above 1.1~$\mu$m, covering our energy range of interest.

\textbf{Optical thickness---} Signal photons can get re-absorbed and subsequently quenched if the gas sample is optically thick. For a low-density gas of two-level molecules with an electric-dipole transition moment $\mu_{1,0} = |\langle 1 |\vect{\mu}_e|0\rangle|$, the light intensity falls off exponentially as $\propto e^{-\alpha l}$, with $l$ the optical path length and $\alpha(\omega)$ the attenuation coefficient for light with an angular frequency $\omega$ near the molecular transition energy $\omega_0$:
\begin{align}
\alpha(\omega)  =~&\frac{2  \omega_0 n p_0 \mu_{1,0}^2}{\gamma_\text{col}} \frac{1}{1+\frac{(\omega-\omega_0)^2}{\gamma_\text{col}^2}},\label{eq:optdepth}\\
\underset{\omega = \omega_0}{ \approx}~& \frac{1}{0.09~\text{mm}}\left( \frac{\omega_0}{\eV}\right)  \left(\frac{\mu_{1,0}}{10^{-2}ea_0}\right)^2  \left( \frac{p_0}{1/10} \right).\nonumber
\end{align}
Here, we employ the collisionally-broadened lineshape, though Doppler broadening can also be important at sufficiently low densities. The inverse $\alpha(\omega)^{-1}$ is the mean absorption depth, which can be quite short for $|\omega - \omega_0| < \gamma_\text{col}$, as the second line of eq.~\ref{eq:optdepth} shows.
For this reason, a Bulk detector will rely on fluorescent decays to intermediate levels which are not thermally occupied. It follows from eq.~\ref{eq:optdepth} that the associated photons from those decays will travel macroscopic distances, because levels that could resonantly absorb them have exponentially small occupation probabilities $p_i \propto e^{-E_i/T}$, while off-resonant absorption is highly suppressed by the Lorentzian line-shape factor, so $\eta_\text{thick} \approx 1$. 

Even for thin Stack detectors, the on-resonance absorption depth $\alpha(\omega_0)^{-1}$ can be smaller than the proposed integrated thickness, $D = 1~\text{mm}$ for Phase I and $D = 100~\text{mm}$ for Phase II. However, in the regime of eq.~\ref{eq:stackineq} where cooperative effects for DM absorption are in action, the radiated photons themselves will also interact cooperatively with the molecular medium. Dicke's seminal work on superradiance showed that due to a phase-matching effect, a photon plane wave can experience coherent scattering in the direction of the incoming wave~\cite{d1954}. This coherent forward scattering in extended volumes was further developed in refs.~\cite{arecchi1970cooperative,rehler1971superradiance}, showing that for large-Fresnel-number samples (such as a slab), photons are re-emitted in a narrow ``lobes'' primarily in the forward direction. Recent theoretical~\cite{scully2006directed} and experimental~\cite{bromley2016collective} work shows that this effect is also obeyed for a \emph{single} photon absorbed by a sufficiently strong dipole transition. Single-photon superradiance in extended volumes is a complicated subject of intense recent study~\cite{araujo2016superradiance,roof2016observation}, and beyond the scope of this work. Further work is needed to show how far a photon will travel in this regime, and to what extent the directional information is preserved after multiple coherent scattering events. This will be crucial for a Phase~II version of the Stack configuration. 

\textbf{Photodetector performance---} The sensitivity of our proposed setups is ultimately limited by photodetector parameters. The key specifications not only include the quantum detection efficiency $\eta_\text{det}$, but also the dark count rate (DCR), photosensitive area ($A_\text{det}$), energy range ($\omega_\text{min}$---$\omega_\text{max}$), energy resolution ($\Delta \omega$), timing jitter ($\Delta t$), and operating temperature. Reviews on these aspects of single-photon counting detectors can be found in refs.~\cite{hadfield2009single,eisaman2011invited}.

The Bulk configuration requires a large photosensitive area to keep $\eta_\text{refl} \approx 1$. In a prototype Phase~I, we propose utilizing off-the-shelf photomultiplier tubes (PMT) with photosensitive area $A_\text{det} = (30~\text{cm})^2$. These allow for (near-)room-temperature operation at visible and near-infrared wavelengths with detection efficiency $\eta_\text{det} \approx 40~\%$, and $\text{DCR} \approx 1~\text{Hz}$ (when they are cooled a few degrees below freezing)~\cite{hamamatsu1,hamamatsu2,hamamatsu3,hamamatsu4}. Timing jitter $\Delta t$ is sub-nanosecond, aiding in timing-based rejection of environmental backgrounds, while the intrinsic energy resolution is quite poor.

For the Phase~II Bulk setup, we assume state-of-the-art photodetector arrays based on microwave kinetic inductance detectors (MKID)~\cite{day2003broadband}. These cryogenic photodetectors can be operated as single-photon counters between ultraviolet wavelengths of 100~nm, all the way to mid-infrared wavelengths greater than 5~$\mu$m, while retaining energy resolution of order $\Delta \omega \sim 0.1~\text{eV}$ as well as good timing resolution $\Delta t \sim 10^{-6}~\text{s}$, with further improvements on the horizon~\cite{mazin2012superconducting}. These devices have essentially no intrinsic dark counts at energies $\omega$ a few times above their energy resolution; any non-signal counts must be due to true environmental photons. Their quantum efficiency is already good even for mid-infrared photons ($\eta_\text{det} > 0.2$), and excellent for wavelengths above 500~nm ($\eta_\text{det} > 0.5$).

Crucially for our purposes, it is the first cryogenic photodetector technology to have been multiplexed into arrays of $2\times 10^4$ ``pixels''~\cite{mazin2013arcons,meeker2015design}, already yielding photosensitive areas as large as $A_\text{det} \sim (1~\text{cm})^2$; larger arrays are already under development~\cite{mazin2015science}. (Most cryogenic single-photon counters rely on the increase in temperature in a superconducting volume from the impinging photon's energy, so any one such volume must necessarily be microscopic.) Anticipating that MKID technology matures even further, we assume a photosensitive area of $A_\text{det}\sim (10~\text{cm})^2$ for the future BII setup volume of $V = (2~\text{m})^3$, where it would occupy a $10^{-3}$ fraction of the container area and thus require tuned dielectric coatings with $1-\bar{\mathcal{R}} \sim 10^{-3}$. 

In the Stack configuration, the photons can be focused onto a small area of order $10^{-6}$ the transverse area of the molecular container. As such, there is no need for particularly large arrays even in the larger Phase~II version, which would require a minimal photosensitive area $A_\text{det} \sim \mathcal{O}(\text{mm}^2)$. MKIDs would perform even better as the smaller phase space of the signal photons allows them to make use of microlenses to steer the photons on their inductive elements~\cite{mazin2012superconducting}. Over such relatively small areas, transition edge sensors (TES)~\cite{irwin1995application} and other cryogenic detectors~\cite{hadfield2009single,eisaman2011invited,zmuidzinas2012superconducting} may also be employed, with similar and potentially better specifications.


\subsection{Environmental backgrounds}\label{sec:backgrounds}
Assuming that DCR can be controlled down to the desired levels specified above, we expect three primary sources of background in our energy range: black body radiation (BBR), natural and cosmogenic radioactivity, and cosmic rays.


\textbf{Blackbody radiation---}
The molecules in the detector volume have to be kept in the gaseous phase in order to retain their resonant absorption characteristics. For all but a few diatomic molecules suitable for our setup, this corresponds to temperatures larger than $100~\text{K}$. The accompanying blackbody radiation (BBR) at a temperature $T$ in thermal equilibrium results in an irreducible background photon rate onto the photosensitive area $A_\text{det}$ of:
\begin{eqnarray}
\Gamma_\text{BBR} \simeq \Delta \omega\frac{\omega^2}{\pi^2} e^{-\omega /T}  A_\text{det}, \label{eq:GammaBBR}
\end{eqnarray} 
in a band $\Delta \omega$ around any frequency $\omega$. Fake counts from BBR will dominate over dark counts in the photodetector whenever $\eta_\text{det} \Gamma_\text{BBR} \gtrsim \text{DCR}$, cfr.~eq.~\ref{eq:deltaGammadet}. Given the exponential tail in eq.~\ref{eq:GammaBBR}, this turnover point occurs at an energy $\omega$ that is not very sensitive to the other experimental specifications. We foresee blackbody radiation to be the dominant background for $\omega < 1.5~\text{eV}$ in any setup at 300~K (BI), and for $\omega < 0.5~\text{eV}$ in a 100~K detector (BII, SI, SII). These turnover points are clearly visible as the ``kinks'' in our sensitivity estimates in e.g.~fig.~\ref{fig:reachhf}.


\textbf{Natural and cosmogenic radioactivity---}
Background photons in the energy range of interest can also originate from radioactive decays in the surrounding material or even the molecular gas itself. Care must be taken to passively mitigate and, in the Phase~II versions of our setups, to actively veto this background. Gamma rays will be the main culprit due to their large penetration length, which is about one and three orders of magnitude larger than that of beta and alpha particles, respectively, at $1~\text{MeV}$. Gamma rays in the surrounding material can dislodge surface electrons via the photoelectric effect or directly ionize the molecules in the bulk (among other channels), which can cause molecular excitations in the frequency band of interest, or even directly trigger the photodetector. 

A later-stage detector proposal should include a detailed detector simulation of these effects; here we provide parametric estimates showing the feasibility of passive and active mechanisms to keep the rate of radioactive decays $\Gamma_\text{RD}$ below the inverse shot time or the DCR, whichever is larger. When $\Gamma_\text{RD} \lesssim \text{DCR} + t_\text{shot}^{-1}$, radioactive backgrounds are subdominant. The attenuation coefficient for gammas at 1~MeV is $\alpha_\gamma \approx 6\times 10^{-2}~\text{cm}^2 \text{g}^{-1}$, giving an attenuation length of about $6~\text{cm}$ in quartz and $1.5~\text{cm}$ in lead for example. The main sources of naturally-occurring radioactivity are ${}^{238}$U, ${}^{232}$Th, and ${}^{40}$K, each giving roughly similar contributions, so we will only estimate contributions from the first. The natural mass-fraction abundance of ${}^{238}$U is $f_m \approx 10^{-6}\text{g}/\text{g}$, while its half-life is $t_{1/2} \approx 4.5~\times 10^9~\text{y}$. Supposing that the container is a cubic volume with boundary area $A_{\delta V} = 6 (2~\text{m})^3$ surrounded by a material of quartz's density, we find a typical radioactive decay rate of $\mathcal{O}(10^4~\text{Hz})$ in a boundary layer with thickness of order the penetration length. The same estimate for a high-purity lead shield with a specific density $\rho_\text{Pb} \approx 11 \text{g/cm}^3$ and a $f_m \sim 10^{-12}$ mass fraction of ${}^{238}$U, similar to those used by other experiments~\cite{leonard2008systematic}, gives a rate of 
\begin{align}
\Gamma_\text{RD}^{{}^{238}\text{U}} \sim \frac{\rho_\text{Pb} f_m A_{\delta V}}{\alpha_\gamma m({}^{238}\text{U})} \frac{\ln 2}{t_{1/2}} \sim \mathcal{O}(10^{-2}~\text{Hz}).
\end{align}

Besides naturally-occurring radioactive isotopes with lifetimes on the order of billions of years, there will generally also be trace concentrations of cosmogenically-activated radiaoactive isotopes in the molecular medium or the container materials. Even though these contaminants occur in much smaller mass fractions, their decay rates are larger, so that they may compete with radioactivity from ${}^{238}$U, ${}^{232}$Th, and ${}^{40}$K. A particularly dangerous isotope is ${}^{14}$C, produced by cosmic-ray collisions high up in the atmosphere, and consequently present at a fractional number density of $10^{-12}$ relative to that of ${}^{12}$C in all organic material not buried for longer than its half-life of $t_{1/2} \approx 5730~\text{y}$. For example, if the molecular medium contains carbon atoms (e.g.~CO), then one can expect a cosmogenic radioactivity contribution of:
\begin{align}
\Gamma_\text{RD}^{{}^{14}\text{C}} = n({}^{14}\text{C}) V \frac{\ln{2}}{ t_{1/2}} \approx 10^{2}~\text{Hz} \left(\frac{n({}^{14}\text{C})}{10^{-12}n_\text{st}}\right).
\end{align} 
Carbon-containing molecules derived from fossil fuels can have greatly depleted ${}^{14}$C content. Certain petroleum reservoirs have been shown to contain $10^{-18}$ fractional concentration of the radioactive isotope~\cite{arpesella2002measurements}, making this background completely subdominant even at large pressures. We foresee that similar provisions can be taken for other cosmogenically-activated radioactive isotopes.

Even a background radioactive rate of $\Gamma_\text{RD} \sim 10^{-2}~\text{Hz}$, potentially achievable with a high-purity shield and keeping radioactive contaminants at a minimum, poses a huge challenge to a typical DM absorption detector in the $\omega \sim \text{eV}$ range. Other detector proposals~\cite{battaglieri2017us, bunting2017magnetic,hochberg2016detecting,hochberg2017absorption, derenzo2017direct,bloch2017searching,Hochberg:2017wce} employ bulk target volumes based on nonresonant absorption onto liquids or solids, and aim to be signal-count-limited at kg-year exposures, requiring powerful active veto methods for large volumes. Our Bulk and Stack configurations each have characteristic properties that provide additional passive mitigation mechanisms not available to the aforementioned nonresonant absorption targets.

The Bulk configuration in scanning mode has three distinct advantages regarding passive radioactive background mitigation. Firstly, by dividing the full integration time over an $\mathcal{O}(1)$ bandwidth into many different shots with independent, narrow-band DM response functions, the radioactive background becomes already negligible when $\Gamma_\text{RD} \lesssim t_\text{shot}^{-1}$---a much looser criterion than $\Gamma_\text{RD} \lesssim t_\text{int}^{-1}$. 

Secondly, one can look for DM signals $\Gamma_\text{det}$ smaller than $\Gamma_\text{RD}$ even when $\Gamma_\text{RD} > t_\text{shot}^{-1}$ (as we have implicitly assumed in eq.~\ref{eq:deltaGammadet}) because there is a natural way to modulate the DM signal rate while keeping the background rate constant. For example, when in neighboring shots one has average total background counts of $\Gamma_\text{bckg} t_\text{shot}\gtrsim 1$, one is sensitive to signal rates as low as $\Gamma_\text{det} \sim \sqrt{\Gamma_\text{bckg} / t_\text{shot}}$ (assuming $\eta_\text{det} = 1$ for simplicity). Such averaging effects and differential response tests are not typically available in most bulk detectors, which do not have a natural way to ``turn off'' the signal while keeping the background rate constant. 

Thirdly, a Bulk detector in scanning mode can be operated at such low density that a typical fast electron only has a small probability $P_\text{ex}^e$ of exciting a molecular state in the frequency band of interest, since that process does not receive a resonant enhancement factor:
\begin{align}
P_\text{ex}^e \simeq n \sigma_\text{ex}^e L \sim 10^{-3} \left(\frac{n}{10^{-2} n_\text{st}}\right) \left( \frac{\sigma_\text{ex}^e}{10^{-4} \alpha a_0^2} \right)\left(\frac{L}{2~\text{m}} \right),
\end{align}
approximately valid for $n \sigma_\text{ex}^e L \lesssim 1$, and $L$ the typical linear size of the detector volume. Above, $\sigma_\text{ex}^e$ should be the (velocity-averaged) cross-section for a typical electron to excite molecular states that give rise to fluorescence photons within a detector's bandwidth around the frequency of interest, for which we have chosen a plausible value for an E1-allowed transition in the numerical estimate~\cite[\S 148]{landau1958quantum}. Forbidden transitions have even lower $\sigma_\text{ex}^e$ that scale like $\alpha^3 a_0^2$ or even higher powers of the fine structure constant. When $N_e P_\text{ex}^e \ll 1$, where $N_e$ is the typical number of electrons produced in a radioactive decay, the radioactive background in the band of interest is further suppressed by this factor.

The Stack configuration has the remarkable property that 84\% of the signal is emitted in a cone of opening angle $2v_0 \approx 5.4~\text{arcmin}$ and solid angle $\Delta \Omega = \pi v_0^2$, a $1.5 \times 10^{-7}$ fraction of the full solid angle $\Delta \Omega = 4\pi$. Radioactive backgrounds from the molecular volume can thus likely be ignored entirely, reducing the problem of radioactive contamination only to the photodetector material and mount, a dramatically smaller volume $V \sim \mathcal{O}(\text{mm}^3)$. 

Furthermore, an active veto system consisting of scintillating material and PMTs surrounding the detector volume may be employed. Radioactive decays give rise to many high-energy particles. When they trigger the scintillating material, any ``signal'' in the photodetector in a short time span around the trigger time in the photodetector can be vetoed. We note that a typical radioactive decay chain releases dozens of primary gammas and betas, as well as a bunch of secondaries, so the trigger efficiency on any \emph{one} photon or electron need not be high. As long as the photodetector has a sufficiently short jitter time $\Delta t$ and the molecular medium a sufficiently fast relaxation rate to thermal equilibrium (through radiative and nonradiative channels), the resulting dead time can be made negligible. Roughly, one requires
\begin{align}
\Gamma_\text{RD} \ll \min \left\lbrace(\Delta t)^{-1}, \gamma_0 + \sum_i \gamma_i+ \gamma_\text{quench} \right\rbrace \label{eq:RDrequirement}
\end{align}
to leave the duty cycle of the experiment essentially unaffected. The timing jitter requirements are easily satisfied by many orders of magnitude. Figure~\ref{fig:rate1} shows that the detector relaxation is sufficiently fast even for low-lying vibrational levels, if they have dipole-allowed radiative decays. Only for vibrational states that have exclusively E1-forbidden radiative decay channels, does eq.~\ref{eq:RDrequirement} become hard to satisfy. 
Given the wealth of passive mechanisms on offer to keep the radioactive background under control, the setups under consideration do not require extremely efficient active veto systems, especially in comparison to other proposed experiments in this energy range. The Phase~I prototypes can likely forego a large-scale active veto system altogether.

\textbf{Cosmic rays---}
Cosmic muons can also make up a considerable fraction of the background. At sea level, the vertical muon flux density is about $70~\text{m}^{-2}\text{s}^{-1}\text{sr}^{-1}$~\cite{de1993absolute,grieder2001cosmic}. The cosmic muon flux is much lower underground, falling to an integrated vertical flux density of $10^{-3} ~\text{m}^{-2}\text{s}^{-1}$ about 1~km deep in standard rock (2.65~km water-equivalent), and $10^{-5} ~\text{m}^{-2}\text{s}^{-1}$ at 2~km depth~\cite{andreev1987proc,aglietta1995neutrino,ambrosio1995vertical,berger1989experimental,waltham2001through}. Underground operation in a mine of moderate depth automatically ensures that the cosmic muon background is subdominant to that of radioactivity. Phase~I prototypes can likely get away with surface-level operation when outfitted with a modest muon veto system that has a rejection power of $10^2$, one that can possibly work in conjunction with a radioactivity veto.


\subsection{Signal discrimination strategies} \label{sec:discrimination}

From the discussion so far, a signal discrimination strategy emerges. Radioactive and cosmic background events can be identified by the fact that they result in multiple photon counts and ionized electrons. All environmental and detector backgrounds will be distributed over a broad energy range. In contrast, the near-monochromatic DM signal can only excite transitions at \emph{one} particular transition energy, and cause a single fluorescence photon at one frequency (or at most a handful of frequencies, if there are decay channels to several intermediate states). Below, we outline that more detailed follow-up studies can \emph{unequivocally verify the DM origin of any potential signal}, and furthermore determine properties such as mass, spin, interaction type, and 3D velocity with pinpoint precision. The spectacular discrimination power of this type of detector relies on three different ``handles'': energy response, selection rules, and spatial coherence.

\textbf{Energy response---}
The highly resonant response of the detector at low number densities can be used to confine the DM signal to better than $10^{-6}$ fractional frequency precision, and even to perform precision studies of its lineshape.
A cryogenic photodetector with energy resolution of $\Delta \omega \lesssim 0.1~\text{eV}$ can resolve both electronic splittings and vibrational fine structure in most molecules. Temperature modulation will change the relative populations of rotational levels, and thus the absorption rates in the rotational fine structure. By varying the pressure, one can determine how far off resonance the signal is. Scanning with external electromagnetic fields and use of different molecular species---and isotopes of the same species---can be employed to further hone in on the true energy of the signal, i.e.~the DM mass.

What is the ultimate energy resolution? Collisional broadening rates can be made arbitrarily small by lowering the pressure, while the fractional radiative linewidth is $(\gamma_0 + \sum_i \gamma_i)/\omega_0 \ll \alpha^3 \approx 4\times 10^{-7}$ and typically several orders of magnitude smaller even for dipole-allowed decays. For photon absorption, the Doppler width due to molecular motion often determines the minimal fractional linewidth of the signal at low number densities. 

Doppler broadening of the absorption line for a \emph{nonrelativistic} particle is less pronounced than for a (necessarily relativistic) photon. Using momentum and energy conservation for an inelastic collision with initial (final) molecular velocity $\vect{v}_1$ ($\vect{v}_2$) during which a dark matter particle with mass $m$, energy $\omega$, and velocity $\vect{v}$ is absorbed onto an internal molecular state with transition energy $\omega_0$, we find:
\begin{gather}
M_\text{mol} \vect{v}_1 + m \vect{v} = M_\text{mol} \vect{v}_2; ~~\, \frac{M_\text{mol} v_1^2}{2} + \omega = \omega_0 + \frac{M_\text{mol} v_2^2}{2} \nonumber \\
\Rightarrow \omega_0 = \omega - m \vect{v}_1 \cdot \vect{v} - \frac{1}{2} \frac{m^2}{M_\text{mol}} v^2.
\end{gather}
The third term on the RHS is quantitatively subdominant, and also does not depend on the molecule's initial velocity. Approximating the dark matter mass with its total energy $\omega$, and expanding in small velocities, we then finally arrive at the condition for absorption:
\begin{align}
\omega = \omega_0 \left( 1+ \vect{v}_1 \cdot \vect{v} \right) \label{eq:dopplercondition}
\end{align}
Because the molecules in the gas move at different velocities, they can absorb dark matter particles of differing energies $\omega$. Defining the molecular speed in the dark matter's velocity direction as $v_{1,\parallel} \equiv \vect{v}_1 \cdot \vect{v}/|\vect{v}|$, we can express the probability of finding a molecule between $v_{1,\parallel}$ and $v_{1,\parallel} + d v_{1,\parallel}$ as proportional to the Boltzmann-weighted $\exp\lbrace - M_\text{mol} v_{1,\parallel}^2/2T\rbrace dv_{1,\parallel}$. Translating this to a normalized molecular transition energy distribution via eq.~\ref{eq:dopplercondition}, we find
\begin{align}
g_0^\text{Dop}(\omega_0,\omega_0') &= \frac{1}{\sqrt{2\pi \Delta^2}} \exp\left[-\frac{(\omega_0 - \omega_0')^2}{2\Delta(v)^2}\right], \label{eq:g0Dop}\\
 \Delta(v) &\equiv \omega_0 v \sqrt{\frac{T}{M_\text{mol}}} ,
\end{align}
with a fractional width of $\Delta(v) / \omega_0 \approx 4 \times 10^{-9} (v/10^{-3})$ for $\text{H}_2$ gas at room temperature, and even lower for heavier molecules and/or colder temperatures. This width $\Delta(v)$ is a factor of $v \sim 10^{-3}$ smaller than the equivalent Doppler width for photon absorption. In case Doppler broadening does dominate over both radiative and collisional broadening, when $\Delta(v) \gg \gamma$, then the result of eq.~\ref{eq:Gammaabscol}  should be convoluted with the line shape of eq.~\ref{eq:g0Dop} to give:
\begin{align}
\Gamma_\text{abs}^\text{Dop}(\omega,\omega_0) \simeq N \Omega^2 \Delta^{-1} \sqrt{\frac{\pi}{8}} \exp\left[-\frac{(\omega - \omega_0)^2}{2\Delta(v)^2}\right], \label{eq:Gammadop}
\end{align}
at least for $|\omega - \omega_0|/\Delta$ not too large. The off-resonance tails are more accurately described by the Voigt profile, a convolution of both the Lorentzian and Gaussian line shapes.

The narrow fractional width allows for extremely narrow spectroscopic studies of the signal line shape at low pressures (when collisional broadening can be ignored), and would be a great signal discriminant should anomalous fluorescence be seen in the experiment. A single velocity component of the dark matter field ensemble has an energy $\omega = m (1 + v^2/2)$ in terms of its square velocity in the lab frame.  The DM's 3D velocity distribution $f(\vect{v})$ can be expected to closely resemble the virialized distribution of eq.~\ref{eq:fv}, yielding a fractional underlying signal frequency width of order $v_0^2/2 \approx 3 \times 10^{-7}$. Because $\Delta(v)/\omega_0 v_0^2 \ll 1$, the molecular resonance can resolve the kinetic energy distribution of DM! In this ``resolved'' regime, the absorption rate as a function of DM mass $m$ is :
\begin{align}
&\Gamma_\text{abs}^\text{DM}(m,\omega_0) \simeq \int d^3\vect{v}\, f(\vect{v}) \Gamma_\text{abs}^\text{Dop}\left(m (1+{v^2}/{2}),\omega_0\right) \nonumber \\
& \simeq \frac{N \Omega^2}{2^{3/2}\pi v_0^3} \int d^3\vect{v}\, \frac{\exp\left\lbrace{-\frac{(\vect{v}-\vect{v}_\text{lab})^2}{v_0^2} - \frac{[m(1+v^2/2) - \omega_0]^2}{2\Delta(v)^2}}\right\rbrace}{\Delta(v)}. \label{eq:GammaDM}
\end{align}
By tuning $\omega_0$ and at large enough statistics, one could determine the DM mass $m$, the velocity dispersion $v_0$, and the magnitude of the relative velocity $|\vect{v_\text{lab}}|$ (using diurnal/annual modulation).  One could possibly even discern finer details of the velocity distribution $f(\vect{v})$, such as the Galactic escape velocity (not taken into account by the virialized distribution in the second line of eq.~\ref{eq:GammaDM}). Note that the integrand in the second line of eq.~\ref{eq:GammaDM} is technically only valid for velocities where $\Delta(v) \gg \gamma$; one should use the full Voigt absorption line shape for the low velocities $v$ where this is no longer a good approximation.

\textbf{Selection rules---}
If a near-monochromatic signal were to be detected at some frequency $\omega$ near a transition frequency $\omega_0$, then one can determine the properties (such as the quantum numbers) of the initial and final states. This is possible because small polyatomic systems are simple enough to have been well-characterized both experimentally and theoretically, as is evident from secs.~\ref{sec:states}~and~\ref{sec:transitions}. 

Any nonthermal SM background must come from interactions with external photons or charged particles, which to leading order interact with a Hamiltonian proportional to the electric dipole moment $\vect{\mu}_e$, giving $\Omega \propto |\vect{\mu}_{1,0}|$. This fact leads to the well-known dipole selection rules for the leading-strength transitions.

DM absorption does not necessarily obey these rules. A scalar DM particle may leave the angular state of the molecule unaffected via a ``monopole'' transition ($\Delta J = 0$), while a pseudoscalar  DM particle may cause ``spin-dipole'' transitions  from a spin-singlet ground state to a spin-triplet excited state ($\Delta S \neq 0$), both at leading order. These processes are normally highly forbidden in small polyatomic systems. Even when the DM particle primarily causes ordinary dipole transitions, one could test if rather than coupling to the dipole moment of electric charge, it instead couples to the dipole moment of e.g.~baryon number, lepton number, or any linear combination thereof. Although we do not discuss it in this work, dark matter candidates with spin $\ge 2$ would dominantly cause higher multipole transitions. 

The modularity of the proposed detector allows for targeted studies with different types of molecules, each having a transition energy $\omega_0$ near a potential candidate signal at $\omega$. Should a DM signal be seen in a broad frequency search, these targeted studies can in principle determine the form of the interaction Hamiltonian $\delta H$.

\textbf{Spatial coherence---}
The nonrelativistic velocity distribution $f(\vect{v})$ leads to a characteristic spatial (and temporal) coherence of the perturbing wave that is imprinted onto the molecular emission via the correlation function in eq.~\ref{eq:gdef}, in turn leading to the dramatic focusing effect for a slab-like container discussed in sec.~\ref{sec:cooperation}. This emission pattern cannot be mimicked by any standard SM background. A robust prediction for such a detector shape is that the center of the emission cones, with off-set from the normal direction proportional to $\vect{v}_\text{lab}$ (see eqs.~\ref{eq:center1}~and~\ref{eq:center2}), would precess on a diurnal and annual basis due to the Earth's spin and orbit around the Sun, respectively, with known phases, directions, and amplitudes. In addition, these measurements of $\vect{v}_\text{lab}$ would have to agree with the magnitude $|\vect{v}_\text{lab}|$ derived from eq.~\ref{eq:GammaDM}; likewise, $v_0$ as determined by the opening angle of the emission cone must agree with the fractional frequency linewidth in eq.~\ref{eq:GammaDM}. These observations can also be compared against astrophysical inferences of the DM's velocity distribution. Finally, a precision line study might reveal additional information not otherwise attainable, such as the existence of DM streams and the relative rotation of the DM halo and the Galactic disk.

\section{Dark matter sensitivity}\label{sec:sensitivity}

In this section, we present sensitivity projections of the proposed setups to the parameter space of specific DM models, after briefly reviewing the chief interactions of each DM candidate. We have classified the models in terms of spin and parity of the DM boson, starting off with spin-1 vectors in sec.~\ref{sec:vectors}, and then continuing with parity-even, spin-0 scalars in sec.~\ref{sec:scalars} and parity-odd, spin-0 pseudoscalars in sec.~\ref{sec:pseudoscalars}. We leave a treatment of DM particles with spin 2 and higher to future work. We summarize the dark matter candidates and couplings, as well as the corresponding transitions they can mediate, in table~\ref{tab:transitions}.

\bgroup
\def\arraystretch{1.7}
\begin{table*}[t]
\resizebox{\textwidth}{!}{%
\begin{tabular}{l l l l l l}
\hline
  \multicolumn{2}{l }{DM type} & Interaction Hamiltonian $\delta H$ ~~& \multicolumn{2}{l}{Transition type and selection rules~~} & $\Omega \left[ \rad~\s^{-1}\right]$ \\ \hline
   & \multirow{8}{*}{parity-even}  & $(d_{m_e} + d_e) \tilde{\phi} k_e R_e R $ & vib & $\Delta v = 1, \Delta J = 0$ & $5.5 \times 10^{-9}  \frac{d_{m_e}}{10^{6}}  $\\ \cline{3-6}
  & & $  (3 d_{m_e} + 4 d_e)\tilde{\phi}\frac{k_e}{2} (R-R_e)^2$ & vib & $\Delta v = 2, \Delta J = 0$ & $7.1 \times 10^{-10}  \frac{d_{m_e}}{10^6}  $ \\ \cline{3-6}
  & & $(d_g + Q_{\hat{m}_q} d_{\hat{m}_q})\tilde{\phi} \frac{\vect{\nabla}_N^2}{2M}  $  & vib & $\Delta v = 2, \Delta J = 0$ & $2.4 \times 10^{-11}  \frac{d_{\hat{m}_q}}{10^6} \frac{Q_{\hat{m}_q}}{0.1}  $\\ \cline{3-6}
  & & \multirow{2}{*}{$(\Delta Q_i d_i) M (\vect{\nabla} \tilde{\phi} \cdot \vect{R}) $} & \textbf{vib} & $\Delta v = 1, \Delta J = \pm 1$ & $3.0 \times 10^{-10}  \frac{d_i}{10^{6}} \frac{\Delta Q_i}{10^{-2}} $ \\ 
  & & & \textbf{rot} & $\Delta J = 1$ & $4.1 \times 10^{-13} \frac{d_i}{10^2} \frac{\Delta Q_i}{10^{-2}} $\\ \cline{3-6}
   & & $(d_{m_e}+d_e)  \tilde{\phi} \frac{\vect{\nabla}_e^2}{2m_e} $ & el & $\Delta \Lambda = 0, \Delta i = 0$ & $9.5 \times 10^{-10}  \frac{d_{m_e}}{10^6}  $ \\ \cline{3-6}
  \multirow{2}{*}{spin-0} & & $d_{m_e} m_e  \vect{\nabla} \tilde{\phi} \cdot \vect{r}_e$ & \textbf{el} & $|\Delta \Lambda| \le 1, \Delta i = 1$ & $7.5 \times 10^{-11} \frac{d_{m_e}}{10^6}$ \\ \cline{2-6}
  & \multirow{5}{*}{parity-odd}  & \multirow{2}{*}{$G_{aNN} \, \partial_t a \, \vect{\sigma}_N \cdot \frac{-i\vect{\nabla}_N}{M}$} & \textbf{vib} & $\Delta v = 1, \Delta J = \pm 1,  |\Delta S_N|  \le 1~$ & $1.7 \times 10^{-10}  \frac{G_{aNN}}{10^{-8}/\GeV}  $ \\ 
  & & & \textbf{rot} & $\Delta J = 1, |\Delta S_N| = 1$& $2.5 \times 10^{-11} \frac{G_{aNN}}{10^{-8}/\GeV}$ \\ \cline{3-6}
  & & \multirow{2}{*}{$\frac{d_\theta}{f_a} a \, \vect{\sigma}_N \cdot \vect{E}$} & \textbf{vib} & $\Delta v = 1, \Delta J = \pm 1,  |\Delta S_N|  \le 1$ & $4.0 \times 10^{-12} \frac{10^{8}~\GeV}{f_a} $\\
  & & & \textbf{rot} & $\Delta J = 1, |\Delta S_N|  \le 1$ & $5.8 \times 10^{-13} \frac{10^{8}~\GeV}{f_a} $ \\ \cline{3-6}
  & &  $G_{aee} \, \partial_t a \, \vect{\sigma}_e \cdot \frac{\vect{-i \nabla}_e}{m_e}$ & \textbf{el} & $|\Delta \Lambda| \le 1, \Delta i = 1, |\Delta S_e| \le 1$ & $4.0 \times 10^{-10}  \frac{G_{aee}}{ 10^{-10}/\GeV}  $ \\ \cline{1-6}
  & 	&  & \textbf{el} & $|\Delta \Lambda| \le 1, \Delta i = 1$ & $1.5 \times 10^{-6} \frac{\epsilon}{10^{-14}}  $ \\ 
  & \multirow{1}{*}{kinetic mixing~~} & \multirow{1}{*}{$\epsilon \vect{\mu}_e \cdot \vect{E}'$} & \textbf{vib}	& $\Delta v = 1, \Delta J = \pm 1$ &  $1.3 \times 10^{-5} \frac{\epsilon}{10^{-12}}  $\\ 
   \multirow{2}{*}{spin-1} & & & \textbf{rot}	& $\Delta J = 1$ & $1.5 \times 10^{-2} \frac{\epsilon}{10^{-10}}  $\\ \cline{2-6}		
  &  &  & \textbf{el} & $|\Delta \Lambda| \le 1, \Delta i = 1$ & $5.0 \times 10^{-6} \frac{g}{10^{-14}}$ \\ 
  & \multirow{1}{*}{$B-L$ charge} & \multirow{1}{*}{$\vect{\mu}_{B-L} \cdot \vect{E}_{B-L}$} & \textbf{vib	} & $\Delta v = 1, \Delta J = \pm 1$ & $4.3 \times 10^{-7} \frac{g}{10^{-14}}$ \\ 
  & 	& & \textbf{rot} & $\Delta J = 1$ & $5.0 \times 10^{-10} \frac{g}{10^{-18}}$\\ \cline{1-6}
\end{tabular}}
\caption{Dark matter candidates classified by their spin, parity, and interaction Hamiltonian $\delta H$, along with the  types and strengths of transitions they can induce. Transition types considered include those of the electronic (el), vibrational (vib), and rotational (rot) kind; bold face is used whenever the transition type can be E1-allowed and thus used in the Stack configuration. The fifth column contains the diatomic-molecule selection rules on molecular-axis angular momentum projection $\Lambda$ and inversion $i$ for electronic transitions, and vibrational quantum number $v$ and total angular momentum $J$ for vibrational and rotational transitions. Unless otherwise noted, the total nuclear spin $S_N$ and electronic spin $S_e$ do not change in absence of spin-orbit coupling. Rabi frequencies $\Omega$ are quoted at $\omega_0 = 5~\text{meV}, 0.4~\eV, 0.8~\eV, 8~\eV$ for rotational, $\Delta v=1$, $\Delta v=2$ vibrational, and electronic transitions, respectively, using experimentally allowed benchmark values of the coupling at $m = \omega_0$. Numerical estimates use FC factors of $10^{-2}$, $\Delta J = 1$ rotational matrix elements of $1/\sqrt{3}$, $ R_e  = a_0$, $\delta_e = \delta_{e,1} = \delta_{e,2} = 1$, $\Delta q_{B-L} = 1$, and matrix element estimates from eqs.~\ref{eq:Mat1},\ref{eq:Mat2},\ref{eq:Mat3},\ref{eq:paramest},\ref{eq:Hmonvir},\ref{eq:dipmatvir},\ref{eq:Mat4},\ref{eq:spindipmatel}.} \label{tab:transitions}
\end{table*}
\egroup


\subsection{Vectors}\label{sec:vectors}

We will start with two vector dark matter candidates, namely a hidden photon kinetically mixed with the usual electromagnetic field, and a new photon that couples to baryon-minus-lepton number $B-L$, both having a St\"uckelberg mass $m_{\gamma'}$. We focus on these two cases because of pedagogy (simple comparisons can be made to the interactions of the normal photon), and because they embody simple extensions of the Standard Model that are theoretically consistent up to very high energy scales. Other types of vectors are certainly possible, but most of the models include other states, like Higgs-like scalars or anomalons, whose interactions are often independently constrained; we ignore those theories only for the sake of brevity.

Weakly-coupled, massive vectors can be produced in the early Universe.  A natural and calculable relic abundance can arise from vector fluctuations during the inflationary era (if the St\"uckelberg mass is ``on'' then), with a present-day relic energy density of  
\begin{align}
\rho_{\gamma'} \approx \rho_\text{DM} \left(\frac{m_{\gamma'}}{1~\text{eV}} \right)^{1/2} \left( \frac{H_I}{5 \times 10^{12}~\text{GeV}}\right)^2,
\end{align}
where $H_I$ is the Hubble scale during the last few e-folds of inflation~\cite{graham2016vector}. We see that the vector can make up all of the DM in the mass range of interest if the Hubble scale is between $10^{12}~\GeV$ and $10^{13}~\GeV$. The field misalignment mechanism, on the other hand, is not effective unless large interactions with curvature invariants are present~\cite{arias2012wispy}.


\textbf{Kinetically mixed photon.---} If a light vector particle is a low-energy remnant of a sector that is coupled to the SM at high energies, the associated vector field $A'_\mu$ can and will generically have effective operators coupling it to the SM even if none of the SM fields are charged under the new U(1) gauge symmetry~\cite{holdom1986two,okun1982limits}. The lowest-dimensional such operator, one that can be expected to capture the dominant effective interactions with the SM at low energies, is the kinetic mixing term $F_{\mu\nu}F^{\prime \mu \nu}$ wherein the ``hidden'' field strength $F_{\mu \nu}' \equiv \partial_\mu A'_\nu - \partial_\nu A'_\mu$ couples to the equivalent quantity $F_{\mu \nu}$ of the SM electromagnetic field $A_\mu$. The strength of this mixing is conventionally quantified by a dimensionless parameter $\epsilon$ in the Lagrangian:
\begin{align}
\mathcal{L}_\text{gauge} = & -\frac{1}{4} F_{\mu\nu}F^{\mu \nu} -\frac{1}{4} F^{\prime}_{\mu\nu}F^{\prime \mu \nu} + \frac{1}{2}\epsilon F_{\mu\nu}F^{\prime \mu \nu} \nonumber \\
 & +\frac{1}{2} m_{\gamma'}^2 A'_\mu A^{\prime \mu} - e A_\mu J^{\mu}_\text{EM}, \label{eq:photgauge}
\end{align}
with $m_{\gamma'}$ the hidden photon's St\"uckelberg mass, and $J^\mu_\text{EM} = \sum_\psi q_\psi \bar{\psi} \gamma^\mu \psi$ the electromagnetic vector current. The above Lagrangian, written in the ``gauge basis'', can be made to have diagonal kinetic and mass terms with a field redefinition, transforming it into the so-called physical basis:
\begin{align}
\mathcal{L}_\text{physical} = &-\frac{1}{4} F_{\mu\nu}F^{\mu \nu} -\frac{1}{4} F^{\prime}_{\mu\nu}F^{\prime \mu \nu} +\frac{1}{2} m_{\gamma'}^2 A'_\mu A^{\prime \mu} \nonumber \\
& - e (A_\mu + \epsilon A'_\mu) J^{\mu}_\text{EM}, \label{eq:photphysical}
\end{align}
now rewritten in terms of redefined fields $A_\mu$ and $A'_\mu$. Diagonalizing the terms that govern the propagation in vacuum comes at the cost of introducing an interaction of the massive photon state with the EM vector current, suppressed by $\epsilon$ relative to that of the massless photon.
In the physical basis, electromagnetic charges interact with a specific linear combination of fields $A_\mu^\text{eff} = A_\mu + \epsilon A_\mu'$, so matrix elements for interactions of $A_\mu'$ with the molecule can be found simply by rescaling those of the photon $A_\mu$ by a factor of $\epsilon$. (This is strictly only true for electric dipole (E1) transitions. For magnetic dipole (M1), electric quadrupole (E2), and even higher-order transitions, whose matrix elements all involve factors of $\vect{k} \cdot \vect{x}$, the matrix elements of a non-relativistic $A_\mu'$ are further suppressed by factors of velocity $v = |\vect{k}|/m_{\gamma'}$ relative to those of the necessarily relativistic photon $A_\mu$.)

The propagation of the two vector fields is qualitatively different, however. In vacuum, the $A_\mu' = (\phi', \vect{A}')$ mass eigenstate obeys the massive wave equation $(\partial_t^2 - \vect{\nabla}^2 + m_{\gamma'}^2 ) A_\mu' = 0$, which means the field can support nonrelativistic solutions as well as a longitudinal modes, which have a vector potential aligned with the propagation velocity $\vect{v}$ (i.e.~$\vect{A}' \cdot \vect{v} \neq 0$). The dark matter state is expected to be well-described by a nonrelativistic, classical solution of this wave equation; a single momentum component of the whole DM ensemble has a vector potential of the form:
\begin{align}
\vect{A}'(t,\vect{x})= A_0' \hat{\vect{n}} \cos\left[m_{\gamma'} (1+v^2/2)t - m_{\gamma'} \vect{v}\cdot \vect{ x} + \alpha_{\vect{v}} \right].\label{eq:Avec}
\end{align}
The full state of the field should be regarded as a mixed state composed out of many of these different velocity components drawn from a probability distribution $f(\vect{v})$, each with random phases $\alpha_{\vect{v}}$ and possibly also random directions $\hat{\vect{n}}$. We assume the velocity distribution to be close to that of eq.~\ref{eq:fv}, and the direction $\hat{\vect{n}}$ of the vector potential to have a coherence time at least as long as $1/mv_0^2$ but shorter than the time scale of the experiment. 
For reasons outlined in appendix~\ref{sec:fullquantum}, all of our main results remain valid even when the classical approximation breaks down, at hidden photon masses $m_{\gamma'} \gtrsim 15~\eV$. In this regime, the local dark-matter field occupation numbers become so low that e.g.~$\langle |\vect{A}'|^2 \rangle \gg \langle |\vect{A}'| \rangle ^2$.

In Lorenz gauge ($\partial_\mu A^{\prime \mu} = 0$), the hidden electric scalar potential $\phi'$ can be determined from the relation $\partial_t \phi' = - \vect{\nabla} \cdot \vect{A}'$, from which it follows that it is suppressed relative to the hidden vector potential, as typically $\phi' \sim v|\vect{A}'|$. The stress-energy tensor contribution 
$T^{\mu \nu} = F^{\prime\mu}_\lambda F^{\prime \lambda \nu} + (1/4) g^{\mu \nu} F^{\prime}_{\lambda \sigma} F^{\prime \lambda \sigma} + (1/2) m_{\gamma'}^2 A^{\prime\mu} A^{\prime\nu}$ from the massive photon state can be evaluated on the solution of eq.~\ref{eq:Avec}, and contributes as an effective fluid with nearly zero average pressure, and energy density:
\begin{align}
\rho_{\gamma'} \equiv T^{00} \simeq \frac{1}{2} m_{\gamma'}^2 \left|A^{\prime}_0\right|^2.
\end{align}
Expressed as fields rather than potentials, we see that the hidden magnetic field $\vect{B}' = \vect{\nabla} \times \vect{A}'$ is velocity suppressed relative to the hidden electric field $\vect{E}' = -\vect{\nabla} \phi' - \partial_t \vect{A}'$, which oscillates with an amplitude of
\begin{align}
\left|\vect{E}'_0\right| \simeq m_{\gamma'} \left|A'_0 \right| \simeq \sqrt{2 \rho_{\gamma'}} \approx 3.3 \times 10^3~\text{V}/\text{m} \left(\frac{\rho_{\gamma'}}{\rho_\text{DM}} \right)^{1/2}. \label{eq:rhogamma}
\end{align}
along the direction of the $\vect{A}'$ potential, up to velocity-suppressed corrections. Because of the enormous size of this hidden electric field, which in some ways is equivalent to shining a kilowatt-class ``hidden'' laser with beam waist of order the detector size of 30~cm (and approaching megawatt power levels for a 3-meter detector size), it is possible to have appreciable event rates in absorptive media even for tiny values of the mixing parameter $\epsilon$, as we showed already in fig.~\ref{fig:rate1}.

A particle $\psi$ with mass $m_\psi$, electromagnetic charge $q_\psi$ coupled to the two photons as in eq.~\ref{eq:photphysical} has nonrelativistic dynamics dictated by the Hamiltonian
\begin{align}
H_\psi &= \frac{1}{2 m_\psi}\left(-i\vect{\nabla}_\psi - e q_\psi \vect{A}_\text{eff} \right)^2 + e q_\psi \phi_\text{eff} \label{eq:Hphot1} \\
  &\simeq \left[-\frac{1}{2m_\psi} \vect{\nabla}_\psi^2 + V \right] + \left[i\frac{q_\psi \epsilon e }{m_\psi} \vect{A}' \cdot \vect{\nabla}_\psi \right] + \dots, \nonumber
\end{align}
where we have ignored spin-orbit coupling, and have assumed in the second line that the particle moves in an electrostatic potential well $V = e q_\psi \phi$ of other nearby particles, as well as in a massive hidden photon wave of the form~\ref{eq:Avec}, with the ellipsis representing terms of $\mathcal{O}(\epsilon^2)$ and $\mathcal{O}(v)$. An ambient $\vect{A}'$ wave will also interact with any surrounding conductive elements, including the reflective coating of the vapor cell. The resulting screening currents will set the interacting linear combination of vector potentials, $\vect{A}_\text{eff} = \vect{A} + \epsilon \vect{A}'$, to near-zero within a hidden photon's Compton wavelength $m_{\gamma'}^{-1}$ away from the container wall~\cite{chaudhuri2015radio}. However, $\vect{A}$ and $\vect{A}'$ propagate at different speeds, so for a sufficiently large container volume  $V \gg m_{\gamma'}^{-3}$ they will oscillate in and out phase in most of the bulk interior such that we have $|\vect{A}_\text{eff}| \simeq A'_0$ to a very good approximation.

In a molecule, we can thus separate the full Hamiltonian into the Hamiltonian $H_0$ from eq.~\ref{eq:H0} and the interaction Hamiltonian $\delta H = \sum_\psi (i q_\psi e \epsilon/m_\psi) \vect{A}' \cdot \vect{\nabla}_\psi$, with the sum running over all particles in the molecule. Using the identity $\vect{\nabla}_\psi = - m_\psi [H_0,\vect{r}_\psi]$, we can rewrite the off-diagonal matrix elements---the only ones that matter for transitions---of $\delta H$ between an initial state $|i\rangle$ and a final state $|f\rangle$ as:
\begin{align}
\langle f | \sum_\psi i\frac{q_\psi \epsilon e }{m_\psi} \vect{A}' \cdot \vect{\nabla}_\psi  | i \rangle \simeq - \langle f | \sum_\psi  q_\psi \epsilon e \vect{E}'\cdot \vect{r}_\psi | i \rangle.
\end{align}
In the second equality, we have made the approximation $i \omega_0 \vect{A}' \simeq \vect{E}'$, which holds insofar as the hidden photon is on resonance with the transition, i.e.~$m_{\gamma'} \simeq \omega_0$. We also assumed that to leading order we can take $\vect{A}(\vect{x})$ and $\vect{E}(\vect{x})$ to be spatially constant, i.e.~independent of $\vect{x}$, over the spatial extent of the molecule. This approximation is especially precise for a \emph{nonrelativistic} wave.

To leading order, we can thus think of transitions being caused by the familiar operator
\begin{align}
\delta H = -  \epsilon \vect{\mu}_{e} \cdot \vect{E}' = - \sum_\psi  q_\psi \epsilon e \vect{E}'\cdot \vect{r}_\psi \label{eq:dHEdip}
\end{align}
of an electric dipole $\vect{\mu}_e$ in an effective electric field $\vect{E}_\text{eff} = \epsilon \vect{E}'$. It can induce transitions in many systems, including vibrational and electronic transitions in diatomic molecules. For vibrational transitions, we can integrate out the electronic motion and regard a neutral diatomic molecule as a spring with charges $\pm \delta_e(R)$ attached to the ends, where we take the charges $\delta_e$ to depend on the internuclear separation $R$. (The dependence on $R$ is easy to see: e.g.~as $R \to \infty$, it must be that $\delta_e(R) \to 0$ if the molecule dissociates into neutral atoms.) This function $\delta_e(R)$ can be calculated from first principles or indirectly measured in absorption spectra. In this simplistic view of the molecule, the transition operator in eq.~\ref{eq:dHEdip} reduces to
\begin{align}
\delta H & =  - \epsilon e \vect{E}'\cdot \vect{R} \delta_e(R) = - \epsilon e \vect{E}'\cdot \hat{\vect{R}} \Big \lbrace R_e \delta_e(R_e) \label{eq:dHEdipvib} \\
& \hspace{5.5em}+ (R-R_e)\underbrace{\left[\delta_e(R_e) + R_e \delta_e'(R_e)  \right]}_{\equiv \delta_{e,1}} +\dots \Big\rbrace. \nonumber
\end{align}
The first term within curly brackets acts trivially on the vibrational state (it only induces rotational transitions), but the second term can induce $\Delta v = \pm 1$ transitions as discussed in sec.~\ref{sec:dipoletransitions}, as long as the factor in square brackets $\delta_{e,1}$ is nonzero. In general, $\delta_{e,1}$ is roughly of order the electric dipole moment of the molecule divided by $R_e$ for heteronuclear diatomics, and zero by symmetry for homonuclear diatomics. The third term and higher-order terms in the Taylor expansion contribute mostly to higher-harmonic transition matrix elements, and can usually be ignored to leading order, which we shall do so for this discussion. Finally, we find the angle-averaged squared Rabi frequency from hidden-photon dark matter to be
\begin{align}
\Omega^2  & =  \left| \langle v_f=1, J_f | \delta H |v_i = 0, J_i\rangle \right|_\text{avg}^2 \label{eq:omegahfvib} \\
 & = \epsilon^2 e^2 \delta_{e,1}^2\left| \langle  v_f = 1 | R-R_e |v_i = 0\rangle \langle  J_f | \vect{E}'_0 \cdot  \hat{\vect{R}} | J_i\rangle \right|_\text{avg}^2 \nonumber \\
 & =  \epsilon^2 e^2 \delta_{e,1}^2 \frac{\rho_\text{DM}}{M \omega_e} \left| \langle  J_f | \hat{\vect{E}}'_0 \cdot \hat{\vect{R}} |, J_i\rangle \right|_\text{avg}^2 \nonumber
\end{align} 
for vibrational transitions from the ground vibrational state $|v_i = 0\rangle$ to the first excited state $|v_f = 1\rangle$. To get to the second line, we used eq.~\ref{eq:rhogamma} with $\rho_{\gamma'} = \rho_\text{DM}$ and eq.~\ref{eq:Mat1} with $M$ the reduced mass of the diatomic and $\omega_e$ the vibrational splitting; the angle-averaged rotational matrix elements are given below eq.~\ref{eq:dipmatrot}. For $\Delta v = 2$ transitions, the squared Rabi frequency is reduced relative to that for $\Delta v= 1$ by the factor $\omega_e x_e / 8 \omega_e$, cfr.~eq.~\ref{eq:Mat1b}.

For electronic transitions, we can repeat the same exercise to find:
\begin{align}
\Omega^2 \simeq~&  \left| \langle \chi^\text{el}_f, v_f', J_f | \delta H |\chi^\text{el}_i, v_i'', J_i\rangle \right|_\text{avg}^2  \label{eq:omegahfel} \\
=~& 
 2 \epsilon^2 \Big|\langle \chi^\text{el}_f|\sum_n  e \vect{r}_{e,n} | \chi^\text{el}_i \rangle \langle v_f' | v_i'' \rangle \langle  J_f | \hat{\vect{E}}'_0 \cdot \hat{\vect{R}} |, J_i\rangle \Big|_\text{avg}^2 \nonumber \\
\equiv~& 2 \epsilon^2 e^2 \delta_{e,2}^2 R_e^2 \rho_\text{DM} \left|\langle v_f' | v_i'' \rangle\right|^2 \left| \langle  J_f | \hat{\vect{E}}'_0 \cdot \hat{\vect{R}} |, J_i\rangle \right|_\text{avg}^2 \nonumber
\end{align}
where $|\langle v_f' | v_i'' \rangle|^2$ is the Franck-Condon factor from eq.~\ref{eq:FC}, and where we have parametrized the electronic transition moment $|\langle \chi^\text{el}_{f}|\sum_n  e \vect{r}_{e,n} | \chi^\text{el}_{i}| \equiv \delta_{e,2} e R_e$ in terms of the dimensionless number $\delta_{e,2}$.

\begin{figure*}
\includegraphics[width=0.75\textwidth]{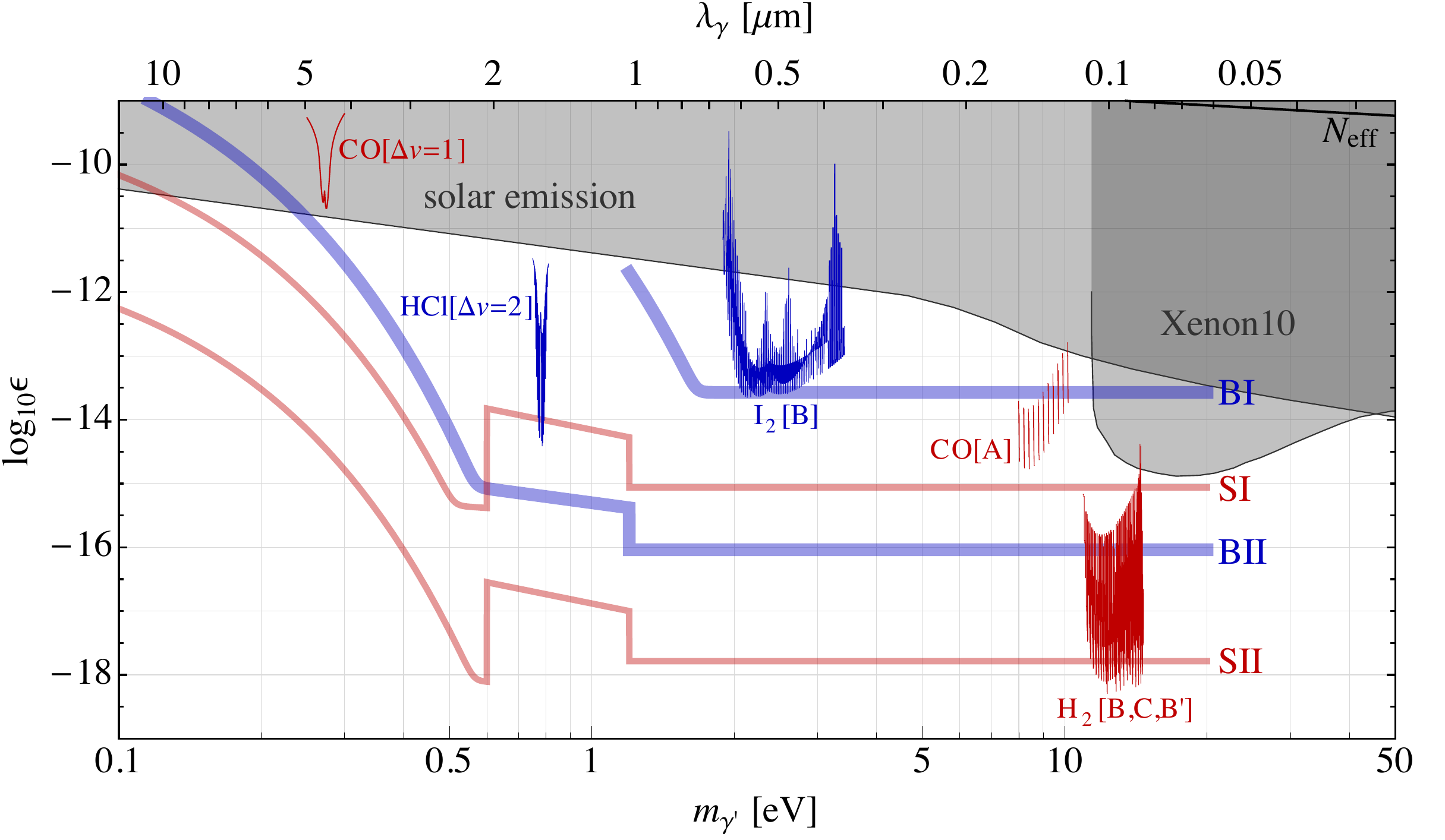}
\caption{Reach for the kinetic mixing parameter $\epsilon$ of a hidden photon with mass $m_{\gamma'}$. The $\text{SNR}=1$ sensitivity estimates for the Bulk configurations (BI~{\&}~BII) are shown as thick blue bands, and those of the Stack configurations (SI~{\&}~SII) as thin red bands. Sensitivity curves are also shown for a single molecular species with the following assumptions: $\text{I}_2$ (blue) in BI, HCl (blue) in BII, CO (red) in SI, and $\text{H}_2$ (red) in SII. Solar emission constraints and DM-induced ionization limits in Xenon10 are shown as gray regions. Above the black line in the top right, hidden photon DM runs afoul of contraints on $N_\text{eff}$.}\label{fig:reachhf}
\end{figure*}

In figure~\ref{fig:reachhf}, we plot the estimated $\epsilon$ sensitivity for the Bulk~I~{\&}~II configurations in thick blue bands, and for the Stack~I~{\&}~II configurations in thin red bands. For these estimates, we used the Rabi frequency reach of eq.~\ref{eq:deltaRabi} using the configuration parameters of table~\ref{tab:configs} in conjunction with the matrix elements of eqs.~\ref{eq:omegahfvib}~and~\ref{eq:omegahfel}. We also assumed $\delta_{e,1} \sim 1$ and the parametric estimates of eq.~\ref{eq:paramest} for vibrational transitions, and a electronic transition moment with $\delta_{e,2} R_e \sim a_0$ and a typical FC factor of $|\langle v_f' | v_i'' \rangle|^2 \sim 10^{-2}$ for electronic transitions. Squared rotational matrix elements were conservatively taken to be $1/6$.

The BI prototype, the least aggressive design, is seen to be already capable of probing new parameter space for electronic transitions above 1.2~eV. The BII configuration would be a drastic step up in reach for the same electronic transitions, and would also extend the reach to lower masses via $\Delta v = 2$ transitions. Transitions with $\Delta v = 1$ would yield larger absorption rates, but would produce trapped and subsequently quenched fluorescence photons due to the high optical thickness of such a Bulk detector, as we discussed around eq.~\ref{eq:optdepth}. 
 
The Stack configurations are also optimally operated with molecules exhibiting strongly-allowed electronic transitions at high energies, and $\Delta v= 2$ vibrational transitions between 0.6~eV and 1.2~eV. They can avoid the quenching issue for $\Delta v = 1$ transitions because of their optically thin planar design, and thus have access to larger matrix elements at energies below 0.6~eV. They also pick up less BBR due to their smaller photosensitive area, with the result that even the  SI prototype will likely outperform the much larger BII detector at low masses.
 
In figure~\ref{fig:reachhf}, we also show a few exemplary sensitivity curves for the proposed experiments.
In blue, we show the reach around $2~\eV$--$3~\eV$ from the famous visible-light absorption band B in ${}^{127}\text{I}_2$, which is both so wide and dense in frequency space that it nearly covers an octave contiguously at standard atmospheric conditions (without need for scanning), making it an attractive molecular candidate for a proof-of-principle prototype experiment. The blue $\text{I}_2$ curve assumes a vapor pressure of $0.25~\text{bar}$, which occurs in equilibrium around room temperature, and an integration time of $10^5~\text{s}$, and otherwise Bulk Phase~I parameters. Molecular information on iodine was taken from refs.~\cite{mulliken1971iodine,lamrini1994electronic,brewer1963radiative, zare1964calculation,paisner1974rotational,tellinghuisen1978intensity}.
Also in blue, at around 0.8~eV, we plot the estimated sensitivity for $\Delta v = 2$ absorption in ${}^1\text{H}{}^{35}\text{Cl}$ at $P = 0.25~\text{bar}$ after a single shot of $10^3~\text{s}$ in the Bulk Phase~II experiment.
We also depict the reach with ${}^{12}\text{C}{}^{16}\text{O}$ in a Stack Phase~I experiment at $P = 5~\text{bar}$, displaying both its infrared $\Delta v = 1$ vibrational transition (see also fig.~\ref{fig:rate2}) and its first allowed electronic transition $\text{X}\to\text{A}$ in the ultraviolet. Molecular data on this electronic transition can be found in refs.~\cite{cooper1981theoretical,chackerian1976electric,halmann1966isotope, krupenie1966band,dotchin1973radiative,tilford1972atlas}.
Finally, we depict in red the sensitivity to absorption of hidden photons heavier than 11~eV onto the first three E1-allowed electronic transitions $\text{X}\to \text{A,B,B'}$ in ${}^1\text{H}_2$~\cite{fantz2006franck} in the Phase~II version of the Stack configuration.
 
The gray exclusion regions in fig.~\ref{fig:reachhf} depict 95\%~CL constraints on $\epsilon$ from null observations by Xenon10~\cite{angle2008first} of hidden photons from the Galactic DM halo~\cite{an2015direct} (``Xenon10'') and of hidden photons emitted by the Sun~\cite{an2013dark} (``solar emission''). Although beyond the scope of this work, it would be interesting to work out the detection prospects of this solar emission component in our proposed setups, even though the resonant detector response is likely not optimal for a thermal emission spectrum. Finally, the black line labeled ``$N_\text{eff}$'' indicates an upper bound on $\epsilon$ derived from the effects that evaporation of hidden photon DM into the photon bath would otherwise have on the effective number of abundant neutrino species in the early Universe~\cite{arias2012wispy}.


\textbf{$\bm{B-L}$ photon.---}
Another possible set of interactions of a new vector particle are couplings to baryon number $B$ and lepton number $L$. Here we shall focus on the anomaly-free linear combination of $B-L$ charges of SM matter charged under a new massive U(1), under which the proton, neutron, and electron have respective charges:
\begin{align}
q_{B-L,p} = q_{B-L,n} = +1, \quad q_{B-L,e} = -1. \label{eq:qBL}
\end{align}
The small number parametrizing the strength of the DM vector interaction is the gauge coupling $g$. For simplicity of presentation, we assume the new vector does \emph{not} kinetically mix with the SM photon, and write its Lagrangian as:
\begin{align}
\mathcal{L} = -\frac{1}{4} F^{\prime}_{\mu\nu}F^{\prime \mu \nu} +\frac{1}{2} m_{\gamma'}^2 A'_\mu A^{\prime \mu} - g\epsilon A'_\mu  J^{\mu}_{B-L}, \label{eq:LagBL}
\end{align}
where the $B-L$ vector current is defined as $J^{\mu}_{B-L} \equiv \sum_\psi q_{B-L,\psi} \bar\psi \gamma^\mu \psi$. Analogously to the analysis done for the (kinetically mixed) photon in eqs.~\ref{eq:Hphot1}--\ref{eq:dHEdip}, we can find that the relevant transition operator
\begin{align}
\delta H & = -\vect{\mu}_{B-L} \cdot \vect{E}'
\end{align}
is that of a $B-L$ dipole moment $\vect{\mu}_{B-L} \equiv g \sum_\psi q_{B-L,\psi} \vect{r}_\psi$ coupled to the electric component of the $B-L$ field strength $\vect{E}'$.

We can write the $B-L$ current $J^{\mu}_{B-L}$ as the sum of the electromagnetic current $J^{\mu}_\text{EM}$ of eq.~\ref{eq:photgauge} and the neutron vector current, as per the charge assignments of eq.~\ref{eq:qBL}. Since nuclear motion can be taken to be ``frozen'' for electronic transitions to leading order in the Born-Oppenheimer approximation, a $B-L$ vector causes the same transition phenomenology as a kinetically mixed hidden photon provided we make the replacement
\begin{align}
\epsilon e  \leftrightarrow g. \label{eq:replacement1}
\end{align}
The separability of the $B-L$ current means that we can decompose the dipole moment as $\vect{\mu}_{B-L} = (g/e)\vect{\mu}_{e}+g \vect{\mu}_{n}$ with $\vect{\mu}_n$ the neutron number dipole moment. Hence, vibrational transitions in a diatomic molecule are caused by the effective operator:
\begin{align}
\delta H &  = - g \left(\frac{\vect{\mu}_{e}}{e}+\vect{\mu}_{n} \right) \cdot \vect{E}' \\
& = g \vect{E}' \cdot \vect{R} \left[\delta_e(R) + \frac{(A_2 - Z_2)M_1 - (A_1 - Z_1) M_2}{M_1+M_2} \right] \nonumber
\end{align}
In the first line, we used the same $\delta_e(R)$ as in eq.~\ref{eq:dHEdipvib}, and took $Z_i$, $A_i$, and $\vect{R}_i$ to be the atomic number, mass number, and position vector for the $i$th nucleus, such that the second two terms represent the interaction with the neutron number current. In the second line, we isolated the component of these terms that depends on the internuclear separation $\vect{R} \equiv \vect{R}_2 - \vect{R}_1$, and neglected terms acting on the molecular center-of-mass position.
So again, we find that, to leading order, vibrational absorption rates of a $B-L$ vector are exactly analogous to those of a kinetically mixed vector, provided we make the replacement:
\begin{align}
\epsilon e \delta_{e,1} \leftrightarrow g \left[\delta_{e,1} + \frac{(A_2 - Z_2)M_1 - (A_1 - Z_1) M_2}{M_1+M_2} \right].
\label{eq:replacement2}
\end{align}
This means that one can expect any E1 transition to be sensitive to absorption of a $B-L$ vector, barring accidental cancellations. In addition, even electric-dipole-forbidden transitions, with $\delta_{e,1} = 0$, may give appreciable absorption rates. For example, the molecule ${}^1$H${}^2$H has no electric dipole transition moment (by symmetry), but has $[(A_2 - Z_2)M_1 - (A_1 - Z_1) M_2]/(M_1+M_2) \approx 2/3$.

\begin{figure}
\includegraphics[width=0.48\textwidth]{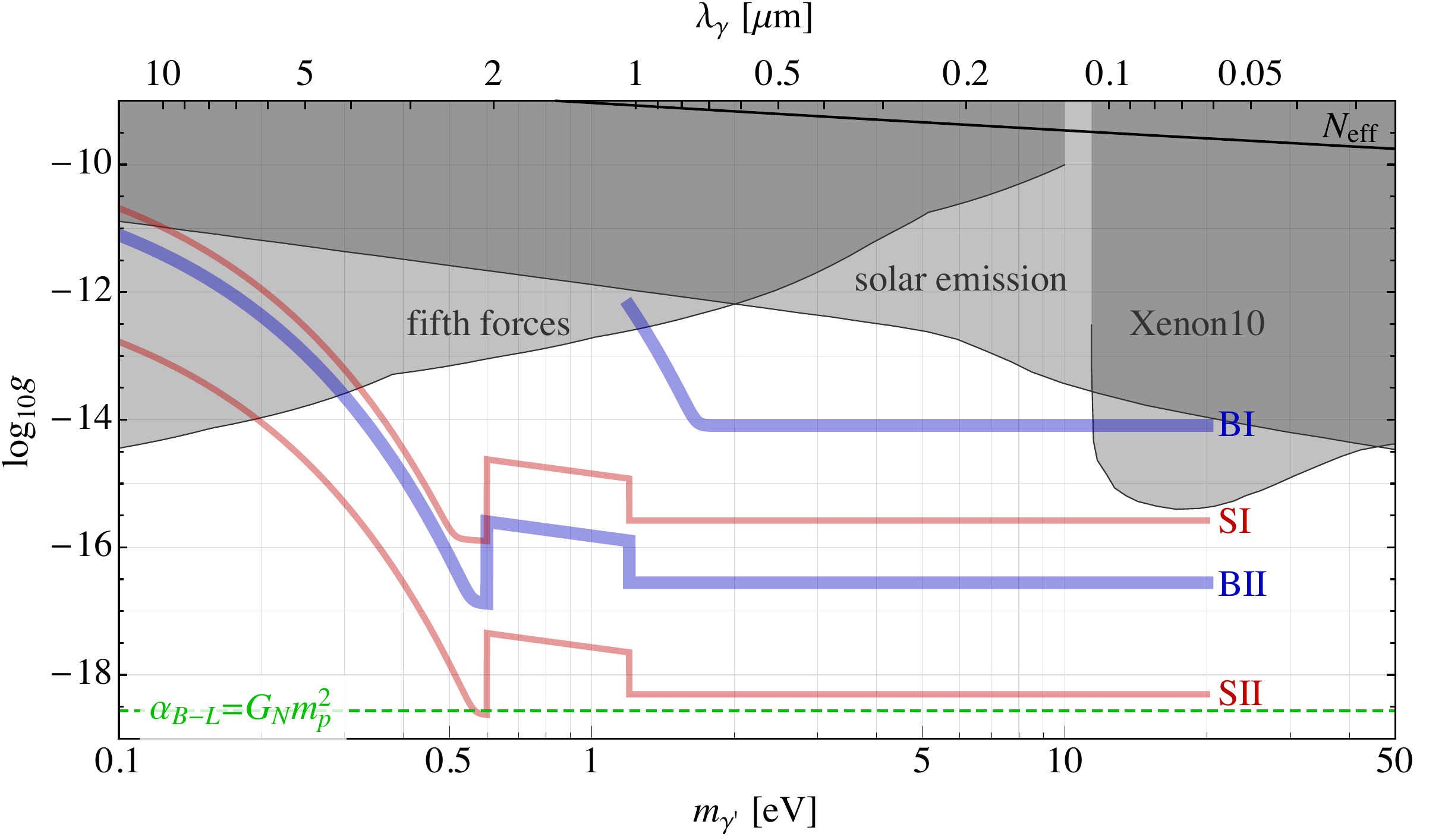}
\caption{Reach for the gauge coupling $g$ of a $B-L$ vector with mass $m_{\gamma'}$. Gross sensitivity projections and exclusion regions are as in fig.~\ref{fig:reachhf}, with the addition of an exclusion region coming from searches for short-range fifth forces. We also indicated in dashed green the $B-L$ gauge coupling that would yield a Coulomb force between nucleons with a strength equal to that of gravity.}\label{fig:reachbl}
\end{figure}

In figure~\ref{fig:reachbl}, we plot the reach estimates for our four proposed configurations. They are exactly analogous to those of a kinetically mixed photon with the replacements of eqs.~\ref{eq:replacement1}~and~\ref{eq:replacement2}, except now a Bulk detector can potentially also look for E1-forbidden $\Delta v = 1$ transitions with low quenching rates. The exclusion regions from the hidden photon apply in exactly the same way, modulo the rescaling of eq.~\ref{eq:replacement1}. In addition, a $B-L$ vector mediates a finite-range ``Coulomb'' force between electrically neutral bodies: the electrons and proton charges cancel each other out, leaving a like-charge repulsive force between neutron pairs. Null results from short-distance gravity tests~\cite{adelberger2003tests} place significant constraints at low masses, indicated by the gray exclusion region labeled ``fifth forces'' in fig.~\ref{fig:reachbl}.


\subsection{Scalars}\label{sec:scalars}

In this section, we first summarize the basics of scalar dark matter, which is analogous to that of vector dark matter in sec.~\ref{sec:vectors} and nearly identical to that of pseudoscalar dark matter in sec.~\ref{sec:pseudoscalars}. Afterwards, we will demonstrate how the proposed experiment can be sensitive to scalar couplings to electrons, photons, quarks, and gluons

If a light scalar (or pseudoscalar) field exists in the spectrum of the theory, then cold ensembles of scalar particles are naturally produced in the early Universe via the field misalignment mechanism~\cite{preskill1983cosmology,dine1983not,abbott1983cosmological}. As in the massive vector case, an abundance of nonrelativistic, weakly coupled scalars form an effectively pressureless fluid, and are thus excellent dark matter candidates (see~\cite{marsh2016axion} for a review of the relevant cosmology). With a Lagrangian of the form
\begin{align}
\mathcal{L} = \frac{1}{2} \partial_\mu \phi \partial^\mu \phi - \frac{1}{2} m_\phi^2 \phi^2 + \delta \mathcal{L}, \label{eq:Lspin0}
\end{align}
and sufficiently weak couplings of $\phi$ in $\delta \mathcal{L}$, scalar dark matter is expected to be a mixed-state superposition of plane waves of the form
\begin{align}
 \phi(t,\vect{x}) = \phi_0  \cos\left[m_\phi(1+v^2/2)t - m_\phi \vect{v}\cdot \vect{x} + \alpha_{\vect{v}} \right] \label{eq:phifield}
\end{align}
drawn from a velocity distribution $f(\vect{v})$ like that of eq.~\ref{eq:fv} with random phases $\alpha_{\vect{v}}$.
The amplitude $\phi_0$ can be thought of as the typical magnitude of the field oscillation; more precisely, we take it to be $\phi_0 = \sqrt{2 \langle \phi(t, \vect{x})^2 \rangle_{\vect{v},\alpha} \rangle }$. The field configuration  then carries an energy density of approximately $\rho = m_\phi^2 \phi_0^2/2$. If it makes up all of the local DM energy density, the field amplitude is expected to be:
\begin{align}
\tilde{\phi}_0 \equiv \sqrt{4\pi G_N} \phi_0 \approx 6 \times 10^{-31} \left(\frac{1~\eV}{m_\phi}\right),\label{eq:phiamplitude}
\end{align}
where we have normalized the amplitude in Planck units ($G_N$ is Newton's gravitational constant). We will use the same notation for the dimensionless field $\tilde{\phi} \equiv \sqrt{4\pi G_N} \phi$.

A light, parity-even scalar may couple to parity-even matter operators (in which case the scalar is often called a modulus or dilaton-like field) of the form:
\begin{align}
\delta \mathcal{L} = ~&  \tilde{\phi} \Big[-d_{m_e} m_e \bar{\psi_e} \psi_e + \frac{d_e}{4} F_{\mu\nu}F^{\mu\nu} \label{eq:modcoupling}\\
&- \hspace{-0.3em}\sum_{q = u,d,s} \left( d_{m_q} + \gamma_{q} d_g\right) m_q \bar{\psi}_q \psi_q - \frac{d_g \beta_3}{2g_3}G_{\mu \nu}^A G^{A \mu \nu}  \Big]. \nonumber
\end{align}
Above, $m_e$ is the electron mass, $\psi_e$ is the electron field, $F_{\mu\nu}$ is the electromagnetic field strength, $\gamma_{q}$ is the anomalous dimension of the quark field $\psi_q$ (up $\psi_u$, down $\psi_d$, and strange $\psi_s$) with mass $m_q$, $\beta_3$ and $g_3$ are the QCD beta function and gauge coupling, and $G^A_{\mu\nu}$ is the QCD field strength. The dimensionless couplings $d_{m_e}$, $d_e$, $d_{m_i}$, $d_g$ parametrize the strength of the leading linear couplings of $\phi$ to electrons, photons, quarks, and gluons, respectively, here written in an effective Lagrangian at a scale just above the QCD confinement scale $\Lambda_3$. In most cases, one can ignore the effects of other higher-dimensional operators of SM fields, and couplings quadratic and higher-order in $\phi$. (One exception is when the linear couplings are absent or highly suppressed, e.g.~via a parity symmetry under $\phi \to - \phi$. In this case, all of our results apply with a straightforward replacement of $\phi \leftrightarrow \phi^2$, $\omega \simeq 2 m_\phi$, and an appropriate field rescaling, while constraints from other experiments typically change qualitatively.)

\begin{figure*}[t]
\includegraphics[width=0.7\textwidth,trim={0cm 0cm 0 0},clip]{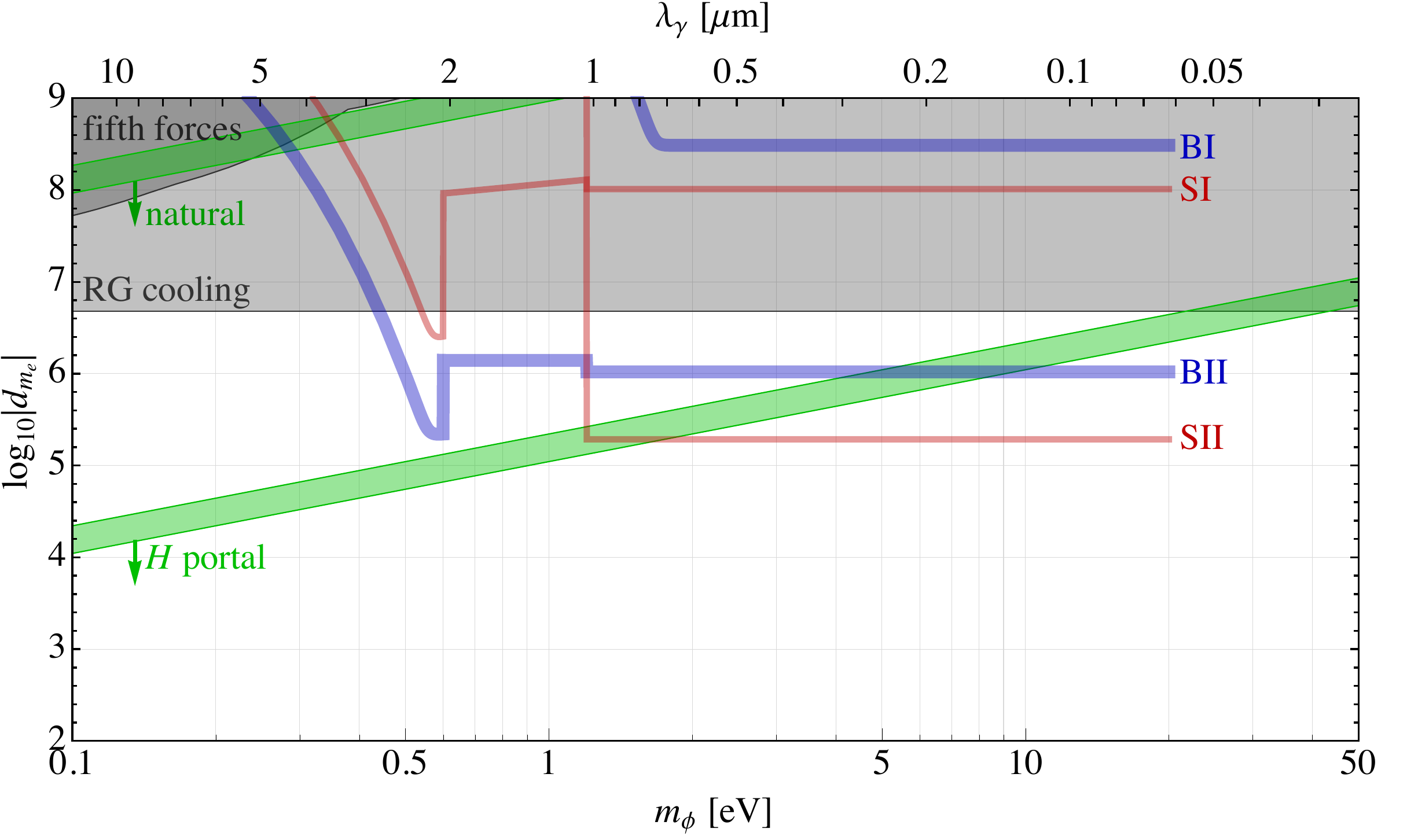}\\
\caption{Reach for the scalar dark matter coupling $d_{m_e}$ to electrons as a function of scalar mass $m_\phi$. Sensitivity projections for the Bulk (BI, BII) and Stack (SI, SII) configurations are shown with assumptions as listed in table~\ref{tab:scalaroperators}. Gray exclusion regions depict constraints from stellar cooling processes in red giants (RG) and from short-range fifth force searches. Parameter space below the upper green band is technically natural (i.e.~not fine tuned) for a 10~TeV cutoff, while the $H$ portal model (see eqs.~\ref{eq:higgsportal}~and~\ref{eq:higgsportalf}) populates the region below the lower green band.} \label{fig:reachscalar1}
\end{figure*}

\begin{figure}[t]
\includegraphics[width=0.48\textwidth,trim={0cm 1cm 0 0cm},clip]{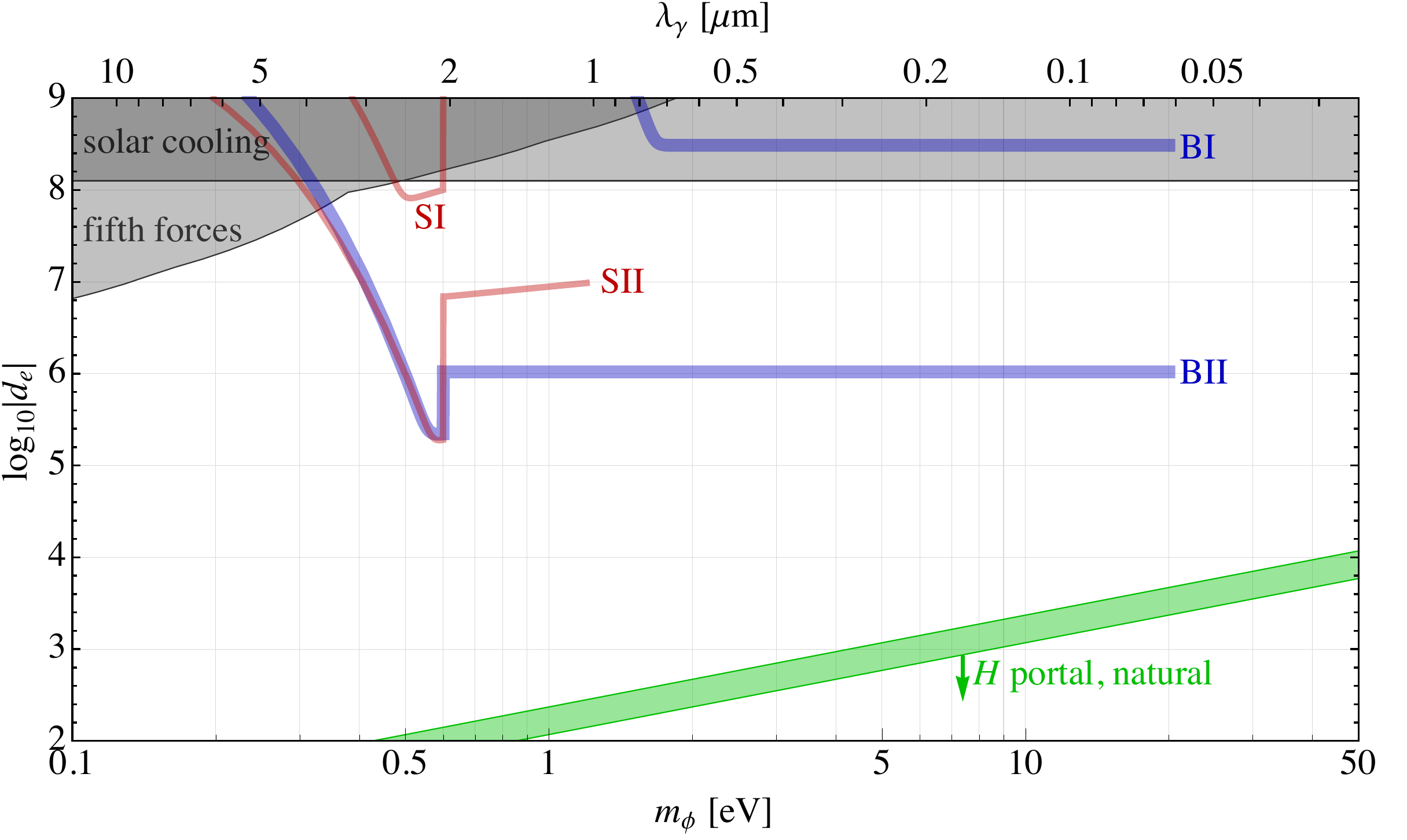}\\
\includegraphics[width=0.48\textwidth,trim={0cm 1cm 0 1cm},clip]{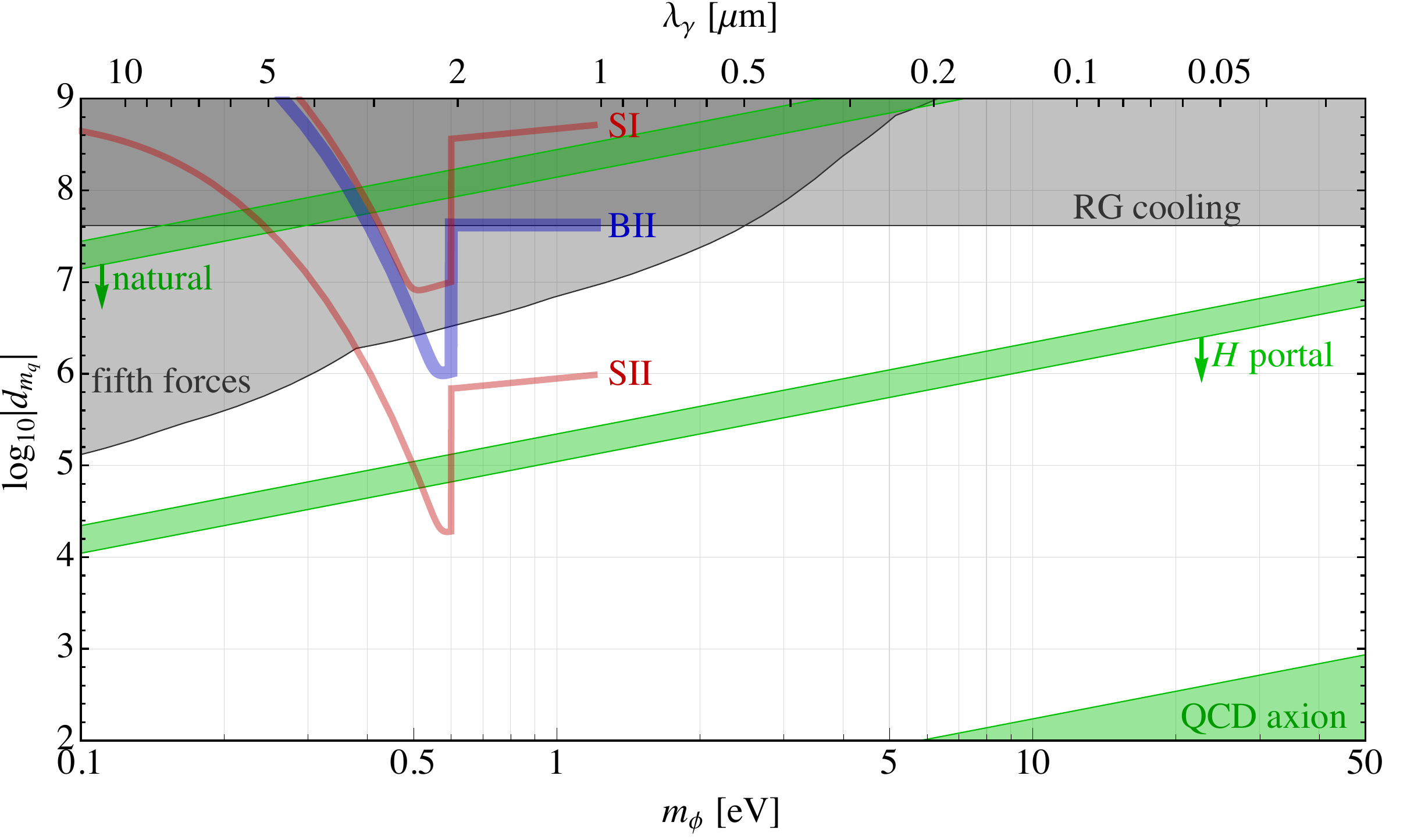}\\
\includegraphics[width=0.48\textwidth,trim={0cm 0cm 0 1cm},clip]{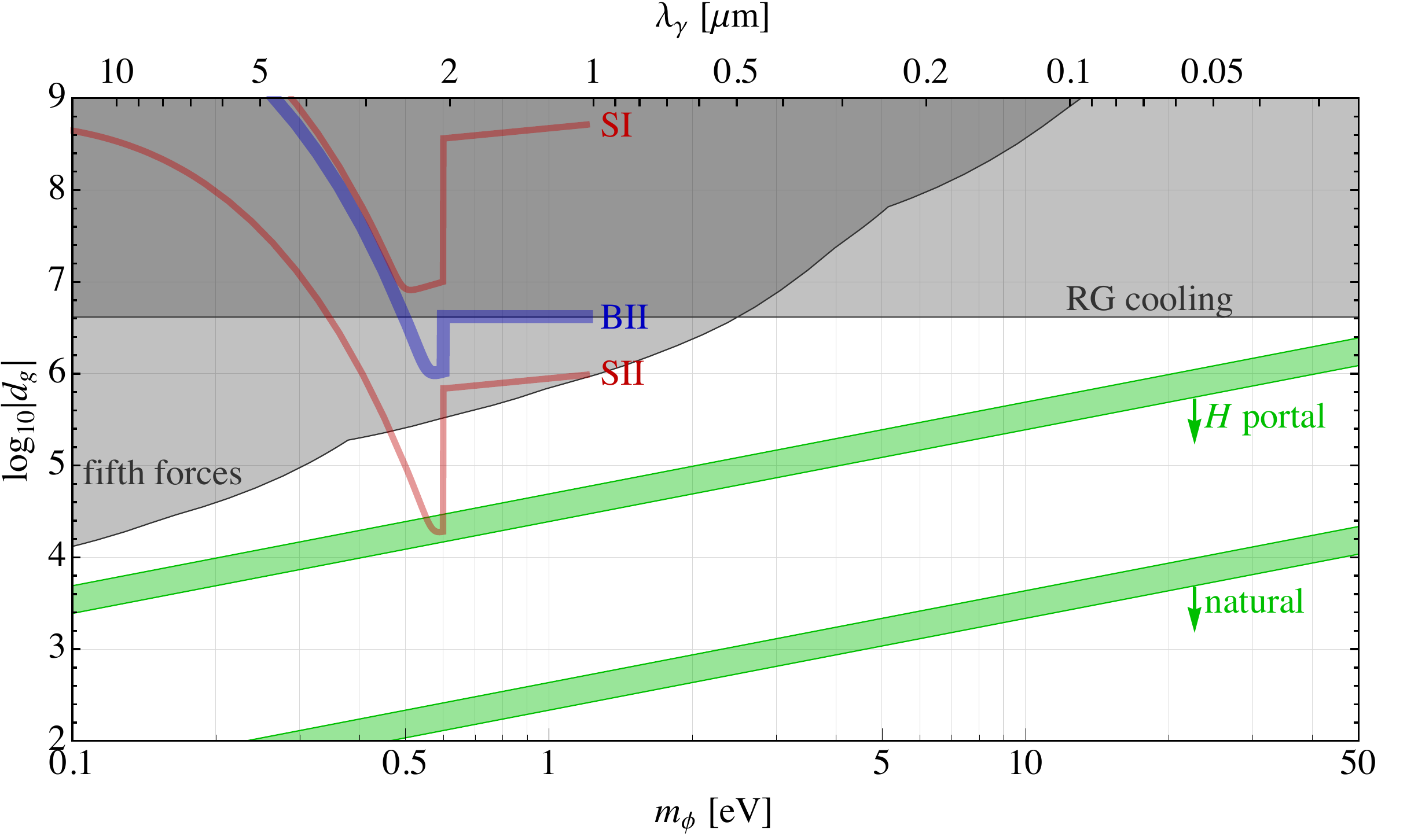}
\caption{Reach for scalar dark matter couplings to photons ($d_e$), light quarks ($d_{\hat{m}_q}$), and gluons ($d_g$) as a function of scalar mass. Labels are as in fig.~\ref{fig:reachscalar1}, except the cooling bound on $d_e$ comes from cooling processes in the Sun rather than red giants. We also show the experimentally allowed region for the scalar coupling $d_{\hat{m_q}}$ of the QCD axion.} \label{fig:reachscalar2}
\end{figure}

We have written the couplings of the scalar field as a low-energy effective theory at the GeV scale, only having included the most relevant operators while remaining agnostic about the theoretical origins of the scalar in the ultraviolet. The simplest UV completion of the couplings in eq.~\ref{eq:modcoupling} is that of the linear Higgs portal~\cite{piazza2010sub}
\begin{align}
\delta \mathcal{L} = b \phi H^\dagger H, \label{eq:higgsportal}
\end{align}
with $b$ a dimension-1 coupling and $H$ the SM Higgs field. At energies far below the Higgs mass, $\phi$ inherits part of the couplings of the Higgs to other SM fields due to the small mixing term that eq.~\ref{eq:higgsportal} induces. Fermion couplings, including to electrons and quarks, are of order:
\begin{align}
d_{m_e} = d_{m_q} = \frac{b}{\sqrt{4\pi G_N} m_h^2} \approx 2.2 \times 10^5 \left(\frac{b}{\text{eV}}\right)\label{eq:higgsportalf}
\end{align} 
with $m_h \approx 125~\GeV$ the Higgs boson mass, while couplings to gauge bosons, $d_e$ and $d_g$, are suppressed~\cite{piazza2010sub}. We indicate on the plots in figs.~\ref{fig:reachscalar1}~and~\ref{fig:reachscalar2} the induced couplings for the Higgs portal model for $b < m_\phi$. (Note that the \emph{relative} couplings to light SM fields in the Higgs portal model are correlated; in particular, their ratios are fixed, e.g.~as in eq.~\ref{eq:higgsportalf}.) Larger couplings naively would destabilize the scalar potential, barring any other mechanism or fine tuning.

Light scalar fields with parity-even couplings can also arise in theories with an extended gravitatational sector, such as the dilaton in string theory~\cite{damour1994string,taylor1988dilaton} or a radion in a theory with extra spatial dimensions~\cite{arkani2000solving,burgess2011naturalness,cicoli2011anisotropic}. In addition, theories with spontaneously broken supersymmetry and/or flavor symmetries usually abound in light moduli fields~\cite{dimopoulos1996macroscopic}. All of the above examples have extra structure beyond that indicated in eq.~\ref{eq:modcoupling}. If new fields associated with this new dynamics come in at a scale $\Lambda$, then $\phi$ generically receives a mass-squared correction of order:
\begin{align}
\Delta m_\phi^2 \sim \frac{G_N \Lambda^4}{4\pi} \left[\frac{d_{m_e}^2 y_e^2}{(4\pi)^2} + \sum_q \frac{d_{m_q}^2 y_q^2}{(4\pi)^2}  +  d_e^2 + d_g^2\right],\label{eq:naturalness}
\end{align}
where $y_e$ and $y_q$ are the electron and quark Yukawa couplings to the Higgs field, and we have assumed $\Lambda \gg m_h$. Without accidental cancellations, one would expect $m_\phi^2 \gtrsim \Delta m_\phi^2$, an inequality we plot for each of the couplings individually in figs.~\ref{fig:reachscalar1}~and~\ref{fig:reachscalar2} for $\Lambda = 10~\text{TeV}$, an energy scale not yet directly explored in collider physics. We stress that parameter space above this ``natural'' region is by no means excluded or unattainable. The scalar may be regarded as a composite particle (not unlike the pion in QCD) at a much lower scale in some theories; see~\cite{arvanitaki2016small} for a notable radion construction along these lines. Alternatively, there may be anthropic pressures for tuning the mass of the DM particle.

In a CP-violating background such as a QCD vacuum with nonzero $\theta$ angle, the QCD axion $a$ will also pick up parity-even couplings to mesons and nucleons~\cite{pospelov1998cp}, which can be expressed in terms of an equivalent quark coupling bounded to the interval:
\begin{align}
\frac{10^{-16}}{\sqrt{4\pi G_N} f_a} \lesssim d_{\hat{m}_q} \lesssim \frac{3\times 10^{-11}}{\sqrt{4\pi G_N} f_a}
\end{align}
with $f_a$ the QCD axion decay constant. The upper bound comes from null results in searches for a neutron electric dipole moment, constraining $|\theta| \lesssim 10^{-10}$, while the lower ``bound'' is the minimum natural size predicted in the Standard Model. At higher loop order, one would also expect the other parity-even couplings, but here we focus on the quark coupling only.

The effective Lagrangian in eq.~\ref{eq:modcoupling} is written in such a way~\cite{damour2010equivalence} that low-energy masses and couplings have the simple functional dependence on the background value of $\phi$:
\begin{align}
m_e \left[\phi(t,\vect{x}) \right] & = m_e \left(1 + d_{m_e} \tilde{\phi}(t,\vect{x}) \right); \label{eq:varcoupling1}\\
\alpha \left[\phi(t,\vect{x}) \right] & = \alpha \left(1 + d_{e} \tilde{\phi}(t,\vect{x}) \right); \label{eq:varcoupling2}\\
m_q\left[\phi(t,\vect{x}) \right] & = m_q \left(1 + d_{m_q} \tilde{\phi}(t,\vect{x}) \right); \label{eq:varcoupling3}\\
\Lambda_3\left[\phi(t,\vect{x}) \right] & = \Lambda_3 \left(1 + d_{g} \tilde{\phi}(t,\vect{x}) \right).\label{eq:varcoupling4}
\end{align}
We will often focus on the symmetric combination of the light quark masses, which has a dependence
\begin{align}
\hat{m}_q\left[\phi(t,\vect{x}) \right] & = \hat{m}_q \left(1 + d_{\hat{m}_q} \tilde{\phi}(t,\vect{x}) \right),
\end{align}
with $\hat{m}_q \equiv (m_u + m_d)/2$ and $d_{\hat{m}_q} \equiv (d_{m_u} m_u + d_{m_d} m_d)/(m_u + m_d)$. A neutral atom with nucleon number $A$ and atomic number $Z$ has a mass $M$ that scales nearly linearly with $\Lambda_3$ but also functionally depends on the pion mass (and thus the sum of quark masses $m_u + m_d$) and the fine-structure constant due to binding energy effects, as well as the electron mass:
\begin{align}
M[\phi(t,\vect{x})] \simeq  M\big\lbrace 1  +  \tilde{\phi}(t,\vect{x})  [&(d_g + (d_{\hat{m}_q} - d_g) Q_{\hat{m}_q} \\
& + (d_e - d_g) Q_e + d_{m_e} Q_{m_e} ]\big\rbrace,\nonumber
\end{align}
where we left out subdominant terms coming from e.g.~the strange quark mass dependence of the nucluear mass, or its dependence on the light-quark mass difference $m_d - m_u$. The ``dilaton charges'' $Q_{\hat{m}_q}$, $Q_e$ and $Q_{m_e}$ have been worked out in ref.~\cite{damour2010equivalence}, and roughly obey the following empirical formulae across the periodic table:
\begin{align}
Q_{\hat{m}_q} \approx & +9.3 \times 10^{-2} - 3.6 \times 10^{-2} \frac{1}{A^{1/3}} \label{eq:dilaton1}\\
& - 2.0 \times 10^{-2} \frac{(A-2Z)^2}{A^2} - 1.4 \times 10^{-4} \frac{Z(Z-1)}{A^{4/3}}, \nonumber \\
Q_e  \approx ~&-1.4 \times 10^{-4} + 8.2 \times 10^{-4} \frac{Z}{A} \label{eq:dilaton2} \\
 & + 7.7 \times 10^{-4} \frac{Z(Z-1)}{A^{4/3}}, \nonumber\\
Q_{m_e} \approx & + 5.5 \times 10^{-4} \frac{Z}{A}\label{eq:dilaton3}.
\end{align}
Spatial and temporal field oscillations in $\phi$ such as those in eq.~\ref{eq:phifield} give rise to fractional variations in e.g.~the electron mass with amplitude given by the coupling constant $d_{m_e}$ times the field amplitude in eq.~\ref{eq:phiamplitude}.



The vibrational Hamiltonian of a diatomic molecule (having integrated out electronic motion) in the presence of a modulus field is:
\begin{align}
H =~& \frac{-\vect{\nabla}_{\vect{R}}^2}{2 M[\phi(t,\vect{x})]} + \frac{k_e[\phi(t,\vect{x})]}{2} (R-R_e[\phi(t,\vect{x})])^2 \nonumber \\
 =~& H_0 + \delta H_\text{I}^{0} + \delta H_\text{II}^{0} + \delta H_\text{III}^{0}+\dots \label{eq:Hscalarvib1} ;\\
\delta H_\text{I}^{0} =~&\tilde{\phi}(t,\vect{x}) (d_{m_e} + d_e) k_e R_e (R-R_e), \label{eq:HI0}\\
\delta H_\text{II}^{0} =~&\tilde{\phi}(t,\vect{x}) (3 d_{m_e} + 4 d_e) \frac{k_e}{2}(R - R_e)^2, \label{eq:HII0}\\
\delta H_\text{III}^{0} =~&\tilde{\phi}(t,\vect{x})(d_g + Q_{\hat{m}_q,\text{eff}} d_{\hat{m}_q}) \frac{\vect{\nabla}_{\vect{R}}^2}{2M} . \label{eq:HIII0}
\end{align}
We defined by $Q_{\hat{m}_q,\text{eff}} \equiv (Q_{\hat{m}_q,1} M_2 + Q_{\hat{m}_q,2} M_1)/(M_1 + M_2)$ the effective dilaton quark charge of the reduced mass $M \equiv M_1 M_2/(M_1 + M_2)$ of the diatomic molecule. In the second line, we have only kept $\tilde{\phi}$ terms with off-diagonal matrix elements, and see that all monopole operators listed in eq.~\ref{eq:HMon1} are generated. The first interaction from $\delta H_\text{I}^{0}$ is the quantum-mechanical operator corresponding to the classical effect of fractional oscillations in the equilibrium size of atoms, upon which the modulus dark matter searches of \cite{arvanitaki2016sound,branca2016search} are based. The other two terms are reminiscent of classical parametric resonance, in that they primarily cause $\Delta v = 2$ transitions when the driving field oscillates at \emph{twice} the harmonic oscillator frequency.

To leading order, we can ignore the spatial dependence in $\phi(t,\vect{x})$ in the matrix elements of the operators in eq.~\ref{eq:Hscalarvib1}. However, we have neglected a tidal force on the diatomic molecule that comes about from the field-induced spatial variation of the nuclear masses: 
\begin{align}
M_N\left[\vect{R}_N\right] \simeq M_N \left(1 +  d_j Q_{j,N}\vect{R}_N \cdot \vect{\nabla} \tilde{\phi}(t,\vect{x})  \right),\label{eq:Mspatial}
\end{align}
Here, $N = 1,2$ runs over the nuclear labels, and $j$ runs over the dilaton charges, the two most important of which are listed in eqs.~\ref{eq:dilaton1}~and~\ref{eq:dilaton2}. This spatial variation leads to a dipole Hamiltonian that acts on the internal state of the molecules:
\begin{align}
\delta H_\text{I}^{1} & = \sum_{N = 1}^2 M_N\left[\vect{R}_N\right] \simeq  d_j  \Delta Q_{j} M \vect{R} \cdot \vect{\nabla}\tilde{\phi}(t,\vect{x}) +\dots , \label{eq:HI1} 
\end{align}
with $\Delta Q_{j} \equiv Q_{j,2} - Q_{j,1}$ the difference in atomic dilaton charges. The ellipsis indicate terms acting on the center-of-mass degrees of freedom.
Heteronuclear diatomics with significantly different dilaton charges may thus experience dipole vibrational transitions as well.
Amusingly, the interaction in eq.~\ref{eq:HI1} is the spectroscopic analog of the differential force macroscopic bodies experience in a dark matter background of very light moduli~\cite{arvanitaki2015searching,graham2016dark}.

Finally, modulus couplings to electrons and photons can cause monopole electronic transitions of the type anticipated in eq.~\ref{eq:HMon2}.
In the presence of a modulus field, and with the nuclear motion frozen, the electronic part of the Hamiltonian in eq.~\ref{eq:H0} becomes:
\begin{align}
H  =~& \sum_{n=1}^{Z_1 + Z_2} \frac{-\vect{\nabla}_{e_n}^2}{2 m_e[\phi(t,\vect{x})]} + \alpha[\phi(t,\vect{x})] \label{eq:Hscalare1}\\
& \hspace{0.5em} \times \left[ \sum_{n<m}^{Z_1 + Z_2} \frac{1}{|\vect{r}_{e_n}-\vect{r}_{e_m}|} - \sum_{n=1}^{Z_1+Z_2} \sum_{N=1}^2 \frac{Z_N}{|\vect{r}_{e_i}-\vect{R}_N|} \right]. \nonumber
\end{align}
Extracting the $\phi$-dependent parts, and using the identity of eq.~\ref{eq:Hmonvir}, we find a transition operator
\begin{align}
\delta H^{0}_\text{IV} = \tilde{\phi}(t,\vect{x}) (d_{m_e} + d_e) \sum_{n=1}^{Z_1 + Z_2} \frac{-\vect{\nabla}_{e_n}^2}{2 m_e},\label{eq:HIV0}
\end{align}
whose transition matrix elements have to be estimated or calculated from first principles. As a first approximation, we use the NDA estimate of eq.~\ref{eq:Hmonvir}. 

The spatial variation in the electron mass  $m_e[\phi(t,\vect{x})]$ can also lead to dipole transitions via the operator
\begin{align}
\delta H_\text{II}^{1} \simeq  d_{m_e} m_e  \vect{\nabla} \tilde{\phi}(t,\vect{x}) \cdot \sum_{n = 1}^{Z_1 + Z_2} \vect{r}_{e,n} \label{eq:HII1}
\end{align}
whose transition matrix elements are suppressed by a factor of $v_\text{lab}/\alpha$ relative to those of the operator in eq.~\ref{eq:HIV0} however. It is clear from the form of eq.~\ref{eq:HII1} that we may view this interaction as that of an effective electric dipole operator with $[e \vect{E}]_\text{eff} = \sqrt{4\pi G_N} d_{m_e} m_e \vect{\nabla} \phi(t,\vect{x})$. At subleading order in the Born-Oppenheimer approximation, modulus-induced spatial variation in the nuclear mass can also cause electronic transitions and thus extend the sensitivity to $d_g$ and $d_{\hat{m}_q}$ to higher masses, but with greatly reduced transition amplitudes.

In figure~\ref{fig:reachscalar1}, we show the reach of the proposed setups to the scalar electron coupling $d_{m_e}$ as a function of the scalar mass $m_\phi$, while the three panels of fig.~\ref{fig:reachscalar2} show the same for the scalar couplings to photons, gluons, and light quarks, respectively. At any point in those parameter spaces, there are several operators contributing to absorption, although one is typically dominant. In general, the Bulk and Stack setups achieve highest signal rates for different operators even at the same point in parameter space, because the Stack setup is insensitive to E1-forbidden transitions, so it can never probe e.g.~monopole operators. We have summarized in table~\ref{tab:scalaroperators} which operators of eqs.~\ref{eq:HI0}, \ref{eq:HII0}, \ref{eq:HIII0}, \ref{eq:HI1}, \ref{eq:HIV0}, and \ref{eq:HII1} drive the estimated optimal sensitivity reach in figs.~\ref{fig:reachscalar1}~and~\ref{fig:reachscalar2}, and our assumptions for the matrix element sizes. We broke down the parameter space into three regions: $\omega \lesssim 0.6~\eV$ (where both $\Delta v = 1$ and $\Delta v = 2$ transitions can occur), $0.6~\eV \lesssim \omega \lesssim 1.2~\eV$ (only $\Delta v = 2$ transitions), $\omega \gtrsim 1.2~\eV$ (only electronic transitions).

\bgroup
\def\arraystretch{1.6}
\begin{table}[t]
\resizebox{0.48\textwidth}{!}{%
\begin{tabular}{l l  c c c}
\hline
 					&   &~~ $\omega < 0.6~\eV$ &~~$0.6~\eV < \omega <1.2~\eV$~~ & $1.2~\eV < \omega$\\
\hline
\multirow{2}{*}{$d_{m_e},d_e$} 	& B & $\delta H_\text{I}^{0}$ & $\delta H_\text{II}^{0}$ & $\delta H_\text{IV}^{0}$\\
						   		& S & $\delta H_\text{I}^{1}$ & $\delta H_\text{I}^{1}$ & $\delta H_\text{II}^{1}$\footnote{$d_{m_e}$ only}\\
\multirow{2}{*}{$d_{\hat{m}_q},d_g$}	& B & $\delta H_\text{I}^{1}$ & $\delta H_\text{III}^{0}$ & ---\\
						   			& S & $\delta H_\text{I}^{1}$ & $\delta H_\text{I}^{1}$ & ---\\
\hline
\end{tabular}}
\caption{Operators from eqs.~\ref{eq:HI0}, \ref{eq:HII0}, \ref{eq:HIII0}, \ref{eq:HI1}, \ref{eq:HIV0}, \ref{eq:HII1}      driving the optimum sensitivity projections in figs.~\ref{fig:reachscalar1}~and~\ref{fig:reachscalar2} for the Bulk (B) and Stack (S) configurations. To estimate vibrational matrix elements, we made use of eqs.~\ref{eq:Mat1}--\ref{eq:Mat1b} and the parametric estimates of eq.~\ref{eq:paramest}. For $\delta H_\text{I}^{1}$, we assumed differential dilaton charges of $\Delta Q_{\hat{m}_q} \sim 1/30$, $\Delta Q_{e} \sim 1/300$, and $\Delta Q_{m_e} \sim 1/4000$. For electronic matrix elements, we utilized eq.~\ref{eq:Hmonvir} for $\delta H_\text{IV}^{0}$ and eq.~\ref{eq:dipmatvir} for $\delta H_\text{II}^{1}$. We also assumed $v_\text{lab} \sim 10^{-3}$ and $\phi$ makes up all of DM such that eq.~\ref{eq:phiamplitude} holds.}
\label{tab:scalaroperators}
\end{table}
\egroup

Short-distance tests of the inverse-square law for the gravitational force between neutral macroscopic bodies strongly constrain the low-mass end of the parameter space of interest~\cite{adelberger2003tests}. Given typical dilaton charges of neutral bodies as listed in eqs.~\ref{eq:dilaton1}~and~\ref{eq:dilaton2} for the quark and electromagnetic couplings, and assuming $Q_{m_e} \approx m_e/2m_p$ for the typical electron-mass dilaton charge, we recast these ``fifth-force'' tests as constraints on the individual couplings $d_j$. Regions excluded at 95\% CL are shown in gray in figs.~\ref{fig:reachscalar1}~and~\ref{fig:reachscalar2}. Absence of anomalous cooling rates in red giant stars sets 95\%-CL upper bounds of $|d_{m_e}| < 4.8 \times 10^6$,  $|d_{\hat{m}_q}| < 4.1 \times 10^7$, and  $|d_g| < 4.1 \times 10^6$~\cite{hardy2017stellar}, while solar cooling constrains $|d_e| < 1.3 \times 10^8$~\cite{raffelt1996stars}, all also shown in gray.


\subsection{Pseudoscalars}\label{sec:pseudoscalars}
Light bosonic dark matter can also consist of a pseudoscalar field, which we shall denote by $a$. Pseudoscalars differ from the scalars in the previous section only by their transformation under CP, simultaneous charge conjugation and parity (or equivalently time-reversal T by CPT conservation), which we take to be approximately conserved at low energies. Under a CP transformation, we assume $a \to - a$, while $\phi \to + \phi$. Equations~\ref{eq:Lspin0}~and~\ref{eq:phifield} hold just as well for a pseudoscalar with the obvious relabeling $\phi \leftrightarrow a$, while many of the early-Universe production mechanisms for spin-0 fields carry over too, in particular the misalignment mechanism. However, its odd transformation under CP symmetry means that $a$ must couple to CP-odd operators; we parametrize the interaction terms at energies below the QCD confinement scale as:
\begin{align}
\mathcal{L} = - \sum_{f = p,n,e} \frac{G_{aff}}{2} \partial_\mu a \bar{\psi}_f \gamma^\mu \gamma_5 \psi_f + \frac{G_{a\gamma\gamma}}{4} a F_{\mu\nu} \tilde{F}^{\mu\nu} \label{eq:Laxioncoupling} ,
\end{align}
with $\tilde{F}^{\mu\nu} \equiv (1/2) \epsilon^{\mu \nu \rho \sigma} F_{\rho \sigma}$. All of the above interactions  obey a shift symmetry in $a$, i.e.~they are invariant under $a \to a + c$ for any constant $c$. The shift symmetry implies that the interactions of eq.~\ref{eq:Laxioncoupling} do not renormalize the mass, so there is no analogous naturalness preference for the mass $m_a$ of $a$ as there was for that of $\phi$, cfr.~eq.~\ref{eq:naturalness}. 

The most famous light pseudoscalar particle is the QCD axion~\cite{weinberg1978new,wilczek1978problem}, whose presence in the theory would explain the observed smallness of the neutron's electic dipole moment $\vect{d}_n$ by a symmetry~\cite{peccei1977cp}, thus highly motivating its existence in Nature (in addition to being a possible DM candidate). The QCD axion, by construction, must couple to the Lagrangian operator
\begin{align}
\mathcal{L} \supset \frac{a}{f_a} \frac{\alpha_3}{8\pi}G_{\mu \nu} \tilde{G}^{\mu\nu},\label{eq:GGdual}
\end{align}
where $\alpha_3 = g_3^2/4\pi$, $f_a$ is the axion decay constant, and $\tilde{G}_{\mu \nu} \equiv \frac{1}{2} \epsilon_{\mu\nu\rho\sigma} G^{\rho \sigma}$ is the dual to the gluon field strength $G_{\mu \nu}$. Nonperturbative effects involving this term break the shift symmetry, and induce a scalar potential for $a$, including a mass-squared term with mass:
\begin{align}
m_a  \simeq \frac{\sqrt{m_u m_d}}{m_u + m_d} \frac{m_\pi f_\pi}{f_a} \approx 0.57~\eV \left(\frac{10^7~\GeV}{f_a}\right),\label{eq:mafa}
\end{align}
with $m_\pi$ and $f_\pi$ the pion's mass and decay constant respectively. The minimum of this potential occurs at a place where the coefficient of $G\tilde{G}$ is (very nearly) zero~\cite{vafa1984parity}, i.e.~at an axion field value where the QCD theta angle is zero and the strong CP problem is solved. However, if the axion energy density $\rho$ makes up part or all of the DM energy density, it will locally oscillate in its potential with a typical amplitude relative to $f_a$ of:
\begin{align}
\frac{a_0}{f_a} = \frac{\sqrt{2 \rho_a}}{m_a f_a} \approx 3.6 \times 10^{-19} \sqrt{\frac{\rho_a}{\rho_\text{DM}}},
\end{align}
the exact analogue of eq.~\ref{eq:phiamplitude}.
In other words, the neutron EDM is no longer zero (or constant), but takes on the field-dependent value~\cite{graham2013new}:
\begin{align}
\vect{d}_n[a(t,\vect{x})] = d_\theta \vect{\sigma}_n \frac{a(t,\vect{x})}{f_a} ,
\end{align}
with a coefficient $d_\theta \approx 2.4 \times 10^{-16} e\text{~cm}$ carrying a 40\% fractional uncertainty~\cite{pospelov1999theta}.

The QCD axion operator of eq.~\ref{eq:GGdual} also gives irreducible, low-energy contributions to the pseudoscalar operators of eq.~\ref{eq:Laxioncoupling}. In particular, one has $G_{a\gamma\gamma}\approx - 1.92(4)\alpha/2\pi f_a$ primarily from mixing with the pion, as well as $G_{a pp} \approx 0.47(3)/f_a$, $G_{a nn} \approx 0.02(3)/f_a$, and a two-loop suppressed $G_{aee}$~\cite{di2016qcd} for the purely hadronic QCD axion in the KSVZ benchmark model~\cite{kim1979weak,shifman1980can}. The couplings in eq.~\ref{eq:Laxioncoupling} are UV-dependent: e.g.~in the DFSZ benchmark model~\cite{dine1981simple,Zhitnitsky:1980tq}, one expects $G_{a\gamma\gamma}\approx 0.74(4)\alpha/2\pi f_a$, and $G_{aff} \sim \mathcal{O}(1)/f_a$ (depending on a continuous angle) for all SM fermions~\cite{di2016qcd}.

Axions also generically emerge out of string theory with exponentially suppressed masses~\cite{svrcek2006axions,arvanitaki2010string}, and other light axion-like particles naturally arise as pseudo-Nambu-Goldstone bosons of global symmetries broken at a high scale $f_a$, with low-energy masses and couplings typically scaling inversely proportional to $f_a$.

The first set of derivative interactions of $a$ with the proton $p$, the neutron $n$, and the electron $e$ in eq.~\ref{eq:Laxioncoupling} lead to the nonrelativistic interaction Hamiltonian (for single particles)
\begin{align}
\delta H = + G_{aff} \vect{\sigma}_f \cdot \left[ (\vect{\nabla}a) + (\partial_t a) \frac{-i\vect{\nabla}_f}{m_f}  \right], \label{eq:windops}
\end{align}
with $f = p,n,e$.
The first term in square brackets has a trivial action on the molecular wavefunction, and can only generate spin flips; it is the basis for cosmic axion search proposals at much lower masses~\cite{barbieri1989axion,graham2013new,sikivie2014axion,barbieri2017searching}, as well as searches for axion-mediated monopole-dipole and dipole-dipole forces~\cite{arvanitaki2014resonantly,moody1984new,tullney2013constraints,raffelt2012limits}. The second term in square brackets can excite (spin-)dipole transitions in molecules at much higher energies, and is the one we focus on here.
By NDA in nonrelativistic molecules, it can be seen that the pseudoscalar coupling to photons, the second term in eq.~\ref{eq:Laxioncoupling}, causes transitions subleading in strength as compared to the nuclear and electronic coupling in typical models, so we will not discuss it here.

The coupling to protons and neutrons generates a coupling to nuclei of the form $\delta H = G_{aNN} (\partial_t a) \vect{\sigma}_N \cdot \frac{-i\vect{\nabla}_N}{M_N}$, where we assume for brevity that $G_{aNN} = G_{app} \sim G_{ann}$ is the same for every nucleus with spin. In general, the coefficient will depend on the nuclear species $N$ as $G_{aNN} = c_{N,p} G_{app} + c_{N,n} G_{ann}$ with $c_{N,p}$ and $c_{N,n}$ coefficients of $\mathcal{O}(1)$, but we will ignore this complication. In a diatomic molecule with a spinless nucleus $N=2$, and in an unpolarized spin state for the $N=1$ nucleus, we find the following square Rabi frequency for $\Delta v = 1$ vibrational transitions:
\begin{align}
\Omega^2  & =  \Big| \langle v_f=1, J_f |G_{aNN} (\partial_t a)\vect{\sigma}_1 \cdot \frac{-i\vect{\nabla}_1}{M_1} |v_i = 0, J_i\rangle \Big|_\text{avg}^2 \nonumber \\
& \simeq \frac{G_{aNN}^2}{M_1^2} \rho_\text{DM} {M \omega_e} \left| \langle  J_f | \vect{\sigma}_1 \cdot \hat{\vect{R}} |, J_i\rangle \right|_\text{avg}^2 . \label{eq:OmegaGannvib}
\end{align}
To get to the second line, we used the vibrational amplitude of eqs.~\ref{eq:psigmaaction}~and~\ref{eq:Mat4}, and assumed that the pseudoscalar is all of the DM. The averaged rotational amplitudes are given below eq.~\ref{eq:dipmatrot}. Generalizations to amplitudes with multiple nuclear spins, polarized spin states, or nonharmonic $\Delta v = 2$ transitions are straightforward.

The irreducible neutron EDM operator of the QCD axion also gives rise to the operator
\begin{align}
\delta H = - \sum_{N=1}^2 \vect{d}_N \cdot \vect{E}(\vect{R_N}) = \frac{d_\theta}{e} \frac{a}{f_a}\sum_{N=1}^2 \frac{\vect{\sigma}_N}{Z_N} \cdot \left[ \vect{\nabla}_N, H_0\right] \label{eq:EDMop}
\end{align}
with $\vect{E}(\vect{R_N})$ the internal electric field of the molecule evaluated at the nuclear position $\vect{R}_N$. The second equality follows from  $\vect{E}(\vect{R_N}) = - \vect{\nabla}_N V(\vect{R_N})/ (eZ_N)$ with $V$ the potential energy terms of $H_0$ in eq.~\ref{eq:H0} and using canonical commutation relations. Famously, this operator has vanishing diagonal matrix elements due to Schiff's theorem~\cite{schiff1963measurability}, but its off-diagonal matrix elements do not vanish. Assuming that the nuclear EDM is similar to that of the neutron, $d_N \sim d_n$, and again starting with a diatomic with a single, unpolarized spin $\vect{\sigma}_1$, we find a square Rabi frequency:
\begin{align}
\Omega^2  & =  \Big|\langle v_f=1, J_f | \frac{d_\theta}{e} \frac{a}{f_a}\frac{\vect{\sigma}_1}{Z_1} \cdot \left[ \vect{\nabla}_1, H_0\right] |v_i = 0, J_i\rangle \Big|_\text{avg}^2 \nonumber \\
& \simeq \frac{d_\theta^2}{Z_1^2 e^2 f_a^2} \rho_\text{DM} M \omega_e \left| \langle  J_f | \vect{\sigma}_1 \cdot \hat{\vect{R}} |, J_i\rangle \right|_\text{avg}^2 . \label{eq:OmegaEDMvib}
\end{align}
We see that the nuclear pseudoscalar operator from eq.~\ref{eq:windops} generates exactly the same transitions as the EDM operator of eq.~\ref{eq:EDMop}. The latter's Rabi frequency is always smaller though, by the fraction $d_\theta M_1 / Z_1 e \simeq 2 d_\theta m_p / e \approx 2.2 \times 10^{-2}$ for $G_{aNN} f_a = 1$. Since its only effect is to give a subleading contribution to transitions already caused by the $G_{app}$ and $G_{ann}$ couplings---which are always generated as well---we will not make separate sensitivity projections for the EDM operator.
If the spin-dipole transition acts trivially on the nuclear spin state, then the transition can be E1-allowed; if the transition is associated with a combined spin flip in one or more nuclei, then it is always strongly forbidden.

Electronic transitions may be excited via the pseudoscalar coupling to electrons in a similar way:
\begin{align}
&\langle 1 | \delta H | 0 \rangle = i G_{aee} (\partial_t a) \omega_0  \langle 1 | \sum^{Z_1+Z_2}_{n=1} \vect{\sigma}_{e,n} \cdot \vect{r}_{e,n} | 0 \rangle;\\
&\Omega^2   =  G_{aee}^2 2 \rho_\text{DM} \delta_{e,2}^2 \omega_0 R_e^2 \left|\langle v_f'|v_i''\rangle \langle  J_f | \hat{\vect{\sigma}}_e \cdot \hat{\vect{R}} |, J_i\rangle \right|_\text{avg}^2. \label{eq:OmegaGaeevib}
\end{align}
We have used the matrix element identity of eq.~\ref{eq:spindipmatel} and defined its strength with $\delta_{e,2}$ as in eq.~\ref{eq:omegahfel}. 
We have notationally suppressed spin degrees of freedom. In a magnetic field, $\Delta \Sigma = \pm 1$ spin-flip transitions lines will receive Zeeman splittings relative to the spin-preserving ones $\Delta \Sigma = 0$, of size given in eq.~\ref{eq:Zeeman}. This Zeeman tuning can be used in the Bulk configuration to achieve efficient and uniform frequency coverage, as all lines receive a uniform shift by the same absolute amount. In absence of significant spin-orbit coupling, $\Delta \Sigma \neq 0$ emission rates are suppressed, so this scanning method is less suitable for use in the Stack setup with small molecules.

\begin{figure*}
\includegraphics[width=0.75\textwidth,trim={0cm 0cm 0 0},clip]{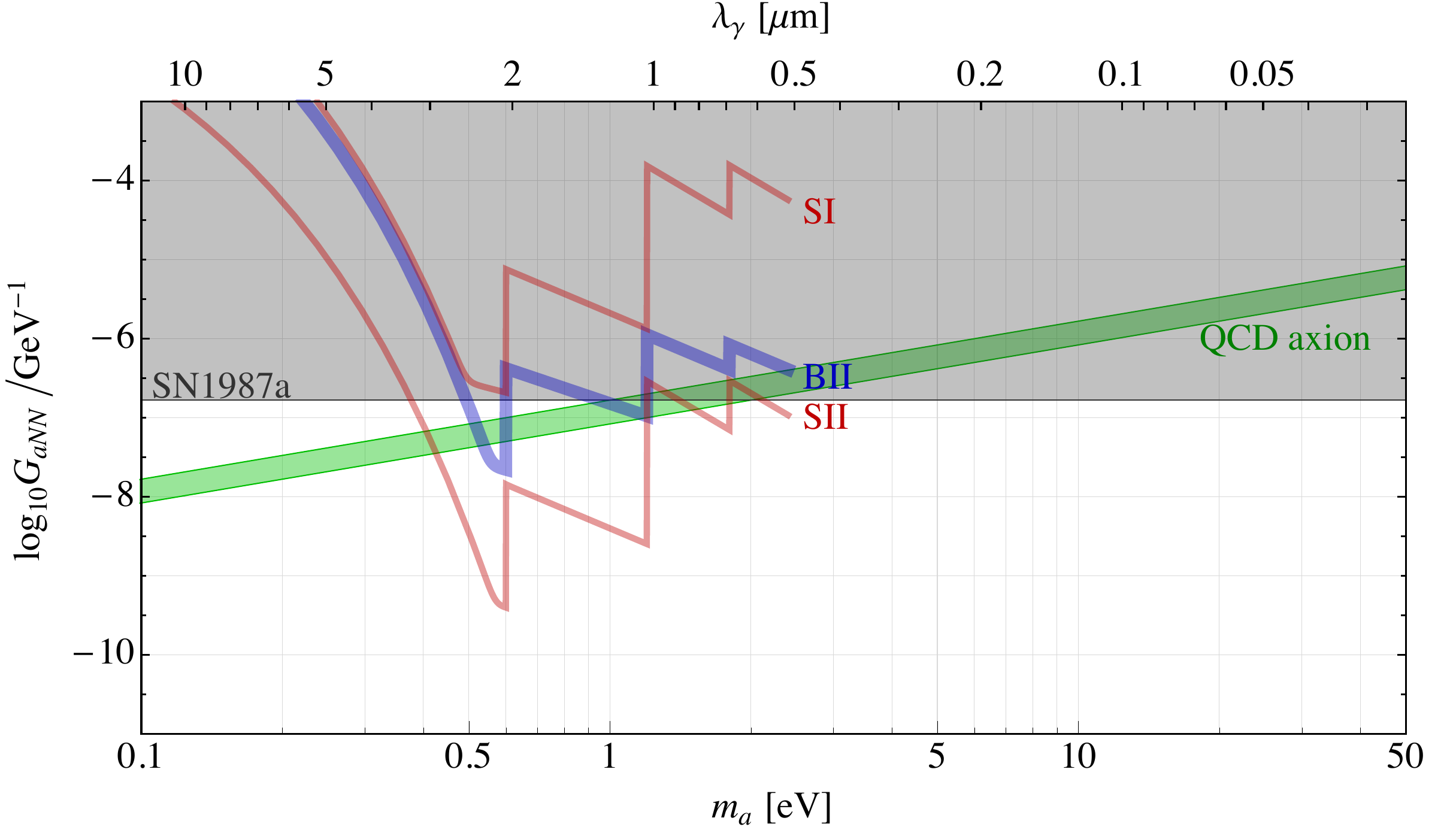}
\caption{Reach for the pseudoscalar coupling $G_{aNN}$ to nucleons as a function of pseudoscalar mass $m_a$, for the configurations BII (thick blue band), SI, and SII (thin red bands). The green band indicates the typical mass-coupling relation between $m_a$ and $G_{aNN} \sim 1/f_a$ for the QCD axion. The gray exclusion region is a ``robust'' bound set by a combination of the neutrino burst duration of SN1987a, and null observations of axion scattering events in water \v{C}erenkov detectors after the same supernova. 
}\label{fig:reachaxion1}
\end{figure*}

\begin{figure}
\includegraphics[width=0.48\textwidth,trim={0cm 0 0 0cm},clip]{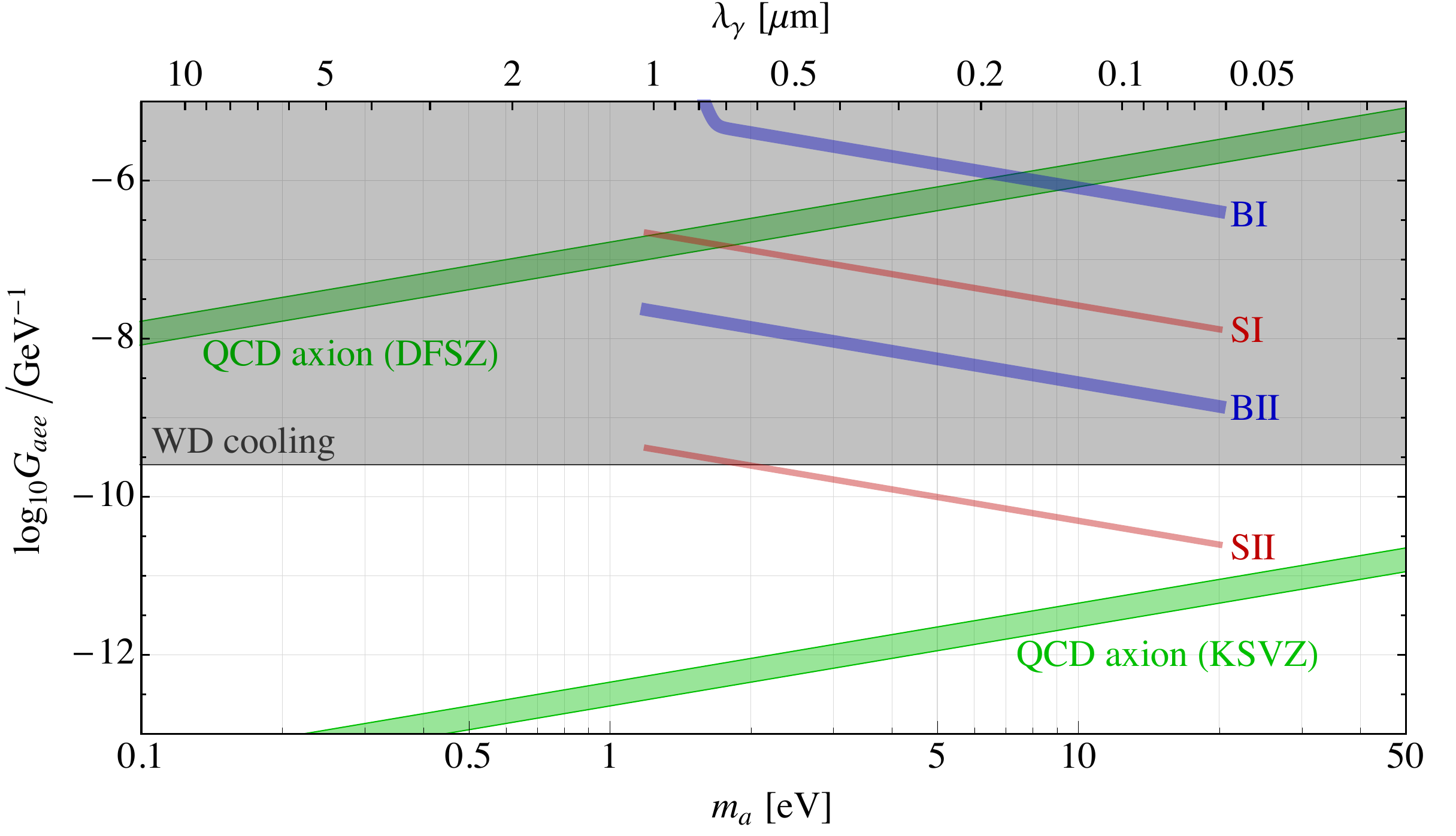}
\caption{Reach for the pseudoscalar coupling $G_{aee}$ to electrons as a function of pseudoscalar mass $m_a$ for the BI~{\&}~BII configurations (thick blue) and the SI~{\&}~II configurations (thin red). The green bands indicate the predicted coupling in two benchmark QCD axion models, namely DFSZ (top) and KSVZ (bottom). The gray exclusion region is excluded by indirect determinations of white dwarf cooling rates.}\label{fig:reachaxion2}
\end{figure}

In figure~\ref{fig:reachaxion1}, we plot gross sensitivity projections to the nuclear coupling $G_{aNN}$ with the four proposed setups for the $\Delta v = 1$ vibrational transitions calculated in eq.~\ref{eq:OmegaGannvib}, as well as for the first higher-harmonic $\Delta v= 2$ absorption. Figure~\ref{fig:reachaxion2} contains similar curves for the electronic transitions of eq.~\ref{eq:OmegaGaeevib} induced by $G_{aee}$. We also plot the typical mass-coupling relation---$f_a^{-1}/2 < G_{aNN} <  f_a^{-1}$ with $m_a$ and $f_a$ related as in eq.~\ref{eq:mafa}---for the QCD axion. Barring accidental fine tuning in \emph{two} separate couplings---$G_{app}$ and $G_{ann}$---any QCD axion model must lie in or above the green band in fig.~\ref{fig:reachaxion1}. Figure~\ref{fig:reachaxion2} depicts by green bands the relation between the electron coupling $G_{aee}$ and the axion mass $m_a$ for the DFSZ and KSVZ benchmark models of the QCD axion. The KSVZ band is roughly the smallest electron coupling an untuned model can exhibit.

In these sensitivity plots, we make the minimal but optimistic assumption that the axion or axion-like particle makes up all of the dark matter energy density. If the QCD axion is to make up \emph{all} of the DM, most cosmological histories would predict or strongly favor axion masses below 1~meV if the cosmological abundance arises due to the misalignment mechanism~\cite{wantz2010axion}, and below 4~meV (with large uncertainties on this bound) if the axions can be produced from decays of topological defects~\cite{ringwald2016axion,sikivie2008axion,wantz2010axion,hiramatsu2011improved,hiramatsu2012production,kawasaki2015axion}. In a standard thermal history, the QCD axion abundance scales as $\Omega_a \propto m_a^{1.19}$ in the dilute instanton gas approximation. Therefore, the standard production mechanisms would predict a fractional axion DM abundance of $\rho_a/\rho_\text{DM} \sim \mathcal{O}(10^{-3})$ if the QCD axion were to exist with a mass $m_a \sim 1~\eV$. One could conceivably construct models wherein these large-mass axions do constitute all of the DM, but we will not attempt to do so here. We note that even if the QCD axion only makes up such a small subcomponent of the DM, the SII setup would still be capable of detecting this component between $0.4~\eV$ and $1.2~\eV$, as the coupling sensitivity only scales as the square root of the energy density $\delta G_{aNN} \propto \sqrt{\rho_a}$. 

The dominant astrophysical bounds on the pseudoscalar couplings to nucleons and electrons come from supernova and white dwarf cooling, respectively. The observed duration of the neutrino burst originating from SN1987a indirectly constrains other efficient cooling mechanisms like the emission of light pseudoscalars. A simple energy loss argument was long thought to set a rough bound of $G_{aNN} \lesssim 2.5\times 10^{-9}~\GeV^{-1}$~\cite{raffelt2008astrophysical}. 
A more detailed analysis, correcting the omission of several physical effects in the early literature and folding in progenitor uncertainties, finds a much weaker robust exclusion bound of $G_{aNN} \lesssim 1.7\times 10^{-7}~\GeV^{-1}$~\cite{samrouven}, which is the one plotted in fig.~\ref{fig:reachaxion1}.
Observations of drifts in light-pulsation periods in white dwarfs provide an indirect measure of their cooling rate; the most stringent constraint, shown in fig.~\ref{fig:reachaxion2}, is $G_{aee} < 2.5 \times 10^{-10}~\GeV^{-1}$ at 95\% CL~\cite{raffelt2008astrophysical,corsico2001potential,isern2003white}.


\section{Discussion}\label{sec:discussion}

We have presented a DM detection scheme based on resonant absorption in molecules. Our proposed setup is sensitive to a wide variety for bosonic DM candidates with masses between 0.2~eV and 20~eV, including axions, dark photons and moduli, and can achieve several orders of magnitude improvement in coupling on current limits over the energy range under consideration. The detector concept may be regarded as a hybrid between low-energy macroscopic oscillators---circuits, cavities, mechanical resonators, electron and nuclear spin resonance systems---and high-energy absorption onto high-density targets. On the one hand, molecules can be viewed as some of the smallest and highest-frequency electromechanical resonators that exist in Nature. Each molecule is tiny and extremely weakly coupled to DM, but a gas of them  contains an enormous number of essentially identical copies, boosting the signal rate. On the other hand, our setup is in some ways similar to proposals aiming to look for DM absorption onto bulk targets~\cite{battaglieri2017us,bunting2017magnetic,hochberg2016detecting, hochberg2017absorption,derenzo2017direct,bloch2017searching,Hochberg:2017wce}. The crucial difference is that instead of looking for DM-induced excitations in a continuum---such as the free-particle continuum, a conduction band, or a phonon spectrum---here we advocate looking for DM absorption into a resolved discretuum of states with a long lifetime $T_1$ and phase-coherence time $T_2$.

A resonant approach comes with numerous advantages, and one major challenge: frequency coverage. Resonant absorption onto a transition line at $\omega \simeq \omega_0$ only yields appreciable event rates in a narrow bandwidth of order $\Delta \omega \sim 1/T_2$ around any one nominal transition frequency. Fortunately, small polyatomic molecules are multimode resonators due to their wealth of electronic levels, each with their own vibrational and rotational fine structure. In a narrow band around each of the lines in this ``forest'', the absorption cross-section is resonantly enhanced by a factor of $\omega_0 T_2$. This enhancement in the absorbing power  allows for excellent DM sensitivity at a discrete set of energies with exposures as small as $10^{-5}~\text{kg~year}$.

Most of the advantages of our proposal boil down to one key feature of molecular spectroscopy: \emph{control}. Small polyatomics exhibit a resolved discretuum of states with spectra, dynamics, selection rules, optical properties, and response to external variables that are well understood both experimentally and theoretically. As such, a gas of molecules has many ``knobs and handles'' to control the susceptibility to any DM absorption signal, unlike most bulk absorption targets. Even if a candidate signal is first seen in a different detector, our proposed setup is ideally suited to perform precision follow-up studies, so in this sense it is complementary to other detectors.

Our detector setup allows for excellent background rejection and signal discrimination. Environmental backgrounds can be naturally suppressed due to the low density of the target material, and on otherwise forbidden transitions. The differential energy response of the molecular sample allows for detectable signal rates \emph{lower} than background rates upon averaging. In addition, background events already become negligible if they occur once per ``shot'', rather than once over the lifetime of the experiment. Active veto systems may also be employed, as the overall detector has a fast response and relaxation time. Finally, the spatial coherence of the DM particles leads to a dramatic focusing effect of the signal photons in the Stack detector, offering a factor of greater than one million in directional isolation from environmental backgrounds.

Should a signal be found, the combination of great intrinsic energy resolution and energy response control with external variables---pressure, temperature, electromagnetic fields, molecular species and isotope---means that the DM mass can be pinpointed with extreme accuracy and precision, easily resolving even its line shape. By using a variety of molecules with a transition line near this energy, detailed information can be gleaned about the DM's interaction Hamiltonian and selection rules. A dedicated array of Stack detectors could be built to exploit and learn about the kinematic and directional properties of DM. Any signal can be unambiguously identified to have a DM origin. Moreover, a positive signal in this energy range would open up a field of observational DM astronomy given the setup's resolution to both the energy and momentum vector of the DM particles.

We see ample research opportunities for the near and far future. First and foremost, it is imperative that prototypes similar to the proposed Phase~I setups get off the ground to demonstrate the feasibility of the experimental strategies outlined in this work. Experimental research and development should include: optimization of MKIDs to deal with isotropic fluorescence photons in the Bulk configuration or other forms of low-noise calorimetric photon detection; manufacturing methods for a physical or artificial set of slabs in the Stack configuration; identification of optimal detector elements, including shield and container materials, and reflective and anti-reflective coatings. On the theoretical front, a significant effort should be devoted to mapping out which molecules are most promising to cover the two decades of energy with their discretuum of transition lines. In addition, optical thickness issues and the effects of multiple forward scatterings of emission photons need to be studied in the context of a Phase~II version of the Stack configuration. There could be other uses of our detector concept, such as inelastic scattering of DM, or detection of particles from other cosmic or astrophysical sources. We have considered absorption via a large---but incomplete---set of DM candidates and couplings; notable omissions include the pseudoscalar coupling to photons, and spin-2 DM candidates.

Molecular-gas-based DM searches define an entirely novel class of detectors without the need for novel or exotic material. We have shown that resonant absorption onto molecular transitions is a promising avenue for detection of well-motivated DM candidates, and a new application of molecular spectroscopy and low-noise photodetectors in fundamental physics. In forthcoming work, we will study how variants on the techniques presented here can potentially extend the reach down to DM masses as low as $10^{-4}~\eV$.


\begin{acknowledgments}
We are greatly indebted to Leo Hollberg for his contributions to many of the experimental aspects of this work, as well as to the conception of the general detection strategy.
We also thank Masha Baryakhtar and Xinlu Huang for their collaboration in an earlier incarnation of this research project. Kent Irwin and Ben Mazin explained to us the capabilities of cryogenic photodetectors. We are grateful for fruitful discussions with Rouven Essig, Andrew Geraci, Roni Harnik, Jason Hogan, Junwu Huang, David E Kaplan, Robert Lasenby, Edward Marti, Samuel McDermott, David Moore, Mariangela Lisanti, Duccio Pappadopulo, Georg Raffelt, Anna-Maria Taki, Scott Thomas, and Neal Weiner during the course of this work. KVT is supported by a Schmidt Fellowship funded by the generosity of Eric and Wendy Schmidt. AA acknowledges the support of NSERC and the Stavros Niarchos Foundation. SD is grateful for support from the National Science Foundation under Grant No.~PHYS-1720397, and would like to thank Perimeter Institute for their hospitality during the completion of this work. Research at Perimeter Institute is supported by the Government of Canada through Industry Canada and by the Province of Ontario through the Ministry of Economic Development \& Innovation.
\end{acknowledgments}

\appendix


\section{Fully-quantized rate calculation}\label{sec:fullquantum}

Throughout the main text, we have been assuming that the interaction with the dark matter bosons can be treated semiclassically, parametrizing the bosonic dark matter as a background field with average ambient energy density $\langle \rho \rangle$. One may question the (degree of) validity of this approximation, especially because quantum discreteness effects should become noticeable at higher dark-matter masses. Indeed, the expected particle occupation number $\langle \hat{\mathcal{N}} \rangle$ within a volume of linear size equal to the spatial coherence length $2/m v_0$ is given by the number density $\langle \rho \rangle/m$ times this volume:
\begin{align}
\langle \hat{\mathcal{N}} \rangle \sim \frac{8 \langle \rho \rangle}{m^4 v_0^3} \approx \left(\frac{15~\eV}{m}\right)^4,
\end{align}
where we have taken the local DM energy density $\langle \rho \rangle \approx 0.4~\GeV/\text{cm}^3$ and velocity dispersion $v_0 \approx 235~\text{km/s}$. We thus find that $\langle \hat{\mathcal{N}} \rangle$ is close to unity for the upper end of the parameter space under consideration in this work. The semiclassical approximation can be expected to be reasonably accurate at low masses, while for $m \gtrsim 15~\eV$ the number density becomes so low that a more appropriate representation of the DM interaction is that of individual DM particles impinging on molecules with a small absorption cross-section. Below, we sketch out a fully quantized treatment that encompasses both regimes. The main result will be that all semiclassical results presented in this paper are valid---even when $ \langle \hat{\mathcal{N}} \rangle \lesssim 1$---for the integration times under consideration. 

We can write the interaction Hamiltonian a molecular system with a bosonic field mode of energy $\omega$ and annihilation (creation) operator $a$ ($a^\dagger$) as
\begin{align}
\delta H'(t)= \widetilde{\Omega} e^{-i\omega t} e^{i\omega_0 t} b^\dagger a + \text{h.c.} \label{eq:quantumint}
\end{align}
The molecule is approximated by a two-level system consisting of states $|0\rangle,|1\rangle$ with energy splitting $\omega_0$ and interaction-picture annihilation (creation) operators $b e^{-i\omega_0 t}$ ($b^\dagger e^{+i\omega_0 t}$) as before. In eq.~\ref{eq:quantumint}, we have (for now) ignored spatial dependence of the local interaction, and absorbed all phases and other numerical constants into $\tilde{\Omega}$. Generalizations to interactions with multiple bosonic field modes is straightforward. 

We are interested in calculating the molecular transition rate of the process $|0\rangle \to |1\rangle$ given an initial dark-matter state $|\text{DM}\rangle$ of the bosonic field mode under consideration. To this end, we compute the partial rate amplitudes $\langle 1; n| \int_0^t\delta H'(t') dt' | 0; \text{DM}\rangle$ for a combined transition of $|0\rangle \to |1\rangle$ in the molecule and $|\text{DM}\rangle \to |n\rangle$ in the DM field mode, as an absorption event will in general also affect the state of the dark-matter field. We take $|n\rangle$ to be members of an orthonormal basis (e.g.~Fock states). The expected absorption probability $P_\text{abs}$ per molecule is found by summing over the squares of all partial amplitudes with different DM final states $|n\rangle$, which to first order in perturbation theory then gives, at short times:
\begin{align}
P_\text{abs} & = \sum_n \left|\left\langle 1; n\left| e^{-i\int_0^t\delta H'(t') dt'} \right| 0;\text{DM} \right\rangle \right|^2\\
& \simeq |\widetilde{\Omega}|^2 \left| \int_0^t \text{d}t'\, e^{-i(\omega_0 - \omega)t'} \langle 1 | b^\dagger |0 \rangle\right|^2 \sum_n |\langle n | a |\text{DM} \rangle|^2 \nonumber \\
& = |\widetilde{\Omega}|^2 \left| \int_0^t \text{d}t'\, e^{-i(\omega_0 - \omega)t'} \right|^2 \langle \text{DM} | a^\dagger a | \text{DM} \rangle. \nonumber
\end{align}
To get to the third line, we have used the fact that the off-diagonal matrix element of $b^\dagger$ is unity, and that $\sum_n |n\rangle \langle n |$ is the unit operator by construction. Matching to the semiclassical result of eq.~\ref{eq:rabi2} can be done with the identification:
\begin{align}
\Omega^2 = |\widetilde{\Omega}|^2 \langle \text{DM} | a^\dagger a | \text{DM} \rangle \equiv |\widetilde{\Omega}|^2 {\rm Tr}\lbrace \hat{\rho}_\text{DM}  \hat{\mathcal{N}} \rbrace.
\end{align}
In the second equality, we defined the number operator $\hat{\mathcal{N}} \equiv a^\dagger a$, and wrote the expectation value as an operator trace weighted by the DM density matrix $\hat{\rho}_\text{DM}$ (not to be confused with the DM energy density $\rho$) to allow for the possibility of mixed states.

The identification of the expectation value $ |\widetilde{\Omega}|^2 \langle \hat{\mathcal{N}} \rangle$ with a semiclassical perturbation of strength $\Omega^2$ does not quite capture all of the physics, for the operator $\hat{\mathcal{N}}$ has quantum fluctuations of its own. To see this, one could compute the moments of $\hat{\mathcal{N}}$, and see that its variance $\langle \hat{\mathcal{N}}^2 \rangle - \langle \hat{\mathcal{N}} \rangle^2$ becomes large compared to its squared expectation value $\langle \hat{\mathcal{N}} \rangle^2$ at low mode occupation numbers. However, these fluctuations---due to the ``particle discreteness'' of the DM state---average down to negligible levels over the long integration times and macroscopic detector volumes under consideration in this work.

\bibliographystyle{apsrev4-1-etal}
\bibliography{resonant_absorption}

\begin{thebibliography}{126}%
\makeatletter
\providecommand \@ifxundefined [1]{%
 \@ifx{#1\undefined}
}%
\providecommand \@ifnum [1]{%
 \ifnum #1\expandafter \@firstoftwo
 \else \expandafter \@secondoftwo
 \fi
}%
\providecommand \@ifx [1]{%
 \ifx #1\expandafter \@firstoftwo
 \else \expandafter \@secondoftwo
 \fi
}%
\providecommand \natexlab [1]{#1}%
\providecommand \enquote  [1]{``#1''}%
\providecommand \bibnamefont  [1]{#1}%
\providecommand \bibfnamefont [1]{#1}%
\providecommand \citenamefont [1]{#1}%
\providecommand \href@noop [0]{\@secondoftwo}%
\providecommand \href [0]{\begingroup \@sanitize@url \@href}%
\providecommand \@href[1]{\@@startlink{#1}\@@href}%
\providecommand \@@href[1]{\endgroup#1\@@endlink}%
\providecommand \@sanitize@url [0]{\catcode `\\12\catcode `\$12\catcode
  `\&12\catcode `\#12\catcode `\^12\catcode `\_12\catcode `\%12\relax}%
\providecommand \@@startlink[1]{}%
\providecommand \@@endlink[0]{}%
\providecommand \url  [0]{\begingroup\@sanitize@url \@url }%
\providecommand \@url [1]{\endgroup\@href {#1}{\urlprefix }}%
\providecommand \urlprefix  [0]{URL }%
\providecommand \Eprint [0]{\href }%
\providecommand \doibase [0]{http://dx.doi.org/}%
\providecommand \selectlanguage [0]{\@gobble}%
\providecommand \bibinfo  [0]{\@secondoftwo}%
\providecommand \bibfield  [0]{\@secondoftwo}%
\providecommand \translation [1]{[#1]}%
\providecommand \BibitemOpen [0]{}%
\providecommand \bibitemStop [0]{}%
\providecommand \bibitemNoStop [0]{.\EOS\space}%
\providecommand \EOS [0]{\spacefactor3000\relax}%
\providecommand \BibitemShut  [1]{\csname bibitem#1\endcsname}%
\let\auto@bib@innerbib\@empty
\bibitem [{\citenamefont {Kimble}\ and\ \citenamefont {Mandel}(1976)}]{km1976}%
  \BibitemOpen
  \bibfield  {author} {\bibinfo {author} {\bibfnamefont {H.~J.}\ \bibnamefont
  {Kimble}}\ and\ \bibinfo {author} {\bibfnamefont {L.}~\bibnamefont
  {Mandel}},\ }\href {\doibase 10.1103/PhysRevA.13.2123} {\bibfield  {journal}
  {\bibinfo  {journal} {Phys. Rev. A}\ }\textbf {\bibinfo {volume} {13}},\
  \bibinfo {pages} {2123} (\bibinfo {year} {1976})}\BibitemShut {NoStop}%
\bibitem [{\citenamefont {Landau}\ and\ \citenamefont
  {Lifshitz}(1980)}]{landau1980statistical}%
  \BibitemOpen
  \bibfield  {author} {\bibinfo {author} {\bibfnamefont {L.}~\bibnamefont
  {Landau}}\ and\ \bibinfo {author} {\bibfnamefont {E.}~\bibnamefont
  {Lifshitz}},\ }\href@noop {} {\bibfield  {journal} {\bibinfo  {journal}
  {Course of theoretical physics}\ }\textbf {\bibinfo {volume} {30}} (\bibinfo
  {year} {1980})}\BibitemShut {NoStop}%
\bibitem [{\citenamefont {Corney}(1978)}]{corney1978atomic}%
  \BibitemOpen
  \bibfield  {author} {\bibinfo {author} {\bibfnamefont {A.}~\bibnamefont
  {Corney}},\ }\href@noop {} {\emph {\bibinfo {title} {Atomic and laser
  spectroscopy}}}\ (\bibinfo  {publisher} {Clarendon Press Oxford},\ \bibinfo
  {year} {1978})\BibitemShut {NoStop}%
\bibitem [{\citenamefont {Loudon}(2000)}]{loudon2000quantum}%
  \BibitemOpen
  \bibfield  {author} {\bibinfo {author} {\bibfnamefont {R.}~\bibnamefont
  {Loudon}},\ }\href@noop {} {\emph {\bibinfo {title} {The quantum theory of
  light}}}\ (\bibinfo  {publisher} {OUP Oxford},\ \bibinfo {year}
  {2000})\BibitemShut {NoStop}%
\bibitem [{\citenamefont {Dicke}(1954)}]{d1954}%
  \BibitemOpen
  \bibfield  {author} {\bibinfo {author} {\bibfnamefont {R.~H.}\ \bibnamefont
  {Dicke}},\ }\href {\doibase 10.1103/PhysRev.93.99} {\bibfield  {journal}
  {\bibinfo  {journal} {Phys. Rev.}\ }\textbf {\bibinfo {volume} {93}},\
  \bibinfo {pages} {99} (\bibinfo {year} {1954})}\BibitemShut {NoStop}%
\bibitem [{\citenamefont {McMillan}\ and\ \citenamefont
  {Binney}(2010)}]{mcmillan2010uncertainty}%
  \BibitemOpen
  \bibfield  {author} {\bibinfo {author} {\bibfnamefont {P.~J.}\ \bibnamefont
  {McMillan}}\ and\ \bibinfo {author} {\bibfnamefont {J.~J.}\ \bibnamefont
  {Binney}},\ }\href@noop {} {\bibfield  {journal} {\bibinfo  {journal}
  {Monthly Notices of the Royal Astronomical Society}\ }\textbf {\bibinfo
  {volume} {402}},\ \bibinfo {pages} {934} (\bibinfo {year}
  {2010})}\BibitemShut {NoStop}%
\bibitem [{\citenamefont {Kerr}\ and\ \citenamefont
  {Lynden-Bell}(1986)}]{kerr1986review}%
  \BibitemOpen
  \bibfield  {author} {\bibinfo {author} {\bibfnamefont {F.~J.}\ \bibnamefont
  {Kerr}}\ and\ \bibinfo {author} {\bibfnamefont {D.}~\bibnamefont
  {Lynden-Bell}},\ }\href@noop {} {\bibfield  {journal} {\bibinfo  {journal}
  {Monthly Notices of the Royal Astronomical Society}\ }\textbf {\bibinfo
  {volume} {221}},\ \bibinfo {pages} {1023} (\bibinfo {year}
  {1986})}\BibitemShut {NoStop}%
\bibitem [{\citenamefont {Reid}\ \emph {et~al.}(2009)\citenamefont {Reid},
  \citenamefont {Menten}, \citenamefont {Zheng}, \citenamefont {Brunthaler},
  \citenamefont {Moscadelli}, \citenamefont {Xu}, \citenamefont {Zhang},
  \citenamefont {Sato}, \citenamefont {Honma}, \citenamefont {Hirota} \emph
  {et~al.}}]{reid2009trigonometric}%
  \BibitemOpen
  \bibfield  {author} {\bibinfo {author} {\bibfnamefont {M.}~\bibnamefont
  {Reid}}, \bibinfo {author} {\bibfnamefont {K.}~\bibnamefont {Menten}},
  \bibinfo {author} {\bibfnamefont {X.}~\bibnamefont {Zheng}}, \bibinfo
  {author} {\bibfnamefont {A.}~\bibnamefont {Brunthaler}}, \bibinfo {author}
  {\bibfnamefont {L.}~\bibnamefont {Moscadelli}}, \bibinfo {author}
  {\bibfnamefont {Y.}~\bibnamefont {Xu}}, \bibinfo {author} {\bibfnamefont
  {B.}~\bibnamefont {Zhang}}, \bibinfo {author} {\bibfnamefont
  {M.}~\bibnamefont {Sato}}, \bibinfo {author} {\bibfnamefont {M.}~\bibnamefont
  {Honma}},  \emph {et~al.},\ }\href@noop {} {\bibfield  {journal} {\bibinfo
  {journal} {The Astrophysical Journal}\ }\textbf {\bibinfo {volume} {700}},\
  \bibinfo {pages} {137} (\bibinfo {year} {2009})}\BibitemShut {NoStop}%
\bibitem [{\citenamefont {Sch{\"o}nrich}\ \emph {et~al.}(2010)\citenamefont
  {Sch{\"o}nrich}, \citenamefont {Binney},\ and\ \citenamefont
  {Dehnen}}]{schonrich2010local}%
  \BibitemOpen
  \bibfield  {author} {\bibinfo {author} {\bibfnamefont {R.}~\bibnamefont
  {Sch{\"o}nrich}}, \bibinfo {author} {\bibfnamefont {J.}~\bibnamefont
  {Binney}}, \ and\ \bibinfo {author} {\bibfnamefont {W.}~\bibnamefont
  {Dehnen}},\ }\href@noop {} {\bibfield  {journal} {\bibinfo  {journal}
  {Monthly Notices of the Royal Astronomical Society}\ }\textbf {\bibinfo
  {volume} {403}},\ \bibinfo {pages} {1829} (\bibinfo {year}
  {2010})}\BibitemShut {NoStop}%
\bibitem [{\citenamefont {Mignard}(2000)}]{mignard2000local}%
  \BibitemOpen
  \bibfield  {author} {\bibinfo {author} {\bibfnamefont {F.}~\bibnamefont
  {Mignard}},\ }\href@noop {} {\bibfield  {journal} {\bibinfo  {journal}
  {Astronomy and Astrophysics}\ }\textbf {\bibinfo {volume} {354}},\ \bibinfo
  {pages} {522} (\bibinfo {year} {2000})}\BibitemShut {NoStop}%
\bibitem [{\citenamefont {Smith}\ \emph {et~al.}(2007)\citenamefont {Smith},
  \citenamefont {Ruchti}, \citenamefont {Helmi}, \citenamefont {Wyse},
  \citenamefont {Fulbright}, \citenamefont {Freeman}, \citenamefont {Navarro},
  \citenamefont {Seabroke}, \citenamefont {Steinmetz}, \citenamefont {Williams}
  \emph {et~al.}}]{smith2007rave}%
  \BibitemOpen
  \bibfield  {author} {\bibinfo {author} {\bibfnamefont {M.~C.}\ \bibnamefont
  {Smith}}, \bibinfo {author} {\bibfnamefont {G.~R.}\ \bibnamefont {Ruchti}},
  \bibinfo {author} {\bibfnamefont {A.}~\bibnamefont {Helmi}}, \bibinfo
  {author} {\bibfnamefont {R.~F.}\ \bibnamefont {Wyse}}, \bibinfo {author}
  {\bibfnamefont {J.~P.}\ \bibnamefont {Fulbright}}, \bibinfo {author}
  {\bibfnamefont {K.~C.}\ \bibnamefont {Freeman}}, \bibinfo {author}
  {\bibfnamefont {J.~F.}\ \bibnamefont {Navarro}}, \bibinfo {author}
  {\bibfnamefont {G.~M.}\ \bibnamefont {Seabroke}}, \bibinfo {author}
  {\bibfnamefont {M.}~\bibnamefont {Steinmetz}},  \emph {et~al.},\ }\href@noop
  {} {\bibfield  {journal} {\bibinfo  {journal} {Monthly Notices of the Royal
  Astronomical Society}\ }\textbf {\bibinfo {volume} {379}},\ \bibinfo {pages}
  {755} (\bibinfo {year} {2007})}\BibitemShut {NoStop}%
\bibitem [{\citenamefont {Derevianko}(2016)}]{derevianko2016detecting}%
  \BibitemOpen
  \bibfield  {author} {\bibinfo {author} {\bibfnamefont {A.}~\bibnamefont
  {Derevianko}},\ }\href@noop {} {\bibfield  {journal} {\bibinfo  {journal}
  {arXiv preprint arXiv:1605.09717}\ } (\bibinfo {year} {2016})}\BibitemShut
  {NoStop}%
\bibitem [{\citenamefont {Struve}(1989)}]{struve1989fundamentals}%
  \BibitemOpen
  \bibfield  {author} {\bibinfo {author} {\bibfnamefont {W.~S.}\ \bibnamefont
  {Struve}},\ }\href@noop {} {\emph {\bibinfo {title} {Fundamentals of
  molecular spectroscopy}}}\ (\bibinfo  {publisher} {Wiley New York},\ \bibinfo
  {year} {1989})\BibitemShut {NoStop}%
\bibitem [{\citenamefont {Huber}\ and\ \citenamefont
  {Herzberg}(1979)}]{huber1979g}%
  \BibitemOpen
  \bibfield  {author} {\bibinfo {author} {\bibfnamefont {K.}~\bibnamefont
  {Huber}}\ and\ \bibinfo {author} {\bibfnamefont {G.}~\bibnamefont
  {Herzberg}},\ }\href@noop {} {\emph {\bibinfo {title} {Constants of diatomic
  molecules}}}\ (\bibinfo  {publisher} {Van Nostrand Reinhold, New York},\
  \bibinfo {year} {1979})\BibitemShut {NoStop}%
\bibitem [{\citenamefont {Massey}(1949)}]{massey1949collisions}%
  \BibitemOpen
  \bibfield  {author} {\bibinfo {author} {\bibfnamefont {H.}~\bibnamefont
  {Massey}},\ }\href@noop {} {\bibfield  {journal} {\bibinfo  {journal}
  {Reports on Progress in Physics}\ }\textbf {\bibinfo {volume} {12}},\
  \bibinfo {pages} {248} (\bibinfo {year} {1949})}\BibitemShut {NoStop}%
\bibitem [{\citenamefont {Landau}\ and\ \citenamefont
  {Teller}(1936)}]{landauteller}%
  \BibitemOpen
  \bibfield  {author} {\bibinfo {author} {\bibfnamefont {L.}~\bibnamefont
  {Landau}}\ and\ \bibinfo {author} {\bibfnamefont {E.}~\bibnamefont
  {Teller}},\ }\href@noop {} {\bibfield  {journal} {\bibinfo  {journal} {Physik
  Z. Sowjetunion}\ }\textbf {\bibinfo {volume} {10}} (\bibinfo {year}
  {1936})}\BibitemShut {NoStop}%
\bibitem [{\citenamefont {Millikan}\ and\ \citenamefont
  {White}(1963)}]{millikan1963systematics}%
  \BibitemOpen
  \bibfield  {author} {\bibinfo {author} {\bibfnamefont {R.~C.}\ \bibnamefont
  {Millikan}}\ and\ \bibinfo {author} {\bibfnamefont {D.~R.}\ \bibnamefont
  {White}},\ }\href@noop {} {\bibfield  {journal} {\bibinfo  {journal} {The
  Journal of chemical physics}\ }\textbf {\bibinfo {volume} {39}},\ \bibinfo
  {pages} {3209} (\bibinfo {year} {1963})}\BibitemShut {NoStop}%
\bibitem [{\citenamefont {Arecchi}\ and\ \citenamefont
  {Courtens}(1970)}]{arecchi1970cooperative}%
  \BibitemOpen
  \bibfield  {author} {\bibinfo {author} {\bibfnamefont {F.}~\bibnamefont
  {Arecchi}}\ and\ \bibinfo {author} {\bibfnamefont {E.}~\bibnamefont
  {Courtens}},\ }\href@noop {} {\bibfield  {journal} {\bibinfo  {journal}
  {Physical Review A}\ }\textbf {\bibinfo {volume} {2}},\ \bibinfo {pages}
  {1730} (\bibinfo {year} {1970})}\BibitemShut {NoStop}%
\bibitem [{\citenamefont {Rehler}\ and\ \citenamefont
  {Eberly}(1971)}]{rehler1971superradiance}%
  \BibitemOpen
  \bibfield  {author} {\bibinfo {author} {\bibfnamefont {N.~E.}\ \bibnamefont
  {Rehler}}\ and\ \bibinfo {author} {\bibfnamefont {J.~H.}\ \bibnamefont
  {Eberly}},\ }\href@noop {} {\bibfield  {journal} {\bibinfo  {journal}
  {Physical Review A}\ }\textbf {\bibinfo {volume} {3}},\ \bibinfo {pages}
  {1735} (\bibinfo {year} {1971})}\BibitemShut {NoStop}%
\bibitem [{\citenamefont {Scully}\ \emph {et~al.}(2006)\citenamefont {Scully},
  \citenamefont {Fry}, \citenamefont {Ooi},\ and\ \citenamefont
  {W{\'o}dkiewicz}}]{scully2006directed}%
  \BibitemOpen
  \bibfield  {author} {\bibinfo {author} {\bibfnamefont {M.~O.}\ \bibnamefont
  {Scully}}, \bibinfo {author} {\bibfnamefont {E.~S.}\ \bibnamefont {Fry}},
  \bibinfo {author} {\bibfnamefont {C.~R.}\ \bibnamefont {Ooi}}, \ and\
  \bibinfo {author} {\bibfnamefont {K.}~\bibnamefont {W{\'o}dkiewicz}},\
  }\href@noop {} {\bibfield  {journal} {\bibinfo  {journal} {Physical review
  letters}\ }\textbf {\bibinfo {volume} {96}},\ \bibinfo {pages} {010501}
  (\bibinfo {year} {2006})}\BibitemShut {NoStop}%
\bibitem [{\citenamefont {Bromley}\ \emph {et~al.}(2016)\citenamefont
  {Bromley}, \citenamefont {Zhu}, \citenamefont {Bishof}, \citenamefont
  {Zhang}, \citenamefont {Bothwell}, \citenamefont {Schachenmayer},
  \citenamefont {Nicholson}, \citenamefont {Kaiser}, \citenamefont {Yelin},
  \citenamefont {Lukin} \emph {et~al.}}]{bromley2016collective}%
  \BibitemOpen
  \bibfield  {author} {\bibinfo {author} {\bibfnamefont {S.~L.}\ \bibnamefont
  {Bromley}}, \bibinfo {author} {\bibfnamefont {B.}~\bibnamefont {Zhu}},
  \bibinfo {author} {\bibfnamefont {M.}~\bibnamefont {Bishof}}, \bibinfo
  {author} {\bibfnamefont {X.}~\bibnamefont {Zhang}}, \bibinfo {author}
  {\bibfnamefont {T.}~\bibnamefont {Bothwell}}, \bibinfo {author}
  {\bibfnamefont {J.}~\bibnamefont {Schachenmayer}}, \bibinfo {author}
  {\bibfnamefont {T.~L.}\ \bibnamefont {Nicholson}}, \bibinfo {author}
  {\bibfnamefont {R.}~\bibnamefont {Kaiser}}, \bibinfo {author} {\bibfnamefont
  {S.~F.}\ \bibnamefont {Yelin}},  \emph {et~al.},\ }\href@noop {} {\bibfield
  {journal} {\bibinfo  {journal} {Nature communications}\ }\textbf {\bibinfo
  {volume} {7}} (\bibinfo {year} {2016})}\BibitemShut {NoStop}%
\bibitem [{\citenamefont {Ara{\'u}jo}\ \emph {et~al.}(2016)\citenamefont
  {Ara{\'u}jo}, \citenamefont {Kre{\v{s}}i{\'c}}, \citenamefont {Kaiser},\ and\
  \citenamefont {Guerin}}]{araujo2016superradiance}%
  \BibitemOpen
  \bibfield  {author} {\bibinfo {author} {\bibfnamefont {M.~O.}\ \bibnamefont
  {Ara{\'u}jo}}, \bibinfo {author} {\bibfnamefont {I.}~\bibnamefont
  {Kre{\v{s}}i{\'c}}}, \bibinfo {author} {\bibfnamefont {R.}~\bibnamefont
  {Kaiser}}, \ and\ \bibinfo {author} {\bibfnamefont {W.}~\bibnamefont
  {Guerin}},\ }\href@noop {} {\bibfield  {journal} {\bibinfo  {journal}
  {Physical review letters}\ }\textbf {\bibinfo {volume} {117}},\ \bibinfo
  {pages} {073002} (\bibinfo {year} {2016})}\BibitemShut {NoStop}%
\bibitem [{\citenamefont {Roof}\ \emph {et~al.}(2016)\citenamefont {Roof},
  \citenamefont {Kemp}, \citenamefont {Havey},\ and\ \citenamefont
  {Sokolov}}]{roof2016observation}%
  \BibitemOpen
  \bibfield  {author} {\bibinfo {author} {\bibfnamefont {S.}~\bibnamefont
  {Roof}}, \bibinfo {author} {\bibfnamefont {K.}~\bibnamefont {Kemp}}, \bibinfo
  {author} {\bibfnamefont {M.}~\bibnamefont {Havey}}, \ and\ \bibinfo {author}
  {\bibfnamefont {I.}~\bibnamefont {Sokolov}},\ }\href@noop {} {\bibfield
  {journal} {\bibinfo  {journal} {Physical review letters}\ }\textbf {\bibinfo
  {volume} {117}},\ \bibinfo {pages} {073003} (\bibinfo {year}
  {2016})}\BibitemShut {NoStop}%
\bibitem [{\citenamefont {Hadfield}(2009)}]{hadfield2009single}%
  \BibitemOpen
  \bibfield  {author} {\bibinfo {author} {\bibfnamefont {R.~H.}\ \bibnamefont
  {Hadfield}},\ }\href@noop {} {\bibfield  {journal} {\bibinfo  {journal}
  {Nature photonics}\ }\textbf {\bibinfo {volume} {3}},\ \bibinfo {pages} {696}
  (\bibinfo {year} {2009})}\BibitemShut {NoStop}%
\bibitem [{\citenamefont {Eisaman}\ \emph {et~al.}(2011)\citenamefont
  {Eisaman}, \citenamefont {Fan}, \citenamefont {Migdall},\ and\ \citenamefont
  {Polyakov}}]{eisaman2011invited}%
  \BibitemOpen
  \bibfield  {author} {\bibinfo {author} {\bibfnamefont {M.}~\bibnamefont
  {Eisaman}}, \bibinfo {author} {\bibfnamefont {J.}~\bibnamefont {Fan}},
  \bibinfo {author} {\bibfnamefont {A.}~\bibnamefont {Migdall}}, \ and\
  \bibinfo {author} {\bibfnamefont {S.~V.}\ \bibnamefont {Polyakov}},\
  }\href@noop {} {\bibfield  {journal} {\bibinfo  {journal} {Review of
  scientific instruments}\ }\textbf {\bibinfo {volume} {82}},\ \bibinfo {pages}
  {071101} (\bibinfo {year} {2011})}\BibitemShut {NoStop}%
\bibitem [{\citenamefont {Photonics}(2017{\natexlab{a}})}]{hamamatsu1}%
  \BibitemOpen
  \bibfield  {author} {\bibinfo {author} {\bibfnamefont {H.}~\bibnamefont
  {Photonics}},\ }\href@noop {} {\enquote {\bibinfo {title} {{Photomultiplier
  Tube R1527P}},}\ }\bibinfo {howpublished}
  {\url{https://www.hamamatsu.com/resources/pdf/etd/R1527_R1527P_TPMS1007E.pdf}}
  (\bibinfo {year} {2017}{\natexlab{a}}),\ \bibinfo {note} {[Online; accessed
  9-Aug-2017]}\BibitemShut {NoStop}%
\bibitem [{\citenamefont {Photonics}(2017{\natexlab{b}})}]{hamamatsu2}%
  \BibitemOpen
  \bibfield  {author} {\bibinfo {author} {\bibfnamefont {H.}~\bibnamefont
  {Photonics}},\ }\href@noop {} {\enquote {\bibinfo {title} {{Photomultiplier
  Tube R4220P}},}\ }\bibinfo {howpublished}
  {\url{https://www.hamamatsu.com/resources/pdf/etd/R4220_R4220P_TPMS1003E.pdf}}
  (\bibinfo {year} {2017}{\natexlab{b}}),\ \bibinfo {note} {[Online; accessed
  9-Aug-2017]}\BibitemShut {NoStop}%
\bibitem [{\citenamefont {Photonics}(2017{\natexlab{c}})}]{hamamatsu3}%
  \BibitemOpen
  \bibfield  {author} {\bibinfo {author} {\bibfnamefont {H.}~\bibnamefont
  {Photonics}},\ }\href@noop {} {\enquote {\bibinfo {title} {{Photomultiplier
  Tube R943-02}},}\ }\bibinfo {howpublished}
  {\url{https://www.hamamatsu.com/resources/pdf/etd/R943-02_TPMH1115E.pdf}}
  (\bibinfo {year} {2017}{\natexlab{c}}),\ \bibinfo {note} {[Online; accessed
  9-Aug-2017]}\BibitemShut {NoStop}%
\bibitem [{\citenamefont {Photonics}(2017{\natexlab{d}})}]{hamamatsu4}%
  \BibitemOpen
  \bibfield  {author} {\bibinfo {author} {\bibfnamefont {H.}~\bibnamefont
  {Photonics}},\ }\href@noop {} {\enquote {\bibinfo {title} {{Photomultiplier
  Tube R12421P}},}\ }\bibinfo {howpublished}
  {\url{https://www.hamamatsu.com/resources/pdf/etd/R12421_H12690_TPMH1347E.pdf}}
  (\bibinfo {year} {2017}{\natexlab{d}}),\ \bibinfo {note} {[Online; accessed
  9-Aug-2017]}\BibitemShut {NoStop}%
\bibitem [{\citenamefont {Day}\ \emph {et~al.}(2003)\citenamefont {Day},
  \citenamefont {LeDuc}, \citenamefont {Mazin}, \citenamefont {Vayonakis},\
  and\ \citenamefont {Zmuidzinas}}]{day2003broadband}%
  \BibitemOpen
  \bibfield  {author} {\bibinfo {author} {\bibfnamefont {P.~K.}\ \bibnamefont
  {Day}}, \bibinfo {author} {\bibfnamefont {H.~G.}\ \bibnamefont {LeDuc}},
  \bibinfo {author} {\bibfnamefont {B.~A.}\ \bibnamefont {Mazin}}, \bibinfo
  {author} {\bibfnamefont {A.}~\bibnamefont {Vayonakis}}, \ and\ \bibinfo
  {author} {\bibfnamefont {J.}~\bibnamefont {Zmuidzinas}},\ }\href@noop {}
  {\bibfield  {journal} {\bibinfo  {journal} {Nature}\ }\textbf {\bibinfo
  {volume} {425}},\ \bibinfo {pages} {817} (\bibinfo {year}
  {2003})}\BibitemShut {NoStop}%
\bibitem [{\citenamefont {Mazin}\ \emph {et~al.}(2012)\citenamefont {Mazin},
  \citenamefont {Bumble}, \citenamefont {Meeker}, \citenamefont {O’Brien},
  \citenamefont {McHugh},\ and\ \citenamefont
  {Langman}}]{mazin2012superconducting}%
  \BibitemOpen
  \bibfield  {author} {\bibinfo {author} {\bibfnamefont {B.~A.}\ \bibnamefont
  {Mazin}}, \bibinfo {author} {\bibfnamefont {B.}~\bibnamefont {Bumble}},
  \bibinfo {author} {\bibfnamefont {S.~R.}\ \bibnamefont {Meeker}}, \bibinfo
  {author} {\bibfnamefont {K.}~\bibnamefont {O’Brien}}, \bibinfo {author}
  {\bibfnamefont {S.}~\bibnamefont {McHugh}}, \ and\ \bibinfo {author}
  {\bibfnamefont {E.}~\bibnamefont {Langman}},\ }\href@noop {} {\bibfield
  {journal} {\bibinfo  {journal} {Optics express}\ }\textbf {\bibinfo {volume}
  {20}},\ \bibinfo {pages} {1503} (\bibinfo {year} {2012})}\BibitemShut
  {NoStop}%
\bibitem [{\citenamefont {Mazin}\ \emph {et~al.}(2013)\citenamefont {Mazin},
  \citenamefont {Meeker}, \citenamefont {Strader}, \citenamefont {Szypryt},
  \citenamefont {Marsden}, \citenamefont {van Eyken}, \citenamefont {Duggan},
  \citenamefont {Walter}, \citenamefont {Ulbricht}, \citenamefont {Johnson}
  \emph {et~al.}}]{mazin2013arcons}%
  \BibitemOpen
  \bibfield  {author} {\bibinfo {author} {\bibfnamefont {B.}~\bibnamefont
  {Mazin}}, \bibinfo {author} {\bibfnamefont {S.~R.}\ \bibnamefont {Meeker}},
  \bibinfo {author} {\bibfnamefont {M.}~\bibnamefont {Strader}}, \bibinfo
  {author} {\bibfnamefont {P.}~\bibnamefont {Szypryt}}, \bibinfo {author}
  {\bibfnamefont {D.}~\bibnamefont {Marsden}}, \bibinfo {author} {\bibfnamefont
  {J.}~\bibnamefont {van Eyken}}, \bibinfo {author} {\bibfnamefont
  {G.}~\bibnamefont {Duggan}}, \bibinfo {author} {\bibfnamefont
  {A.}~\bibnamefont {Walter}}, \bibinfo {author} {\bibfnamefont
  {G.}~\bibnamefont {Ulbricht}},  \emph {et~al.},\ }\href@noop {} {\bibfield
  {journal} {\bibinfo  {journal} {Publications of the Astronomical Society of
  the Pacific}\ }\textbf {\bibinfo {volume} {125}},\ \bibinfo {pages} {1348}
  (\bibinfo {year} {2013})}\BibitemShut {NoStop}%
\bibitem [{\citenamefont {Meeker}\ \emph {et~al.}(2015)\citenamefont {Meeker},
  \citenamefont {Mazin}, \citenamefont {Jensen-Clem}, \citenamefont {Walter},
  \citenamefont {Szypryt}, \citenamefont {Strader},\ and\ \citenamefont
  {Bockstiegel}}]{meeker2015design}%
  \BibitemOpen
  \bibfield  {author} {\bibinfo {author} {\bibfnamefont {S.}~\bibnamefont
  {Meeker}}, \bibinfo {author} {\bibfnamefont {B.}~\bibnamefont {Mazin}},
  \bibinfo {author} {\bibfnamefont {R.}~\bibnamefont {Jensen-Clem}}, \bibinfo
  {author} {\bibfnamefont {A.}~\bibnamefont {Walter}}, \bibinfo {author}
  {\bibfnamefont {P.}~\bibnamefont {Szypryt}}, \bibinfo {author} {\bibfnamefont
  {M.}~\bibnamefont {Strader}}, \ and\ \bibinfo {author} {\bibfnamefont
  {C.}~\bibnamefont {Bockstiegel}},\ }in\ \href@noop {} {\emph {\bibinfo
  {booktitle} {Adaptive Optics for Extremely Large Telescopes 4--Conference
  Proceedings}}},\ Vol.~\bibinfo {volume} {1}\ (\bibinfo {year}
  {2015})\BibitemShut {NoStop}%
\bibitem [{\citenamefont {Mazin}\ \emph {et~al.}(2015)\citenamefont {Mazin},
  \citenamefont {Becker}, \citenamefont {France}, \citenamefont {Fraser},
  \citenamefont {Howell}, \citenamefont {Jones}, \citenamefont {Meeker},
  \citenamefont {O'Brien}, \citenamefont {Prochaska}, \citenamefont {Siana}
  \emph {et~al.}}]{mazin2015science}%
  \BibitemOpen
  \bibfield  {author} {\bibinfo {author} {\bibfnamefont {B.~A.}\ \bibnamefont
  {Mazin}}, \bibinfo {author} {\bibfnamefont {G.}~\bibnamefont {Becker}},
  \bibinfo {author} {\bibfnamefont {K.}~\bibnamefont {France}}, \bibinfo
  {author} {\bibfnamefont {W.}~\bibnamefont {Fraser}}, \bibinfo {author}
  {\bibfnamefont {D.~A.}\ \bibnamefont {Howell}}, \bibinfo {author}
  {\bibfnamefont {T.}~\bibnamefont {Jones}}, \bibinfo {author} {\bibfnamefont
  {S.}~\bibnamefont {Meeker}}, \bibinfo {author} {\bibfnamefont
  {K.}~\bibnamefont {O'Brien}}, \bibinfo {author} {\bibfnamefont {J.~X.}\
  \bibnamefont {Prochaska}},  \emph {et~al.},\ }\href@noop {} {\bibfield
  {journal} {\bibinfo  {journal} {arXiv preprint arXiv:1506.03458}\ } (\bibinfo
  {year} {2015})}\BibitemShut {NoStop}%
\bibitem [{\citenamefont {Irwin}(1995)}]{irwin1995application}%
  \BibitemOpen
  \bibfield  {author} {\bibinfo {author} {\bibfnamefont {K.}~\bibnamefont
  {Irwin}},\ }\href@noop {} {\bibfield  {journal} {\bibinfo  {journal} {Applied
  Physics Letters}\ }\textbf {\bibinfo {volume} {66}},\ \bibinfo {pages} {1998}
  (\bibinfo {year} {1995})}\BibitemShut {NoStop}%
\bibitem [{\citenamefont {Zmuidzinas}(2012)}]{zmuidzinas2012superconducting}%
  \BibitemOpen
  \bibfield  {author} {\bibinfo {author} {\bibfnamefont {J.}~\bibnamefont
  {Zmuidzinas}},\ }\href@noop {} {\bibfield  {journal} {\bibinfo  {journal}
  {Annu. Rev. Condens. Matter Phys.}\ }\textbf {\bibinfo {volume} {3}},\
  \bibinfo {pages} {169} (\bibinfo {year} {2012})}\BibitemShut {NoStop}%
\bibitem [{\citenamefont {Leonard}\ \emph {et~al.}(2008)\citenamefont
  {Leonard}, \citenamefont {Grinberg}, \citenamefont {Weber}, \citenamefont
  {Baussan}, \citenamefont {Djurcic}, \citenamefont {Keefer}, \citenamefont
  {Piepke}, \citenamefont {Pocar}, \citenamefont {Vuilleumier}, \citenamefont
  {Vuilleumier} \emph {et~al.}}]{leonard2008systematic}%
  \BibitemOpen
  \bibfield  {author} {\bibinfo {author} {\bibfnamefont {D.}~\bibnamefont
  {Leonard}}, \bibinfo {author} {\bibfnamefont {P.}~\bibnamefont {Grinberg}},
  \bibinfo {author} {\bibfnamefont {P.}~\bibnamefont {Weber}}, \bibinfo
  {author} {\bibfnamefont {E.}~\bibnamefont {Baussan}}, \bibinfo {author}
  {\bibfnamefont {Z.}~\bibnamefont {Djurcic}}, \bibinfo {author} {\bibfnamefont
  {G.}~\bibnamefont {Keefer}}, \bibinfo {author} {\bibfnamefont
  {A.}~\bibnamefont {Piepke}}, \bibinfo {author} {\bibfnamefont
  {A.}~\bibnamefont {Pocar}}, \bibinfo {author} {\bibfnamefont {J.-L.}\
  \bibnamefont {Vuilleumier}},  \emph {et~al.},\ }\href@noop {} {\bibfield
  {journal} {\bibinfo  {journal} {Nuclear Instruments and Methods in Physics
  Research Section A: Accelerators, Spectrometers, Detectors and Associated
  Equipment}\ }\textbf {\bibinfo {volume} {591}},\ \bibinfo {pages} {490}
  (\bibinfo {year} {2008})}\BibitemShut {NoStop}%
\bibitem [{\citenamefont {Arpesella}\ \emph {et~al.}(2002)\citenamefont
  {Arpesella}, \citenamefont {Back}, \citenamefont {Balata}, \citenamefont
  {Beau}, \citenamefont {Bellini}, \citenamefont {Benziger}, \citenamefont
  {Bonetti}, \citenamefont {Brigatti}, \citenamefont {Buck}, \citenamefont
  {Caccianiga} \emph {et~al.}}]{arpesella2002measurements}%
  \BibitemOpen
  \bibfield  {author} {\bibinfo {author} {\bibfnamefont {C.}~\bibnamefont
  {Arpesella}}, \bibinfo {author} {\bibfnamefont {H.}~\bibnamefont {Back}},
  \bibinfo {author} {\bibfnamefont {M.}~\bibnamefont {Balata}}, \bibinfo
  {author} {\bibfnamefont {T.}~\bibnamefont {Beau}}, \bibinfo {author}
  {\bibfnamefont {G.}~\bibnamefont {Bellini}}, \bibinfo {author} {\bibfnamefont
  {J.}~\bibnamefont {Benziger}}, \bibinfo {author} {\bibfnamefont
  {S.}~\bibnamefont {Bonetti}}, \bibinfo {author} {\bibfnamefont
  {A.}~\bibnamefont {Brigatti}}, \bibinfo {author} {\bibfnamefont
  {C.}~\bibnamefont {Buck}},  \emph {et~al.},\ }\href@noop {} {\bibfield
  {journal} {\bibinfo  {journal} {Astroparticle Physics}\ }\textbf {\bibinfo
  {volume} {18}},\ \bibinfo {pages} {1} (\bibinfo {year} {2002})}\BibitemShut
  {NoStop}%
\bibitem [{\citenamefont {Battaglieri}\ \emph {et~al.}(2017)\citenamefont
  {Battaglieri}, \citenamefont {Belloni}, \citenamefont {Chou}, \citenamefont
  {Cushman}, \citenamefont {Echenard}, \citenamefont {Essig}, \citenamefont
  {Estrada}, \citenamefont {Feng}, \citenamefont {Flaugher}, \citenamefont
  {Fox} \emph {et~al.}}]{battaglieri2017us}%
  \BibitemOpen
  \bibfield  {author} {\bibinfo {author} {\bibfnamefont {M.}~\bibnamefont
  {Battaglieri}}, \bibinfo {author} {\bibfnamefont {A.}~\bibnamefont
  {Belloni}}, \bibinfo {author} {\bibfnamefont {A.}~\bibnamefont {Chou}},
  \bibinfo {author} {\bibfnamefont {P.}~\bibnamefont {Cushman}}, \bibinfo
  {author} {\bibfnamefont {B.}~\bibnamefont {Echenard}}, \bibinfo {author}
  {\bibfnamefont {R.}~\bibnamefont {Essig}}, \bibinfo {author} {\bibfnamefont
  {J.}~\bibnamefont {Estrada}}, \bibinfo {author} {\bibfnamefont {J.~L.}\
  \bibnamefont {Feng}}, \bibinfo {author} {\bibfnamefont {B.}~\bibnamefont
  {Flaugher}},  \emph {et~al.},\ }\href@noop {} {\bibfield  {journal} {\bibinfo
   {journal} {arXiv preprint arXiv:1707.04591}\ } (\bibinfo {year}
  {2017})}\BibitemShut {NoStop}%
\bibitem [{\citenamefont {Bunting}\ \emph {et~al.}(2017)\citenamefont
  {Bunting}, \citenamefont {Gratta}, \citenamefont {Melia},\ and\ \citenamefont
  {Rajendran}}]{bunting2017magnetic}%
  \BibitemOpen
  \bibfield  {author} {\bibinfo {author} {\bibfnamefont {P.~C.}\ \bibnamefont
  {Bunting}}, \bibinfo {author} {\bibfnamefont {G.}~\bibnamefont {Gratta}},
  \bibinfo {author} {\bibfnamefont {T.}~\bibnamefont {Melia}}, \ and\ \bibinfo
  {author} {\bibfnamefont {S.}~\bibnamefont {Rajendran}},\ }\href@noop {}
  {\bibfield  {journal} {\bibinfo  {journal} {Physical Review D}\ }\textbf
  {\bibinfo {volume} {95}},\ \bibinfo {pages} {095001} (\bibinfo {year}
  {2017})}\BibitemShut {NoStop}%
\bibitem [{\citenamefont {Hochberg}\ \emph {et~al.}(2016)\citenamefont
  {Hochberg}, \citenamefont {Lin},\ and\ \citenamefont
  {Zurek}}]{hochberg2016detecting}%
  \BibitemOpen
  \bibfield  {author} {\bibinfo {author} {\bibfnamefont {Y.}~\bibnamefont
  {Hochberg}}, \bibinfo {author} {\bibfnamefont {T.}~\bibnamefont {Lin}}, \
  and\ \bibinfo {author} {\bibfnamefont {K.~M.}\ \bibnamefont {Zurek}},\
  }\href@noop {} {\bibfield  {journal} {\bibinfo  {journal} {Physical Review
  D}\ }\textbf {\bibinfo {volume} {94}},\ \bibinfo {pages} {015019} (\bibinfo
  {year} {2016})}\BibitemShut {NoStop}%
\bibitem [{\citenamefont {Hochberg}\ \emph
  {et~al.}(2017{\natexlab{a}})\citenamefont {Hochberg}, \citenamefont {Lin},\
  and\ \citenamefont {Zurek}}]{hochberg2017absorption}%
  \BibitemOpen
  \bibfield  {author} {\bibinfo {author} {\bibfnamefont {Y.}~\bibnamefont
  {Hochberg}}, \bibinfo {author} {\bibfnamefont {T.}~\bibnamefont {Lin}}, \
  and\ \bibinfo {author} {\bibfnamefont {K.~M.}\ \bibnamefont {Zurek}},\
  }\href@noop {} {\bibfield  {journal} {\bibinfo  {journal} {Physical Review
  D}\ }\textbf {\bibinfo {volume} {95}},\ \bibinfo {pages} {023013} (\bibinfo
  {year} {2017}{\natexlab{a}})}\BibitemShut {NoStop}%
\bibitem [{\citenamefont {Derenzo}\ \emph {et~al.}(2017)\citenamefont
  {Derenzo}, \citenamefont {Essig}, \citenamefont {Massari}, \citenamefont
  {Soto},\ and\ \citenamefont {Yu}}]{derenzo2017direct}%
  \BibitemOpen
  \bibfield  {author} {\bibinfo {author} {\bibfnamefont {S.}~\bibnamefont
  {Derenzo}}, \bibinfo {author} {\bibfnamefont {R.}~\bibnamefont {Essig}},
  \bibinfo {author} {\bibfnamefont {A.}~\bibnamefont {Massari}}, \bibinfo
  {author} {\bibfnamefont {A.}~\bibnamefont {Soto}}, \ and\ \bibinfo {author}
  {\bibfnamefont {T.-T.}\ \bibnamefont {Yu}},\ }\href@noop {} {\bibfield
  {journal} {\bibinfo  {journal} {Physical Review D}\ }\textbf {\bibinfo
  {volume} {96}},\ \bibinfo {pages} {016026} (\bibinfo {year}
  {2017})}\BibitemShut {NoStop}%
\bibitem [{\citenamefont {Bloch}\ \emph {et~al.}(2017)\citenamefont {Bloch},
  \citenamefont {Essig}, \citenamefont {Tobioka}, \citenamefont {Volansky},\
  and\ \citenamefont {Yu}}]{bloch2017searching}%
  \BibitemOpen
  \bibfield  {author} {\bibinfo {author} {\bibfnamefont {I.~M.}\ \bibnamefont
  {Bloch}}, \bibinfo {author} {\bibfnamefont {R.}~\bibnamefont {Essig}},
  \bibinfo {author} {\bibfnamefont {K.}~\bibnamefont {Tobioka}}, \bibinfo
  {author} {\bibfnamefont {T.}~\bibnamefont {Volansky}}, \ and\ \bibinfo
  {author} {\bibfnamefont {T.-T.}\ \bibnamefont {Yu}},\ }\href@noop {}
  {\bibfield  {journal} {\bibinfo  {journal} {Journal of High Energy Physics}\
  }\textbf {\bibinfo {volume} {2017}},\ \bibinfo {pages} {1} (\bibinfo {year}
  {2017})}\BibitemShut {NoStop}%
\bibitem [{\citenamefont {Hochberg}\ \emph
  {et~al.}(2017{\natexlab{b}})\citenamefont {Hochberg}, \citenamefont {Kahn},
  \citenamefont {Lisanti}, \citenamefont {Zurek}, \citenamefont {Grushin},
  \citenamefont {Ilan}, \citenamefont {Griffin}, \citenamefont {Liu},
  \citenamefont {Weber},\ and\ \citenamefont {Neaton}}]{Hochberg:2017wce}%
  \BibitemOpen
  \bibfield  {author} {\bibinfo {author} {\bibfnamefont {Y.}~\bibnamefont
  {Hochberg}}, \bibinfo {author} {\bibfnamefont {Y.}~\bibnamefont {Kahn}},
  \bibinfo {author} {\bibfnamefont {M.}~\bibnamefont {Lisanti}}, \bibinfo
  {author} {\bibfnamefont {K.~M.}\ \bibnamefont {Zurek}}, \bibinfo {author}
  {\bibfnamefont {A.}~\bibnamefont {Grushin}}, \bibinfo {author} {\bibfnamefont
  {R.}~\bibnamefont {Ilan}}, \bibinfo {author} {\bibfnamefont {S.~M.}\
  \bibnamefont {Griffin}}, \bibinfo {author} {\bibfnamefont {Z.-F.}\
  \bibnamefont {Liu}}, \bibinfo {author} {\bibfnamefont {S.~F.}\ \bibnamefont
  {Weber}}, \ and\ \bibinfo {author} {\bibfnamefont {J.~B.}\ \bibnamefont
  {Neaton}},\ }\href@noop {} {\  (\bibinfo {year} {2017}{\natexlab{b}})},\
  \Eprint {http://arxiv.org/abs/1708.08929} {arXiv:1708.08929 [hep-ph]}
  \BibitemShut {NoStop}%
\bibitem [{\citenamefont {Landau}\ and\ \citenamefont
  {Lifshitz}(1958)}]{landau1958quantum}%
  \BibitemOpen
  \bibfield  {author} {\bibinfo {author} {\bibfnamefont {L.~D.}\ \bibnamefont
  {Landau}}\ and\ \bibinfo {author} {\bibfnamefont {E.~M.}\ \bibnamefont
  {Lifshitz}},\ }\href@noop {} {\  (\bibinfo {year} {1958})}\BibitemShut
  {NoStop}%
\bibitem [{\citenamefont {De~Pascale}\ \emph {et~al.}(1993)\citenamefont
  {De~Pascale}, \citenamefont {Morselli}, \citenamefont {Picozza},
  \citenamefont {Golden}, \citenamefont {Grimani}, \citenamefont {Kimbell},
  \citenamefont {Stephens}, \citenamefont {Stochaj}, \citenamefont {Webber},
  \citenamefont {Basini} \emph {et~al.}}]{de1993absolute}%
  \BibitemOpen
  \bibfield  {author} {\bibinfo {author} {\bibfnamefont {M.}~\bibnamefont
  {De~Pascale}}, \bibinfo {author} {\bibfnamefont {A.}~\bibnamefont
  {Morselli}}, \bibinfo {author} {\bibfnamefont {P.}~\bibnamefont {Picozza}},
  \bibinfo {author} {\bibfnamefont {R.}~\bibnamefont {Golden}}, \bibinfo
  {author} {\bibfnamefont {C.}~\bibnamefont {Grimani}}, \bibinfo {author}
  {\bibfnamefont {B.}~\bibnamefont {Kimbell}}, \bibinfo {author} {\bibfnamefont
  {S.}~\bibnamefont {Stephens}}, \bibinfo {author} {\bibfnamefont
  {S.}~\bibnamefont {Stochaj}}, \bibinfo {author} {\bibfnamefont
  {W.}~\bibnamefont {Webber}},  \emph {et~al.},\ }\href@noop {} {\bibfield
  {journal} {\bibinfo  {journal} {Journal of Geophysical Research: Space
  Physics}\ }\textbf {\bibinfo {volume} {98}},\ \bibinfo {pages} {3501}
  (\bibinfo {year} {1993})}\BibitemShut {NoStop}%
\bibitem [{\citenamefont {Grieder}(2001)}]{grieder2001cosmic}%
  \BibitemOpen
  \bibfield  {author} {\bibinfo {author} {\bibfnamefont {P.~K.}\ \bibnamefont
  {Grieder}},\ }\href@noop {} {\emph {\bibinfo {title} {Cosmic rays at
  Earth}}}\ (\bibinfo  {publisher} {Gulf Professional Publishing},\ \bibinfo
  {year} {2001})\BibitemShut {NoStop}%
\bibitem [{\citenamefont {Andreev}\ \emph {et~al.}(1987)\citenamefont
  {Andreev}, \citenamefont {Gurentzov},\ and\ \citenamefont
  {Kogai}}]{andreev1987proc}%
  \BibitemOpen
  \bibfield  {author} {\bibinfo {author} {\bibfnamefont {Y.~M.}\ \bibnamefont
  {Andreev}}, \bibinfo {author} {\bibfnamefont {V.}~\bibnamefont {Gurentzov}},
  \ and\ \bibinfo {author} {\bibfnamefont {I.}~\bibnamefont {Kogai}},\
  }\href@noop {} {\bibfield  {journal} {\bibinfo  {journal} {Moscow}\ }\textbf
  {\bibinfo {volume} {6}},\ \bibinfo {pages} {200} (\bibinfo {year}
  {1987})}\BibitemShut {NoStop}%
\bibitem [{\citenamefont {Aglietta}\ \emph {et~al.}(1995)\citenamefont
  {Aglietta}, \citenamefont {Alpat}, \citenamefont {Alyea}, \citenamefont
  {Antonioli}, \citenamefont {Badino}, \citenamefont {Ban}, \citenamefont
  {Bari}, \citenamefont {Basile}, \citenamefont {Benelli}, \citenamefont
  {Berezinsky} \emph {et~al.}}]{aglietta1995neutrino}%
  \BibitemOpen
  \bibfield  {author} {\bibinfo {author} {\bibfnamefont {M.}~\bibnamefont
  {Aglietta}}, \bibinfo {author} {\bibfnamefont {B.}~\bibnamefont {Alpat}},
  \bibinfo {author} {\bibfnamefont {E.}~\bibnamefont {Alyea}}, \bibinfo
  {author} {\bibfnamefont {P.}~\bibnamefont {Antonioli}}, \bibinfo {author}
  {\bibfnamefont {G.}~\bibnamefont {Badino}}, \bibinfo {author} {\bibfnamefont
  {Y.}~\bibnamefont {Ban}}, \bibinfo {author} {\bibfnamefont {G.}~\bibnamefont
  {Bari}}, \bibinfo {author} {\bibfnamefont {M.}~\bibnamefont {Basile}},
  \bibinfo {author} {\bibfnamefont {A.}~\bibnamefont {Benelli}},  \emph
  {et~al.},\ }\href@noop {} {\bibfield  {journal} {\bibinfo  {journal}
  {Astroparticle Physics}\ }\textbf {\bibinfo {volume} {3}},\ \bibinfo {pages}
  {311} (\bibinfo {year} {1995})}\BibitemShut {NoStop}%
\bibitem [{\citenamefont {Ambrosio}\ \emph {et~al.}(1995)\citenamefont
  {Ambrosio}, \citenamefont {Antolini}, \citenamefont {Auriemma}, \citenamefont
  {Baker}, \citenamefont {Baldini}, \citenamefont {Barbarino}, \citenamefont
  {Barish}, \citenamefont {Battistoni}, \citenamefont {Bellotti}, \citenamefont
  {Bemporad} \emph {et~al.}}]{ambrosio1995vertical}%
  \BibitemOpen
  \bibfield  {author} {\bibinfo {author} {\bibfnamefont {M.}~\bibnamefont
  {Ambrosio}}, \bibinfo {author} {\bibfnamefont {R.}~\bibnamefont {Antolini}},
  \bibinfo {author} {\bibfnamefont {G.}~\bibnamefont {Auriemma}}, \bibinfo
  {author} {\bibfnamefont {R.}~\bibnamefont {Baker}}, \bibinfo {author}
  {\bibfnamefont {A.}~\bibnamefont {Baldini}}, \bibinfo {author} {\bibfnamefont
  {G.}~\bibnamefont {Barbarino}}, \bibinfo {author} {\bibfnamefont
  {B.}~\bibnamefont {Barish}}, \bibinfo {author} {\bibfnamefont
  {G.}~\bibnamefont {Battistoni}}, \bibinfo {author} {\bibfnamefont
  {R.}~\bibnamefont {Bellotti}},  \emph {et~al.},\ }\href@noop {} {\bibfield
  {journal} {\bibinfo  {journal} {Physical Review D}\ }\textbf {\bibinfo
  {volume} {52}},\ \bibinfo {pages} {3793} (\bibinfo {year}
  {1995})}\BibitemShut {NoStop}%
\bibitem [{\citenamefont {Berger}\ \emph {et~al.}(1989)\citenamefont {Berger},
  \citenamefont {Fr{\"o}hlich}, \citenamefont {M{\"o}nch}, \citenamefont
  {Nisius}, \citenamefont {Raupach}, \citenamefont {Schleper}, \citenamefont
  {Benadjal}, \citenamefont {Blum}, \citenamefont {Bourdarious}, \citenamefont
  {Dudelzak} \emph {et~al.}}]{berger1989experimental}%
  \BibitemOpen
  \bibfield  {author} {\bibinfo {author} {\bibfnamefont {C.}~\bibnamefont
  {Berger}}, \bibinfo {author} {\bibfnamefont {M.}~\bibnamefont
  {Fr{\"o}hlich}}, \bibinfo {author} {\bibfnamefont {H.}~\bibnamefont
  {M{\"o}nch}}, \bibinfo {author} {\bibfnamefont {R.}~\bibnamefont {Nisius}},
  \bibinfo {author} {\bibfnamefont {F.}~\bibnamefont {Raupach}}, \bibinfo
  {author} {\bibfnamefont {P.}~\bibnamefont {Schleper}}, \bibinfo {author}
  {\bibfnamefont {Y.}~\bibnamefont {Benadjal}}, \bibinfo {author}
  {\bibfnamefont {D.}~\bibnamefont {Blum}}, \bibinfo {author} {\bibfnamefont
  {C.}~\bibnamefont {Bourdarious}},  \emph {et~al.},\ }\href@noop {} {\bibfield
   {journal} {\bibinfo  {journal} {Physical Review D}\ }\textbf {\bibinfo
  {volume} {40}},\ \bibinfo {pages} {2163} (\bibinfo {year}
  {1989})}\BibitemShut {NoStop}%
\bibitem [{\citenamefont {Waltham}\ \emph {et~al.}(2001)\citenamefont
  {Waltham}, \citenamefont {Collaboration} \emph
  {et~al.}}]{waltham2001through}%
  \BibitemOpen
  \bibfield  {author} {\bibinfo {author} {\bibfnamefont {C.}~\bibnamefont
  {Waltham}}, \bibinfo {author} {\bibfnamefont {S.}~\bibnamefont
  {Collaboration}},  \emph {et~al.},\ }in\ \href@noop {} {\emph {\bibinfo
  {booktitle} {International Cosmic Ray Conference}}},\ Vol.~\bibinfo {volume}
  {3}\ (\bibinfo {year} {2001})\ p.\ \bibinfo {pages} {991}\BibitemShut
  {NoStop}%
\bibitem [{\citenamefont {Graham}\ \emph
  {et~al.}(2016{\natexlab{a}})\citenamefont {Graham}, \citenamefont {Mardon},\
  and\ \citenamefont {Rajendran}}]{graham2016vector}%
  \BibitemOpen
  \bibfield  {author} {\bibinfo {author} {\bibfnamefont {P.~W.}\ \bibnamefont
  {Graham}}, \bibinfo {author} {\bibfnamefont {J.}~\bibnamefont {Mardon}}, \
  and\ \bibinfo {author} {\bibfnamefont {S.}~\bibnamefont {Rajendran}},\
  }\href@noop {} {\bibfield  {journal} {\bibinfo  {journal} {Physical Review
  D}\ }\textbf {\bibinfo {volume} {93}},\ \bibinfo {pages} {103520} (\bibinfo
  {year} {2016}{\natexlab{a}})}\BibitemShut {NoStop}%
\bibitem [{\citenamefont {Arias}\ \emph {et~al.}(2012)\citenamefont {Arias},
  \citenamefont {Cadamuro}, \citenamefont {Goodsell}, \citenamefont {Jaeckel},
  \citenamefont {Redondo},\ and\ \citenamefont {Ringwald}}]{arias2012wispy}%
  \BibitemOpen
  \bibfield  {author} {\bibinfo {author} {\bibfnamefont {P.}~\bibnamefont
  {Arias}}, \bibinfo {author} {\bibfnamefont {D.}~\bibnamefont {Cadamuro}},
  \bibinfo {author} {\bibfnamefont {M.}~\bibnamefont {Goodsell}}, \bibinfo
  {author} {\bibfnamefont {J.}~\bibnamefont {Jaeckel}}, \bibinfo {author}
  {\bibfnamefont {J.}~\bibnamefont {Redondo}}, \ and\ \bibinfo {author}
  {\bibfnamefont {A.}~\bibnamefont {Ringwald}},\ }\href@noop {} {\bibfield
  {journal} {\bibinfo  {journal} {Journal of Cosmology and Astroparticle
  Physics}\ }\textbf {\bibinfo {volume} {2012}},\ \bibinfo {pages} {013}
  (\bibinfo {year} {2012})}\BibitemShut {NoStop}%
\bibitem [{\citenamefont {Holdom}(1986)}]{holdom1986two}%
  \BibitemOpen
  \bibfield  {author} {\bibinfo {author} {\bibfnamefont {B.}~\bibnamefont
  {Holdom}},\ }\href@noop {} {\bibfield  {journal} {\bibinfo  {journal}
  {Physics Letters B}\ }\textbf {\bibinfo {volume} {166}},\ \bibinfo {pages}
  {196} (\bibinfo {year} {1986})}\BibitemShut {NoStop}%
\bibitem [{\citenamefont {Okun}(1982)}]{okun1982limits}%
  \BibitemOpen
  \bibfield  {author} {\bibinfo {author} {\bibfnamefont {L.~B.}\ \bibnamefont
  {Okun}},\ }\href@noop {} {\bibfield  {journal} {\bibinfo  {journal} {Zhurnal
  Eksperimental'noi i Teoreticheskoi Fiziki}\ }\textbf {\bibinfo {volume}
  {83}},\ \bibinfo {pages} {892} (\bibinfo {year} {1982})}\BibitemShut
  {NoStop}%
\bibitem [{\citenamefont {Chaudhuri}\ \emph {et~al.}(2015)\citenamefont
  {Chaudhuri}, \citenamefont {Graham}, \citenamefont {Irwin}, \citenamefont
  {Mardon}, \citenamefont {Rajendran},\ and\ \citenamefont
  {Zhao}}]{chaudhuri2015radio}%
  \BibitemOpen
  \bibfield  {author} {\bibinfo {author} {\bibfnamefont {S.}~\bibnamefont
  {Chaudhuri}}, \bibinfo {author} {\bibfnamefont {P.~W.}\ \bibnamefont
  {Graham}}, \bibinfo {author} {\bibfnamefont {K.}~\bibnamefont {Irwin}},
  \bibinfo {author} {\bibfnamefont {J.}~\bibnamefont {Mardon}}, \bibinfo
  {author} {\bibfnamefont {S.}~\bibnamefont {Rajendran}}, \ and\ \bibinfo
  {author} {\bibfnamefont {Y.}~\bibnamefont {Zhao}},\ }\href@noop {} {\bibfield
   {journal} {\bibinfo  {journal} {Physical Review D}\ }\textbf {\bibinfo
  {volume} {92}},\ \bibinfo {pages} {075012} (\bibinfo {year}
  {2015})}\BibitemShut {NoStop}%
\bibitem [{\citenamefont {Mulliken}(1971)}]{mulliken1971iodine}%
  \BibitemOpen
  \bibfield  {author} {\bibinfo {author} {\bibfnamefont {R.~S.}\ \bibnamefont
  {Mulliken}},\ }\href@noop {} {\bibfield  {journal} {\bibinfo  {journal} {The
  Journal of Chemical Physics}\ }\textbf {\bibinfo {volume} {55}},\ \bibinfo
  {pages} {288} (\bibinfo {year} {1971})}\BibitemShut {NoStop}%
\bibitem [{\citenamefont {Lamrini}\ \emph {et~al.}(1994)\citenamefont
  {Lamrini}, \citenamefont {Bacis}, \citenamefont {Cerny}, \citenamefont
  {Churassy}, \citenamefont {Crozet},\ and\ \citenamefont
  {Ross}}]{lamrini1994electronic}%
  \BibitemOpen
  \bibfield  {author} {\bibinfo {author} {\bibfnamefont {M.}~\bibnamefont
  {Lamrini}}, \bibinfo {author} {\bibfnamefont {R.}~\bibnamefont {Bacis}},
  \bibinfo {author} {\bibfnamefont {D.}~\bibnamefont {Cerny}}, \bibinfo
  {author} {\bibfnamefont {S.}~\bibnamefont {Churassy}}, \bibinfo {author}
  {\bibfnamefont {P.}~\bibnamefont {Crozet}}, \ and\ \bibinfo {author}
  {\bibfnamefont {A.}~\bibnamefont {Ross}},\ }\href@noop {} {\bibfield
  {journal} {\bibinfo  {journal} {The Journal of chemical physics}\ }\textbf
  {\bibinfo {volume} {100}},\ \bibinfo {pages} {8780} (\bibinfo {year}
  {1994})}\BibitemShut {NoStop}%
\bibitem [{\citenamefont {Brewer}\ \emph {et~al.}(1963)\citenamefont {Brewer},
  \citenamefont {Berg},\ and\ \citenamefont
  {Rosenblatt}}]{brewer1963radiative}%
  \BibitemOpen
  \bibfield  {author} {\bibinfo {author} {\bibfnamefont {L.}~\bibnamefont
  {Brewer}}, \bibinfo {author} {\bibfnamefont {R.~A.}\ \bibnamefont {Berg}}, \
  and\ \bibinfo {author} {\bibfnamefont {G.~M.}\ \bibnamefont {Rosenblatt}},\
  }\href@noop {} {\bibfield  {journal} {\bibinfo  {journal} {The Journal of
  Chemical Physics}\ }\textbf {\bibinfo {volume} {38}},\ \bibinfo {pages}
  {1381} (\bibinfo {year} {1963})}\BibitemShut {NoStop}%
\bibitem [{\citenamefont {Zare}(1964)}]{zare1964calculation}%
  \BibitemOpen
  \bibfield  {author} {\bibinfo {author} {\bibfnamefont {R.}~\bibnamefont
  {Zare}},\ }\href@noop {} {\bibfield  {journal} {\bibinfo  {journal} {The
  Journal of Chemical Physics}\ }\textbf {\bibinfo {volume} {40}},\ \bibinfo
  {pages} {1934} (\bibinfo {year} {1964})}\BibitemShut {NoStop}%
\bibitem [{\citenamefont {Paisner}\ and\ \citenamefont
  {Wallenstein}(1974)}]{paisner1974rotational}%
  \BibitemOpen
  \bibfield  {author} {\bibinfo {author} {\bibfnamefont {J.}~\bibnamefont
  {Paisner}}\ and\ \bibinfo {author} {\bibfnamefont {R.}~\bibnamefont
  {Wallenstein}},\ }\href@noop {} {\bibfield  {journal} {\bibinfo  {journal}
  {The Journal of Chemical Physics}\ }\textbf {\bibinfo {volume} {61}},\
  \bibinfo {pages} {4317} (\bibinfo {year} {1974})}\BibitemShut {NoStop}%
\bibitem [{\citenamefont {Tellinghuisen}(1978)}]{tellinghuisen1978intensity}%
  \BibitemOpen
  \bibfield  {author} {\bibinfo {author} {\bibfnamefont {J.}~\bibnamefont
  {Tellinghuisen}},\ }\href@noop {} {\bibfield  {journal} {\bibinfo  {journal}
  {Journal of Quantitative Spectroscopy and Radiative Transfer}\ }\textbf
  {\bibinfo {volume} {19}},\ \bibinfo {pages} {149} (\bibinfo {year}
  {1978})}\BibitemShut {NoStop}%
\bibitem [{\citenamefont {Cooper}\ and\ \citenamefont
  {Langhoff}(1981)}]{cooper1981theoretical}%
  \BibitemOpen
  \bibfield  {author} {\bibinfo {author} {\bibfnamefont {D.~M.}\ \bibnamefont
  {Cooper}}\ and\ \bibinfo {author} {\bibfnamefont {S.~R.}\ \bibnamefont
  {Langhoff}},\ }\href@noop {} {\bibfield  {journal} {\bibinfo  {journal} {The
  Journal of Chemical Physics}\ }\textbf {\bibinfo {volume} {74}},\ \bibinfo
  {pages} {1200} (\bibinfo {year} {1981})}\BibitemShut {NoStop}%
\bibitem [{\citenamefont {Chackerian~Jr}(1976)}]{chackerian1976electric}%
  \BibitemOpen
  \bibfield  {author} {\bibinfo {author} {\bibfnamefont {C.}~\bibnamefont
  {Chackerian~Jr}},\ }\href@noop {} {\bibfield  {journal} {\bibinfo  {journal}
  {The Journal of Chemical Physics}\ }\textbf {\bibinfo {volume} {65}},\
  \bibinfo {pages} {4228} (\bibinfo {year} {1976})}\BibitemShut {NoStop}%
\bibitem [{\citenamefont {Halmann}\ and\ \citenamefont
  {Laulicht}(1966)}]{halmann1966isotope}%
  \BibitemOpen
  \bibfield  {author} {\bibinfo {author} {\bibfnamefont {M.}~\bibnamefont
  {Halmann}}\ and\ \bibinfo {author} {\bibfnamefont {I.}~\bibnamefont
  {Laulicht}},\ }\href@noop {} {\bibfield  {journal} {\bibinfo  {journal} {The
  Astrophysical Journal Supplement Series}\ }\textbf {\bibinfo {volume} {12}},\
  \bibinfo {pages} {307} (\bibinfo {year} {1966})}\BibitemShut {NoStop}%
\bibitem [{\citenamefont {Krupenie}(1966)}]{krupenie1966band}%
  \BibitemOpen
  \bibfield  {author} {\bibinfo {author} {\bibfnamefont {P.~H.}\ \bibnamefont
  {Krupenie}},\ }\href@noop {} {\emph {\bibinfo {title} {The band spectrum of
  carbon monoxide}}},\ \bibinfo {type} {Tech. Rep.}\ (\bibinfo  {institution}
  {NATIONAL STANDARD REFERENCE DATA SYSTEM},\ \bibinfo {year}
  {1966})\BibitemShut {NoStop}%
\bibitem [{dot()}]{dotchin1973radiative}%
  \BibitemOpen
  \href@noop {} {\ }\BibitemShut {NoStop}%
\bibitem [{\citenamefont {Tilford}\ and\ \citenamefont
  {Simmons}(1972)}]{tilford1972atlas}%
  \BibitemOpen
  \bibfield  {author} {\bibinfo {author} {\bibfnamefont {S.}~\bibnamefont
  {Tilford}}\ and\ \bibinfo {author} {\bibfnamefont {J.}~\bibnamefont
  {Simmons}},\ }\href@noop {} {\bibfield  {journal} {\bibinfo  {journal}
  {Journal of Physical and Chemical Reference Data}\ }\textbf {\bibinfo
  {volume} {1}},\ \bibinfo {pages} {147} (\bibinfo {year} {1972})}\BibitemShut
  {NoStop}%
\bibitem [{\citenamefont {Fantz}\ and\ \citenamefont
  {W{\"u}nderlich}(2006)}]{fantz2006franck}%
  \BibitemOpen
  \bibfield  {author} {\bibinfo {author} {\bibfnamefont {U.}~\bibnamefont
  {Fantz}}\ and\ \bibinfo {author} {\bibfnamefont {D.}~\bibnamefont
  {W{\"u}nderlich}},\ }\href@noop {} {\bibfield  {journal} {\bibinfo  {journal}
  {Atomic Data and Nuclear Data Tables}\ }\textbf {\bibinfo {volume} {92}},\
  \bibinfo {pages} {853} (\bibinfo {year} {2006})}\BibitemShut {NoStop}%
\bibitem [{\citenamefont {Angle}\ \emph {et~al.}(2008)\citenamefont {Angle},
  \citenamefont {Aprile}, \citenamefont {Arneodo}, \citenamefont {Baudis},
  \citenamefont {Bernstein}, \citenamefont {Bolozdynya}, \citenamefont
  {Brusov}, \citenamefont {Coelho}, \citenamefont {Dahl}, \citenamefont
  {DeViveiros} \emph {et~al.}}]{angle2008first}%
  \BibitemOpen
  \bibfield  {author} {\bibinfo {author} {\bibfnamefont {J.}~\bibnamefont
  {Angle}}, \bibinfo {author} {\bibfnamefont {E.}~\bibnamefont {Aprile}},
  \bibinfo {author} {\bibfnamefont {F.}~\bibnamefont {Arneodo}}, \bibinfo
  {author} {\bibfnamefont {L.}~\bibnamefont {Baudis}}, \bibinfo {author}
  {\bibfnamefont {A.}~\bibnamefont {Bernstein}}, \bibinfo {author}
  {\bibfnamefont {A.}~\bibnamefont {Bolozdynya}}, \bibinfo {author}
  {\bibfnamefont {P.}~\bibnamefont {Brusov}}, \bibinfo {author} {\bibfnamefont
  {L.}~\bibnamefont {Coelho}}, \bibinfo {author} {\bibfnamefont
  {C.}~\bibnamefont {Dahl}},  \emph {et~al.},\ }\href@noop {} {\bibfield
  {journal} {\bibinfo  {journal} {Physical Review Letters}\ }\textbf {\bibinfo
  {volume} {100}},\ \bibinfo {pages} {021303} (\bibinfo {year}
  {2008})}\BibitemShut {NoStop}%
\bibitem [{\citenamefont {An}\ \emph {et~al.}(2015)\citenamefont {An},
  \citenamefont {Pospelov}, \citenamefont {Pradler},\ and\ \citenamefont
  {Ritz}}]{an2015direct}%
  \BibitemOpen
  \bibfield  {author} {\bibinfo {author} {\bibfnamefont {H.}~\bibnamefont
  {An}}, \bibinfo {author} {\bibfnamefont {M.}~\bibnamefont {Pospelov}},
  \bibinfo {author} {\bibfnamefont {J.}~\bibnamefont {Pradler}}, \ and\
  \bibinfo {author} {\bibfnamefont {A.}~\bibnamefont {Ritz}},\ }\href@noop {}
  {\bibfield  {journal} {\bibinfo  {journal} {Physics Letters B}\ }\textbf
  {\bibinfo {volume} {747}},\ \bibinfo {pages} {331} (\bibinfo {year}
  {2015})}\BibitemShut {NoStop}%
\bibitem [{\citenamefont {An}\ \emph {et~al.}(2013)\citenamefont {An},
  \citenamefont {Pospelov},\ and\ \citenamefont {Pradler}}]{an2013dark}%
  \BibitemOpen
  \bibfield  {author} {\bibinfo {author} {\bibfnamefont {H.}~\bibnamefont
  {An}}, \bibinfo {author} {\bibfnamefont {M.}~\bibnamefont {Pospelov}}, \ and\
  \bibinfo {author} {\bibfnamefont {J.}~\bibnamefont {Pradler}},\ }\href@noop
  {} {\bibfield  {journal} {\bibinfo  {journal} {Physical review letters}\
  }\textbf {\bibinfo {volume} {111}},\ \bibinfo {pages} {041302} (\bibinfo
  {year} {2013})}\BibitemShut {NoStop}%
\bibitem [{\citenamefont {Adelberger}\ \emph {et~al.}(2003)\citenamefont
  {Adelberger}, \citenamefont {Heckel},\ and\ \citenamefont
  {Nelson}}]{adelberger2003tests}%
  \BibitemOpen
  \bibfield  {author} {\bibinfo {author} {\bibfnamefont {E.~G.}\ \bibnamefont
  {Adelberger}}, \bibinfo {author} {\bibfnamefont {B.~R.}\ \bibnamefont
  {Heckel}}, \ and\ \bibinfo {author} {\bibfnamefont {A.}~\bibnamefont
  {Nelson}},\ }\href@noop {} {\bibfield  {journal} {\bibinfo  {journal} {Annual
  Review of Nuclear and Particle Science}\ }\textbf {\bibinfo {volume} {53}},\
  \bibinfo {pages} {77} (\bibinfo {year} {2003})}\BibitemShut {NoStop}%
\bibitem [{\citenamefont {Preskill}\ \emph {et~al.}(1983)\citenamefont
  {Preskill}, \citenamefont {Wise},\ and\ \citenamefont
  {Wilczek}}]{preskill1983cosmology}%
  \BibitemOpen
  \bibfield  {author} {\bibinfo {author} {\bibfnamefont {J.}~\bibnamefont
  {Preskill}}, \bibinfo {author} {\bibfnamefont {M.~B.}\ \bibnamefont {Wise}},
  \ and\ \bibinfo {author} {\bibfnamefont {F.}~\bibnamefont {Wilczek}},\
  }\href@noop {} {\bibfield  {journal} {\bibinfo  {journal} {Physics Letters
  B}\ }\textbf {\bibinfo {volume} {120}},\ \bibinfo {pages} {127} (\bibinfo
  {year} {1983})}\BibitemShut {NoStop}%
\bibitem [{\citenamefont {Dine}\ and\ \citenamefont
  {Fischler}(1983)}]{dine1983not}%
  \BibitemOpen
  \bibfield  {author} {\bibinfo {author} {\bibfnamefont {M.}~\bibnamefont
  {Dine}}\ and\ \bibinfo {author} {\bibfnamefont {W.}~\bibnamefont
  {Fischler}},\ }\href@noop {} {\bibfield  {journal} {\bibinfo  {journal}
  {Physics Letters B}\ }\textbf {\bibinfo {volume} {120}},\ \bibinfo {pages}
  {137} (\bibinfo {year} {1983})}\BibitemShut {NoStop}%
\bibitem [{\citenamefont {Abbott}\ and\ \citenamefont
  {Sikivie}(1983)}]{abbott1983cosmological}%
  \BibitemOpen
  \bibfield  {author} {\bibinfo {author} {\bibfnamefont {L.~F.}\ \bibnamefont
  {Abbott}}\ and\ \bibinfo {author} {\bibfnamefont {P.}~\bibnamefont
  {Sikivie}},\ }\href@noop {} {\bibfield  {journal} {\bibinfo  {journal}
  {Physics Letters B}\ }\textbf {\bibinfo {volume} {120}},\ \bibinfo {pages}
  {133} (\bibinfo {year} {1983})}\BibitemShut {NoStop}%
\bibitem [{\citenamefont {Marsh}(2016)}]{marsh2016axion}%
  \BibitemOpen
  \bibfield  {author} {\bibinfo {author} {\bibfnamefont {D.~J.}\ \bibnamefont
  {Marsh}},\ }\href@noop {} {\bibfield  {journal} {\bibinfo  {journal} {Physics
  Reports}\ }\textbf {\bibinfo {volume} {643}},\ \bibinfo {pages} {1} (\bibinfo
  {year} {2016})}\BibitemShut {NoStop}%
\bibitem [{\citenamefont {Piazza}\ and\ \citenamefont
  {Pospelov}(2010)}]{piazza2010sub}%
  \BibitemOpen
  \bibfield  {author} {\bibinfo {author} {\bibfnamefont {F.}~\bibnamefont
  {Piazza}}\ and\ \bibinfo {author} {\bibfnamefont {M.}~\bibnamefont
  {Pospelov}},\ }\href@noop {} {\bibfield  {journal} {\bibinfo  {journal}
  {Physical Review D}\ }\textbf {\bibinfo {volume} {82}},\ \bibinfo {pages}
  {043533} (\bibinfo {year} {2010})}\BibitemShut {NoStop}%
\bibitem [{\citenamefont {Damour}\ and\ \citenamefont
  {Polyakov}(1994)}]{damour1994string}%
  \BibitemOpen
  \bibfield  {author} {\bibinfo {author} {\bibfnamefont {T.}~\bibnamefont
  {Damour}}\ and\ \bibinfo {author} {\bibfnamefont {A.~M.}\ \bibnamefont
  {Polyakov}},\ }\href@noop {} {\bibfield  {journal} {\bibinfo  {journal}
  {Nuclear Physics B}\ }\textbf {\bibinfo {volume} {423}},\ \bibinfo {pages}
  {532} (\bibinfo {year} {1994})}\BibitemShut {NoStop}%
\bibitem [{\citenamefont {Taylor}\ and\ \citenamefont
  {Veneziano}(1988)}]{taylor1988dilaton}%
  \BibitemOpen
  \bibfield  {author} {\bibinfo {author} {\bibfnamefont {T.}~\bibnamefont
  {Taylor}}\ and\ \bibinfo {author} {\bibfnamefont {G.}~\bibnamefont
  {Veneziano}},\ }\href@noop {} {\bibfield  {journal} {\bibinfo  {journal}
  {Physics Letters B}\ }\textbf {\bibinfo {volume} {213}},\ \bibinfo {pages}
  {450} (\bibinfo {year} {1988})}\BibitemShut {NoStop}%
\bibitem [{\citenamefont {Arkani-Hamed}\ \emph {et~al.}(2000)\citenamefont
  {Arkani-Hamed}, \citenamefont {Hall}, \citenamefont {Smith},\ and\
  \citenamefont {Weiner}}]{arkani2000solving}%
  \BibitemOpen
  \bibfield  {author} {\bibinfo {author} {\bibfnamefont {N.}~\bibnamefont
  {Arkani-Hamed}}, \bibinfo {author} {\bibfnamefont {L.}~\bibnamefont {Hall}},
  \bibinfo {author} {\bibfnamefont {D.}~\bibnamefont {Smith}}, \ and\ \bibinfo
  {author} {\bibfnamefont {N.}~\bibnamefont {Weiner}},\ }\href@noop {}
  {\bibfield  {journal} {\bibinfo  {journal} {Physical Review D}\ }\textbf
  {\bibinfo {volume} {62}},\ \bibinfo {pages} {105002} (\bibinfo {year}
  {2000})}\BibitemShut {NoStop}%
\bibitem [{\citenamefont {Burgess}\ \emph {et~al.}(2011)\citenamefont
  {Burgess}, \citenamefont {Maharana},\ and\ \citenamefont
  {Quevedo}}]{burgess2011naturalness}%
  \BibitemOpen
  \bibfield  {author} {\bibinfo {author} {\bibfnamefont {C.}~\bibnamefont
  {Burgess}}, \bibinfo {author} {\bibfnamefont {A.}~\bibnamefont {Maharana}}, \
  and\ \bibinfo {author} {\bibfnamefont {F.}~\bibnamefont {Quevedo}},\
  }\href@noop {} {\bibfield  {journal} {\bibinfo  {journal} {Journal of High
  Energy Physics}\ }\textbf {\bibinfo {volume} {2011}},\ \bibinfo {pages} {1}
  (\bibinfo {year} {2011})}\BibitemShut {NoStop}%
\bibitem [{\citenamefont {Cicoli}\ \emph {et~al.}(2011)\citenamefont {Cicoli},
  \citenamefont {Burgess},\ and\ \citenamefont
  {Quevedo}}]{cicoli2011anisotropic}%
  \BibitemOpen
  \bibfield  {author} {\bibinfo {author} {\bibfnamefont {M.}~\bibnamefont
  {Cicoli}}, \bibinfo {author} {\bibfnamefont {C.}~\bibnamefont {Burgess}}, \
  and\ \bibinfo {author} {\bibfnamefont {F.}~\bibnamefont {Quevedo}},\
  }\href@noop {} {\bibfield  {journal} {\bibinfo  {journal} {Journal of High
  Energy Physics}\ }\textbf {\bibinfo {volume} {2011}},\ \bibinfo {pages} {1}
  (\bibinfo {year} {2011})}\BibitemShut {NoStop}%
\bibitem [{\citenamefont {Dimopoulos}\ and\ \citenamefont
  {Giudice}(1996)}]{dimopoulos1996macroscopic}%
  \BibitemOpen
  \bibfield  {author} {\bibinfo {author} {\bibfnamefont {S.}~\bibnamefont
  {Dimopoulos}}\ and\ \bibinfo {author} {\bibfnamefont {G.~F.}\ \bibnamefont
  {Giudice}},\ }\href@noop {} {\bibfield  {journal} {\bibinfo  {journal}
  {Physics Letters B}\ }\textbf {\bibinfo {volume} {379}},\ \bibinfo {pages}
  {105} (\bibinfo {year} {1996})}\BibitemShut {NoStop}%
\bibitem [{\citenamefont {Arvanitaki}\ \emph
  {et~al.}(2016{\natexlab{a}})\citenamefont {Arvanitaki}, \citenamefont
  {Dimopoulos}, \citenamefont {Gorbenko}, \citenamefont {Huang},\ and\
  \citenamefont {Van~Tilburg}}]{arvanitaki2016small}%
  \BibitemOpen
  \bibfield  {author} {\bibinfo {author} {\bibfnamefont {A.}~\bibnamefont
  {Arvanitaki}}, \bibinfo {author} {\bibfnamefont {S.}~\bibnamefont
  {Dimopoulos}}, \bibinfo {author} {\bibfnamefont {V.}~\bibnamefont
  {Gorbenko}}, \bibinfo {author} {\bibfnamefont {J.}~\bibnamefont {Huang}}, \
  and\ \bibinfo {author} {\bibfnamefont {K.}~\bibnamefont {Van~Tilburg}},\
  }\href@noop {} {\bibfield  {journal} {\bibinfo  {journal} {arXiv preprint
  arXiv:1609.06320}\ } (\bibinfo {year} {2016}{\natexlab{a}})}\BibitemShut
  {NoStop}%
\bibitem [{\citenamefont {Pospelov}(1998)}]{pospelov1998cp}%
  \BibitemOpen
  \bibfield  {author} {\bibinfo {author} {\bibfnamefont {M.}~\bibnamefont
  {Pospelov}},\ }\href@noop {} {\bibfield  {journal} {\bibinfo  {journal}
  {Physical Review D}\ }\textbf {\bibinfo {volume} {58}},\ \bibinfo {pages}
  {097703} (\bibinfo {year} {1998})}\BibitemShut {NoStop}%
\bibitem [{\citenamefont {Damour}\ and\ \citenamefont
  {Donoghue}(2010)}]{damour2010equivalence}%
  \BibitemOpen
  \bibfield  {author} {\bibinfo {author} {\bibfnamefont {T.}~\bibnamefont
  {Damour}}\ and\ \bibinfo {author} {\bibfnamefont {J.~F.}\ \bibnamefont
  {Donoghue}},\ }\href@noop {} {\bibfield  {journal} {\bibinfo  {journal}
  {Physical Review D}\ }\textbf {\bibinfo {volume} {82}},\ \bibinfo {pages}
  {084033} (\bibinfo {year} {2010})}\BibitemShut {NoStop}%
\bibitem [{\citenamefont {Arvanitaki}\ \emph
  {et~al.}(2016{\natexlab{b}})\citenamefont {Arvanitaki}, \citenamefont
  {Dimopoulos},\ and\ \citenamefont {Van~Tilburg}}]{arvanitaki2016sound}%
  \BibitemOpen
  \bibfield  {author} {\bibinfo {author} {\bibfnamefont {A.}~\bibnamefont
  {Arvanitaki}}, \bibinfo {author} {\bibfnamefont {S.}~\bibnamefont
  {Dimopoulos}}, \ and\ \bibinfo {author} {\bibfnamefont {K.}~\bibnamefont
  {Van~Tilburg}},\ }\href@noop {} {\bibfield  {journal} {\bibinfo  {journal}
  {Physical review letters}\ }\textbf {\bibinfo {volume} {116}},\ \bibinfo
  {pages} {031102} (\bibinfo {year} {2016}{\natexlab{b}})}\BibitemShut
  {NoStop}%
\bibitem [{\citenamefont {Branca}\ \emph {et~al.}(2016)\citenamefont {Branca},
  \citenamefont {Bonaldi}, \citenamefont {Cerdonio}, \citenamefont {Conti},
  \citenamefont {Falferi}, \citenamefont {Marin}, \citenamefont {Mezzena},
  \citenamefont {Ortolan}, \citenamefont {Prodi}, \citenamefont {Taffarello}
  \emph {et~al.}}]{branca2016search}%
  \BibitemOpen
  \bibfield  {author} {\bibinfo {author} {\bibfnamefont {A.}~\bibnamefont
  {Branca}}, \bibinfo {author} {\bibfnamefont {M.}~\bibnamefont {Bonaldi}},
  \bibinfo {author} {\bibfnamefont {M.}~\bibnamefont {Cerdonio}}, \bibinfo
  {author} {\bibfnamefont {L.}~\bibnamefont {Conti}}, \bibinfo {author}
  {\bibfnamefont {P.}~\bibnamefont {Falferi}}, \bibinfo {author} {\bibfnamefont
  {F.}~\bibnamefont {Marin}}, \bibinfo {author} {\bibfnamefont
  {R.}~\bibnamefont {Mezzena}}, \bibinfo {author} {\bibfnamefont
  {A.}~\bibnamefont {Ortolan}}, \bibinfo {author} {\bibfnamefont {G.~A.}\
  \bibnamefont {Prodi}},  \emph {et~al.},\ }\href@noop {} {\bibfield  {journal}
  {\bibinfo  {journal} {arXiv preprint arXiv:1607.07327}\ } (\bibinfo {year}
  {2016})}\BibitemShut {NoStop}%
\bibitem [{\citenamefont {Arvanitaki}\ \emph {et~al.}(2015)\citenamefont
  {Arvanitaki}, \citenamefont {Huang},\ and\ \citenamefont
  {Van~Tilburg}}]{arvanitaki2015searching}%
  \BibitemOpen
  \bibfield  {author} {\bibinfo {author} {\bibfnamefont {A.}~\bibnamefont
  {Arvanitaki}}, \bibinfo {author} {\bibfnamefont {J.}~\bibnamefont {Huang}}, \
  and\ \bibinfo {author} {\bibfnamefont {K.}~\bibnamefont {Van~Tilburg}},\
  }\href@noop {} {\bibfield  {journal} {\bibinfo  {journal} {Physical Review
  D}\ }\textbf {\bibinfo {volume} {91}},\ \bibinfo {pages} {015015} (\bibinfo
  {year} {2015})}\BibitemShut {NoStop}%
\bibitem [{\citenamefont {Graham}\ \emph
  {et~al.}(2016{\natexlab{b}})\citenamefont {Graham}, \citenamefont {Kaplan},
  \citenamefont {Mardon}, \citenamefont {Rajendran},\ and\ \citenamefont
  {Terrano}}]{graham2016dark}%
  \BibitemOpen
  \bibfield  {author} {\bibinfo {author} {\bibfnamefont {P.~W.}\ \bibnamefont
  {Graham}}, \bibinfo {author} {\bibfnamefont {D.~E.}\ \bibnamefont {Kaplan}},
  \bibinfo {author} {\bibfnamefont {J.}~\bibnamefont {Mardon}}, \bibinfo
  {author} {\bibfnamefont {S.}~\bibnamefont {Rajendran}}, \ and\ \bibinfo
  {author} {\bibfnamefont {W.~A.}\ \bibnamefont {Terrano}},\ }\href@noop {}
  {\bibfield  {journal} {\bibinfo  {journal} {Physical Review D}\ }\textbf
  {\bibinfo {volume} {93}},\ \bibinfo {pages} {075029} (\bibinfo {year}
  {2016}{\natexlab{b}})}\BibitemShut {NoStop}%
\bibitem [{\citenamefont {Hardy}\ and\ \citenamefont
  {Lasenby}(2017)}]{hardy2017stellar}%
  \BibitemOpen
  \bibfield  {author} {\bibinfo {author} {\bibfnamefont {E.}~\bibnamefont
  {Hardy}}\ and\ \bibinfo {author} {\bibfnamefont {R.}~\bibnamefont
  {Lasenby}},\ }\href@noop {} {\bibfield  {journal} {\bibinfo  {journal}
  {Journal of High Energy Physics}\ }\textbf {\bibinfo {volume} {2017}},\
  \bibinfo {pages} {33} (\bibinfo {year} {2017})}\BibitemShut {NoStop}%
\bibitem [{\citenamefont {Raffelt}(1996)}]{raffelt1996stars}%
  \BibitemOpen
  \bibfield  {author} {\bibinfo {author} {\bibfnamefont {G.~G.}\ \bibnamefont
  {Raffelt}},\ }\href@noop {} {\emph {\bibinfo {title} {Stars as laboratories
  for fundamental physics: The astrophysics of neutrinos, axions, and other
  weakly interacting particles}}}\ (\bibinfo  {publisher} {University of
  Chicago press},\ \bibinfo {year} {1996})\BibitemShut {NoStop}%
\bibitem [{\citenamefont {Weinberg}(1978)}]{weinberg1978new}%
  \BibitemOpen
  \bibfield  {author} {\bibinfo {author} {\bibfnamefont {S.}~\bibnamefont
  {Weinberg}},\ }\href@noop {} {\bibfield  {journal} {\bibinfo  {journal}
  {Physical Review Letters}\ }\textbf {\bibinfo {volume} {40}},\ \bibinfo
  {pages} {223} (\bibinfo {year} {1978})}\BibitemShut {NoStop}%
\bibitem [{\citenamefont {Wilczek}(1978)}]{wilczek1978problem}%
  \BibitemOpen
  \bibfield  {author} {\bibinfo {author} {\bibfnamefont {F.}~\bibnamefont
  {Wilczek}},\ }\href@noop {} {\bibfield  {journal} {\bibinfo  {journal}
  {Physical Review Letters}\ }\textbf {\bibinfo {volume} {40}},\ \bibinfo
  {pages} {279} (\bibinfo {year} {1978})}\BibitemShut {NoStop}%
\bibitem [{\citenamefont {Peccei}\ and\ \citenamefont
  {Quinn}(1977)}]{peccei1977cp}%
  \BibitemOpen
  \bibfield  {author} {\bibinfo {author} {\bibfnamefont {R.~D.}\ \bibnamefont
  {Peccei}}\ and\ \bibinfo {author} {\bibfnamefont {H.~R.}\ \bibnamefont
  {Quinn}},\ }\href@noop {} {\bibfield  {journal} {\bibinfo  {journal}
  {Physical Review Letters}\ }\textbf {\bibinfo {volume} {38}},\ \bibinfo
  {pages} {1440} (\bibinfo {year} {1977})}\BibitemShut {NoStop}%
\bibitem [{\citenamefont {Vafa}\ and\ \citenamefont
  {Witten}(1984)}]{vafa1984parity}%
  \BibitemOpen
  \bibfield  {author} {\bibinfo {author} {\bibfnamefont {C.}~\bibnamefont
  {Vafa}}\ and\ \bibinfo {author} {\bibfnamefont {E.}~\bibnamefont {Witten}},\
  }\href@noop {} {\bibfield  {journal} {\bibinfo  {journal} {Physical Review
  Letters}\ }\textbf {\bibinfo {volume} {53}},\ \bibinfo {pages} {535}
  (\bibinfo {year} {1984})}\BibitemShut {NoStop}%
\bibitem [{\citenamefont {Graham}\ and\ \citenamefont
  {Rajendran}(2013)}]{graham2013new}%
  \BibitemOpen
  \bibfield  {author} {\bibinfo {author} {\bibfnamefont {P.~W.}\ \bibnamefont
  {Graham}}\ and\ \bibinfo {author} {\bibfnamefont {S.}~\bibnamefont
  {Rajendran}},\ }\href@noop {} {\bibfield  {journal} {\bibinfo  {journal}
  {Physical Review D}\ }\textbf {\bibinfo {volume} {88}},\ \bibinfo {pages}
  {035023} (\bibinfo {year} {2013})}\BibitemShut {NoStop}%
\bibitem [{\citenamefont {Pospelov}\ and\ \citenamefont
  {Ritz}(1999)}]{pospelov1999theta}%
  \BibitemOpen
  \bibfield  {author} {\bibinfo {author} {\bibfnamefont {M.}~\bibnamefont
  {Pospelov}}\ and\ \bibinfo {author} {\bibfnamefont {A.}~\bibnamefont
  {Ritz}},\ }\href@noop {} {\bibfield  {journal} {\bibinfo  {journal} {Physical
  review letters}\ }\textbf {\bibinfo {volume} {83}},\ \bibinfo {pages} {2526}
  (\bibinfo {year} {1999})}\BibitemShut {NoStop}%
\bibitem [{\citenamefont {di~Cortona}\ \emph {et~al.}(2016)\citenamefont
  {di~Cortona}, \citenamefont {Hardy}, \citenamefont {Vega},\ and\
  \citenamefont {Villadoro}}]{di2016qcd}%
  \BibitemOpen
  \bibfield  {author} {\bibinfo {author} {\bibfnamefont {G.~G.}\ \bibnamefont
  {di~Cortona}}, \bibinfo {author} {\bibfnamefont {E.}~\bibnamefont {Hardy}},
  \bibinfo {author} {\bibfnamefont {J.~P.}\ \bibnamefont {Vega}}, \ and\
  \bibinfo {author} {\bibfnamefont {G.}~\bibnamefont {Villadoro}},\ }\href@noop
  {} {\bibfield  {journal} {\bibinfo  {journal} {Journal of High Energy
  Physics}\ }\textbf {\bibinfo {volume} {2016}},\ \bibinfo {pages} {34}
  (\bibinfo {year} {2016})}\BibitemShut {NoStop}%
\bibitem [{\citenamefont {Kim}(1979)}]{kim1979weak}%
  \BibitemOpen
  \bibfield  {author} {\bibinfo {author} {\bibfnamefont {J.~E.}\ \bibnamefont
  {Kim}},\ }\href@noop {} {\bibfield  {journal} {\bibinfo  {journal} {Physical
  Review Letters}\ }\textbf {\bibinfo {volume} {43}},\ \bibinfo {pages} {103}
  (\bibinfo {year} {1979})}\BibitemShut {NoStop}%
\bibitem [{\citenamefont {Shifman}\ \emph {et~al.}(1980)\citenamefont
  {Shifman}, \citenamefont {Vainshtein},\ and\ \citenamefont
  {Zakharov}}]{shifman1980can}%
  \BibitemOpen
  \bibfield  {author} {\bibinfo {author} {\bibfnamefont {M.~A.}\ \bibnamefont
  {Shifman}}, \bibinfo {author} {\bibfnamefont {A.}~\bibnamefont {Vainshtein}},
  \ and\ \bibinfo {author} {\bibfnamefont {V.~I.}\ \bibnamefont {Zakharov}},\
  }\href@noop {} {\bibfield  {journal} {\bibinfo  {journal} {Nuclear Physics
  B}\ }\textbf {\bibinfo {volume} {166}},\ \bibinfo {pages} {493} (\bibinfo
  {year} {1980})}\BibitemShut {NoStop}%
\bibitem [{\citenamefont {Dine}\ \emph {et~al.}(1981)\citenamefont {Dine},
  \citenamefont {Fischler},\ and\ \citenamefont {Srednicki}}]{dine1981simple}%
  \BibitemOpen
  \bibfield  {author} {\bibinfo {author} {\bibfnamefont {M.}~\bibnamefont
  {Dine}}, \bibinfo {author} {\bibfnamefont {W.}~\bibnamefont {Fischler}}, \
  and\ \bibinfo {author} {\bibfnamefont {M.}~\bibnamefont {Srednicki}},\
  }\href@noop {} {\bibfield  {journal} {\bibinfo  {journal} {Physics letters
  B}\ }\textbf {\bibinfo {volume} {104}},\ \bibinfo {pages} {199} (\bibinfo
  {year} {1981})}\BibitemShut {NoStop}%
\bibitem [{\citenamefont {Zhitnitsky}(1980)}]{Zhitnitsky:1980tq}%
  \BibitemOpen
  \bibfield  {author} {\bibinfo {author} {\bibfnamefont {A.~R.}\ \bibnamefont
  {Zhitnitsky}},\ }\href@noop {} {\bibfield  {journal} {\bibinfo  {journal}
  {Sov. J. Nucl. Phys.}\ }\textbf {\bibinfo {volume} {31}},\ \bibinfo {pages}
  {260} (\bibinfo {year} {1980})},\ \bibinfo {note} {[Yad.
  Fiz.31,497(1980)]}\BibitemShut {NoStop}%
\bibitem [{\citenamefont {Svrcek}\ and\ \citenamefont
  {Witten}(2006)}]{svrcek2006axions}%
  \BibitemOpen
  \bibfield  {author} {\bibinfo {author} {\bibfnamefont {P.}~\bibnamefont
  {Svrcek}}\ and\ \bibinfo {author} {\bibfnamefont {E.}~\bibnamefont
  {Witten}},\ }\href@noop {} {\bibfield  {journal} {\bibinfo  {journal}
  {Journal of High Energy Physics}\ }\textbf {\bibinfo {volume} {2006}},\
  \bibinfo {pages} {051} (\bibinfo {year} {2006})}\BibitemShut {NoStop}%
\bibitem [{\citenamefont {Arvanitaki}\ \emph {et~al.}(2010)\citenamefont
  {Arvanitaki}, \citenamefont {Dimopoulos}, \citenamefont {Dubovsky},
  \citenamefont {Kaloper},\ and\ \citenamefont
  {March-Russell}}]{arvanitaki2010string}%
  \BibitemOpen
  \bibfield  {author} {\bibinfo {author} {\bibfnamefont {A.}~\bibnamefont
  {Arvanitaki}}, \bibinfo {author} {\bibfnamefont {S.}~\bibnamefont
  {Dimopoulos}}, \bibinfo {author} {\bibfnamefont {S.}~\bibnamefont
  {Dubovsky}}, \bibinfo {author} {\bibfnamefont {N.}~\bibnamefont {Kaloper}}, \
  and\ \bibinfo {author} {\bibfnamefont {J.}~\bibnamefont {March-Russell}},\
  }\href@noop {} {\bibfield  {journal} {\bibinfo  {journal} {Physical Review
  D}\ }\textbf {\bibinfo {volume} {81}},\ \bibinfo {pages} {123530} (\bibinfo
  {year} {2010})}\BibitemShut {NoStop}%
\bibitem [{\citenamefont {Barbieri}\ \emph {et~al.}(1989)\citenamefont
  {Barbieri}, \citenamefont {Cerdonio}, \citenamefont {Fiorentini},\ and\
  \citenamefont {Vitale}}]{barbieri1989axion}%
  \BibitemOpen
  \bibfield  {author} {\bibinfo {author} {\bibfnamefont {R.}~\bibnamefont
  {Barbieri}}, \bibinfo {author} {\bibfnamefont {M.}~\bibnamefont {Cerdonio}},
  \bibinfo {author} {\bibfnamefont {G.}~\bibnamefont {Fiorentini}}, \ and\
  \bibinfo {author} {\bibfnamefont {S.}~\bibnamefont {Vitale}},\ }\href@noop {}
  {\bibfield  {journal} {\bibinfo  {journal} {Physics Letters B}\ }\textbf
  {\bibinfo {volume} {226}},\ \bibinfo {pages} {357} (\bibinfo {year}
  {1989})}\BibitemShut {NoStop}%
\bibitem [{\citenamefont {Sikivie}(2014)}]{sikivie2014axion}%
  \BibitemOpen
  \bibfield  {author} {\bibinfo {author} {\bibfnamefont {P.}~\bibnamefont
  {Sikivie}},\ }\href@noop {} {\bibfield  {journal} {\bibinfo  {journal}
  {Physical review letters}\ }\textbf {\bibinfo {volume} {113}},\ \bibinfo
  {pages} {201301} (\bibinfo {year} {2014})}\BibitemShut {NoStop}%
\bibitem [{\citenamefont {Barbieri}\ \emph {et~al.}(2017)\citenamefont
  {Barbieri}, \citenamefont {Braggio}, \citenamefont {Carugno}, \citenamefont
  {Gallo}, \citenamefont {Lombardi}, \citenamefont {Ortolan}, \citenamefont
  {Pengo}, \citenamefont {Ruoso},\ and\ \citenamefont
  {Speake}}]{barbieri2017searching}%
  \BibitemOpen
  \bibfield  {author} {\bibinfo {author} {\bibfnamefont {R.}~\bibnamefont
  {Barbieri}}, \bibinfo {author} {\bibfnamefont {C.}~\bibnamefont {Braggio}},
  \bibinfo {author} {\bibfnamefont {G.}~\bibnamefont {Carugno}}, \bibinfo
  {author} {\bibfnamefont {C.}~\bibnamefont {Gallo}}, \bibinfo {author}
  {\bibfnamefont {A.}~\bibnamefont {Lombardi}}, \bibinfo {author}
  {\bibfnamefont {A.}~\bibnamefont {Ortolan}}, \bibinfo {author} {\bibfnamefont
  {R.}~\bibnamefont {Pengo}}, \bibinfo {author} {\bibfnamefont
  {G.}~\bibnamefont {Ruoso}}, \ and\ \bibinfo {author} {\bibfnamefont
  {C.}~\bibnamefont {Speake}},\ }\href@noop {} {\bibfield  {journal} {\bibinfo
  {journal} {Physics of the Dark Universe}\ }\textbf {\bibinfo {volume} {15}},\
  \bibinfo {pages} {135} (\bibinfo {year} {2017})}\BibitemShut {NoStop}%
\bibitem [{\citenamefont {Arvanitaki}\ and\ \citenamefont
  {Geraci}(2014)}]{arvanitaki2014resonantly}%
  \BibitemOpen
  \bibfield  {author} {\bibinfo {author} {\bibfnamefont {A.}~\bibnamefont
  {Arvanitaki}}\ and\ \bibinfo {author} {\bibfnamefont {A.~A.}\ \bibnamefont
  {Geraci}},\ }\href@noop {} {\bibfield  {journal} {\bibinfo  {journal}
  {Physical review letters}\ }\textbf {\bibinfo {volume} {113}},\ \bibinfo
  {pages} {161801} (\bibinfo {year} {2014})}\BibitemShut {NoStop}%
\bibitem [{\citenamefont {Moody}\ and\ \citenamefont
  {Wilczek}(1984)}]{moody1984new}%
  \BibitemOpen
  \bibfield  {author} {\bibinfo {author} {\bibfnamefont {J.}~\bibnamefont
  {Moody}}\ and\ \bibinfo {author} {\bibfnamefont {F.}~\bibnamefont
  {Wilczek}},\ }\href@noop {} {\bibfield  {journal} {\bibinfo  {journal}
  {Physical Review D}\ }\textbf {\bibinfo {volume} {30}},\ \bibinfo {pages}
  {130} (\bibinfo {year} {1984})}\BibitemShut {NoStop}%
\bibitem [{\citenamefont {Tullney}\ \emph {et~al.}(2013)\citenamefont
  {Tullney}, \citenamefont {Allmendinger}, \citenamefont {Burghoff},
  \citenamefont {Heil}, \citenamefont {Karpuk}, \citenamefont {Kilian},
  \citenamefont {Knappe-Gr{\"u}neberg}, \citenamefont {M{\"u}ller},
  \citenamefont {Schmidt}, \citenamefont {Schnabel} \emph
  {et~al.}}]{tullney2013constraints}%
  \BibitemOpen
  \bibfield  {author} {\bibinfo {author} {\bibfnamefont {K.}~\bibnamefont
  {Tullney}}, \bibinfo {author} {\bibfnamefont {F.}~\bibnamefont
  {Allmendinger}}, \bibinfo {author} {\bibfnamefont {M.}~\bibnamefont
  {Burghoff}}, \bibinfo {author} {\bibfnamefont {W.}~\bibnamefont {Heil}},
  \bibinfo {author} {\bibfnamefont {S.}~\bibnamefont {Karpuk}}, \bibinfo
  {author} {\bibfnamefont {W.}~\bibnamefont {Kilian}}, \bibinfo {author}
  {\bibfnamefont {S.}~\bibnamefont {Knappe-Gr{\"u}neberg}}, \bibinfo {author}
  {\bibfnamefont {W.}~\bibnamefont {M{\"u}ller}}, \bibinfo {author}
  {\bibfnamefont {U.}~\bibnamefont {Schmidt}},  \emph {et~al.},\ }\href@noop {}
  {\bibfield  {journal} {\bibinfo  {journal} {Physical review letters}\
  }\textbf {\bibinfo {volume} {111}},\ \bibinfo {pages} {100801} (\bibinfo
  {year} {2013})}\BibitemShut {NoStop}%
\bibitem [{\citenamefont {Raffelt}(2012)}]{raffelt2012limits}%
  \BibitemOpen
  \bibfield  {author} {\bibinfo {author} {\bibfnamefont {G.}~\bibnamefont
  {Raffelt}},\ }\href@noop {} {\bibfield  {journal} {\bibinfo  {journal}
  {Physical Review D}\ }\textbf {\bibinfo {volume} {86}},\ \bibinfo {pages}
  {015001} (\bibinfo {year} {2012})}\BibitemShut {NoStop}%
\bibitem [{\citenamefont {Schiff}(1963)}]{schiff1963measurability}%
  \BibitemOpen
  \bibfield  {author} {\bibinfo {author} {\bibfnamefont {L.}~\bibnamefont
  {Schiff}},\ }\href@noop {} {\bibfield  {journal} {\bibinfo  {journal}
  {Physical Review}\ }\textbf {\bibinfo {volume} {132}},\ \bibinfo {pages}
  {2194} (\bibinfo {year} {1963})}\BibitemShut {NoStop}%
\bibitem [{\citenamefont {Wantz}\ and\ \citenamefont
  {Shellard}(2010)}]{wantz2010axion}%
  \BibitemOpen
  \bibfield  {author} {\bibinfo {author} {\bibfnamefont {O.}~\bibnamefont
  {Wantz}}\ and\ \bibinfo {author} {\bibfnamefont {E.}~\bibnamefont
  {Shellard}},\ }\href@noop {} {\bibfield  {journal} {\bibinfo  {journal}
  {Physical Review D}\ }\textbf {\bibinfo {volume} {82}},\ \bibinfo {pages}
  {123508} (\bibinfo {year} {2010})}\BibitemShut {NoStop}%
\bibitem [{\citenamefont {Ringwald}\ and\ \citenamefont
  {Saikawa}(2016)}]{ringwald2016axion}%
  \BibitemOpen
  \bibfield  {author} {\bibinfo {author} {\bibfnamefont {A.}~\bibnamefont
  {Ringwald}}\ and\ \bibinfo {author} {\bibfnamefont {K.}~\bibnamefont
  {Saikawa}},\ }\href@noop {} {\bibfield  {journal} {\bibinfo  {journal}
  {Physical Review D}\ }\textbf {\bibinfo {volume} {93}},\ \bibinfo {pages}
  {085031} (\bibinfo {year} {2016})}\BibitemShut {NoStop}%
\bibitem [{\citenamefont {Sikivie}(2008)}]{sikivie2008axion}%
  \BibitemOpen
  \bibfield  {author} {\bibinfo {author} {\bibfnamefont {P.}~\bibnamefont
  {Sikivie}},\ }\href@noop {} {\bibfield  {journal} {\bibinfo  {journal}
  {Axions}\ ,\ \bibinfo {pages} {19}} (\bibinfo {year} {2008})}\BibitemShut
  {NoStop}%
\bibitem [{\citenamefont {Hiramatsu}\ \emph {et~al.}(2011)\citenamefont
  {Hiramatsu}, \citenamefont {Kawasaki}, \citenamefont {Sekiguchi},
  \citenamefont {Yamaguchi},\ and\ \citenamefont
  {Yokoyama}}]{hiramatsu2011improved}%
  \BibitemOpen
  \bibfield  {author} {\bibinfo {author} {\bibfnamefont {T.}~\bibnamefont
  {Hiramatsu}}, \bibinfo {author} {\bibfnamefont {M.}~\bibnamefont {Kawasaki}},
  \bibinfo {author} {\bibfnamefont {T.}~\bibnamefont {Sekiguchi}}, \bibinfo
  {author} {\bibfnamefont {M.}~\bibnamefont {Yamaguchi}}, \ and\ \bibinfo
  {author} {\bibfnamefont {J.}~\bibnamefont {Yokoyama}},\ }\href@noop {}
  {\bibfield  {journal} {\bibinfo  {journal} {Physical Review D}\ }\textbf
  {\bibinfo {volume} {83}},\ \bibinfo {pages} {123531} (\bibinfo {year}
  {2011})}\BibitemShut {NoStop}%
\bibitem [{\citenamefont {Hiramatsu}\ \emph {et~al.}(2012)\citenamefont
  {Hiramatsu}, \citenamefont {Kawasaki}, \citenamefont {Saikawa},\ and\
  \citenamefont {Sekiguchi}}]{hiramatsu2012production}%
  \BibitemOpen
  \bibfield  {author} {\bibinfo {author} {\bibfnamefont {T.}~\bibnamefont
  {Hiramatsu}}, \bibinfo {author} {\bibfnamefont {M.}~\bibnamefont {Kawasaki}},
  \bibinfo {author} {\bibfnamefont {K.}~\bibnamefont {Saikawa}}, \ and\
  \bibinfo {author} {\bibfnamefont {T.}~\bibnamefont {Sekiguchi}},\ }\href@noop
  {} {\bibfield  {journal} {\bibinfo  {journal} {Physical Review D}\ }\textbf
  {\bibinfo {volume} {85}},\ \bibinfo {pages} {105020} (\bibinfo {year}
  {2012})}\BibitemShut {NoStop}%
\bibitem [{\citenamefont {Kawasaki}\ \emph {et~al.}(2015)\citenamefont
  {Kawasaki}, \citenamefont {Saikawa},\ and\ \citenamefont
  {Sekiguchi}}]{kawasaki2015axion}%
  \BibitemOpen
  \bibfield  {author} {\bibinfo {author} {\bibfnamefont {M.}~\bibnamefont
  {Kawasaki}}, \bibinfo {author} {\bibfnamefont {K.}~\bibnamefont {Saikawa}}, \
  and\ \bibinfo {author} {\bibfnamefont {T.}~\bibnamefont {Sekiguchi}},\
  }\href@noop {} {\bibfield  {journal} {\bibinfo  {journal} {Physical Review
  D}\ }\textbf {\bibinfo {volume} {91}},\ \bibinfo {pages} {065014} (\bibinfo
  {year} {2015})}\BibitemShut {NoStop}%
\bibitem [{\citenamefont {Raffelt}(2008)}]{raffelt2008astrophysical}%
  \BibitemOpen
  \bibfield  {author} {\bibinfo {author} {\bibfnamefont {G.~G.}\ \bibnamefont
  {Raffelt}},\ }in\ \href@noop {} {\emph {\bibinfo {booktitle} {Axions}}}\
  (\bibinfo  {publisher} {Springer},\ \bibinfo {year} {2008})\ pp.\ \bibinfo
  {pages} {51--71}\BibitemShut {NoStop}%
\bibitem [{\citenamefont {Chang}\ \emph {et~al.}(2017)\citenamefont {Chang},
  \citenamefont {Essig},\ and\ \citenamefont {McDermott}}]{samrouven}%
  \BibitemOpen
  \bibfield  {author} {\bibinfo {author} {\bibfnamefont {J.}~\bibnamefont
  {Chang}}, \bibinfo {author} {\bibfnamefont {R.}~\bibnamefont {Essig}}, \ and\
  \bibinfo {author} {\bibfnamefont {S.}~\bibnamefont {McDermott}},\ }\href@noop
  {} {\bibfield  {journal} {\bibinfo  {journal} {to appear}\ } (\bibinfo {year}
  {2017})}\BibitemShut {NoStop}%
\bibitem [{\citenamefont {Corsico}\ \emph {et~al.}(2001)\citenamefont
  {Corsico}, \citenamefont {Benvenuto}, \citenamefont {Althaus}, \citenamefont
  {Isern},\ and\ \citenamefont {Garc{\i}a-Berro}}]{corsico2001potential}%
  \BibitemOpen
  \bibfield  {author} {\bibinfo {author} {\bibfnamefont {A.~H.}\ \bibnamefont
  {Corsico}}, \bibinfo {author} {\bibfnamefont {O.~G.}\ \bibnamefont
  {Benvenuto}}, \bibinfo {author} {\bibfnamefont {L.~G.}\ \bibnamefont
  {Althaus}}, \bibinfo {author} {\bibfnamefont {J.}~\bibnamefont {Isern}}, \
  and\ \bibinfo {author} {\bibfnamefont {E.}~\bibnamefont {Garc{\i}a-Berro}},\
  }\href@noop {} {\bibfield  {journal} {\bibinfo  {journal} {New Astronomy}\
  }\textbf {\bibinfo {volume} {6}},\ \bibinfo {pages} {197} (\bibinfo {year}
  {2001})}\BibitemShut {NoStop}%
\bibitem [{\citenamefont {Isern}\ and\ \citenamefont
  {Garcia-Berro}(2003)}]{isern2003white}%
  \BibitemOpen
  \bibfield  {author} {\bibinfo {author} {\bibfnamefont {J.}~\bibnamefont
  {Isern}}\ and\ \bibinfo {author} {\bibfnamefont {E.}~\bibnamefont
  {Garcia-Berro}},\ }\href@noop {} {\bibfield  {journal} {\bibinfo  {journal}
  {Nuclear Physics B-Proceedings Supplements}\ }\textbf {\bibinfo {volume}
  {114}},\ \bibinfo {pages} {107} (\bibinfo {year} {2003})}\BibitemShut
  {NoStop}%
\end{thebibliography}%

\end{document}